\documentclass[a4paper,fleqn,usenatbib]{mnras}

\usepackage{amsmath}	
\usepackage{txfonts}

\usepackage[T1]{fontenc}
\usepackage{ae,aecompl}

\usepackage{amsfonts}
\usepackage{ctable}
\usepackage{longtable}
\usepackage{subcaption}
\usepackage[normalem]{ulem}
\title[Multiphase Analysis of QSO Absorbers]{The COS Absorption Survey of Baryon Harbors: Unveiling the Physical Conditions of Circumgalactic Gas through Multiphase Bayesian Ionization Modeling}

\author[K. J. Haislmaier et al.]{%
Karl J. Haislmaier$^{1}$\thanks{E-mail: khaislmaier@astro.umass.edu},
Todd M. Tripp$^{1}$,
Neal Katz$^{1}$,
J. Xavier Prochaska$^{2}$,
\newauthor
Joseph N. Burchett$^{3}$,
John M. O'Meara$^{4}$,
Jessica K. Werk$^{5}$
\\
$^{1}$Department of Astronomy, University of Massachusetts, Amherst, MA 01003\\
$^{2}$Department of Astronomy, New Mexico State University, Las Cruces, NM 88001\\
$^{3}$UCO/Lick Observatory, Department of Astronomy and Astrophysics, University of California, Santa Cruz, CA 95064\\
$^{4}$W. M. Keck Observatory, Kamuela, HI 96743\\
$^{5}$Department of Astronomy, University of Washington, Seattle, WA 98195-1580
}

\date{Accepted XXX. Received YYY; in original form ZZZ}

\pubyear{2017}

\begin{document}
\label{firstpage}
\pagerange{\pageref{firstpage}--\pageref{lastpage}}
\maketitle

\begin{abstract}
Quasar absorption systems encode a wealth of information about the abundances, ionization structure, and physical conditions in intergalactic and circumgalactic media. Simple (often single-phase) photoionization models are frequently used to decode such data. Using five discrete absorbers from the COS Absorption Survey of Baryon Harbors (CASBaH) that exhibit a wide range of detected ions (e.g., Mg~\textsc{ii}, \textsc{S~ii} -- \textsc{S~vi}, \textsc{O~ii} -- \textsc{O~vi}, Ne~\textsc{viii}), we show several examples where single-phase ionization models cannot reproduce the full set of measured column densities. To explore models that can self-consistently explain the measurements and kinematic alignment of disparate ions, we develop a Bayesian multiphase ionization modeling framework that characterizes discrete phases by their unique physical conditions and also investigates variations in the shape of the UV flux field, metallicity, and relative abundances. Our models require at least two (but favor three) distinct ionization phases ranging from $T \approx 10^{4}$ K photoionized gas to warm-hot phases at $T \lesssim 10^{5.8}$ K. For some ions, an apparently single absorption ``component'' includes contributions from more than one phase, and up to 30\% of the \textsc{H~i} is not from the lowest ionization phase. If we assume that all of the phases are photoionized, we cannot find solutions in thermal pressure equilibrium.  By introducing hotter, collisionally ionized phases, however, we can achieve balanced pressures.  The best models indicate moderate metallicities, often with sub-solar N/$\alpha$, and, in two cases, ionizing flux fields that are softer and brighter than the fiducial Haardt \& Madau UV background model.

\end{abstract}





\section{Introduction}

	An understanding of gas physics in the circumgalactic medium (CGM) of galaxies is vital for several areas of research.  The spatial cycling of baryons and the fueling of star formation, the impact of active galactic nuclei (AGN) and stellar feedback on galaxy evolution, and the outcome of cosmological simulations all depend on the physical conditions in the CGM, and measurements of the abundances and characteristics of the CGM and IGM can provide insight into a variety of processes.  Since halo gas is very challenging to study in emission with current facilities, ultraviolet (UV) absorption features imprinted by circumgalactic/intergalactic gases on the spectra of background QSOs provide the most practical means for probing the CGM in a wide variety of contexts with statistically useful samples \citep[e.g.,][]{Tumlinson17,Rudie19}.  Quasar absorption-line systems trace the dominant component of the baryon distribution at high redshifts \citep{Weinberg97,Rauch97}, and current evidence indicates that QSO absorbers reveal important baryon harbors at low redshifts as well \citep[e.g.,][]{Shull96,Tripp00,Penton04,Tumlinson11,Werk14,Burchett19}.  In most CGM absorbers, the gas is optically thin and the hydrogen is $>$90\% ionized, so ionization modeling is essential for deriving the chemical and thermodynamic properties of the gas/plasma, which are in turn critical variables for understanding its role in galaxy evolution.

	To gain further insights there are several physical properties that are important to measure, including the gas density, temperature, pressure, and composition, as well as the ambient ionizing radiation field.  The first three are related to the physical structure of clouds in the CGM, including aspects such as the cloud size, pressure balance, mass, and energy content.  The gas metallicity and \textit{relative} abundance ratios (e.g., the abundance of nitrogen vs. $\alpha -$group elements such as oxygen) are age-sensitive quantities, and hence trace the origin and evolution of gas and/or the presence of dust in the CGM, but interpreting abundances is complicated by the uncertain physics that govern the transport and mixing of galaxy ejecta into the CGM \citep[see, e.g.,][]{Frye19}.  The ionizing radiation field is significant cosmologically, but also for its systematic effect on the photoionization rate for all ions.  However, typical ionization models only vary the overall metallicity and ionization parameter (a ratio between the number densities of ionizing photons and hydrogen atoms) and typically assume a solar gas composition with a single fixed incident radiation field from available models such as \citet{Haardt12}, but it is well known that changes in the shape of the ionizing radiation field can significantly alter model results \citep[e.g.,][]{Crighton15,Chen17,Khaire19a,Wotta19}.  The primary driver of the usual assumptions and simplifications of the models is not rooted in physics but rather in the inadequate number of absorption line diagnostics and model constraints that are typically recorded in a single absorption system.  However, as we summarize in the following section, we are conducting a new survey (Tripp et al. 2021, \emph{in prep.}) that provides a significant improvement in the number and quality of constraints available for individual QSO absorption systems. This larger number of constraints enables more detailed studies of absorber ionization models.  In this paper we present a Bayesian Markov-Chain Monte-Carlo method for modeling these more observationally constrained absorbers, and we discuss the implications of these models for three absorption systems in particular, which present a variety of constraints and typify the data from our new program.  

\subsection{The COS Absorption Survey of Baryon Harbors}
\label{sec:survey}

	To investigate the physical nature and evolution of the CGM over a broad redshift range, we have initiated a program to study QSO absorption systems from $z = 0$ to $z \approx$ 1.5 using ultraviolet spectra recorded with the Cosmic Origins Spectrograph \citep[COS,][]{Green12} and the Space Telescope Imaging Spectrograph \citep[STIS,][]{Woodgate98} on the \textit{Hubble Space Telescope} (HST) as well as optical echelle spectra obtained with HIRES on Keck \citep{Vogt94}.  Our program --- the COS Absorption Survey of Baryon Harbors (CASBaH, P.I. T. M. Tripp) --- offers some unique advantages for probing the ionization, abundances, and physics of halo gas.  This section reviews some of the motivations and strengths of CASBaH for the study of QSO absorption systems and the baryon cycle; additional information about the survey can be found in \citet{Burchett19}, \citet{Prochaska19}, and Tripp et al. (2021, \emph{in prep.}).

\begin{figure}
\centering
\includegraphics[width=4.9in]{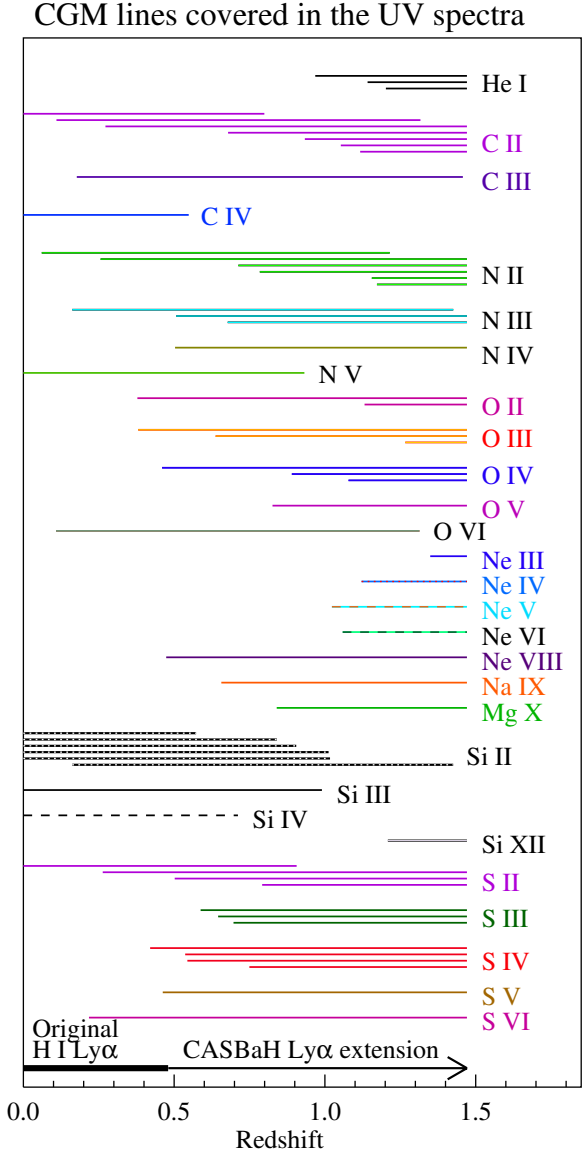}
\caption{Schematic summary of the redshift range over which detectable resonance lines of various species can be detected in intervening absorbers in the CASBaH spectra. The species shown are labeled on the right side of the diagram. For each species, a horizontal bar indicates the $z$ range that is covered for a specific transition; the lower bound indicates the redshift at which the transition is redshifted into the COS FUV band (observed $\lambda >$ 1152 \AA), and the upper bound indicates the redshift where the line is redshifted beyond the long-wavelength end of the CASBaH NUV data; with one exception (see text), the NUV spectra all extend to at least 2400 \AA .  For example, we cover the \textsc{C~ii} 1036.34 \AA\ line from $z_{\rm abs}$ = 0.1097 to at least 1.3168.  However, for some of the sight lines, the NUV spectrum extends to even longer wavelengths, so that the upper redshift cutoff can be higher than this minimum range.  The highest redshift that our survey can cover is set by the redshift of our highest-redshift QSO, PG1630+377 ($z = 1.4789$) . For some species, we can access several resonance lines.  For example, we have the ability to detect the \textsc{C~ii} lines at 543.26, 560.24, 594.80, 687.05, 903.63, 903.96, 1036.34, and 1334.53 \AA . The coverage of each transition is indicated by a separate line unless the lines are quite close in wavelength (e.g., 903.63 and 903.96 \AA ), in which case a single line indicates the coverage.  \label{fig:casbah_lines}}
\end{figure}

	The CASBaH program observed nine quasars with redshifts ranging from $z_{\rm QSO}$ = 0.91573 to $z_{\rm QSO}$ = 1.47895, and the spectra extend from observed wavelengths $\lambda _{\rm ob}$ = 1152 \AA\ to roughly the redshifted wavelength of the \textsc{H~i} Ly$\alpha$ line at the quasar redshift [i.e., $(1 + z_{\rm QSO}) \times 1215.67$].  The UV spectra in the full CASBaH dataset span \textit{rest-frame} wavelengths extending from 465 to $>3000$ \AA, and the optical HIRES data extend the coverage farther into the red.  As noted by \cite{Verner94} and illustrated in Figure~\ref{fig:casbah_lines}, the UV rest-frame wavelength range below 912 \AA\ includes a rich suite of resonance absorption lines \citep[see also Figure 2 in][]{Tripp13}, and access to this wavelength range provides unique diagnostics such as hot-gas tracers (e.g., Ne~\textsc{viii}) and, more importantly for this work, various banks of adjacent ions such as \textsc{O~i}/\textsc{O~ii}/\textsc{O~iii}/\textsc{O~iv}/\textsc{O~v}/\textsc{O~vi}, \textsc{S~i}/\textsc{S~ii}/\textsc{S~iii}/\textsc{S~iv}/\textsc{S~v}/\textsc{S~vi}, etc. that provide greatly improved constraints on ionization and physical condition models (see \citealt{Verner96b} for a compilation of absorption line data $<<$ 912 \AA). For comparison, most previous studies of oxygen in QSO absorbers have covered only \textsc{O~i} and \textsc{O~vi}.  Without any information on \textsc{O~ii} -- \textsc{O~v}, it is difficult to draw precise conclusions about the gas ionization and modeling assumptions, which can lead to misleading results (see further discussion below). In this paper, we demonstrate the advantages gained in QSO absorber studies by 1) pushing deep into the far ultraviolet (FUV) and 2) using a Bayesian Markov-chain Monte-Carlo (MCMC) method for exploring the implications of the diagnostics afforded by the FUV coverage.  In a follow-up paper, we will apply our methodology to the full set of CASBaH systems.

\subsubsection{Challenges and Limitations of Typical Absorber Ionization Models} 
\label{sec:modeling_challenges}

	Since many studies are limited by the number of ions detected in an absorber (especially in the low-density CGM), necessary simplifications could lead to significant systematic errors.   For example, in many studies the detected CGM lines are limited to species such as \textsc{H~i}, \textsc{C~ii}, \textsc{C~iv}, Si~\textsc{ii}, Si~\textsc{iii}, and Si~\textsc{iv}. At lower redshifts, where there are fewer \textsc{H~i} lines per unit redshift, transitions such as \textsc{N~ii} 1083.99 \AA , \textsc{N~iii} 989.80 \AA , \textsc{C~iii} 977.02 \AA , \textsc{O~vi} 1031.93, 1037.62 can also be accessed. At higher redshifts (e.g., $z\gtrsim2$), however, transitions at observed wavelengths $\lambda _{\rm ob} <$ 1216 \AA\ become increasingly difficult to measure because they are often severely blended with \textsc{H~i} lines in the Ly$\alpha$ forest (or the spectrum is obliterated by an optically thick intervening Lyman limit system).  While the typical sets of ions and elements employed in the majority of QSO absorber studies have established that the CGM is an important component of galaxies with some surprising characteristics \citep[e.g.,][]{Steidel10,Tumlinson11}, many analyses of the ionization and physical conditions suffer from the following limitations.  

\subsubsection{Improved Constraints from Far-Ultraviolet Lines}

	First, important phases of the absorbing entity may be poorly constrained or not constrained at all.   For example, if the only detected species are \textsc{H~i},  \textsc{C~ii}, and \textsc{C~iv}, the lack of knowledge of intermediate- and high-ionization stages can introduce serious systematic errors. Some papers choose to analyze such an ion set as a single-phase absorber modeled with a photoionization code such as \textsc{Cloudy} \citep{Ferland17}, but these systems are not necessarily single-phase ``clouds'', and indeed many papers have presented evidence that QSO absorbers can arise in complex, multiphase media \citep[e.g.,][]{Charlton03,Ding03,Savage05,Tripp08,Tripp11,Burchett15}.  A single-phase photoionization model can typically fit the column densities of \textsc{H~i},  \textsc{C~ii}, and \textsc{C~iv} quite nicely, but if the \textsc{C~iv} actually arises in a separate phase, the assumption of a single-phase model will lead to an erroneous solution.  The addition of species such as \textsc{C~iii}, Si~\textsc{iii}, and \textsc{O~vi} to the analysis is helpful, but these additions can be problematic owing to line saturation and non-solar relative abundances (see below).   Alternatively, it is sometimes assumed that only low-ionization stages can be safely assumed to be cospatial and well mixed with the \textsc{H~i}, and some authors use photoionization models to fit \textsc{H~i}, \textsc{C~ii}, Si~\textsc{ii} and ignore higher ionization stages that are also detected (e.g., \textsc{C~iv}, Si~\textsc{iv}) on the grounds that the higher ions are likely from a separate phase.  This is also problematic because the \textsc{C~iv}- and  Si~\textsc{iv}-bearing phases {\it can} contribute significantly to the total \textsc{H~i} column density, and the assumption that 100\% of the \textsc{H~i} originates in the \textsc{C~ii} and Si~\textsc{ii} phase could lead to incorrect inferences about the physical conditions and metallicities.  This problem is exacerbated by evidence that the absorption lines of intermediate ionization stages such as \textsc{C~iii} and Si~\textsc{iii} come from yet another phase that is distinct from both the \textsc{C~ii} phase and possibly also the \textsc{C~iv} phase (see below).  It is necessary to consider how the results might change if the detected \textsc{H~i} and other ions are distributed amongst two, three, or even more phases. 

	Second, when only a small number of lines are covered, some potentially valuable constraints can be ruined by line saturation or blending.  Column densities of species like \textsc{C~iii} and Si~\textsc{iii} can provide insights on multiphase absorbers, but most observations only record a single line from \textsc{C~iii} and Si~\textsc{iii} and, moreover, these single lines are very strong and prone to strong/severe saturation, which limits their usefulness.   This problem is especially insidious in spectra from the Cosmic Origins Spectrograph on HST because the modest spectral resolution and broad wings of the COS line-spread function \citep{Kriss11} can cause saturated lines to appear to be unsaturated, and the presence or absence of unresolved saturation can be difficult to establish with the coverage of only a single strong line.  Similarly, a single line of an important species can be unmeasurable if it is badly blended with a strong absorption feature from a different redshift.

	Third, many studies assume that the relative abundances of elements follow the solar pattern (e.g., from \citealt{Lodders03} or \citealt{Asplund09}).  However, there is ample evidence of departures from solar abundances.  In various contexts, including QSO absorbers arising in galaxy halos, there are clear indications that nitrogen can be underabundant by various amounts \citep[e.g.,][]{Vilacostas93, vanZee06, Pettini08, Battisti12, Berg12}, the alpha-group elements have well-known non-solar abundances compared to iron-group elements \citep[e.g.,][]{McWilliam97, Prochaska00, Wolfe05}, and the relative carbon abundance exhibits a complex dependence on metallicity in both stars and QSO absorbers \citep{Akerman04, Pettini08, Fabbian09, Penprase10}.  Thus the assumption of solar relative abundances could also introduce substantial systematic errors if intercomparison of species such as \textsc{C~ii}, \textsc{N~ii}, Si~\textsc{ii}, and Fe~\textsc{ii} provide the primary model constraints.

	Finally, ionizing models require other assumptions regarding, for example, the shape of the ionizing flux field impinging on the gas and the basic ionization mechanism (whether the absorber is photoionized or collisionally ionized or both).  Often a single shape for the ionizing flux is assumed, but plausible variations in the shape of the radiation field can significantly change the model outcomes, and the recent controversy regarding the shape and intensity of the UV background light \citep{Kollmeier14,Shull15,Khaire15,Khaire19a,Khaire19b} illustrates the difficulties and uncertainties encountered in models of the extragalactic UV background.  

\begin{figure*}
\includegraphics[width=16cm]{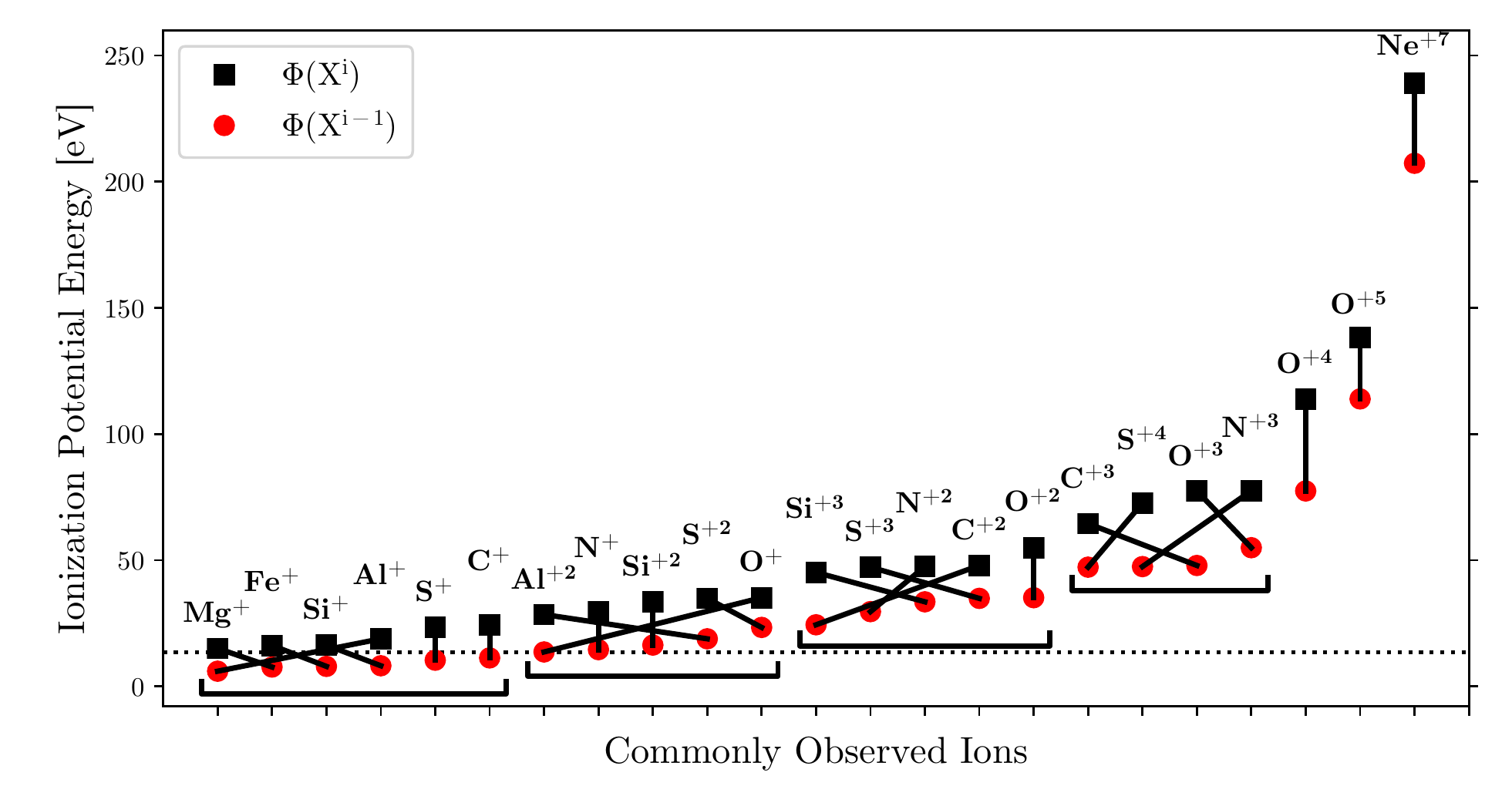}
\includegraphics[width=16cm]{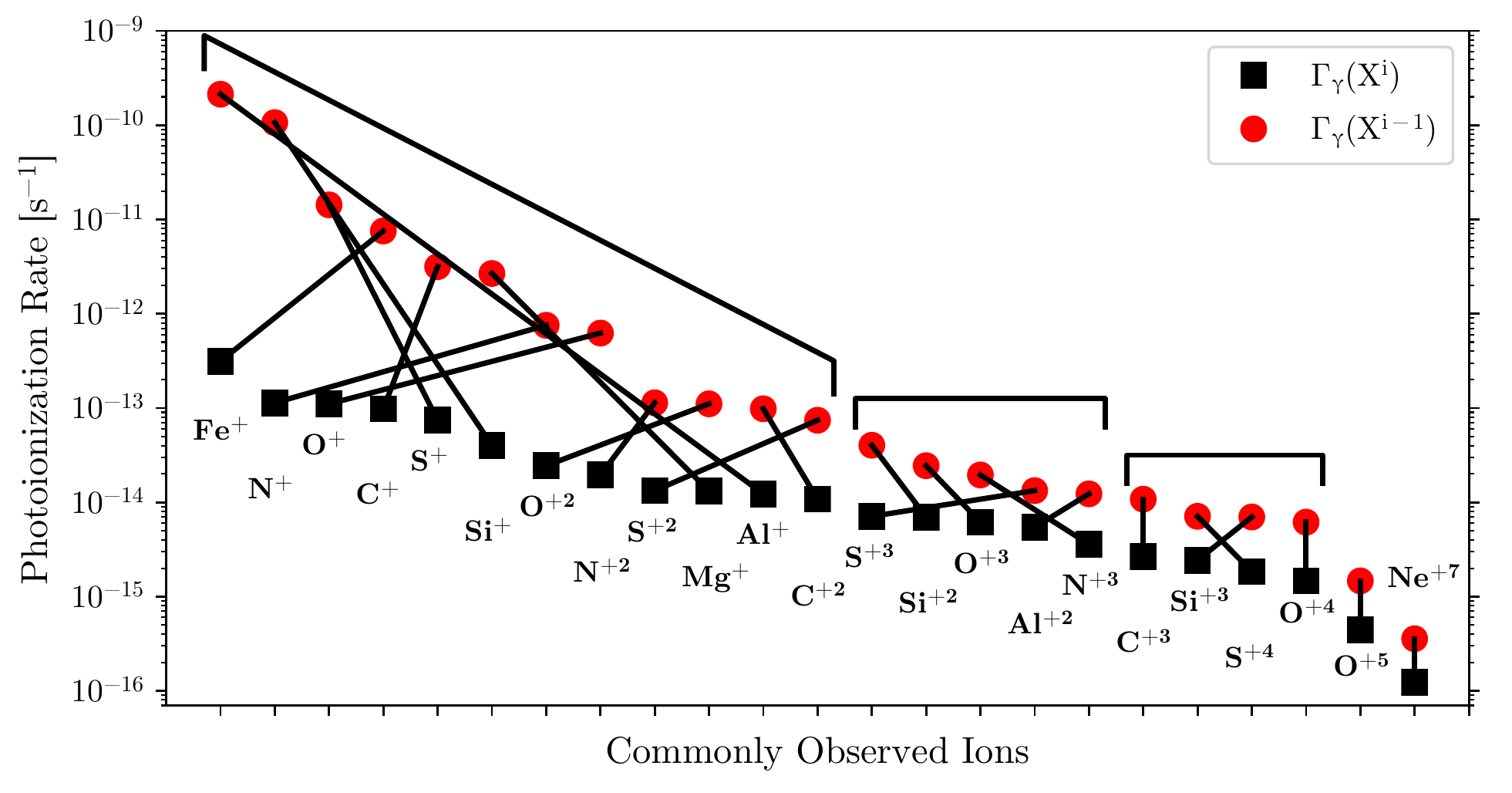}
\caption[]{\emph{Top:} A comparison of ionization potentials (IP) of ions that are commonly detected in CASBaH spectra.  For each element, the black square indicates the IP of the labeled ion (X$^{\rm i}$), which is connected with a black line to a red circle that shows the IP of the next lower stage (X$^{i-1}$). Thus, the red dot indicates the energy required to create the ion, and the black square plots the energy needed to destroy it. The points are plotted in order of increasing ionization energy. 
Horizontal black brackets group species with overlapping ionization creation and destruction energies (overlapping slanting black lines).  Isolated species (i.e., S$^{+}$, C$^{+}$, and O$^{+2}$ in the top panel and C$^{+3}$ and O$^{+4}$ in the bottom panel) were assigned to the bracketed groups with the most similar ionization potentials.  Heuristically these categorize groups of ions that might be expected to coexist in a specific gas phase.  The horizontal dotted line at 1 Rydberg shows the ionization potential of H~\textsc{i}.  \emph{Bottom:}  Photoionization rates ($\Gamma$) of the same ions, calculated for an optically thin gas cloud that is ionized by the UV background from \cite{Haardt12} at redshift $z$=0.5, with photoionization cross sections from \cite{Verner96a}.  As in the top panel, red circles and black squares indicate X$^{{\rm i}-1}$ and X$^{\rm i}$, respectively, and the labeled ions and their adjacent lower ionization stages are connected with a black line.  Here the points are plotted in order of decreasing $\Gamma$.  Note the very low photoionization rates expected for the highest ionization stages.\label{fig:ion_potentials}}
\end{figure*}

\ctable[
	caption={Characteristics of the three absorption systems (five total 
			absorbers) under consideration here.},
	label = {tab:abs_sys_info},
	doinside = \footnotesize,
	star
	]{lllcll}{
\tnote[a]{\footnotesize Throughout this work, we reserve the term ``absorber'' to mean a distinct \emph{velocity component} of a given \emph{absorption system}.  Refer to Section~\ref{sec:implementation} for more details.}
\tnote[b]{\footnotesize The velocity of H~\textsc{i} (from Voigt profile fitting) relative to the systemic redshift of the absorption system.}
\tnote[c]{\footnotesize The column density uncertainties are derived from Voigt profile fits to numerous, high S/N H~\textsc{i} Lyman Series lines simultaneously, hence the high precision.}
\tnote[d]{\footnotesize These lists include lower limits, but exclude upper limits, which are presented in Tables~\ref{tab:z041761_meas}--\ref{tab:z068605_meas}.}
}{															\FL
Absorber\tmark[a] & QSO Sightline & $\textit{z}_{\textrm{abs}}$ & \textit{v}\tmark[b] [km s$^{-1}$] & log\emph{N}\textsubscript{H~\textsc{i}}\tmark[c] [cm\textsuperscript{-2}] & Detected Ions\tmark[d]	\ML
0.4A	&	PG1630+377	&	0.41760	&	0	&	$15.46\pm0.02$	&	H~\textsc{i} (full Lyman series), C~\textsc{ii}/\textsc{iii}, N~\textsc{ii}/\textsc{iii},	\NN
 & & & & & O~\textsc{ii}/\textsc{iii}/\textsc{vi}, Mg~\textsc{ii}, Si~\textsc{ii}/\textsc{iii}, Fe~\textsc{ii}	\NN
0.4B	&	PG1630+377	&	0.41760	&	44	&	$15.33\pm0.01$	&	H~\textsc{i} (full Lyman series), C~\textsc{ii}/\textsc{iii}, N~\textsc{iii},	\NN
 & & & & & O~\textsc{ii}/\textsc{iii}/\textsc{vi}, Mg~\textsc{ii}, Si~\textsc{iii}	\NN
0.5A	&	PG1522+101	&	0.51850	&	3	&	$16.22\pm0.01$	&	H~\textsc{i} (full Lyman series), C~\textsc{ii}/\textsc{iii}, N~\textsc{iii}, Si~\textsc{ii}/\textsc{iii}	\NN
 & & & & & O~\textsc{ii}/\textsc{iii}, Mg~\textsc{ii}, Fe~\textsc{ii}	\NN
0.6A	&	PG1338+416	&	0.68606	&	0	&	$16.52\pm0.02$	&	H~\textsc{i} (full Lyman series), C~\textsc{ii}, N~\textsc{ii}/\textsc{iii}/\textsc{iv}, \NN
 & & & & & O~\textsc{ii}/\textsc{iii}/\textsc{vi}, Mg~\textsc{i}/\textsc{ii}, Ne~\textsc{viii}, Al~\textsc{iii},	\NN
 & & & & & Si~\textsc{ii}, S~\textsc{ii}/\textsc{iii}/\textsc{v}, Fe~\textsc{ii}	\NN
0.6B	&	PG1338+416	&	0.68606	&	46	&	$15.90\pm0.06$	&	H~\textsc{i} (full Lyman series), C~\textsc{ii}/\textsc{iii}, N~\textsc{iii}/\textsc{iv},	\NN
 & & & & & O~\textsc{ii}/\textsc{iii}/\textsc{iv}/\textsc{vi}, Mg~\textsc{ii}, Ne~\textsc{viii}, S~\textsc{iii}/\textsc{v}/\textsc{vi}	\LL
}

	To compactly summarize the ions, elements, and individual transitions that are available at any given redshift in the CASBaH spectra, Figure~\ref{fig:casbah_lines} shows the redshift range over which the UV spectra cover a given \textit{detectable} transition of a given ion. Often, multiple transitions with a range of line strengths can be detected from certain species, which is advantageous for dealing with saturation and blending.\footnote{The species shown in Figure~\ref{fig:casbah_lines} have many more resonance lines than indicated, but the lines that are not shown are too weak to be detected at the typical signal-to-noise (S/N) of CASBaH data or are outside the wavelength range of the spectra.}  The UV spectra extend from $\lambda _{ob}$ = 1152 \AA\ to at least 2400 \AA , and for the higher-redshift targets, the spectra extend to at least 2780 \AA .  Thus, for example, both lines of the \textsc{O~vi} doublet at $\lambda _{r}$ = 1031.93 and 1037.62 \AA\ are covered from $z_{\rm abs}$ = 0.11636 to at least $z_{\rm abs}$ = 1.31299, as indicated in the bar labeled ``\textsc{O~vi} '' in Figure~\ref{fig:casbah_lines} (for many species, the maximum absorber redshift in a given sight line is set by the QSO redshift, not the wavelength range of the data, so some of the sight lines cover less redshift path than shown in the figure).  For some ions (e.g., \textsc{C~ii} and \textsc{N~ii}), many resonance lines are covered, and each line is indicated by a separate horizontal bar (unless the resonance-transition wavelengths are very close, e.g., \textsc{C~ii} 903.63 and 903.96, in which case both lines are indicated with a single bar). 

	A few remarks and caveats about Figure~\ref{fig:casbah_lines} and the survey data are worth noting here.  First, as the absorber redshift ($z_{\rm abs}$) increases, the number of species increases, but many of the ions that are often studied in the QSO absorber literature (e.g., Mg~\textsc{ii} and \textsc{C~iv}) are redshifted beyond the long-wavelength limit of the CASBaH UV spectra at higher $z_{\rm abs}$ (Mg~\textsc{ii} is not shown in Fig.~\ref{fig:casbah_lines} because it is not covered at all in most of the UV spectra).  By design, these favorite lines are often covered in our Keck HIRES data. Second, while some very highly ionized species in Figure~\ref{fig:casbah_lines} such as Na~\textsc{ix} and Mg~\textsc{x} have been detected at high significance in the CASBaH data, these mostly occur in ``proximate'' absorbers with $z_{\rm abs} \approx z_{\rm QSO}$ \citep{Muzahid13,Muzahid16}.  Only \cite{Qu16} have reported the detection of weak Mg~\textsc{x} in one intervening ($z_{\rm abs} << z_{\rm QSO}$) CASBaH absorber.  The Ne~\textsc{viii} doublet is the most highly ionized species frequently detected in the CASBaH spectra \citep[Tripp et al. 2021 \emph{in prep.}]{Meiring13,Burchett19}. Third, while species such as \textsc{C~i}, \textsc{N~i}, and \textsc{O~i} have a large number of resonance transitions in the far-UV including lines with a broad range of strengths (see, e.g., \citealt{Sembach04}), we do not show these species in Figure~\ref{fig:casbah_lines} because they are rarely detected in CASBaH spectra.  This may owe to the ``inner CGM'' bias \citep{Prochaska17} -- the CASBaH targets were selected to cover large paths for species such as Ne~\textsc{viii}, so sight lines with known optically thick Lyman-limit absorbers or low FUV GALEX fluxes were excluded from consideration.

	For guidance in interpreting detections of various ions, Figure~\ref{fig:ion_potentials} graphically portrays the ionization potentials and photoionization rates of species that are commonly detected in CASBaH spectra. In each panel, the black squares represent the labeled species of interest (e.g., \textsc{C~ii}), and the red dot indicates the next lower ionization stage (e.g., \textsc{C~i}).  All of the points are presented in order of increasing ionization potential (upper panel) or photoionization rate (lower panel), and pairs of adjacent ions of a given element are connected with a line. This figure shows some factors that might lead certain species to coexist in the same gas phases.  For example, there is a set of species with ionization potentials just above 1 Ryd (\textsc{N~ii}, \textsc{O~ii}, Al~\textsc{iii}, Si~\textsc{iii}, and \textsc{S~iii}), and similar energies are required to ionize all of these ions. Thus, we might expect these ions to trace similar conditions. The information in Figure~\ref{fig:ion_potentials}  is useful for understanding the outcomes of the various ionization models discussed below.

\section{DATA AND MEASUREMENTS}
\label{sec:data}

	To develop our methods and begin to exploit the rich information provided by the CASBaH program, this paper focuses on three line-rich UV absorption systems found in high-quality QSO spectra from CASBaH in the range 0.4 < \emph{z} < 0.7.  These are taken from a larger (>35) sample of H~\textsc{i}-selected $\log{N_{\text{H I}}} \ge 15$ cm$^{-2}$ CASBaH intervening absorption systems that will be analyzed in a subsequent paper.  Our primary goal is to select systems well-suited for developing and testing the methodology presented herein. To that end, we randomly selected three systems that (1) have robust, but optically thin ($\log{N_{\text{H I}}} \lesssim 17$ cm$^{-2}$), $N_\textrm{H I}$ measurements, (2) are detected in a variety of metals, including C, N, and Fe in order to constrain models probing their relative abundance variations, (3) exhibit species across a range of ionization stages, including one system with only low-ions as well as two featuring both low- and high-ions, and (4) present some variety of component structure (to test how well the software performs in cases with simple vs. complex component structure). Some properties of the selected absorption systems are presented in Table~\ref{tab:abs_sys_info}.  For convenience, we will often refer to the absorbers by the shorthand name in the first column of Table~\ref{tab:abs_sys_info}, which is just the rough redshift of the absorption system plus `A' or `B' to refer to individual components at that redshift. The large number of well-detected lines in these systems (see below) enables us to test the software in detail and to demonstrate some initial results.  The high quality of the data, including high resolution detections of Mg~\textsc{ii} and Fe~\textsc{ii} with the ground-based Keck HIRES spectrograph, allow us to not only model for the physical parameters outlined above, but to do so separately for individual velocity components within each absorption system.  Before proceeding, we emphasize our adopted parlance by which the terms \emph{component} and/or \emph{absorber} denote a group of kinematically aligned absorption lines, while \emph{phases} are spectroscopically unresolved partitions of a given velocity component/absorber corresponding to distinct sets of physical conditions.  

\subsection{Observations}
\label{sec:observations}

	The COS FUV spectra covering 1152$-$1800 \AA\ were taken with the G130M and G160M gratings, which provide a spectral resolution of $\approx$15--20 km\ s$^{-1}$ per resolution element \citep{Fischer19}.  The data were recorded in the time-tagged photon-counting mode, and two grating tilts were used with both gratings to fill in the chip gap in the COS FUV detector (see Table 1.2 in \citealt{Rafelski18}).  In addition, multiple exposures at three or four focal-plane split positions were used at each grating tilt, so the flux at any given wavelength was recorded at six to eight different locations on the detector.  This greatly mitigates detector fixed-pattern noise when the exposures are aligned in wavelength space and coadded.  Initial data reduction steps were completed using the CALCOS pipeline (version 3.1.7) to produce extractions of the one-dimensional spectra (see \citealt{Rafelski18} for details).  The exposures were then aligned using well-detected lines of comparable strength from the ISM or extragalactic systems (e.g., higher \textsc{H~i} Lyman series lines) and coadded as described in \cite{Meiring11} and Tripp et al. (2021, \emph{in prep.}).   Finally, the combined spectra were binned to the Nyquist sampling rate of two pixels per resolution element (by default, the COS pipeline delivers 6 pixels per resolution element).  The coadded and binned FUV spectra of the the three CASBaH QSOs in this paper have signal-to-noise ratios (per resolution element) ranging from $15 - 43$ in the G130M range and $12 - 28$ in the G160M range.\footnote{These S/N ratios apply to the majority of the FUV spectra; the S/N ratios drop in small wavelength ranges at $\lambda_{\rm ob} < 1160$ and $\lambda_{\rm ob} > 1750$ \AA .}

\begin{figure*}
\includegraphics[width=\textwidth]{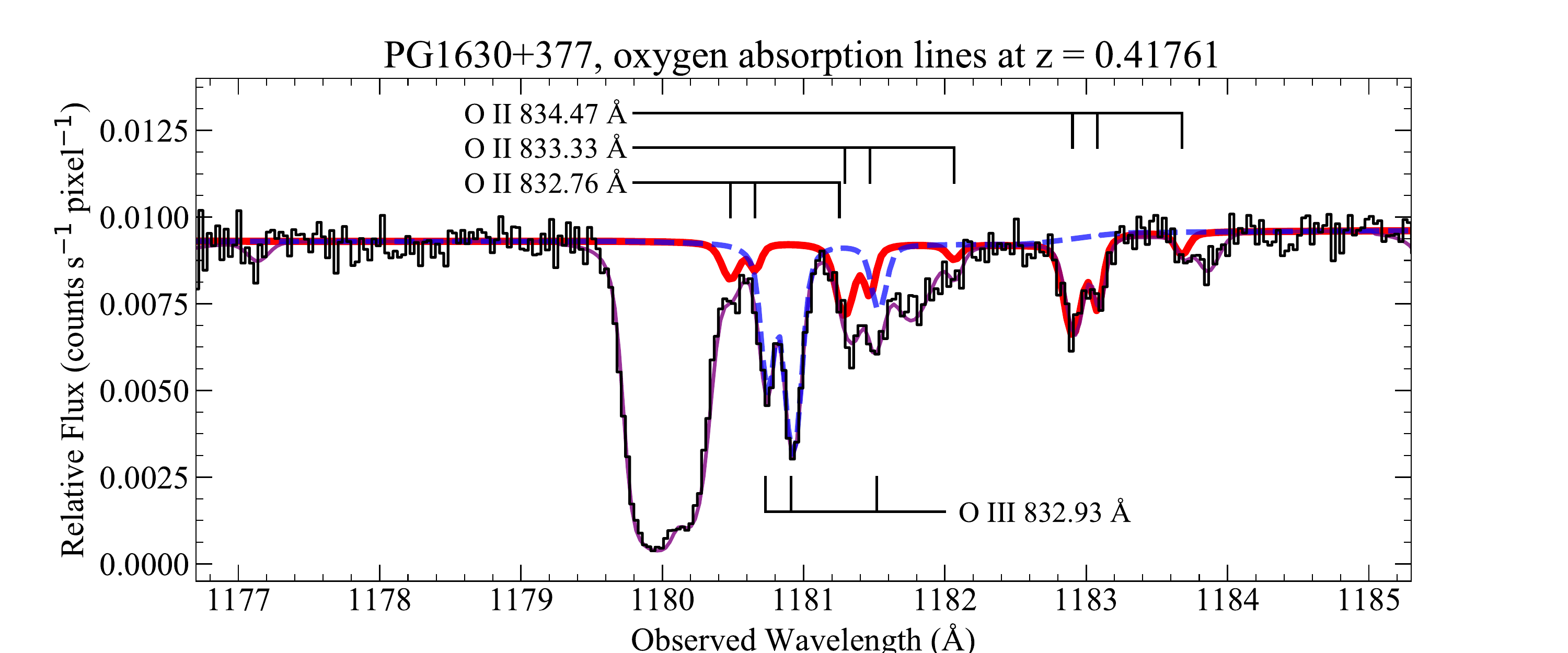}
\caption{A small portion of the far-ultraviolet COS spectrum of PG1630+377 (black histogram) obtained as part of the CASBaH survey. This example shows the CASBaH data quality and complexity and illustrates some of the advantages and challenges presented by the data, as discussed in the text. In this wavelength range, absorption lines from the \textsc{O~ii} multiplet at 832.76, 833.33, and 834.47~\AA\ as well as the \textsc{O~iii} 832.93 line at $z_{\rm abs}$ = 0.41760 are detected but are blended with each other and with absorption lines from other absorption systems at different redshifts.  To extract absorption-line measurements from these data, we have fit Voigt profiles to the entire blend, indicated with a thin purple line.  In these \textsc{O~ii} and \textsc{O~iii} transitions, we find evidence of three components at $v = 0, 43$ and 197 km s$^{-1}$ in the  $z_{\rm abs}$ = 0.41760 system. These absorption components from the \textsc{O~ii} multiplet are shown with a thick red line, and the \textsc{O~iii} absorption components are plotted with a dashed blue line.  The \textsc{O~ii} and \textsc{O~iii} lines of interest are blended with \textsc{H~i} Lyman series lines at $z_{\rm abs}$ = 0.2153, 0.2741, and 0.2780 as well as multicomponent Ne~\textsc{viii} 770.409 \AA\ absorption at $z_{\rm abs}$ = 0.5334.  The identification of the \textsc{H~i} lines, and their detailed contributions to this blend, are highly secure because they are constrained by many other \textsc{H~i} Lyman series lines at the same redshifts.  Likewise, the Ne~\textsc{viii} 770.409 features are corroborated by their kinematical alignment with \textsc{O~vi} and other metals at the same redshift. \label{fig:data_sample}}
\end{figure*}

	To obtain NUV spectra of the targets, we used either the NUV mode of COS or the E230M echelle mode of STIS, depending on the target wavelength range of the spectrum.  At $\lambda_{ob} \gtrsim$ 2200 \AA , we generally preferred to use STIS because it provides somewhat higher spectral resolution and a better, more Gaussian line-spread function; the COS NUV modes provide a spectral resolution of $\approx$15--20 km s$^{-1}$ while the STIS E230M resolution is $\approx$10 km s$^{-1}$. In the 1800$-$2200 \AA\ range the COS NUV mode is much more sensitive and efficient than the corresponding STIS configuration, but this COS mode has two large gaps in the covered wavelengths at any given grating tilt (see Figure 13.17 in \citealt{Fischer19}).  To obtain complete COS NUV spectra without gaps, we used the G185M and G225M gratings with several grating tilts to fill in the gaps (see Tripp et al. 2021, \emph{in prep.}).  The COS NUV data were reduced in a manner similar to the FUV data but with an updated version of CALCOS (version 3.1.8) that was needed to correct a wavelength calibration problem that we encountered when we initially reduced the NUV data. We also used a 9-pixel extraction box rather than the default 57-pixel box.  The smaller extraction box leads to a small loss of flux owing to the broad wings of the COS line-spread function but it significantly improves the S/N of the extracted spectrum. The STIS data were reduced using the method of \cite{Tripp01}, including a weighted coaddition of the overlapping regions of adjacent orders as well as a weighted coaddition of the individual exposures.  The NUV spectra were obtained to detect stronger lines such as \textsc{H~i} Ly$\alpha$ and \textsc{O~vi} doublets redshifted into the NUV range.  Consequently, the NUV spectra were not required to attain as high S/N as the FUV observations.  The final NUV spectra used here have S/N per resolution element ranging from 4 to 19.

As noted above, we also obtained high-resolution optical spectroscopy with the HIRES instrument on the Keck 1 telescope.  Most of the CASBaH targets were observed with the HIRES C1 decker, which provides a spectral resolution of 6 km s$^{-1}$, and were recorded with the 3-chip mosaic CCD detector.\footnote{Two of the targets in the broader CASBaH program were observed with the C5 decker, and those spectra have a somewhat lower resolution of 8 km s$^{-1}$.  However, all of the data presented in this paper employed the higher resolution C1 aperture.}   The data were reduced and continuum normalized using the procedures described by \cite{OMeara15}.  Additional information about these observations can be found in Tripp et al. (2021, \emph{in prep.}).

\begin{figure*}
\includegraphics[width=\textwidth]{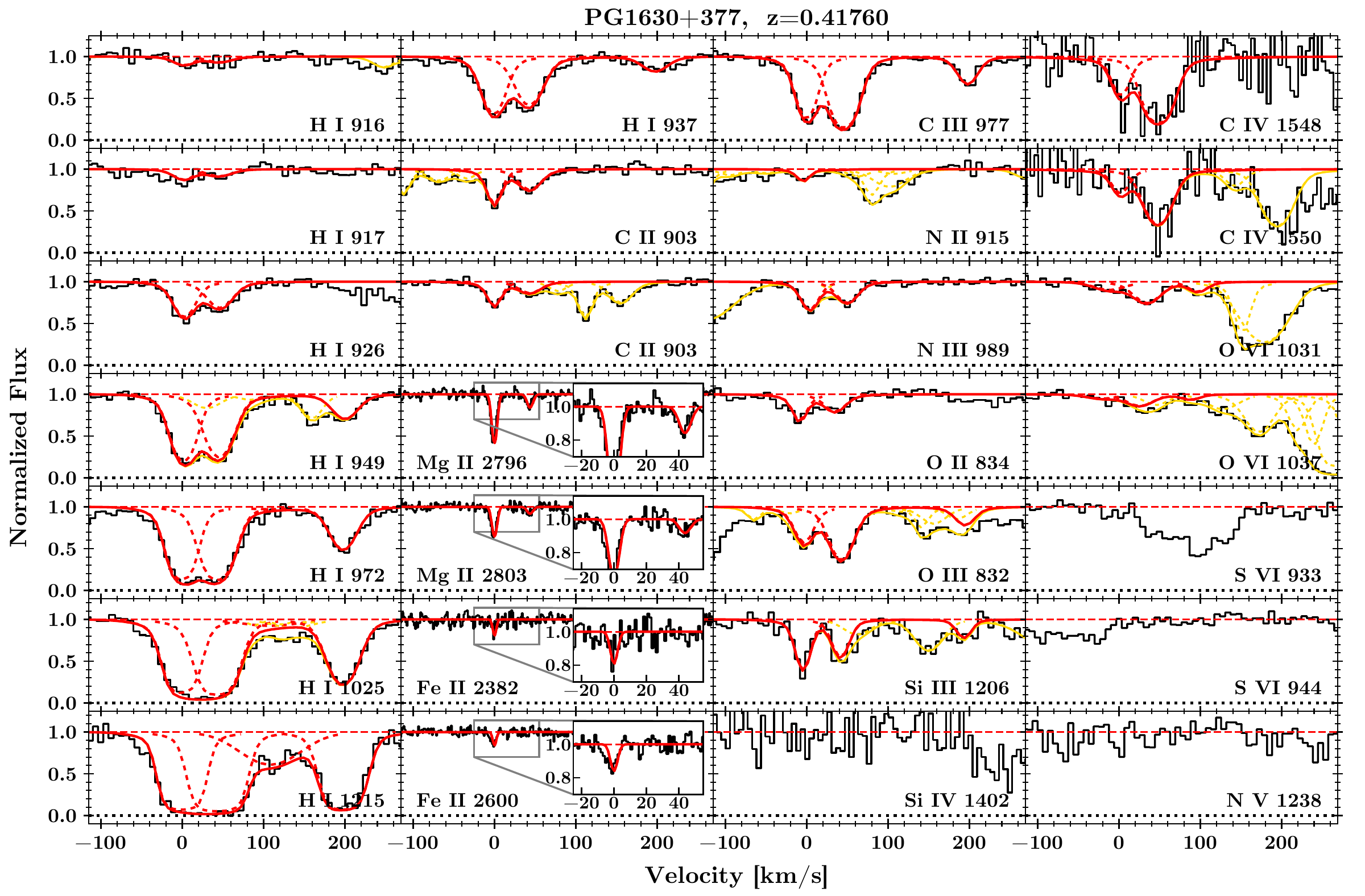}
\caption{Absorption velocity profiles and best-fit Voigt Profiles for detections and selected non-detections at \emph{z}=0.41760 in the PG1630+377 sightline (includes Absorbers~0.4A and~0.4B at velocities $\approx$0 and $\approx$43 km s$^{-1}$, respectively).  These draw from observations using multiple instruments with significant differences in S/N and resolution between them.  The flux has been continuum-normalized as described in the text.  Dashed red (yellow) lines illustrate individual Voigt profiles for the labeled species(unlabeled contamination), with solid red(yellow) lines indicating the total modeled absorption.  The Voigt profile fits to contaminating lines from absorbers at other redshifts (yellow curves) are all reasonably well constrained through co-fitting with another transition (at the contaminating absorber's redshift) of the same or similar species observed in a separate region of the spectrum.}
\label{fig:z041761_vpfits}
\end{figure*}

Figure~\ref{fig:data_sample} shows an example of the CASBaH FUV COS spectra.  This small portion of the PG1630+377 spectrum demonstrates some of the advantages and challenges of high-S/N and high-resolution observations of $z \approx$ 1 quasars.  The absorption feature in Figure~\ref{fig:data_sample} is a complicated blend of lines from various species at various redshifts, and one might reasonably question whether this blend can be usefully decomposed.  However, since there are many resonance transitions available in the far-UV (see Figure~\ref{fig:casbah_lines}), we find that we can obtain meaningful constraints by exploiting all of the information in the data and the broad wavelength coverage of the survey; the velocity centroids and line widths of blended features are often constrained by other transitions of the same species recorded elsewhere in the spectra.  Indeed, this is evident in Figure~\ref{fig:data_sample}: much of the absorption in this example is caused by \textsc{O~ii} and \textsc{O~iii} ions in the absorption system at $z_{\rm abs}$ = 0.41760 (see also Figure~\ref{fig:z041761_vpfits}), and by comparing the optical depths at the expected wavelengths of the various transitions, one finds that the  \textsc{O~ii} and \textsc{O~iii} features are robustly identified and well constrained.


\begin{figure*}
\includegraphics[width=\textwidth]{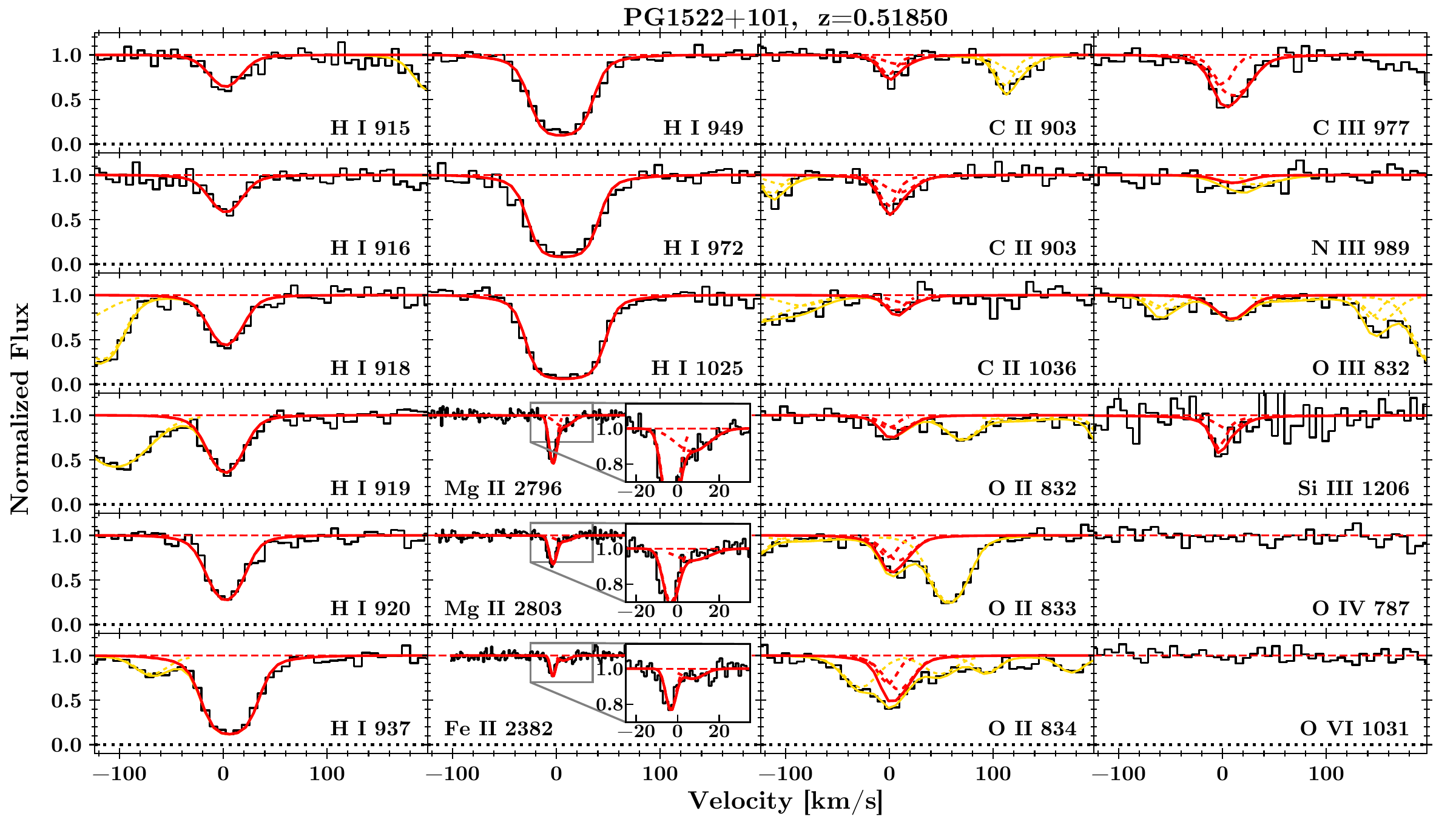}
\caption{Absorption velocity profiles and best-fit Voigt Profiles for detections and selected non-detections at \emph{z}=0.51850 in the PG1522+101 sightline (Absorber~0.5A).  See Figure~\ref{fig:z041761_vpfits} for a description of the colored lines.}
\label{fig:z051850_vpfits}
\end{figure*}

\begin{figure*}
\includegraphics[width=\textwidth]{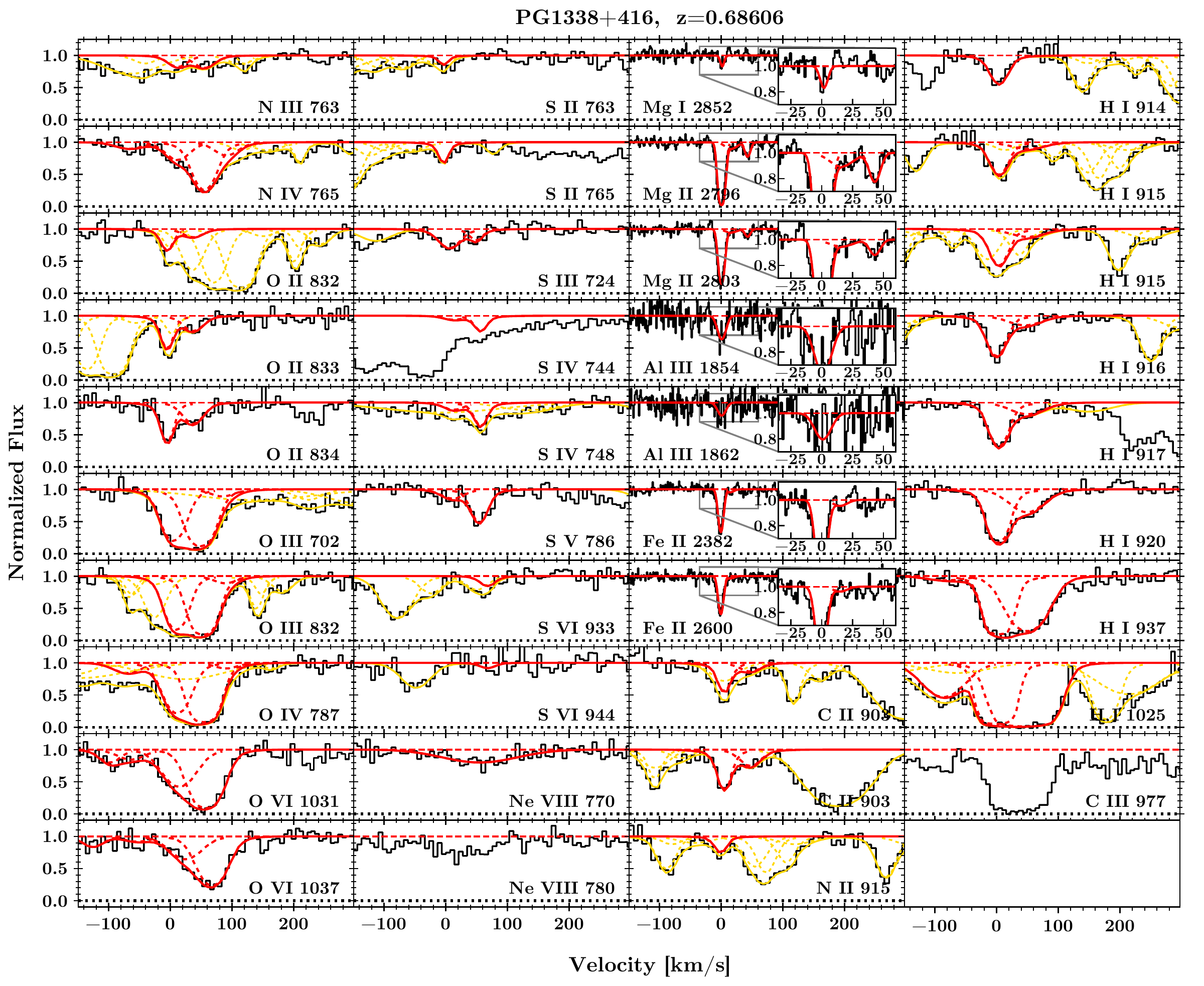}
\caption{Absorption velocity profiles and best-fit Voigt Profiles (VPs) for detections and selected non-detections at \emph{z}=0.68606 in the PG1338+416 sightline (includes Absorbers~0.6A and~0.6B at velocities $\approx$0 and $\approx$43 km s$^{-1}$, respectively).  Ne~\textsc{viii} 780$\textrm{\AA}$ is located in an unconstrained blend and could not be fitted, but is consistent with the fit to Ne~\textsc{viii} 770$\textrm{\AA}$.  The red dashed-line VPs shown for Ne~\textsc{viii} 780$\textrm{\AA}$ are drawn based on best fit parameters from Ne~\textsc{viii} 770$\textrm{\AA}$ for comparison.  The velocities and $b$-values for VP fits to saturated lines were constrained through co-fitting with non-saturated lines following the co-fitting considerations described in the text (note that the column densities listed in Table~\ref{tab:z068605_meas} include those derived from the fits shown here as well as those for conservative upper and lower limits).  See Figure~\ref{fig:z041761_vpfits} for a description of the colored lines.}
\label{fig:z068605_vpfits}
\end{figure*}

\subsection{Column Density Measurements}
\label{sec:measurements}
We identified absorption metal lines associated with the prominent H~\textsc{i} Lyman Series lines of the systems in Table~\ref{tab:abs_sys_info} and then measured their column densities and limits.  Figures~\ref{fig:z041761_vpfits}--\ref{fig:z068605_vpfits} show the absorption profiles in these systems, and the measured line parameters are listed in Tables~\ref{tab:z041761_meas}--\ref{tab:z068605_meas}.  The unusual quality and richness of the CASBaH QSO spectra enhances our ability to detect absorption and useful non-detections with lower uncertainties for a larger number of species compared with average quality QSO spectra (e.g., \emph{z}\textsubscript{QSO}$\approx$0.2, S/N$\lesssim$10). For many ions, the uncertainty is reduced by the presence of multiple resonance transitions (transitions that are unavailable at lower redshift) as well as by the decrease in shot noise.  The improved confidence in our measurements stems from a) an enhanced ability to identify and constrain systematics owing to intervening absorption lines from other redshifts, b) careful decomposition of absorbers' underlying velocity structures, and c) co-fitting multiple components with shared parameters.  Below we relate our insights related to the absorbers in this study and to the analysis of high-quality, line-rich QSO spectra in general.

\subsubsection{Line Identification and Deblending}
	Prior to making line measurements, we identified all absorption systems in our spectra based on their H~\textsc{i} Lyman Series lines and/or corroborating metal line transitions at the same redshift (Tripp et al. 2021 \emph{in prep.}).  A full census of lines in each sightline is essential for identifying and deblending contaminating lines, i.e., absorption features from various redshifts that coincidentally overlap in observed wavelength with an absorption line of interest.  Potentially unidentified contamination, whether from weak Ly$\alpha$ lines (\emph{N}\textsubscript{H I} $\lesssim$ 10\textsuperscript{13.5} cm\textsuperscript{-2}) or metal lines associated with an obscured or weak \textsc{H~i} line (see, e.g., Figure 7 in \citealt{Tripp08} and Figures 4 and 7 in \citealt{Meiring13}) could lead to an overestimation of the contaminated line's absorption strength.  This potential for contamination increases with the redshift of the observed QSO, owing to the steadily increasing density of UV transitions that are redshifted to observable wavelengths as well as the increase in path length along the line-of-sight.

	Additional checks for hidden contamination can be performed for species with an observed coverage of doublet and multiplet transitions by comparing their apparent column density (ACD) profiles \citep{Savage91,Sembach92}, which should match in the absence of contamination.  For species with only one observable transition (e.g., C~\textsc{iii} $\rm{\lambda}$977.02$\textrm{\AA}$, N~\textsc{iv} $\rm{\lambda}$765.15$\textrm{\AA}$, Si~\textsc{iii}  $\rm{\lambda}$1206.05$\textrm{\AA}$, and S~\textsc{v}  $\rm{\lambda}$786.48$\textrm{\AA}$), a comparison of velocity profiles (absorption or ACD) with those of the full set of detected species will often help to confirm a detection by virtue of their similarity, or it may reveal some contaminating absorption line as a spurious bulge in the velocity profiles.  Two examples of this are the high velocity O~\textsc{iii} and S~\textsc{iii} lines at \emph{v} $\approx \ 200$ km s$^{-1}$ in the \emph{z}=0.41760 absorber from the PG1630+377 sightline.  Figure~\ref{fig:z041761_vpfits} shows absorption profiles in red for species at this redshift, including Si~\textsc{iii}, as well as contaminating lines in yellow.  We can confirm Si~\textsc{iii} absorption at \emph{v}$\approx$ 200 km s\textsuperscript{-1} by noting its similarity to the C~\textsc{iii} and H~\textsc{i} absorption features present at that velocity, and we can simultaneously conclude that a strong Si~\textsc{iii} feature at \emph{v}$\approx$145 km s\textsuperscript{-1} is unlikely because it is not corroborated in C~\textsc{iii}, nor in H~\textsc{i}.  O~\textsc{iii}, on the other hand, is apparent in the flux decrement remaining after accounting for all the adjacent, well-constrained contaminating lines.  In the same figure, the O~\textsc{vi} doublet profiles provide an example of contamination (centered at \emph{v}$\approx$30 km s\textsuperscript{-1} in the O~\textsc{vi} $\rm{\lambda}$1037$\textrm{\AA}$ profile) that becomes apparent when comparing multiplet species.  These examples underscore the complexity of the data at moderate redshifts and the importance of high quality spectra, and extensive coverage of the Ly$\alpha$ forest region, for de-blending contamination.  Spectra with insufficient resolution or S/N, or an inadequate wavelength range to cover corroborating lines, would not be nearly as effective at constraining such large numbers of coincident absorption features.

\subsubsection{Continuum Normalization}
	As a precursor to Voigt Profile fitting, we continuum-normalized our COS and STIS spectra, using continuum placements determined by fitting polynomials to arbitrary 10--50$\textrm{\AA}$ windows of continuum pixels (identified by eye) in the spectrum.  We prefer this approach, which directly emphasizes human intuition, versus a more automated approach, which typically combine indirect human intuitions with additional poorly characterized algorithmic systematics.  In general, we expect the continuum placement uncertainty (typically $\approx 0.01$ dex) to have a minimal impact on the final absorption line column density measurement, based on a comparison of line measurements using continuum placements estimated independently by two of the authors (Haislmaier and Tripp).  The precision of our continuum estimates is a consequence of the excellent spectral S/N, which significantly narrows the continuum placement ambiguity.  The HIRES spectra were normalized using the methods of \cite{OMeara15}.

\subsubsection{Voigt Profile Fitting}
\label{sec:vpfitting}
We measured the column density, linewidth, and velocity of all absorption lines associated with our selected H~\textsc{i} absorbers by fitting Voigt Profiles (VP) to the relevant portions of the continuum-normalized spectra; the results are listed in Tables~\ref{tab:z041761_meas} - \ref{tab:z068605_meas}.  In this paper, we chose the redshift of 
the strongest H~\textsc{i} component in each of the three absorption systems as the systemic redshift.  A visual comparison of ACD profiles suggests that all ionization stages are consistent with having a common velocity structure (i.e., the velocity separation between components), a point we discuss further below.  We leverage this information to minimize VP uncertainties by co-fitting velocities for various groupings of ions.

Unfortunately, the absolute kinematic alignment of lines separated by as little as 1 \AA\ can be obscured by highly localized distortions of the wavelength scale caused by the type of detector used by COS \citep{Dixon07}.  For example, N~\textsc{iii} 989\AA, O~\textsc{ii} 834\AA, and O~\textsc{iii} 832\AA\ in Figure~\ref{fig:z041761_vpfits} all have the same two prominent components separated by $\approx$43 km s$^{-1}$, but the velocities of the two O~\textsc{ii} components are offset by $\approx$-6 km s$^{-1}$ from the O~\textsc{iii} components, while N~\textsc{iii} exhibits a $\approx$+4 km s$^{-1}$ shift.  A similar effect also plagues the C~\textsc{iv} and O~\textsc{vi} in the same figure.  While in these examples the offsets could be real, similar offsets are found for different lines of the same species that should have exactly the same centroid.  These distortions are typically largest near the detector edges and the COS detector gap. To model this source of error, we add a free parameter (labeled as $\mathrm{v}_{corr}$ in Tables~\ref{tab:z041761_meas} - \ref{tab:z068605_meas}) for each $\approx$1 \AA\ segment of each spectrum that, in effect, adjusts the wavelength calibration of each segment independently.  $\mathrm{v}_{corr}$ is mathematically redundant with the VP velocity parameter, so we only need it when the VP component velocities, $\mathrm{v}$, are shared or ``co-fit'' between lines in different segments of the spectrum.  While this issue with the COS wavelength scale is somewhat annoying (and possibly impossible to fully rectify), we underscore that these calibration errors are small enough so that they do not profoundly affect the conclusions in this paper.

The fits for C~\textsc{ii} and O~\textsc{ii} at \emph{z}=0.68606 in PG1338+416 are an excellent example of how co-fitting can improve the measurement.  Fitted individually, the stronger C~\textsc{ii} VP converges to an unrealistic column density of \emph{N}\textsubscript{C II}=10\textsuperscript{17.62} cm\textsuperscript{-2} and a spurious broadening parameter value of \emph{b}=0.45 km s\textsuperscript{-1}.  This poor fit primarily results from poorly constrained velocity centroids for C~\textsc{ii}.  A similar problem occurs with O~\textsc{ii}.  Meanwhile, the \emph{v} $\approx$ 0 and \emph{v} $\approx$ 43 km s\textsuperscript{-1} Mg~\textsc{ii} doublet components in this absorption system can be precisely measured in the high resolution HIRES spectrum.  Since all three species are low-ions and follow similar absorption profiles, i.e., strong absorption at \emph{v} $\approx$ 0 km s\textsuperscript{-1} relative to the weaker \emph{v} $\approx$ 43 km s\textsuperscript{-1} component, they likely arise from similar physical processes and thus ought to make good candidates for co-fitting.\footnote{Note that tying velocity centroids together does not necessarily require the respective species to be associated with the same phase (see discussion in Section~\ref{sec:kinematic_alignment}).  The phase structure present in a given velocity component can only be inferred by ionization modeling, such as we perform below.}  With this technique, we manage to recover useful constraints on the C~\textsc{ii} and O~\textsc{ii} column densities (reported in Table~\ref{tab:z068605_meas}).

	In some cases it is also advantageous to fit two species with shared broadening parameters in addition to shared velocities.  This must be done with caution, however, since several factors influence the line-widths.  First, ions produced in different temperature gas phases (e.g., low and high-ions in a multiphase medium) will have distinct thermal broadening solutions.  Additionally, the degree of thermal broadening varies with atomic weight, and this should be considered when comparing different species.  We refer to this as ``thermally-scaled'' broadening.  Second, we must account for the possibility of non-thermal broadening that has the potential to smooth out linewidth differences between different-mass species, what we call ``evenly-scaled'' broadening.  Finally, weak, narrow, and unresolved lines can appear to broaden lines after smearing by the line spread function\footnote{We use the line spread functions provided by the Space Telescope Science Institute for COS \citep{Kriss11} and a 2 pixel Gaussian FWHM for HIRES spectra.}.  It is reasonable to question when it is possible to share broadening parameters in light of these considerations.  In practice, however, as long as one restricts shared broadening to species with similar ionization potentials (i.e., expected to be in the same gas ``phase''), broadening parameters from the co-fitted VPs are predominantly consistent with the results of independent VP fits (but typically with improved uncertanties), and the difference has an even smaller impact on the column density.  The exception to this statement is in the presence of narrow, saturated lines (see the discussion of limits below for an example from the PG1630+377, \emph{v}=0.41760 system).  When co-fitting broadening parameters, we take care to compare fits based on ``thermally-scaled'' and ``evenly-scaled'' broadening.  The bracketed numerical superscripts in Tables~\ref{tab:z041761_meas} - \ref{tab:z068605_meas} mark line groups that have a shared parameter.  The type of shared parameter varies from case-to-case; notes about these measurements are provided in Sections~\ref{sec:column_limits} and \ref{sec:absorber_notes}.

\subsubsection{Upper and Lower Limits}
\label{sec:column_limits}
	Limits were measured for saturated lines and non-detections in various ways.  We adopted 3$\sigma$ upper limits for lines that did not meet a 3$\sigma$ equivalent width (EW) threshold for detection.  In addition, occasionally upper limits were measured from VP fitting for absorption features suspected to be affected by an unidentified blend (e.g., S~\textsc{vi} at \emph{z}=0.68606, \emph{v} $\approx$ 43 km s\textsuperscript{-1} in PG1338+416, Figure~\ref{fig:z068605_vpfits}).  Lower limits to saturated lines were estimated from their equivalent widths.
	In some cases intrinsically narrow, saturated lines can deceptively appear to be optically thin (e.g., the line core is well above the zero-flux level) as a consequence of the COS line spread function's broad wings \citep{Kriss11}.  For example, the C~\textsc{iii} for the \emph{z}=0.41760 system in PG1630+377 is well-fit by two apparently optically thin VPs.  Our ionization modeling (see below) persistently overpredicts the C~\textsc{iii} column density in both components, however, so we explored its absorption more carefully and found that the same lines were well-fit with a separate set of VP parameters having optically thick column densities consistent with those predicted by the ionization model.  This alternative fit for C~\textsc{iii} requires significantly narrower (i.e., smaller broadening parameter) lines than the original fit, however.  To ensure that we adopt a self-consistent set of VP parameters, we re-fit the N~\textsc{iii}, O~\textsc{iii}, and Si~\textsc{iii} lines along with the C~\textsc{iii}, choosing to co-fit their velocities and thermally-scaled broadening parameters for various fixed choices of broadening parameter since all four ions have similar ionization potentials.  In fact, we found in this case that all four ions are consistent with having saturated (or nearly saturated), narrow VPs, and that a double-bounded limit is more accurate to report than a single best-fit.  We report lower limits estimated from the naive fit, i.e., assuming the lines are optically thin, and upper limits from a fit where the broadening parameter was fixed to match the low-ion broadening.

\ctable[
	caption={Column Density Measurements for the \emph{z}=0.41760 Absorption 
			System in PG1630+377 (including Absorbers~0.4A and~0.4B)},
	label = {tab:z041761_meas},
	footerwidth = 3.1in,
	doinside = \scriptsize
	]{l@{~}l@{~}r@{ $\pm$}lr@{$\pm$}l@{~}r@{$\pm$}l@{~}r@{$\pm$}l@{}}{
\tnote[a]{\scriptsize Only atomic transitions used to produce the measurements are listed.  Some transitions covered by the observations had to be excluded as they did not provide meaningful constraints to the fit, e.g., due to severe blending or low signal-to-noise.}
\tnote[b]{\scriptsize Limits preceded by partial inequality signs ($\leq$ or $\geq$) indicate solid detections affected by significant systematics (e.g., saturation or complex covariances) in one direction.  Non-detections with a single numeric (i.e., no extra uncertainty term) were derived from 3$\sigma$ equivalent width upper limits.}
\tnote[c]{\scriptsize The Voigt Profile (VP) broadening parameter, or $b$-value.}
\tnote[d]{\scriptsize All parameters sharing the same bracketed superscript number in a given column derive their value from a single free parameter, i.e., that parameter was shared by several different Voigt profiles.} 
\tnote[e]{\scriptsize Section~\ref{sec:vpfitting} describes the relation between ``$\mathrm{v}$'', the component velocity, and ``$\mathrm{v}_{corr}$'', a wavelength dependent velocity correction.  The systemic redshift is listed in the table header.}
\tnote[f]{\scriptsize These measurements were derived from the equivalent width.}
\tnote[g]{\scriptsize See the text regarding the estimation of these upper limits.}
}{                                                                                                                                                 \FL
$\mathrm{Ion}$ & \multicolumn{1}{c}{$\lambda_{0}\tmark[a]$} & \multicolumn{2}{c}{$\log N(X^{i})\tmark[b]$} & \multicolumn{2}{c}{$\rm{b}$\tmark[c,]\tmark[d]} & \multicolumn{2}{c}{$\mathrm{v}$\tmark[d,]\tmark[e]} & \multicolumn{2}{c}{$\mathrm{v}_{corr}$\tmark[d,]\tmark[e]}                                     \NN
               & \multicolumn{1}{c}{$\textrm{(\AA)}$}       & \multicolumn{2}{c}{$\mathrm{(cm^{-2})}$}            & \multicolumn{2}{c}{$\mathrm{(km\ s^{-2})}$}    & \multicolumn{2}{c}{$\mathrm{(km\ s^{-2})}$} & \multicolumn{2}{c}{$\mathrm{(km\ s^{-2})}$}                                          \ML

\multicolumn{10}{c}{$\rm{Component\ A\ \textrm{(}\textit{v}\approx\textrm{0 km s}^{-1}\textrm{)}}$}                                                \ML
$\rm{HI     }$ & $\rm{Ly\alpha-\epsilon,\eta}$ & $15.46$ & $0.016$ & $13.6$ & $0.4$        & $  0.0$ & $1.0$$^{[1]}$   & $ -1.0$ & $1.3$$^{[1]}$   \NN
$\rm{CII    }$ & $\rm{903,903        }$ &     $13.46$ & $0.059$ & $ 5.6$ & $0.7$$^{[1]}$   & $  0.0$ & $1.0$$^{[1]}$   & $ -0.2$ & $1.1$$^{[2]}$   \NN
$\rm{CIII   }$ & $\rm{977            }$ & $\geq13.52$ & $0.029$ & $11.7$ & $0.2$$^{[2]}$   & $  0.4$ & $0.5$$^{[2]}$   & $ -0.4$ & $0.6$$^{[3]}$   \NN
$\rm{CIII   }$ & $\rm{977\tmark[g]   }$ &    $<14.20$ & $0.000$ & $ 7.5$ & $0.2$$^{[2]}$   & \multicolumn{4}{c}{}                                  \NN
$\rm{CIV    }$ & $\rm{1548,1550      }$ &     $13.48$ & $0.175$ & $11.7$ & $0.2$$^{[2]}$   & $  0.4$ & $0.5$$^{[2]}$   & $  1.3$ & $3.0$$^{[4]}$   \NN
$\rm{NII    }$ & $\rm{915            }$ &     $13.05$ & $0.104$ & $ 5.2$ & $0.7$$^{[1]}$   & $  0.0$ & $1.0$$^{[1]}$   & $ -4.4$ & $2.8$$^{[5]}$   \NN
$\rm{NIII   }$ & $\rm{989            }$ &     $13.62$ & $0.034$ & $10.8$ & $0.2$$^{[2]}$   & $  0.4$ & $0.5$$^{[2]}$   & $  3.4$ & $1.1$$^{[6]}$   \NN
$\rm{OII    }$ & $\rm{832,833,834    }$ &     $13.70$ & $0.051$ & $ 4.8$ & $0.7$$^{[1]}$   & $  0.0$ & $1.0$$^{[1]}$   & $ -3.8$ & $0.0$           \NN
$\rm{OIII   }$ & $\rm{832            }$ & $\geq13.99$ & $0.031$ & $10.1$ & $0.6$           & $  0.4$ & $0.6$           & $ -3.8$ & $0.0$           \NN
$\rm{OIII   }$ & $\rm{832\tmark[g]   }$ &    $<14.11$ & $0.045$ & $ 6.5$ & $0.2$$^{[2]}$   & \multicolumn{4}{c}{}                                  \NN
$\rm{OVI    }$ & $\rm{1031,1037      }$ &     $13.17$ & $0.156$ & $22.8$ & $11.3$          & $  0.4$ & $0.5$$^{[2]}$   & $ -1.5$ & $3.0$$^{[7]}$   \NN
$\rm{MgII   }$ & $\rm{2796,2803      }$ &     $12.17$ & $0.013$ & $ 4.0$ & $0.2$$^{[3]}$   & $  0.0$ & $1.0$$^{[1]}$   & $ -1.3$ & $1.1$$^{[8]}$   \NN
$\rm{SiIII  }$ & $\rm{1206           }$ & $\geq12.79$ & $0.077$ & $ 7.6$ & $0.2$$^{[2]}$   & $  0.4$ & $0.5$$^{[2]}$   & $ -5.5$ & $1.3$$^{[9]}$   \NN
$\rm{SiIII  }$ & $\rm{1206\tmark[g]  }$ &    $<13.05$ & $0.121$ & $ 4.9$ & $0.2$$^{[2]}$   & \multicolumn{4}{c}{}                                  \NN
$\rm{SVI    }$ & $\rm{933\tmark[f]   }$ &    $<12.43$ & $0.117$          & \multicolumn{6}{c}{}                                                    \NN
$\rm{FeII   }$ & $\rm{2344,2382,2600 }$ &     $11.80$ & $0.042$ & $ 3.5$ & $0.5$           & $  0.0$ & $1.0$$^{[1]}$   & $ -1.3$ & $1.1$$^{[8]}$   \ML

\multicolumn{10}{c}{$\rm{Component\ B\ \textrm{(}\textit{v}\approx\textrm{44 km s}^{-1}\textrm{)}}$}                                               \ML
$\rm{HI     }$ & $\rm{Ly\alpha-\epsilon,\eta}$ & $15.33$ & $0.012$ & $17.4$ & $0.5$        & $ 43.7$ & $1.0$$^{[3]}$   & $ -1.0$ & $1.3$$^{[1]}$   \NN
$\rm{CII    }$ & $\rm{903,903        }$ &     $13.16$ & $0.031$ & $14.6$ & $1.7$$^{[4]}$   & $ 43.7$ & $1.0$$^{[3]}$   & $ -0.2$ & $1.1$$^{[2]}$   \NN
$\rm{CIII   }$ & $\rm{977            }$ & $\geq13.83$ & $0.035$ & $15.4$ & $0.2$$^{[5]}$   & $ 46.5$ & $0.4$$^{[4]}$   & $ -0.4$ & $0.6$$^{[3]}$   \NN
$\rm{CIII   }$ & $\rm{977\tmark[g]   }$ &    $<14.40$ & $0.000$ & $11.0$ & $0.2$$^{[5]}$   & \multicolumn{4}{c}{}                                  \NN
$\rm{CIV    }$ & $\rm{1548,1550      }$ &     $14.19$ & $0.136$ & $17.5$ & $4.2$           & $ 46.5$ & $0.4$$^{[4]}$   & $  1.3$ & $3.0$$^{[4]}$   \NN
$\rm{NII    }$ & $\rm{915            }$ &     $11.82$ & $1.872$ & $13.5$ & $1.7$$^{[4]}$   & $ 43.7$ & $1.0$$^{[3]}$   & $ -4.4$ & $2.8$$^{[5]}$   \NN
$\rm{NIII   }$ & $\rm{989            }$ & $\leq13.53$ & $0.038$ & $14.3$ & $0.2$$^{[5]}$   & $ 46.5$ & $0.4$$^{[4]}$   & $  3.4$ & $1.1$$^{[6]}$   \NN
$\rm{NIII   }$ & $\rm{989\tmark[g]   }$ &    $>13.50$ & $0.039$ & $10.2$ & $0.2$$^{[5]}$   & \multicolumn{4}{c}{}                                  \NN
$\rm{NV     }$ & $\rm{1238\tmark[f]  }$ &    $<12.99$ & $0.204$          & \multicolumn{6}{c}{}                                                    \NN
$\rm{OII    }$ & $\rm{832,833,834    }$ &     $13.48$ & $0.048$ & $12.6$ & $1.7$$^{[4]}$   & $ 43.7$ & $1.0$$^{[3]}$   & $ -3.8$ & $0.0$           \NN
$\rm{OIII   }$ & $\rm{832            }$ & $\geq14.35$ & $0.023$ & $13.4$ & $0.6$           & $ 46.5$ & $0.5$           & $ -3.8$ & $0.0$           \NN
$\rm{OIII   }$ & $\rm{832\tmark[g]   }$ &    $<14.52$ & $0.040$ & $ 9.5$ & $0.2$$^{[5]}$   & \multicolumn{4}{c}{}                                  \NN
$\rm{OVI    }$ & $\rm{1031,1037      }$ &     $13.58$ & $0.056$ & $18.6$ & $3.6$           & $ 46.5$ & $0.4$$^{[4]}$   & $ -1.5$ & $3.0$$^{[7]}$   \NN
$\rm{MgII   }$ & $\rm{2796,2803      }$ &     $11.53$ & $0.039$ & $ 4.7$ & $0.6$$^{[6]}$   & $ 43.7$ & $1.0$$^{[3]}$   & $ -1.3$ & $1.1$$^{[8]}$   \NN
$\rm{SiIII  }$ & $\rm{1206           }$ &     $12.56$ & $0.295$ & $10.1$ & $0.2$$^{[5]}$   & $ 46.5$ & $0.4$$^{[4]}$   & $ -2.1$ & $1.3$$^{[9]}$   \NN
$\rm{SiIV   }$ & $\rm{1402\tmark[f]  }$ &    $<13.02$ & $0.200$          & \multicolumn{6}{c}{}                                                    \NN
$\rm{SVI    }$ & $\rm{944\tmark[f]   }$ &    $<12.56$ & $0.180$          & \multicolumn{6}{c}{}                                                    \NN
$\rm{FeII   }$ & $\rm{2344,2382,2600 }$ &     $10.91$ & $0.238$ & $ 3.1$ & $0.6$$^{[6]}$   & $ 43.7$ & $1.0$$^{[3]}$   & $ -1.3$ & $1.1$$^{[8]}$   \ML

\multicolumn{10}{c}{$\rm{Components\ A+B}$}                                                                                                        \ML
$\rm{NV     }$ & $\rm{1238\tmark[f]  }$	& \multicolumn{2}{c}{$<13.35$}   & \multicolumn{6}{c}{}                                                    \NN
$\rm{CIV    }$ & $\rm{1550\tmark[f]  }$	& $\geq14.12$ & $0.052$          & \multicolumn{6}{c}{}                                                    \NN
$\rm{OI     }$ & $\rm{988\tmark[f]   }$	& \multicolumn{2}{c}{$<13.56$}   & \multicolumn{6}{c}{}                                                    \NN
$\rm{OVI    }$ & $\rm{1031\tmark[f]  }$	& $\geq13.66$ & $0.033$          & \multicolumn{6}{c}{}                                                    \NN
$\rm{MgI    }$ & $\rm{2852\tmark[f]  }$	& \multicolumn{2}{c}{$<11.03$}   & \multicolumn{6}{c}{}                                                    \NN
$\rm{AlII   }$ & $\rm{1670\tmark[f]  }$	& \multicolumn{2}{c}{$<12.22$}   & \multicolumn{6}{c}{}                                                    \NN
$\rm{AlIII  }$ & $\rm{1854\tmark[f]  }$	& \multicolumn{2}{c}{$<12.76$}   & \multicolumn{6}{c}{}                                                    \NN
$\rm{SiII   }$ & $\rm{1193\tmark[f]  }$ &    $<12.78$ & $0.070$          & \multicolumn{6}{c}{}                                                    \NN
$\rm{SiIV   }$ & $\rm{1402\tmark[f]  }$	& \multicolumn{2}{c}{$<13.49$}   & \multicolumn{6}{c}{}                                                    \ML

\multicolumn{10}{c}{$\rm{Component\ C\ \textrm{(}\textit{v}\approx\textrm{109 km s}^{-1}\textrm{)}}$}                                              \ML
$\rm{HI     }$ &$\rm{Ly\alpha-\epsilon}$&     $13.51$ & $0.076$ & $45.4$ & $10.6$$^{[7]}$  & $109.4$ & $3.3$$^{[5]}$   & $ -1.7$ & $0.0$           \NN
$\rm{OVI    }$ & $\rm{1031,1037      }$ &     $13.09$ & $0.094$ & $11.4$ & $10.6$$^{[7]}$  & $109.4$ & $3.3$$^{[5]}$   & $ -1.7$ & $0.0$           \ML

\multicolumn{10}{c}{$\rm{Component\ D\ \textrm{(}\textit{v}\approx\textrm{199 km s}^{-1}\textrm{)}}$}                                              \ML
$\rm{HI     }$ &$\rm{Ly\alpha-\epsilon}$&     $14.64$ & $0.011$ & $17.4$ & $0.5$           & $198.5$ & $0.9$$^{[6]}$   & $ -1.0$ & $1.3$$^{[1]}$   \NN
$\rm{CIII   }$ & $\rm{977            }$ &     $12.87$ & $0.029$ & $12.9$ & $1.4$$^{[8]}$   & $198.5$ & $0.9$$^{[6]}$   & $ -0.2$ & $0.6$$^{[3]}$   \NN
$\rm{OIII   }$ & $\rm{832            }$ &     $13.61$ & $0.071$ & $11.2$ & $1.4$$^{[8]}$   & $198.5$ & $0.9$$^{[6]}$   & $ -3.8$ & $0.0$           \NN
$\rm{SiIII  }$ & $\rm{1206           }$ &     $12.12$ & $0.142$ & $ 8.5$ & $1.4$$^{[8]}$   & $198.5$ & $0.9$$^{[6]}$   & $ -2.1$ & $1.3$$^{[9]}$   \LL
}

\ctable[
	caption={Column Density Measurements for the \emph{z}=0.51850 Absorption 
			System in PG1522+101 (Absorber~0.5A)},
	label = {tab:z051850_meas},
	footerwidth = 3.1in,
	doinside = \scriptsize
	]{l@{~}l@{~}r@{ $\pm$}l@{~}r@{$\pm$}l@{~}r@{$\pm$}l@{~}r@{$\pm$}l@{}}{
\tnote[a-f]{\scriptsize Tablenotes \textit{a--f} are given in Table~\ref{tab:z041761_meas}}
\tnote[g]{\scriptsize Components A1 and A2 represent our best attempt to resolve the absorption features into two components.  As discussed in the text, we use the total column densities in component A (i.e., A1 + A2) as constraints on our ionization models.}
}{                                                                                                                                             \FL
$\mathrm{Ion}$ & \multicolumn{1}{c}{$\lambda_{0}\tmark[a]$} & \multicolumn{2}{c}{$\log N(X^{i})\tmark[b]$} & \multicolumn{2}{c}{$\rm{b}$\tmark[c,]\tmark[d]} & \multicolumn{2}{c}{$\mathrm{v}$\tmark[d,]\tmark[e]} & \multicolumn{2}{c}{$\mathrm{v}_{corr}$\tmark[d,]\tmark[e]}                                     \NN
               & \multicolumn{1}{c}{$\textrm{(\AA)}$}       & \multicolumn{2}{c}{$\mathrm{(cm^{-2})}$}            & \multicolumn{2}{c}{$\mathrm{(km\ s^{-2})}$}    & \multicolumn{2}{c}{$\mathrm{(km\ s^{-2})}$} & \multicolumn{2}{c}{$\mathrm{(km\ s^{-2})}$}                                          \ML

\multicolumn{10}{c}{$\rm{Component\ A1\tmark[g]\ \textrm{(}\textit{v}\approx\textrm{-5 km s}^{-1}\textrm{)}}$}                                     \ML
$\rm{CII    }$ & $\rm{903.6,903.9,1036}$& $13.22$ & $0.121$ & $ 6.4$ & $0.3$$^{[1]}$   & $ -4.6$ & $0.2$$^{[1]}$   & $  4.0$ & $1.6$$^{[2]}$   \NN
$\rm{CIII   }$ & $\rm{977            }$ & $12.85$ & $0.289$ & $ 6.4$ & $0.3$$^{[1]}$   & $ -4.6$ & $0.2$$^{[1]}$   & $  4.0$ & $2.2$$^{[1]}$   \NN
$\rm{OII    }$ & $\rm{832,833,834    }$ & $13.67$ & $0.166$ & $ 5.5$ & $0.3$$^{[1]}$   & $ -4.6$ & $0.2$$^{[1]}$   & $  3.2$ & $2.0$$^{[3]}$   \NN
$\rm{MgII   }$ & $\rm{2796,2803      }$ & $12.13$ & $0.036$ & $ 4.5$ & $0.3$$^{[1]}$   & $ -4.6$ & $0.2$$^{[1]}$   & $  0.0$ & $0.0$           \NN
$\rm{SiII   }$ & $\rm{1193,989       }$ & $12.38$ & $0.443$ & $ 4.2$ & $0.3$$^{[1]}$   & $ -4.6$ & $0.2$$^{[1]}$   & $  4.0$ & $2.2$$^{[1]}$   \NN
$\rm{SiIII  }$ & $\rm{1206           }$ & $12.65$ & $0.648$ & $ 4.2$ & $0.3$$^{[1]}$   & $ -4.6$ & $0.2$$^{[1]}$   & $  1.4$ & $4.9$$^{[4]}$   \NN
$\rm{FeII   }$ & $\rm{2382           }$ & $11.89$ & $0.051$ & $ 3.0$ & $0.3$$^{[1]}$   & $ -4.6$ & $0.2$$^{[1]}$   & $  0.0$ & $0.0$           \ML
\multicolumn{10}{c}{$\rm{Component\ A2\tmark[g]\ \textrm{(}\textit{v}\approx\textrm{+6 km s}^{-1}\textrm{)}}$}                                    \ML
$\rm{CII    }$ & $\rm{903.6,903.9,1036}$& $13.04$ & $0.157$ & $16.0$ & $1.8$$^{[2]}$   & $  5.8$ & $1.7$$^{[2]}$   & $  4.0$ & $1.6$$^{[2]}$   \NN
$\rm{CIII   }$ & $\rm{977            }$ & $13.18$ & $0.094$ & $18.7$ & $2.1$$^{[3]}$   & $  5.8$ & $1.7$$^{[2]}$   & $  4.0$ & $2.2$$^{[1]}$   \NN
$\rm{NIII   }$ & $\rm{989            }$ & $13.09$ & $0.220$ & $17.3$ & $2.1$$^{[3]}$   & $  5.8$ & $1.7$$^{[2]}$   & $  4.0$ & $2.2$$^{[1]}$   \NN
$\rm{OII    }$ & $\rm{832,833,834    }$ & $13.82$ & $0.096$ & $13.8$ & $1.8$$^{[2]}$   & $  5.8$ & $1.7$$^{[2]}$   & $  3.2$ & $2.0$$^{[3]}$   \NN
$\rm{OIII   }$ & $\rm{832            }$ & $13.76$ & $0.049$ & $16.2$ & $2.1$$^{[3]}$   & $  5.8$ & $1.7$$^{[2]}$   & $  3.2$ & $2.0$$^{[3]}$   \NN
$\rm{MgII   }$ & $\rm{2796,2803      }$ & $11.79$ & $0.092$ & $11.2$ & $1.8$$^{[2]}$   & $  5.8$ & $1.7$$^{[2]}$   & $  0.0$ & $0.0$           \NN
$\rm{SiIII  }$ & $\rm{1206           }$ & $12.11$ & $0.713$ & $12.2$ & $2.1$$^{[3]}$   & $  5.8$ & $1.7$$^{[2]}$   & $  1.4$ & $4.9$$^{[4]}$   \NN
$\rm{FeII   }$ & $\rm{2382           }$ & $11.59$ & $0.122$ & $ 7.4$ & $1.8$$^{[2]}$   & $  5.8$ & $1.7$$^{[2]}$   & $  0.0$ & $0.0$           \ML
\multicolumn{10}{c}{$\rm{Component\ A = A1 + A2\tmark[g]}$}                                                                                    \ML
$\rm{HI     }$ & $\rm{Ly\beta-\epsilon,\theta-\nu}$&$16.22$ & $0.011$ & $16.9$ & $0.2$ & $  3.2$ & $0.4$           & $  0.0$ & $0.0$           \NN
$\rm{CII    }$ & $\rm{903.6\tmark[f] }$ & $\geq13.34$ & $0.042$        & \multicolumn{6}{c}{}                                                  \NN
$\rm{CIII   }$ & $\rm{977\tmark[f]   }$ & $\geq13.23$ & $0.025$        & \multicolumn{6}{c}{}                                                  \NN
$\rm{CIV    }$ & $\rm{1548\tmark[f]  }$	& \multicolumn{2}{c}{$<13.62$} & \multicolumn{6}{c}{}                                                  \NN
$\rm{NII    }$ & $\rm{915\tmark[f]   }$ & \multicolumn{2}{c}{$<13.29$} & \multicolumn{6}{c}{}                                                  \NN
$\rm{NIV    }$ & $\rm{765\tmark[f]   }$ & $\leq12.97$ & $0.092$        & \multicolumn{6}{c}{}                                                  \NN
$\rm{OI     }$ & $\rm{988\tmark[f]   }$ & \multicolumn{2}{c}{$<13.83$} & \multicolumn{6}{c}{}                                                  \NN
$\rm{OIV    }$ & $\rm{787\tmark[f]   }$ & \multicolumn{2}{c}{$<13.56$} & \multicolumn{6}{c}{}                                                  \NN
$\rm{OVI    }$ & $\rm{1031\tmark[f]  }$	& \multicolumn{2}{c}{$<13.40$} & \multicolumn{6}{c}{}                                                  \NN
$\rm{SiIII  }$ & $\rm{1206\tmark[f]  }$ & $\geq12.42$ & $0.133$        & \multicolumn{6}{c}{}                                                  \NN
$\rm{SiIV   }$ & $\rm{1393\tmark[f]  }$ & \multicolumn{2}{c}{$<13.21$} & \multicolumn{6}{c}{}                                                  \NN
$\rm{SII    }$ & $\rm{765\tmark[f]   }$ & $\leq12.41$ & $0.181$        & \multicolumn{6}{c}{}                                                  \NN
$\rm{SV     }$ & $\rm{786\tmark[f]   }$ & \multicolumn{2}{c}{$<12.48$} & \multicolumn{6}{c}{}                                                  \LL
}

\ctable[
	caption={Column Density Measurements for the \emph{z}=0.68606 Absorption 
			System in PG1338+416 (including Absorbers~0.6A and~0.6B)},
	label = {tab:z068605_meas},
	footerwidth = 3.1in,
	doinside = \scriptsize
	]{l@{~}l@{~}r@{ $\pm$}lr@{$\pm$}l@{}r@{$\pm$}l@{}r@{$\pm$}l@{}}{
\tnote[a-f]{\scriptsize Footnotes \textit{a--f} are given in Table~\ref{tab:z041761_meas}}
\tnote[g]{\scriptsize See the text regarding how and why these upper limits were estimated.}
}{                                                                                                                                                 \FL
$\mathrm{Ion}$ & \multicolumn{1}{c}{$\lambda_{0}\tmark[a]$} & \multicolumn{2}{c}{$\log N(X^{i})\tmark[b]$} & \multicolumn{2}{c}{$\rm{b}$\tmark[c,]\tmark[d]} & \multicolumn{2}{c}{$\mathrm{v}$\tmark[d,]\tmark[e]} & \multicolumn{2}{c}{$\mathrm{v}_{corr}$\tmark[d,]\tmark[e]}                                     \NN
               & \multicolumn{1}{c}{$\textrm{(\AA)}$}       & \multicolumn{2}{c}{$\mathrm{(cm^{-2})}$}            & \multicolumn{2}{c}{$\mathrm{(km\ s^{-2})}$}    & \multicolumn{2}{c}{$\mathrm{(km\ s^{-2})}$} & \multicolumn{2}{c}{$\mathrm{(km\ s^{-2})}$}                                          \ML

\multicolumn{10}{c}{$\rm{Component\ A\ \textrm{(}\textit{v}\approx\textrm{-83 km s}^{-1}\textrm{)}}$}                                              \ML
$\rm{HI     }$ & $\rm{Ly\beta,\epsilon}$&     $14.45$ & $0.048$ & $40.4$ & $6.0$           & $-82.9$ & $4.0$           & $ -0.3$ & $0.0$$^{[1]}$   \NN
$\rm{NIV    }$ & $\rm{765            }$ &     $12.71$ & $0.129$ & $27.2$ & $9.8$$^{[1]}$   & $-62.9$ & $6.1$$^{[1]}$   & $ -0.3$ & $0.0$$^{[1]}$   \NN
$\rm{OIV    }$ & $\rm{787            }$ &     $13.64$ & $0.146$ & $27.2$ & $9.8$$^{[1]}$   & $-62.9$ & $6.1$$^{[1]}$   & $ -4.1$ & $0.0$           \NN
$\rm{OVI    }$ & $\rm{1031,1037      }$ &     $13.42$ & $0.201$ & $27.2$ & $9.8$$^{[1]}$   & $-62.9$ & $6.1$$^{[1]}$   & $ 12.3$ & $0.0$$^{[2]}$   \NN
$\rm{NeVIII }$ & $\rm{770\tmark[f]   }$ &    $<13.40$ & $0.230$ & \multicolumn{6}{c}{}                                                             \ML

\multicolumn{10}{c}{$\rm{Component\ B = B1\ \textrm{(}\textit{v}\approx\textrm{-3 km s}^{-1}\textrm{)} \cup B2\ \textrm{(}\textit{v}\approx\textrm{13 km s}^{-1} \textrm{)}}$} \ML
$\rm{HI     }$ & $\rm{Ly\beta,\epsilon,\iota,\mu-\pi}$& $16.52$ & $0.022$ & $15.3$ & $0.9$ & $  0.0$ & $2.7$           & $ -0.3$ & $0.0$           \NN
$\rm{CII    }$ & $\rm{903.6,903.9    }$ & $\geq13.75$ & $0.093$ & $ 9.2$ & $1.3$$^{[2]}$   & $ -2.9$ & $0.6$$^{[2]}$   & $  7.7$ & $1.2$$^{[3]}$   \NN
$\rm{CII    }$ & $\rm{903.6,903.9\tmark[g]}$&$<14.04$ & $0.106$ & $ 6.1$ & $0.0$$^{[2]}$   & \multicolumn{4}{c}{}                                  \NN
$\rm{NII    }$ & $\rm{915            }$ &     $13.38$ & $0.100$ & $ 9.2$ & $1.3$$^{[2]}$   & $ -2.9$ & $0.6$$^{[2]}$   & $  2.0$ & $3.0$$^{[4]}$   \NN
$\rm{NIII   }$ & $\rm{763            }$ &     $13.71$ & $0.092$ & $17.6$ & $2.4$$^{[3]}$   & $ 12.5$ & $1.9$$^{[3]}$   & $ -0.3$ & $0.0$$^{[1]}$   \NN
$\rm{NIV    }$ & $\rm{765            }$ &     $13.13$ & $0.077$ & $22.0$ & $3.2$$^{[4]}$   & $ 12.5$ & $1.9$$^{[3]}$   & $ -0.3$ & $0.0$$^{[1]}$   \NN
$\rm{OII    }$ & $\rm{832,833,834    }$ & $\geq14.10$ & $0.056$ & $ 9.2$ & $1.3$$^{[2]}$   & $ -2.9$ & $0.6$$^{[2]}$   & $ -2.1$ & $0.0$           \NN
$\rm{OII    }$ & $\rm{832,833,834\tmark[g]}$&$<14.32$ & $0.081$ & $ 6.1$ & $0.0$$^{[2]}$   & \multicolumn{4}{c}{}                                  \NN
$\rm{OIII   }$ & $\rm{702,832        }$ & $\geq14.66$ & $0.071$ & $17.6$ & $2.4$$^{[3]}$   & $ 12.5$ & $1.9$$^{[3]}$   & $ -2.1$ & $0.0$           \NN
$\rm{OIV    }$ & $\rm{787            }$ &     $14.61$ & $0.084$ & $22.0$ & $3.2$$^{[4]}$   & $ 12.5$ & $1.9$$^{[3]}$   & $ -1.1$ & $2.7$$^{[5]}$   \NN
$\rm{OVI    }$ & $\rm{1031,1037      }$ &     $14.19$ & $0.098$ & $31.4$ & $5.4$           & $ 12.5$ & $1.9$$^{[3]}$   & $ 12.3$ & $0.0$$^{[2]}$   \NN
$\rm{NeVIII }$ & $\rm{770\tmark[f]   }$ & $\leq13.42$ & $0.221$ & \multicolumn{6}{c}{}                                                             \NN
$\rm{MgI    }$ & $\rm{2852           }$ &     $11.01$ & $0.097$ & $ 4.0$ & $1.1$           & $ -2.6$ & $0.8$           & $  2.3$ & $0.7$$^{[6]}$   \NN
$\rm{MgII   }$ & $\rm{2796,2803      }$ &     $12.95$ & $0.019$ & $ 5.7$ & $0.1$           & $ -2.9$ & $0.6$$^{[2]}$   & $  2.3$ & $0.7$$^{[6]}$   \NN
$\rm{MgII   }$ & $\rm{2796,2803      }$ &     $11.77$ & $0.121$ & $13.0$ & $4.0$           & $ 12.5$ & $1.9$$^{[3]}$   & $  2.3$ & $0.7$$^{[6]}$   \NN
$\rm{AlIII  }$ & $\rm{1854,1862      }$ &     $12.42$ & $0.071$ & $ 8.9$ & $1.7$           & $ -2.9$ & $0.6$$^{[2]}$   & $  2.3$ & $0.7$$^{[6]}$   \NN
$\rm{SII    }$ & $\rm{763,765        }$ &     $12.71$ & $0.084$ & $ 5.1$ & $0.2$$^{[5]}$   & $ -2.9$ & $0.6$$^{[2]}$   & $ -0.3$ & $0.0$$^{[1]}$   \NN
$\rm{SIII   }$ & $\rm{724            }$ &     $13.38$ & $0.063$ & $18.9$ & $5.1$$^{[6]}$   & $ 12.5$ & $1.9$$^{[3]}$   & $ -6.4$ & $3.4$$^{[7]}$   \NN
$\rm{SIV    }$ & $\rm{748            }$ &     $12.78$ & $0.187$ & $18.9$ & $5.1$$^{[6]}$   & $ 12.5$ & $1.9$$^{[3]}$   & $  0.9$ & $2.5$$^{[8]}$   \NN
$\rm{SV     }$ & $\rm{786            }$ &     $12.44$ & $0.156$ & $22.0$ & $3.2$$^{[4]}$   & $ 12.5$ & $1.9$$^{[3]}$   & $ -1.1$ & $2.7$$^{[5]}$   \NN
$\rm{SVI    }$ & $\rm{944\tmark[f]   }$ &    $<13.00$ & $0.050$ & \multicolumn{6}{c}{}                                                             \NN
$\rm{FeII   }$ & $\rm{2382,2586,2600 }$ &     $12.72$ & $0.014$ & $ 5.1$ & $0.2$$^{[5]}$   & $ -2.9$ & $0.6$$^{[2]}$   & $  1.3$ & $0.7$$^{[9]}$   \NN
$\rm{FeII   }$ & $\rm{2382,2586,2600 }$ & $\leq11.52$ & $0.137$ & $ 7.8$ & $0.0$           & $ 12.5$ & $1.9$$^{[3]}$   & $  1.3$ & $0.7$$^{[9]}$   \ML

\multicolumn{10}{c}{$\rm{Component\ C = C1\ \textrm{(}\textit{v}\approx\textrm{-39 km s}^{-1}\textrm{)} \cup C2\ \textrm{(}\textit{v}\approx\textrm{56 km s}^{-1} \textrm{)}}$} \ML
$\rm{HI     }$ & $\rm{Ly\beta,\epsilon,\iota,\mu-\pi}$& $15.90$ & $0.057$ & $33.5$ & $3.0$ & $ 46.4$ & $5.5$           & $  5.0$ & $0.0$           \NN
$\rm{CII    }$ & $\rm{903.6,903.9    }$ &     $13.24$ & $0.080$ & $19.0$ & $0.0$$^{[7]}$   & $ 39.4$ & $0.8$$^{[4]}$   & $ 13.0$ & $1.2$$^{[3]}$   \NN
$\rm{NIII   }$ & $\rm{763            }$ &     $13.77$ & $0.076$ & $19.0$ & $2.0$$^{[8]}$   & $ 55.8$ & $1.8$$^{[5]}$   & $  5.0$ & $0.0$$^{[1]}$   \NN
$\rm{NIV    }$ & $\rm{765            }$ &     $13.80$ & $0.057$ & $19.0$ & $2.0$$^{[8]}$   & $ 55.8$ & $1.8$$^{[5]}$   & $  5.0$ & $0.0$$^{[1]}$   \NN
$\rm{OII    }$ & $\rm{832,833,834    }$ &     $13.78$ & $0.047$ & $19.0$ & $0.0$$^{[7]}$   & $ 39.4$ & $0.8$$^{[4]}$   & $  3.2$ & $0.0$           \NN
$\rm{OIII   }$ & $\rm{702,832        }$ & $\geq15.02$ & $0.126$ & $19.0$ & $2.0$$^{[8]}$   & $ 55.8$ & $1.8$$^{[5]}$   & $  3.2$ & $0.0$           \NN
$\rm{OIV    }$ & $\rm{787            }$ & $\geq15.26$ & $0.219$ & $19.0$ & $2.0$$^{[8]}$   & $ 55.8$ & $1.8$$^{[5]}$   & $  4.2$ & $2.7$$^{[5]}$   \NN
\NN
$\rm{OVI    }$ & $\rm{1031,1037      }$ &     $14.68$ & $0.053$ & $25.8$ & $2.8$           & $ 55.8$ & $1.8$$^{[5]}$   & $ 17.6$ & $0.0$$^{[2]}$   \NN
$\rm{NeVIII }$ & $\rm{770            }$ &     $14.19$ & $0.062$ & $78.8$ & $14.1$          & $ 55.8$ & $1.8$$^{[5]}$   & $  5.0$ & $0.0$$^{[1]}$   \NN
$\rm{MgII   }$ & $\rm{2796,2803      }$ &     $11.79$ & $0.046$ & $ 5.9$ & $0.8$           & $ 39.4$ & $0.8$$^{[4]}$   & $  7.6$ & $0.7$$^{[6]}$   \NN
$\rm{AlIII  }$ & $\rm{1854           }$ & \multicolumn{2}{c}{$<12.34$} & \multicolumn{6}{c}{}                                                      \NN
$\rm{SII    }$ & $\rm{765\tmark[f]   }$ &    $<13.42$ & $0.225$ & \multicolumn{6}{c}{}                                                             \NN
$\rm{SIII   }$ & $\rm{724            }$ &     $12.96$ & $0.136$ & $ 8.4$ & $3.3$$^{[9]}$   & $ 55.8$ & $1.8$$^{[5]}$   & $ -1.1$ & $3.4$$^{[7]}$   \NN
$\rm{SIV    }$ & $\rm{748            }$ &     $13.27$ & $0.074$ & $ 8.4$ & $3.3$$^{[9]}$   & $ 55.8$ & $1.8$$^{[5]}$   & $  6.2$ & $2.5$$^{[8]}$   \NN
$\rm{SV     }$ & $\rm{786            }$ &     $13.03$ & $0.056$ & $13.2$ & $2.9$$^{[10]}$  & $ 55.8$ & $1.8$$^{[5]}$   & $  4.2$ & $2.7$$^{[5]}$   \NN
$\rm{SVI    }$ & $\rm{933,944        }$ &     $12.76$ & $0.340$ & $13.2$ & $2.9$$^{[10]}$  & $ 55.8$ & $1.8$$^{[5]}$   & $ 16.9$ & $6.9$$^{[10]}$  \NN
$\rm{FeII   }$ & $\rm{2382\tmark[f]  }$ & \multicolumn{2}{c}{$<12.01$} & \multicolumn{6}{c}{}                                                      \ML

\multicolumn{10}{c}{$\rm{Components\ B + C}$}                                                                                                      \ML
$\rm{CIII   }$ & $\rm{977\tmark[f]   }$ & $\geq14.11$ & $0.014$ & \multicolumn{6}{c}{}                                                             \NN
$\rm{OI     }$ & $\rm{877\tmark[f]   }$ &    $<13.42$ & $0.221$ & \multicolumn{6}{c}{}                                                             \NN
$\rm{OIV    }$ & $\rm{787\tmark[f]   }$ & $\geq15.05$ & $0.006$ & \multicolumn{6}{c}{}                                                             \NN
$\rm{OVI    }$ & $\rm{1037\tmark[f]  }$ & $\geq14.75$ & $0.023$ & \multicolumn{6}{c}{}                                                             \NN
$\rm{SVI    }$ & $\rm{944\tmark[f]   }$ & $\leq13.16$ & $0.060$ & \multicolumn{6}{c}{}                                                             \LL
}

\subsection{Notes for Individual Absorption Systems}
\label{sec:absorber_notes}

\subsubsection{\emph{z}=0.41760, PG1630+377 (Absorbers 0.4A and 0.4B)}
\label{sec:v041760}

As shown in Figure~\ref{fig:z041761_vpfits}, in this absorption system we identify four components (Components A--D in Table~\ref{tab:z041761_meas}), including two line-rich components with relatively high \textsc{H~i} column densities at \emph{v}$=0$ and 43 km s$^{-1}$ as well as two weaker components with lower \textit{N}\textsubscript{H I} values at \emph{v}$ = 110$ and 197 km s$^{-1}$ (see Table~\ref{tab:z041761_meas}). Interestingly, the three components with low/mid-ion absorption (i.e., all except the one at \emph{v} $\approx$ 110 km s\textsuperscript{-1}) exhibit a gradient in ionic column densities and in velocity: the degree of ionization in each velocity component appears to be correlated with increasing velocity.  The strongest \textsc{H~i}, C~\textsc{ii}, O~\textsc{ii}, and Mg~\textsc{ii} absorption appears in the lowest velocity component, while C~\textsc{iii}, C~\textsc{iv}, O~\textsc{iii}, and O~\textsc{vi} are stronger in the higher-velocity features.  This ionization gradient is most evident in the \textsc{H~i} lines; the \textsc{H~i} column densities clearly decrease with increasing velocity.  This could simply be a consequence of less gas in the higher-velocity clouds, but the fact that the column densities of more highly ionized metals do not decrease as well suggests that this owes to the gas becoming increasingly more ionized with increasing velocity.

It is also interesting to note that Component C at \emph{v} = 110 km s$^{-1}$ is only detected in \textsc{O~vi} and \textsc{H~i}.  In their blind STIS survey, \cite{Tripp08} report many absorbers only detected in \textsc{O~vi} and \textsc{H~i} (no \textsc{C~iii}), and similarly \cite{Werk16} found a category of absorbers (their ``No-low'' sample) with a similar absence of low-ionization stages, yet having \textsc{O~vi} aligned with moderate \textrm{N}(\textsc{H~i}).  We measure a narrow linewidth for \textsc{O~vi} in Component C, and our best fit to the \textsc{H~i} feature at this velocity returns a doppler broadening parameter $b$(\textsc{H~i}) $\approx \ 4\times \ b$(\textsc{O~vi}).  Both measurements are consistent with the \textsc{O~vi} and \textsc{H~i} arising in a single, thermally-broadened gas phase.  While the best-fit line widths for this component imply a temperature slightly below the $T$ range where \textsc{O~vi} is expected to peak in collisional ionization equilibrium, the line width uncertainties are substantial, and detailed analysis of this feature would require a higher spectral resolution and/or better S/N.

In Component D at \emph{v} $\approx$ 200 km s\textsuperscript{-1}, there is enough degeneracy in the fit to the spectrum to allow for some weak O~\textsc{vi} absorption, but this component is also highly uncertain owing to its location in a complex blend, and we do not address it further in this work.

\subsubsection{\emph{z}=0.51850, PG1522+101 (Absorber 0.5A)}
\label{sec:z051850}
	In this system, plotted in Figure~\ref{fig:z051850_vpfits} and summarized in Table~\ref{tab:z051850_meas}, we mainly detect singly-ionized species, along with C~\textsc{iii}, O~\textsc{iii}, and possibly Si~\textsc{iii}, with no clear evidence for a hot phase.  Indeed, we selected this system for this paper to test how our code would work for absorbers with no affiliated highly ionized species; such absorbers are not uncommon.

	The two species detected in the higher resolution HIRES spectra provide interesting insights into this absorber.  The Mg~\textsc{ii} doublet requires a weaker and broader component at $\Delta$\emph{v} $\approx$ +6 km s\textsuperscript{-1} in addition to the stronger and narrower component at $\approx$ -5 km s\textsuperscript{-1} (Figure~\ref{fig:z051850_vpfits}).  The same component structure is evident in the Fe~\textsc{ii} absorption profile as well, though it does not appear in the remaining detected species, presumably owing to the lower spectral resolution at which they were observed (i.e., with COS).  While a single-component VP fit adequately matches the absorption profiles of the COS detections, a two-component fit encompasses a wider range of plausible solutions.  For example, a two-component fit for C~\textsc{iii}, where both component velocities are co-fitted with the higher resolution Mg~\textsc{ii} lines, includes a narrower, slightly saturated C~\textsc{iii} line that is kinematically aligned with the narrower Mg~\textsc{ii} component.

	Although we report two-component VP parameters in Table~\ref{tab:z051850_meas} for several detected species (see also Figure~\ref{fig:z051850_vpfits}), we only list one component for H~\textsc{i}, N~\textsc{iii}, and O~\textsc{iii}.  In the case of O~\textsc{iii}, a two-component fit would not converge; the column density of the lower-velocity component is pushed to a negligibly low value, effectively leaving a single component fit.  The weak N~\textsc{iii} absorption feature is too noisy and crowded with intervening lines from other redshifts to reasonably constrain a multiple component fit.  Finally, we fit a single component to the H~\textsc{i} Lyman series lines to get a tight measurement of the total \textit{N}\textsubscript{H I}.

	In a later section, we reproduce the measured column densities of the detected species through various ionization models.  We find that the uncertainties of the forced two-component fits to the COS detections are too high to meaningfully constrain a two-component or two-phase ionization model, and that a single-component, single-phase ionization model suffices to explain the total measured column densities.  To be clear, we derive the total column density by summing the measured column densities of each component VP and summing their uncertainties in quadrature.

\subsubsection{\emph{z}=0.68606, PG1338+416 (Absorbers 0.6A and 0.6B)}
\label{sec:v068606}
	This system (Figure~\ref{fig:z068605_vpfits} and Table~\ref{tab:z068605_meas}) is a Ne~\textsc{viii} absorber that exhibits strong absorption in all ionization stages (low though high-ions).  As such, it represents an excellent test for the multiphase modeling discussed below (Section~\ref{sec:results}).  In this system we firmly identify three components at \emph{v} $\approx$ -83, -3, and 39 km s$^{-1}$. There is room for a fourth component at $\approx$150 km s$^{-1}$ in the \textsc{N~iv} and \textsc{O~iv} profiles, though both are also contaminated by blending from other redshifts and we are unable to search for corroborating \textsc{H~i} absorption at this velocity owing to a lack of adequate\footnote{The \textsc{H~i} in this candidate component is likely to be quite weak, so we would need to use the Ly$\alpha$ transition to search for \textsc{H~i} at this velocity. At this $z$, Ly$\alpha$ is redshifted into the NUV.  We do not have NUV COS or STIS spectra of PG1338+416, and the resolution of archival NUV spectra captured by HST's Faint Object Spectrograph is too low to reliably de-blend and identify weak \textsc{H~i} lines.} coverage in the NUV for this particular sightline.  We do not include any VP-fitting results for this potential fourth component in Table~\ref{tab:z068605_meas}, as we focus our analysis on the three well-detected components.

	As with Absorber~0.5A, the higher resolution Mg~\textsc{ii} and Fe~\textsc{ii} spectral profiles reveal sub-components unresolved in the COS spectra.  Component B splits into two subcomponents, separated by $\approx$15 km s$^{-1}$.  Additionally, a comparison in Component C of the HIRES vs. COS spectra suggests the presence of two subcomponents. The low-ions dominate subcomponents B1 and C1, while the mid- and high-ions seem to be mostly concentrated in B2 and C2. It is less likely that these misalignments between low and mid/high-ions are artifacts of COS wavelength calibration issues since we detect these velocity offsets in several pairs of low-ion lines adjacent in the spectrum to observed mid/high-ion lines.  These include O~\textsc{ii} ($\rm{\lambda\lambda}$832, 833, and 834 $\textrm{\AA}$) vs. O~\textsc{iii} ($\rm{\lambda}$832 $\textrm{\AA}$), S~\textsc{ii} ($\rm{\lambda}$765 $\textrm{\AA}$) vs. N~\textsc{iv} ($\rm{\lambda}$765 $\textrm{\AA}$), and H~\textsc{i}\footnote{The small offset between H~\textsc{i} and the low-ions is probed by N~\textsc{ii} ($\rm{\lambda}$915 $\textrm{\AA}$) vs. H~\textsc{i} ($\rm{\lambda}$915.3 and 916 $\textrm{\AA}$).} (Ly$\beta$ $\rm{\lambda}$1025 $\textrm{\AA}$) vs. O~\textsc{vi} ($\rm{\lambda}$1031 $\textrm{\AA}$).  A similar offset is seen between N~\textsc{iv} ($\rm{\lambda}$765 $\textrm{\AA}$) and Ne~\textsc{viii} ($\rm{\lambda}$770 $\textrm{\AA}$), though these ions are not as close in wavelength as the pairs previously mentioned.  As is also the case for Absorber~0.5A, we do not have sufficient constraints from species detected in the higher resolution HIRES spectrum to explore ionization models for each of these subcomponents separately.  Instead, we use the multiphase models developed below to infer the fraction of each species in a particular phase.

	While we focus our ionization modeling efforts in this work on the two main components, there is also a substantial apparent O~\textsc{vi} absorption feature at \emph{v} $\approx$ -63 km s\textsuperscript{-1} that appears to coincide with possible H~\textsc{i}, N~\textsc{iv}, O~\textsc{iv}, and Ne~\textsc{viii} features, although the velocity alignment is not perfect between these species (possibly owing to hidden substructure as evidenced above).  Independent fits of N~\textsc{iv} prefer a significantly narrower feature than O~\textsc{vi} and a velocity offset of $\approx$ +16 km s\textsuperscript{-1}.  O~\textsc{iv} is severely contaminated, but the contaminating lines are positioned such that there is some inexplicable optical depth in the spectrum if O~\textsc{iv} is removed.  Hence, we can obtain reasonable measurements for the O~\textsc{iv} column density by co-fitting its broadening parameter and velocity parameter with N~\textsc{iv} and O~\textsc{vi}.


\section{THE IONIZATION MODELING SCHEME}
\label{sec:scheme}
As discussed in \S~\ref{sec:modeling_challenges}, physical conditions in QSO absorption systems are traditionally inferred from simple (often single-phase) photoionization models that best fit the measured ionic column densities. Motivated by the multiphase kinematic alignment of detected species discussed below, we leverage the CASBaH spectra to explore multiphase models that expand on basic, two/three parameter models in several ways: 1) by parameterizing individually resolved absorption features (i.e., velocity components) as distinct gas clouds, which we refer to henceforth as separate \textit{absorbers} even if they share the same systemic redshift (labels assigned to each are listed in Table~\ref{tab:abs_sys_info}), 2) by allowing for multiple discrete \textit{phases} within each absorber, 3) through the addition of physically motivated parameters for the shape of the ultraviolet (UV) ionizing radiation field and gas-phase metal abundances, and 4) by applying Bayesian models to make probabilistic parametric inferences that incorporate prior knowledge.  We use Markov Chain Monte Carlo (MCMC) sampling to robustly characterize the posterior distributions of the most probable models.

\subsection{Kinematic Alignment of Disparate Ionization Stages}
\label{sec:kinematic_alignment}
All but one (Absorber~0.5A) of the absorption systems examined here exhibit detections of high-ions such as C~\textsc{iv}, S~\textsc{v}, O~\textsc{vi}, and Ne~\textsc{viii} alongside low-ions such as Mg~\textsc{ii}, Fe~\textsc{ii}, and C~\textsc{ii} (as well as many intermediate ionization stages, see Section~\ref{sec:data}).  As summarized graphically in Figure~\ref{fig:ion_potentials}, the energies required to create and destroy these species through ionization span a tremendous range extending from 7.6 to 239.1 eV. Even without any ionization modeling, this wide range of ionization stages on its own implies the presence of a multiphase entity, since no single gas temperature/density state can produce appreciable abundances of low, intermediate, and high ionization stages simultaneously.  However, there is a puzzling aspect of these absorption systems. As we and others have shown in several previous studies \citep[e.g.,][]{Tripp00,Tripp06,Tripp08,Tripp11,Muzahid12,Meiring13,Savage14,Burchett15,Crighton15,Werk16,Rudie19} these disparate ionization stages are often remarkably well aligned in velocity space.  Their column density ratios seem to vary from component to component, but the velocity centroids and line widths of the various ions are quite similar (typically well within the measurement uncertainties).  This kinematic alignment is not naturally expected in a simplistic picture of the CGM.  For example, if the \textsc{O~vi} arises in a large, volume-filling phase in the halo of a galaxy while the Mg~\textsc{ii} originates in tiny clouds distributed throughout the disk and halo of that galaxy, one might expect normal galaxy kinematics to cause these ions to have significant (and easily measured) differences in their velocity centroids and absorption-profile shapes.   

\begin{figure}
\includegraphics[width=8cm]{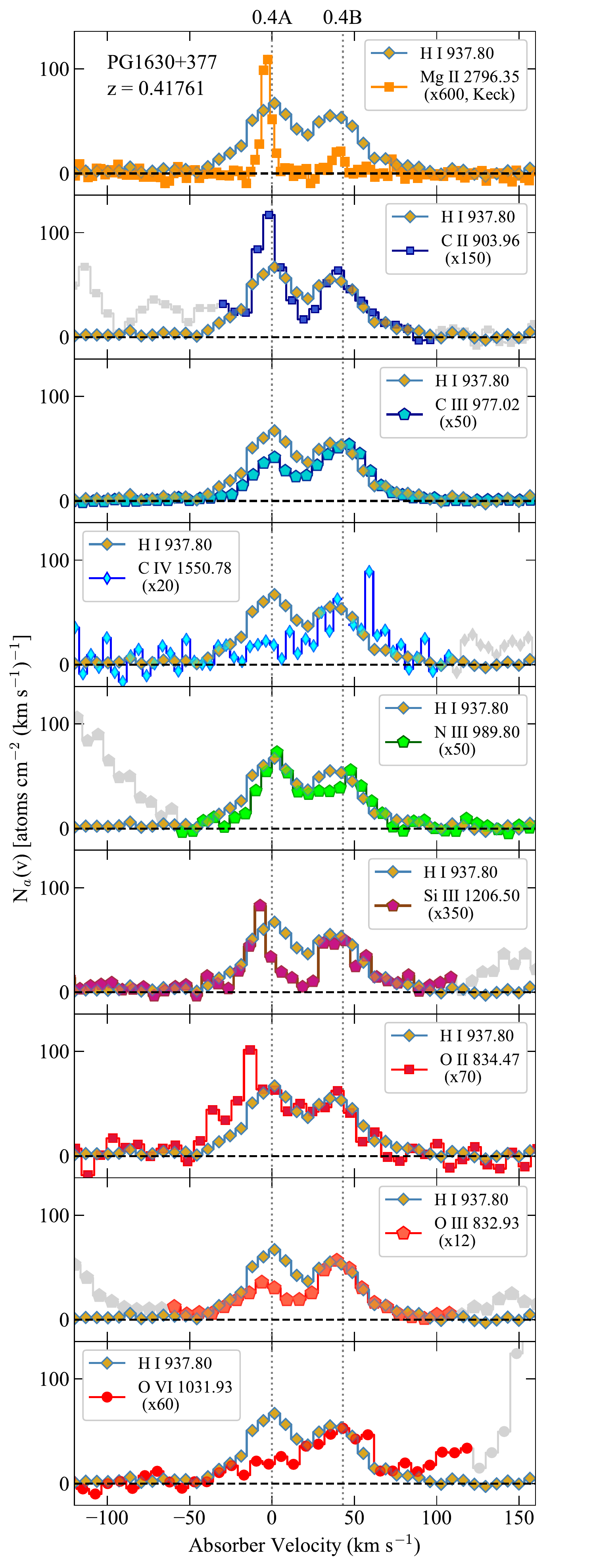}
\caption{Comparison of apparent column density profiles [$N_{\rm a}(\emph{v})$, see text] of several ions detected in the PG1630+377 absorption system at $z_{\rm abs} = 0.41760$. The species that are compared are indicated in the legend of each panel.  In all panels, the \textsc{H~i} 930.804 \AA\ line (gold diamonds) is compared to a metal ion that is scaled up by the factor listed in the legend, and regions contaminated by interloping lines from other redshifts are plotted in gray. Vertical dotted lines mark the velocities of Absorbers 0.4A and 0.4B (see Table~\ref{tab:abs_sys_info}).}
\label{v_align1630}
\end{figure}

\begin{figure}
\includegraphics[width=8cm]{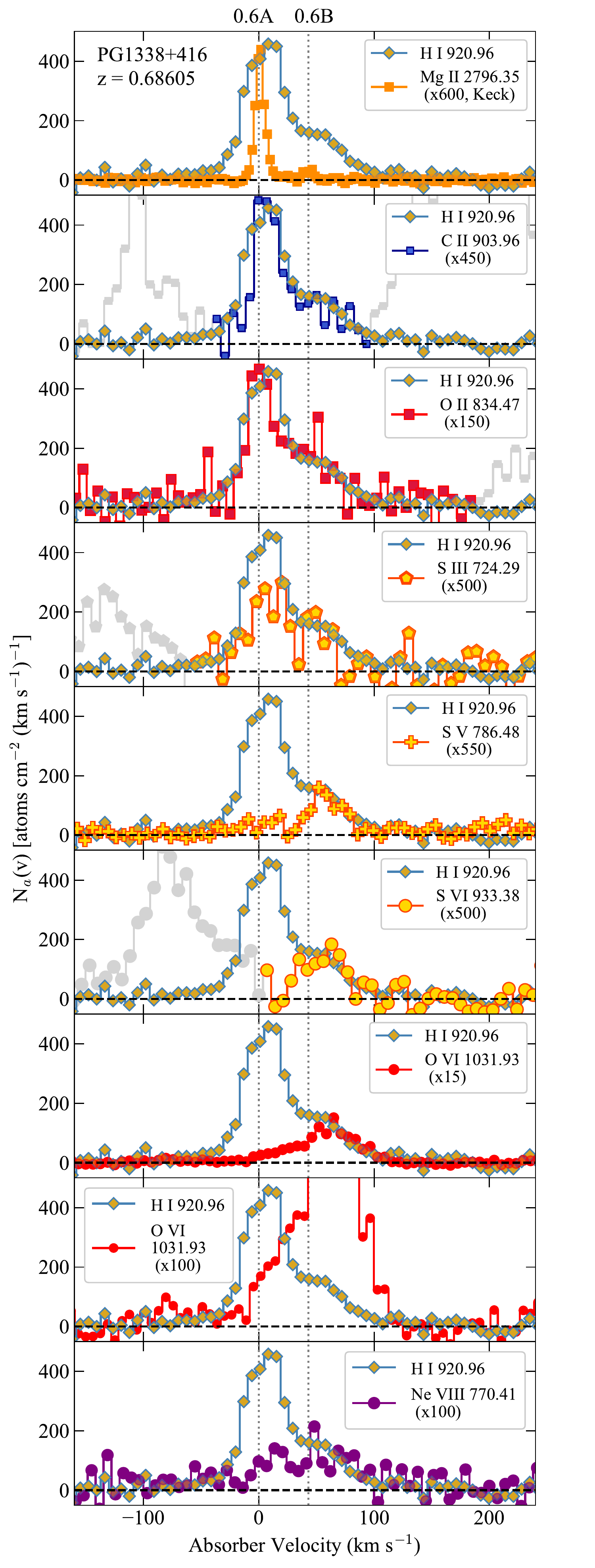}
\caption{Comparison of \textsc{H~i} vs. metal $N_{\rm a}(\emph{v})$ profiles, as in Figure~\ref{v_align1630}, for the absorption system at $z_{\rm abs} = 0.68606$ in the spectrum of PG1338+413.  In this case, the metal profiles are compared to the \textsc{H~i} 920.963 \AA\ line in each panel.  Dotted lines indicate the velocities of Absorbers 0.6A and 0.6B (Table~\ref{tab:abs_sys_info}).}
\label{v_align1338}
\end{figure}

Instead, the profiles of low-, intermediate-, and high-ionization stages often have remarkably similar profiles.  In some instances, one can recognize small differences in centroids and/or line widths when comparing diverse ions, but overall, a wide range of ions often exhibit very similar and distinctive kinematics, which suggests that there is a direct relationship between the various ionization stages. To demonstrate this, Figures~\ref{v_align1630} and \ref{v_align1338} show examples of the kinematic velocity alignment of the wide range of ions detected in absorbers 0.4A, 0.4B, 0.6A, and 0.6B.  To compare the various ion profiles in these figures, we have constructed apparent column density profiles \citep[e.g.,][]{Savage91,Jenkins96}, which provide linear presentations of the absorption.\footnote{The ``raw'' absorption profiles are exponential attenuations of the QSO light and thus are more difficult to directly scale for comparison among different elements and ions, which can have vastly different abundances and thus require some scaling to overplot their profiles.} Briefly, in this method the apparent optical depth in each pixel as a function of velocity, $\tau_{\rm a}(\emph{v})$, is used to calculate the apparent column density per unit velocity, $N_{\rm a}(\emph{v}) = (m_{\rm e}c/\pi e^{2})(f\lambda)^{-1}\tau_{\rm a}(\emph{v})$, where $f$ is the oscillator strength, $\lambda$ is the rest wavelength of the transition, and the other symbols have their usual meanings.  While $N_{\rm a}(\emph{v})$ profiles can be used in various ways  \citep{Savage91,Jenkins96}, our main application in this paper is to use them to directly overlay the profiles of various ions, as we have done in Figures~\ref{v_align1630} and \ref{v_align1338}.

Beginning with the PG1630+377 system in Figure~\ref{v_align1630}, we see that in both Absorbers 0.4A and 0.4B, while the absorption profiles are not identical, there is a striking correspondence of the detected ions.  There is an obvious difference between Mg~\textsc{ii} and the other lines, primarily on account of the higher spectral resolution provided by the Keck HIRES spectrograph that recorded the Mg~\textsc{ii} data, yet it is still useful to compare the Mg~\textsc{ii} centroids to those of the other ions (all other species in Fig.~\ref{v_align1630} were recorded with COS and thus have the same resolution).  Especially in Absorber 0.4B, there is a profound similarity of the line shapes and velocity structure (i.e., the velocity offsets between the centroids of the two dominant components) of species with ionization potentials ranging from 13.6 eV (\textsc{H~i}) to 64.5 eV (\textsc{C~iv}) to 138.1 eV (\textsc{O~vi}).  This begs a question: how can the similarity of the profiles be reconciled with the fact that the physical conditions that will maximize one set of ions should lead to undetectable amounts of other sets of ions at the same velocity in this system?  Moreover, some of the aligned species have significantly different atomic masses, so their lines should have detectably different widths if they arise in the same gas and are thermally broadened \citep[see section 4.1 in][]{Tripp08}. Similar line widths of, e.g., H and O, implies that the lines are \textsl{not} predominantly thermally broadened, which in turn indicates that the gas is relatively cool \citep{Tripp08}.  Could the gas be photoionized?  This is an important question for a variety of reasons. The \textsc{O~vi} ion, which is relatively easy to detect and is often assumed to be collisionally ionized, would have very different implications if it is produced by photoionization.

In Absorber 0.4A we find more indications of complexity.  First, the peaks of several metals appear to have a small (2$-$10 km s$^{-1}$) offset to the blue compared to the \textsc{H~i} peak, and there is marginal evidence that the low-ionization metal peaks are offset by the same amount compared to the intermediate- and high-ionization metals.  This could reflect unresolved component structure \textsl{within} 0.4A, with a lower ionization component on the blue side of 0.4A, a higher ionization component on the red side, and \textsc{H~i} from both components that smears together to give the appearance of a single \textsc{H~i} feature.  Alternatively, these subtle misalignments could be coincidental effects of the aforementioned COS wavelength distortions (Section~\ref{sec:measurements}).  To sort this out would require higher resolution and higher S/N.  Such spectra will not be available in the near future, so we must find a way to interpret the data in hand.  Second, the intermediate- and high-ionizations stages are weaker, relative to the low-ionization stages, in 0.4A.  Nevertheless, the more highly ionized gas is still kinematically aligned with the lower ionization gas, albeit with different $N$(low)/$N$(high) patterns.  That is, there is some type of physical relationship between the low- and high-ionization materials; these are highly unlikely to be random patches of low- and high-ionization gases that have no connection whatsoever.  This is reminiscent of the behavior seen in other complex but kinematically aligned QSO absorption systems \citep[see, e.g., Fig.~3 in][]{Tripp11}.

Absorbers 0.6A and 0.6B in the PG1338+416 system (Figure~\ref{v_align1338}) exhibit very similar patterns.  However, this system is arguably even more interesting because its higher redshift provides access to a greater number of more highly ionized species.  Indeed, we detect \textsc{S~v}, \textsc{S~vi}, and Ne~\textsc{viii} in addition to \textsc{O~vi}, and Figure~\ref{v_align1338} shows that the shapes of the $N_{\rm a}(\emph{v})$ profiles of these high ions are remarkably similar to the shapes of the \textsc{H~i} and lower-ionization metals in 0.6B.  Like the PG1630+377 system, the lower-ionization absorption is relatively stronger in 0.6A than in 0.6B, but the high-ionization species are still present in 0.6A (see also Figure~\ref{fig:z068605_vpfits}).  Again, there are some hints that the component structure may be more complex than just two components, but better information on the component structure must await a future space telescope with greater sensitivity and spectral resolution.
 
The kinematic alignment of different ionization phases makes it impossible to directly determine the fraction of a given ion's absorption in each gas phase \emph{from the spectrum alone}.  While it is often assumed that the low-ion columns originate entirely in the lowest ionization gas phase (i.e., the other, more ionized phases contribute only negligible additional amounts to the low-ion column densities), in fact the ionization potentials of ``low ions'' are often well above 1 Rydberg (see Figure~\ref{fig:ion_potentials}), and this assumption may be spurious unless it can be verified by multiphase ionization modeling that accurately models the full set of observed ions simultaneously.  Indeed, this concern strongly motivates our detailed attempts to formulate a model that consistently accounts for \emph{all} the detected absorption lines.

	In many cases though, it is difficult or even impossible to identify individual velocity components in systems with densely packed component structures, particularly in spectra of moderate signal-to-noise and/or moderate spectral resolution (such as is afforded by the COS G130M and G160M gratings).  Consequently, it is common practice to ignore inter-component variances altogether by integrating all absorption from a given species near a common systemic redshift into a single representative column density.  This practice implicitly assumes that the absorbing medium is homogeneous in composition and physical conditions.  To avoid making this assumption, we require an ionization model (along with sufficient observational constraints) that can account for the component-to-component variances in ion column density ratios.

\subsection{Implementation of Multiple Phases and Components}
\label{sec:implementation}
We designed our ionization modeling scheme to flexibly fit an arbitrary combination of velocity components and ionization phases simultaneously.  The ionic column densities associated with distinct velocity components are measured through absorption line fitting (see Section~\ref{sec:measurements}).  Each measured column density is then resolved probabilistically into one or more phases using a Bayesian multiphase ionization model.  For example, suppose O~\textsc{iii}, O~\textsc{iv}, O~\textsc{v}, and O~\textsc{vi} are all detected at \textit{v} = 0 km s$^{-1}$ in a hypothetical absorption system.  We could assume, for example, that this \textit{v} = 0 km s$^{-1}$ \textit{absorber} consists of two discrete \textit{phases}, including one that produces O~\textsc{iii} and O~\textsc{iv} absorption, and a second (more highly ionized) phase that produces O~\textsc{v} and O~\textsc{vi} absorption.  Unfortunately, this is not a sound assumption because the column density for some species (e.g., O~\textsc{iv}) might include appreciable contributions from each phase, or there might be more than two phases present.  Ostensibly, it is difficult to make correct assumptions about what fraction of a given ion's column density arises in each of the phases, but we can use Bayes' Thereom coupled with a Monte Carlo Markov Chain (MCMC) method to explore the range of physical conditions in each component that are consistent with the data, and we can use metrics such as Bayes factors to judge the relative probability of, e.g., a 2-phase vs. a 3-phase model.

For each absorber in this study (Table~\ref{tab:abs_sys_info}), we compute models with one, two, and three ionization phases and use their Bayes factors to select the most probable model.  While each phase can, in principle, be modeled with a unique set of parameters, it can be advantageous and logical to reduce the dimensionality of the parameter space significantly by assuming that the metallicity and gas composition parameters (e.g., C/$\alpha$, N/$\alpha$, and $\alpha$/Fe) remain uniform across all the phases within a given absorber.  The kinematic alignment of low- and high-ions discussed above lends credence to this idea that multiphase ions arise in gas phases that have some type of physical relationship and are likely to have similar abundances.  We can also assume that the various phases in an (optically thin) absorption system are exposed to the same ionizing radiation field (i.e., in a given system, the optically thin gas is permeated and photoionized by the same UV background and/or the same light escaping from nearby galaxies).  We also explore the effect of parameterizing the UV ionizing radiation field at the level of individual absorbers (one set of UV parameters per \textit{absorber}) \textit{versus} at the systemic level (one set of shared UVB parameters for all components in a given \textit{absorption system}).  After sharing various parameters across phases in this manner, only the gas density, temperature, and H~\textsc{i} column density\footnote{The \textsl{total} H~\textsc{i} column density is well-constrained in these systems, but the way the H I is distributed among the kinematically-aligned phases is unknown and is allowed to vary in our models.} remain as parameters uniquely specified at the level of individual phases.

\subsection{Bayesian Modeling}
\label{sec:formalism}

\subsubsection{Parameteric Inference via Bayes Theorem}
Following similar studies of QSO absorbers \citep{Crighton15,Fumagalli16,Prochaska17}, this work adopts a Bayesian modeling perspective, which offers two key advantages over traditional $\chi^{2}$ fitting techniques: 1) a probabilistic inference of the model parameters, and 2) a natural framework for incorporating prior knowledge.  In the Bayesian paradigm, \textit{p}($\bmath{\theta}$|\textit{\textbf{N},M}) is the probability of the model parameters $\bmath{\theta}$ given the data (here \textit{\textbf{N}} is used to denote column densities) and a particular model \textit{M}.  Bayes Theorem states that \textit{p}($\bmath{\theta}$|\textit{\textbf{N},M}) is proportional to the product of the likelihood distribution \textit{L}(\textit{\textbf{N}}|$\bmath{\theta}$\textit{,M}) and the prior probability of the parameters \textit{p}($\bmath{\theta}$|\textit{M}):
\begin{equation}\label{eqn:bayes}
\textit{p}(\bmath{\theta}|\textit{\textbf{N},M}) = \frac{\textit{L}(\textit{\textbf{N}}|\bmath{\theta},\textit{M}) \textit{p}(\bmath{\theta}|\textit{M})}{\textit{p}(\textit{\textbf{N}}|\textit{M})}.
\end{equation}
where the prior probability of the data \textit{p}(\textit{\textbf{N}}|\textit{M}), also known as the marginal likelihood, is a normalizing constant that depends on the particular model \textit{M}.  In the context of our ionization modeling application, a model $\textit{M} \equiv \textit{M}(\textit{n}_{\textrm{phases}}, \textit{I}_{\textrm{mech}}^{1},...,\textit{I}_{\textrm{mech}}^{\textit{n}_{\textrm{phases}}})$ is specified by the number of model phases $\textit{n}_{\textrm{phases}}$, the ionization mechanism $\textit{I}_{\textrm{mech}}^\textit{i}$ used for the $i$th phase, and the order in which the phases are assigned.  The order of the phases is only important if the multiphase model relies on multiple types of ionization mechanisms, since it affects various ``ranking priors'' that we introduce below (i.e., for the density and temperature parameters).  We condense this information into a compact notation by using the letters ``P'' and ``C'' to refer to photoionization equilibrium (PIE) only and a combination of photoionization \textsl{and} collisional ionization mechanisms (defined below), respectively, and by combining letters so that each phase is represented by a single letter and their order reflects their use in the ionization model (lowest ionization stage first).  Thus, a ``PPC'' model has three phases and only the third (most highly ionized) phase incorporates P+C ionization.  Sometimes, column density measurements that are not kinematically resolved across individual absorbers will necessitate joint modeling of two (or more) absorbers. For example, in the PG1338+416 system,  the  \textsc{C~iii} components from Absorbers~0.6A \&~0.6B are strongly blended and cannot be disentangled (see Fig.~\ref{fig:z068605_vpfits}), but we can still use the \textsc{C~iii} information by jointly modeling the two absorbers.  We refer to absorbers linked in this manner as \textit{component groups}.  We can extend this notation to account for this scenario. For example, ``PPCa-PCb,'' indicates that Absorber~0.6A has a ``PPC'' model and Absorber~0.6B takes a ``PC'' model, specified by the joint posterior \textit{p}($\bmath{\theta}_\textbf{PPC}$,$\bmath{\theta}_\textbf{PC}$|\textit{\textbf{N}}\textsubscript{\textbf{0.6A}},\textit{\textbf{N}}\textsubscript{\textbf{0.6B}},\textit{\textbf{N}}\textsubscript{\textbf{0.6AB}},\textit{M}\textsubscript{PPC},\textit{M}\textsubscript{PC}).

	The parameters inferred using Bayes Theorem are not point estimates, but fully specified probability densities.  One derives the marginal posterior density of individual parameters $\theta_{i,p}$ from the full posterior \textit{p}($\bmath{\theta}$|\textit{\textbf{N},M}) by integrating (i.e., \emph{marginalizing}) over all the other parameters:
\begin{equation}
\textit{p}(\theta_{i}|\textit{\textbf{N}},\textit{M}) = \int \textit{p}(\bmath{\theta}|\textit{\textbf{N}},\textit{M}) d\theta_1 ... d\theta_{i-1} d\theta_{i+1}... d\theta_n.
\end{equation}
Wherever possible, we make use of the full parameter probabilities (i.e., the marginal posterior densities), but in some figures that rely on point estimates for clarity, we represent the corresponding marginal posterior densities by the primary mode of the distribution with uncertainties that encompass the interior 68\% of the distribution in each direction.  For example, a parameter \textit{p} with a distribution resembling the standard log-normal distribution ($\mu$=0, $\sigma$=1) would be represented as \textit{p}=$0.37_{-0.19}^{+1.86}$.  We prefer this notation since it imparts some information about the skew in the underlying distribution.

The Bayesian likelihood distribution describes the probability of observing the data given a set of model parameters.  We treat each measured column density as an independent random variate of a zero-truncated normal (ZTN) distribution (Equation~\ref{eqn:ZTNpdf}) with the mean given by the multiphase ionization model and the variance set by the uncertainty in the corresponding Voigt profile fit.  Empirically, the ZTN distribution appears to best describe simulated measurements of a single, noisy absorption feature across a wide range in column density (from non-detection to saturated) and signal-to-noise ratios.  We model upper and lower limits as lower and upper tail probabilities, respectively, of the corresponding ZTN distribution.

The likelihood function that we employ in this paper is described in detail in Appendix~\ref{app:likelihood}. This likelihood function depends on two key assumptions: 1) that ionic column densities derived from Voigt-profile fitting are essentially independent, and 2) that our adopted error model with fixed variance (per species) adequately reflects the uncertainty in the column density measurements.  These same assumptions have been used in most of the papers in the literature that use Bayesian methodology \citep[e.g.,][]{Crighton15,Fumagalli16,Glidden16,Prochaska17} to model quasar absorption line spectra.  We expect that our likelihood function is at least accurate to first order, especially for ``clean'' (unblended and not saturated) measurements; the measurement of severely overlapping Voigt profiles (e.g., with velocity separations on the order of their linewidths $\delta\textit{v}_{12} \lesssim \textit{b}_1,\textit{b}_2$) are not fully independent.  We are currently investigating the relative significance of these effects in a separate project.  For the present study, we will follow the literature and assume that the two assumptions are accurate to first order.

A key part of any Bayesian model is the prior distribution $\textit{p}(\bmath{\theta}|\textit{M})$.  As the ratio of the number of data points to model parameters decreases, the weight of the prior increases (relative to the likelihood).  In many of our models, the number of parameters is close (within a factor of two) to the number of strong column density constraints, so the prior distribution plays an important role in our inferences.  We strive to incorporate \textit{informative} priors wherever possible since so-called \textit{non-informative} priors can introduce undesired biases.  In the following parameter overview sections, we describe various priors on both the model parameters and derived quantities.  Appendix~\ref{app:priors} formally describes $\textit{p}(\bmath{\theta}|\textit{M})$.

\subsubsection{MCMC Sampling}
We sample the posterior densities of our models using an ensemble Monte Carlo Markov Chain (MCMC) algorithm, specifically the affine invariant ``Parallel Tempered Ensemble Sampler'' (PTES) implemented in the popular Python package \textsc{emcee v2.2.1}\footnote{Note, the PTES has been removed from the most recent version of \textsc{emcee} (\textsc{v3.0.2}) and is now the stand-alone package \textsc{ptemcee}: https://github.com/willvousden/ptemcee.  These program changes did not alter the underlying mathematical algorithm, so it is consistent in both versions.} \citep{ForemanMackey13}.  The core (non-tempered) algorithm initializes an ensemble of ``walkers'', which are similar to Metropolis-Hastings chains with the key difference being that the proposal distribution for a given walker depends on the other walkers' positions.  In our application, the tempered version of the algorithm is faster and more likely to converge than the basic algorithm.  The ensemble proposal distribution also includes a scale parameter, \emph{a}, that 
primarily affects the acceptance rate.  We find that scale parameters in the range \emph{a}=1.2--2 optimally balance the algorithm's convergence rate against its ability to explore parameter space.  The tempered version initializes multiple ensembles of walkers and allows each ensemble to wander through a power-law scaled version of the posterior (including one ensemble dedicated to the true, non-scaled posterior).  These tempered chains can swap positions in a manner that preserves the detailed balance of the MCMC.  This results in a more robust sampling of multimodal posteriors, and frequently a boost in their convergence rate as well.  We initialize the sampler with 70--150 walkers and 2--3 temperings (using the provided default power-law scaling).  Each walker's initial position is uniformly sampled across the allowed parameter volume (i.e., where $\textit{p}(\bmath{\theta}|\textit{M}) > 0$), an initialization technique that we find leads more reliably to convergence for complex models than initializing all the walkers near a single best-guess point in parameter space.

MCMC algorithms are guaranteed to draw samples from the stationary distribution (e.g., the posterior) if run for an infinite duration.  For finite MCMC runs, however, one must rely on heuristics to assess when the chain has moved away from the initial positions and converged on the stationary distribution.  The Gelman-Rubin (GR) statistic \citep{Gelman92,Brooks98} is a commonly used heuristic that tracks how well multiple independent, co-evolved chains have mixed.  Unfortunately, since the PTES walkers are not individually Markovian (each ensemble is, in fact, a single large Markov Chain), it is incorrect to use the GR statistic on the ensemble walkers.  One could, in principle, compute the statistic for several co-evolving ensembles, but this is typically impractical.  Instead, \cite{ForemanMackey13} advocate using the integrated autocorrelation time (IACT) $\tau_\textit{iact}$ of the ensemble to determine when to cease sampling.  The idea behind using the IACT as a stopping criterion is rooted in the observation that Monte Carlo integration errors decrease as a function of the number of samples: $\sigma_\textrm{Monte Carlo} \propto N^{-0.5}$.  The effective sample size is reduced, however, by the IACT between samples: $N^{i}_\textit{eff} = N / \tau^{i}_\textit{iact}$, where $i=1,2,...,n_\textit{par}$ and $n_\textit{par}$ is the number of model parameters.  It makes sense to follow $N^{min}_\textit{eff} = N / \max\{ \tau^{i}_\textit{iact} \}$ for determining convergence.  One may stop the simulation once $N^{min}_\textit{eff}$ reaches a target threshold (we target $N^{min}_\textit{eff}>10^4$).  In practice, $\tau^{i}_\textit{iact}$ can be poorly estimated when $N^{i}_\textit{eff}$ is low.  Hence, we wait until $N^{min}_\textit{eff}>100$ to begin collecting samples.  In practice, we find that the IACT used in this way works well in an empirical sense: the sampled posterior density does not evolve significantly after the IACT has converged and the results are consistently reproducible.

\subsubsection{Model Selection Metrics}
\label{sec:model_selection_metrics}
	In subsequent sections we will seek to quantify the performance of individual models relative to a set of competing models.  We utilize two common (Bayesian) metrics for this purpose: 1) Bayesian evidence ratios or Bayes factors \citep{Kass95}, which express the relative probabilities or ``odds'' of pairs of models, and 2) posterior predictive \textit{p}-values \citep{Gelman96}, which are Bayesian versions of classical \textit{p}-values.

	The Bayes factor \textit{K} for two models \textit{M}\textsubscript{1}, \textit{M}\textsubscript{2} is the ratio of their marginal likelihoods multiplied by the prior odds ratio of the models.  This gives the posterior odds, i.e., the relative probabilities of the two models:
\begin{equation}
\textrm{(posterior odds)} = \textrm{(Bayes factor)} \times \textrm{(prior odds)}
\end{equation}
or equivalently,
\begin{equation}   
\frac{\textit{p}(\textit{M}_2|\textit{\textbf{N}})}{\textit{p}(\textit{M}_1|\textit{\textbf{N}})} = \frac{ \int \textit{p}(\textit{\textbf{N}}|\bmath{\theta}_2,\textit{M}_2) \textit{p}(\bmath{\theta}_2|\textit{M}_2) \mathrm{d}\bmath{\theta}_2 }{ \int \textit{p}(\textit{\textbf{N}}|\bmath{\theta}_1,\textit{M}_1) \textit{p}(\bmath{\theta}_1|\textit{M}_1) \mathrm{d}\bmath{\theta}_1 } \times \frac{ \textit{p}(M_2) }{ \textit{p}(M_1) }
\end{equation}
Lacking information that favors any particular model over the others \textit{a priori}, we assign a factor of unity for the prior odds of all pairs of our models and treat the Bayes factor as the \textit{de facto} posterior odds.  We compute the marginal likelihood by identifying and resampling posterior modes following the approach outlined in \cite{Weinberg13a}.  Details are given in Appendix~\ref{app:marginal_likelihood}.

	By virtue of their definition as probabilities, Bayes factors allow one to consider how strongly the data support a particular model.  Unlike classical hypothesis tests, which only consider evidence \textit{against} the preferred model (the null hypothesis), Bayes factors also consider evidence \textit{in favor of} the model.  Thus, using Bayesian evidence and Bayes factors, one may rank a set of models probabilistically and \textit{accept} those above some ``decisive'' odds ratio threshold rather than merely rejecting those below the threshold.

	In practice, Bayes factors are sensitive to the choice of prior and the decision threshold is somewhat subjective.  Additionally, accurate estimates of the marginal likelihood are often computationally challenging, particularly for models with moderate to high dimensionality and/or multimodal posteriors.  To address these shortcomings, we 1) set informative priors wherever possible, and 2) model mock absorbers to assess the sensitivity and accuracy of our Bayes factor estimates.  Our test results (see Appendix~\ref{app:tests}) indicate a decision threshold of $\textit{K}>10^{1.5}$ suffices to choose one model over another.  Note that this corresponds to ``very strong'' evidence on the oft-invoked interpretation scale of \citet{Jeffreys61}.

	Bayes factors provide a useful metric for comparing two models, but unless one has computed the complete set of competing models, it is impossible to state the \textit{absolute} probability of any given model.  Thus, we would also like some criterion to reject models in an absolute sense when discrepancies between the model and data are unlikely to have arisen by chance.  The posterior predictive \textit{p}-value (PP \textit{p}-value), is a Bayesian version of the classical \textit{p} statistic that probabilistically evaluates (in simulated datasets) a model's ability to replicate some summary statistic \textit{T}(\textit{y\textsuperscript{obs}},$\theta$) of the data \textit{y\textsuperscript{obs}} and model parameters $\theta$.  For example, a PP \textit{p}-value of 0.35 means that, assuming the model is true, we can expect simulated datasets \textit{y\textsuperscript{sim}} to produce a statistic \textit{T}(\textit{y\textsuperscript{sim}},$\theta$)$>$\textit{T}(\textit{y\textsuperscript{obs}},$\theta$) 35\% of the time.  Extending this concept, models that result in PP \textit{p}-values near 0 are deemed unlikely and rejected.  Conversely, it is appropriate to be suspicious about models with PP \textit{p}-values very close to 1 -- this could indicate that such models are overfitting the data, which compromises the insight that they provide.  Models with PP \textit{p}-values $\approx$ 1 should be subjected to extra scrutiny to evaluate whether, e.g., the model has too many parameters. The rejection threshold or significance level $\alpha$ and the choice of summary statistic are subjective and depend on the problem at hand.  We set $\alpha$=0.05 and define \textit{T}(\textit{y},$\theta$) to be the scaled and summed residuals between observed/simulated and modeled column densities.  Appendix~\ref{app:ppp} expounds on the details of this statistic.

\subsection{Overview of Model Parameters}
\label{sec:parameters}

\subsubsection{Ionization Mechanism}
To infer the physical conditions in a given absorber, we need to know the parameters that are most likely to produce the measured ionic column densities.  This necessarily involves differencing the measured and model-predicted columns, where we define the model prediction as the sum of column densities across all ionization phases assigned by the model for each absorber (Equations~\ref{eqn:ZTNpdf}--\ref{eqn:like}).  We use version \textsc{17.00} of \textsc{cloudy}, last described by \citep{Ferland17}, to compute the ionization equilibrium and radiative transfer through the absorbing gas.  Assuming that the gas is optically thin, we can build a basic multiphase model as the composite of several single-phase models, e.g., as layers or shells of a gas cloud.  For a given slab of gas, \textsc{cloudy} iteratively adjusts the thickness until \emph{N}\textsubscript{H I} reaches a prescribed value, while simultaneously solving the radiative transfer equations and balancing the heating and cooling processes.  The resulting \textsc{cloudy} solution reports the abundances of all the ionization states for the first thirty elements (through Zn).  \textsc{cloudy} also has an option to fix the slab temperature to an arbitrary value.  This effectively forces \textsc{cloudy} to function as a collisional ionization model while simultaneously accounting for the energy supplied by a background radiation field.  Hence we label this mode photo- plus collisional-ionization (P+C). Note that in P+C models, heating and cooling are not necessarily balanced.  \textsc{cloudy}'s prevalence in the literature makes it a natural choice as our core ionization model, to enable a straightforward comparison between our multiphase models and prior work.  We note in passing, however, that there are other intriguing alternative ionization models that might benefit from the same multiphase framework we develop in this work \citep[e.g.,][]{Gnat07,Oppenheimer18,Buie20}.

Since \textsc{cloudy} models are relatively expensive to compute on-the-fly, we precomputed large grids of ionization models for both the PIE-only and P+C ionization modes at the grid points specified in Table~\ref{tab:param_info}.  To facilitate arbitrary sampling across the parameter space these grids were then interpolated into continuous functions via 1-D Akima\footnote{1-D Akima interpolation is less prone to strong spurious oscillations with small numbers of data points \citep{Akima70}.} interpolants applied iteratively along each dimension.

\subsubsection{The UV Radiation Field}
\label{sec:UVB}

\begin{figure}
	\begin{subfigure}[b]{0.465\textwidth}
	\centering
	\includegraphics[width=3.2in]{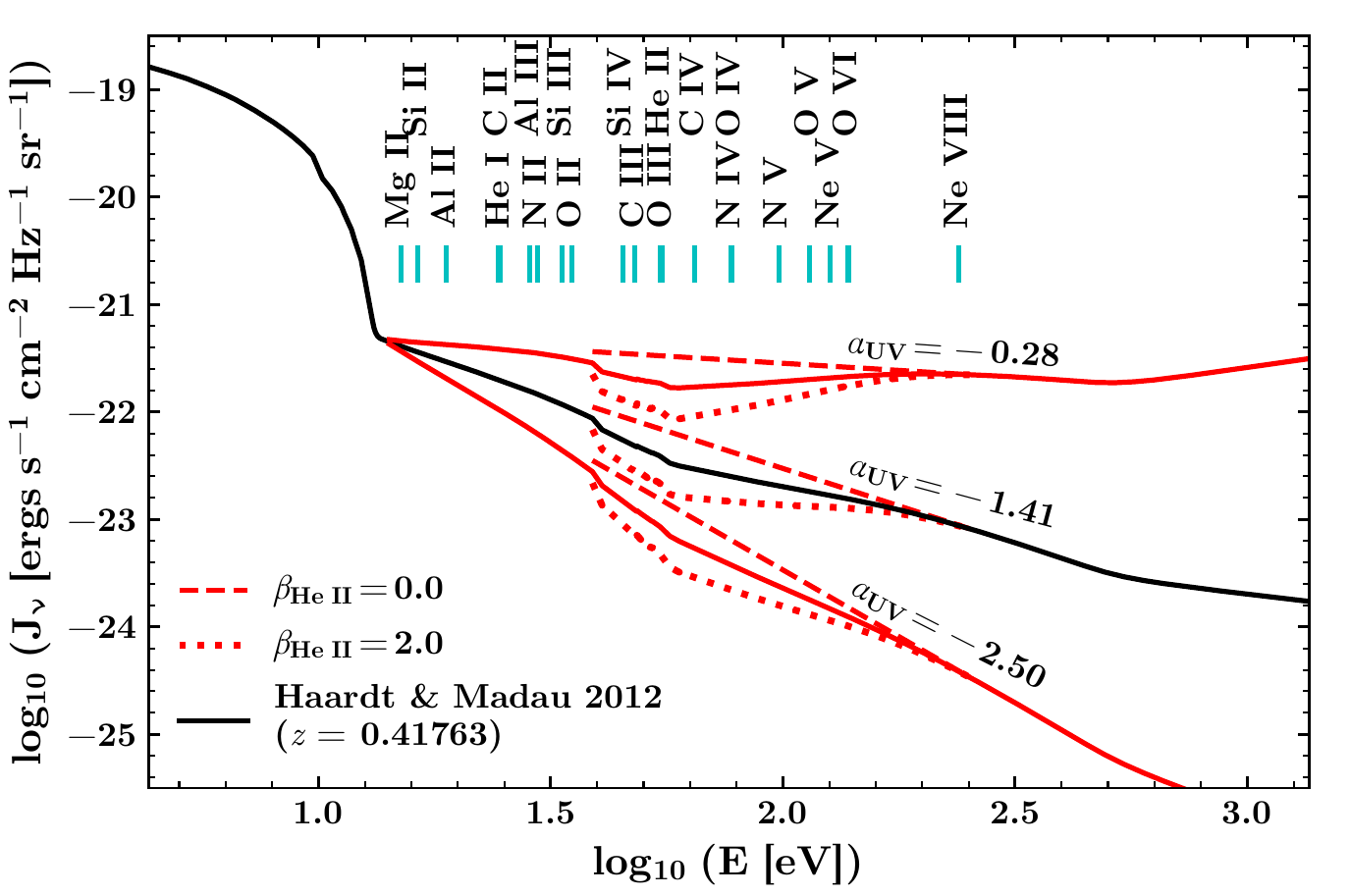}
	\caption{UVB parameterization details}
	\label{fig:UVBplotA}
	\end{subfigure}
	\begin{subfigure}[b]{0.465\textwidth}
	\centering
	\includegraphics[width=3.2in]{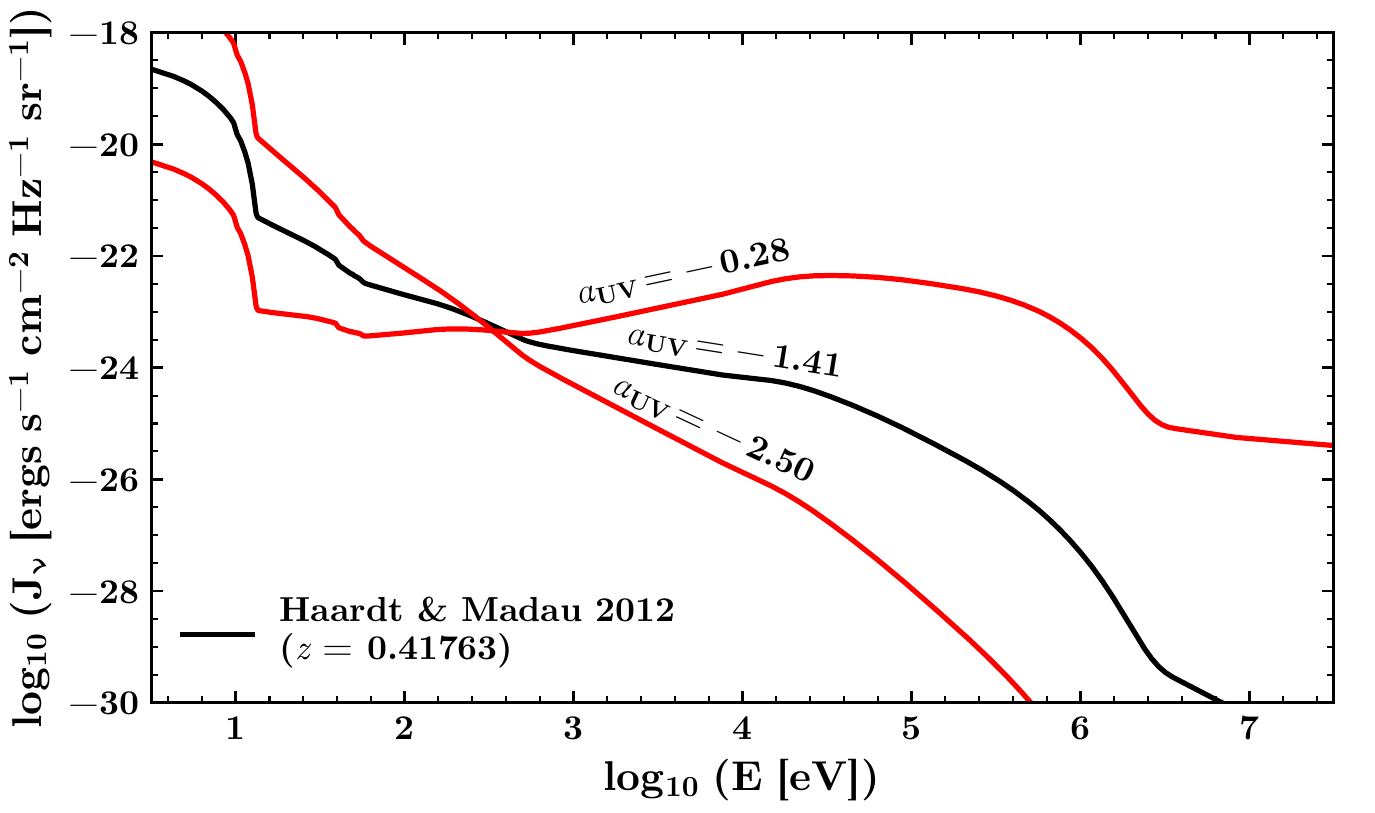}
	\caption{Full ionizing radiation field, normalized by $\Gamma$\textsubscript{H I}}
	\label{fig:UVBplotB} 
	\end{subfigure}
\caption[]{An illustration of the various UV ionizing radiation field parameterizations explored by our models.  (a) In the top panel we vary the spectral slope $\alpha_{\textrm{UV}}$ (with minimal and maximal values illustrated by the red solid curves) and the He~\textsc{ii} Lyman Limit absorption strength $\beta_{\textrm{H I}}$ (curves with different linestyles) relative to the fiducial \cite{Haardt12} UVB (black curve).  Ionization potential energies are marked for some commonly detected ions, to underscore the potential impact of variations in the shape of the UV spectral energy distribution on their photoionization rates.  (b) The bottom panel includes a wider view of the ionizing radiation field, with each curve's intensity normalized to the fiducial H~\textsc{i} photoionization rate $\Gamma_{\textrm{H I}}^0$ as described in the text.}
\label{fig:UVBplot}
\end{figure}

	The impinging ultraviolet light is expected to be the dominant source of ionization in diffuse IGM/CGM absorbers, but local variations in its shape and intensity are poorly characterized.  The UV background (UVB) contributes a substantial fraction toward the UV ionizing photon budget, yet models of the cosmic mean extragalactic radiation field \citep[e.g.,][]{Haardt96,Haardt01,Haardt12, FaucherGiguere09} vary year-to-year in their predictions (e.g., see Figure 11 in \citealt{Fumagalli17}).  Unknown intrinsic uncertainties in the UVB estimates are not the only concern, however, as \cite{Upton-Sanderbeck18} list several ionizing flux sources local to galaxies that may significantly enhance the density of ionizing photons.  Additionally, the ionization in some intervening absorbers is likely impacted by their proximity to AGN (foreground to the target QSO and adjacent to the sight line), which can exhibit a wide variation in UV spectral slopes and luminosity \citep{Scott04,Stevans14}.  This consideration may be more significant given the potential for ``flickering'' AGN \citep{Oppenheimer18} to produce ionization fossils that persist while the AGN is inactive.

	To account for a wide range of physically plausible local radiation fields, we introduce two parameters that adjust the spectral shape of the UV radiation field, illustrated in Figure~\ref{fig:UVBplot}.  First, borrowing from the approach of \cite{Crighton15} \citep[see also][]{Fumagalli16} we vary the approximately powerlaw slope of the fiducial radiation field (in this study, \citealt{Haardt12}) by applying a ``hinge'' at 1 Ryd and modifying the slope of the spectrum on the higher energy side (of the hinge) via the power law spectral index parameter, $\alpha_{\textrm{UV}}$.  These variations are implemented as shown in Equation~\ref{eqn:uvb} through the function $f(E)$.  We apply a truncated normal prior to $\alpha_{\textrm{UV}}$, specifically $p_\textrm{Trunc. Normal}(\alpha_{\textrm{UV}},-1.41,0.7,-2.5,-0.28)$ (see Equation~\ref{eqn:truncnorm}), where the bounds are tied to the range of our precomputed grid of ionization models and the values for $\mu$ and $\sigma$ are derived from the distribution of AGN spectral indices $\alpha^{\nu}$ measured by \cite{Stevans14} at rest-frame FUV (500--1000 $\textrm{\AA}$) (see the blue histogram in their Figure 10).

	Moderately and highly ionized species (see tickmarks in the upper panel of Fig.~\ref{fig:UVBplot}) are also sensitive to the strength of the He~\textsc{ii} Lyman Limit absorption trough ($\lambda_\textrm{rest} $ $\approx$ $228\textrm{\AA}$ corresponding to $\approx$ $10^{1.74}$ eV), and estimates of this feature in models of the cosmic mean UVB have varied significantly over the past two decades. Hence, we introduce a second scaling parameter $\beta_{\textrm{He II}}$ that makes the helium dip deeper or shallower.\footnote{More precisely, it adjusts the total integrated flux decrement from the unabsorbed powerlaw in the vicinity of the He~\textsc{ii} lyman limit, with a scaling of unity returning the fiducial curve, and a maximum scaling reflecting the largest estimate of such a flux decrement found in the literature \citep[i.e.,][]{Haardt96}.}  Given the wide range in $\beta_{\textrm{He II}}$ over the limited set of UVB estimates in the literature, we place a uniform prior $p_\textrm{Uniform}(\beta_{\textrm{He II}},0,3)$ (Equation~\ref{eqn:uniform}) on $\beta_{\textrm{He II}}$ designed to encompass the variations seen in the literature (see Fig.~\ref{fig:UVBplotA}).

Given the two parameters $\alpha_\textrm{UV}$ and $\beta_\textrm{HeII}$, the following equation defines the shape $F_\nu(E)$ of our variable radiation field for energy $E$:
\begin{equation}\label{eqn:uvb}
	\log_{10} F_\nu(E) = g(E, f(E))
\end{equation}
where
\[
	f(E) = 
	\begin{cases}
		H(E), & E \leq E_0\\
		H(E) + (\alpha_\textrm{UV} + 1.4)E, & E>E_0
	\end{cases}
\]
and
\[
	g(E,x) = 
	\begin{cases}
		\beta_\textrm{HeII} x + (1 - \beta_\textrm{HeII}) (mE + b), & 1\text{Ryd} < E < 18.4\text{Ryd}\\
		x, & \text{otherwise}
	\end{cases}
\]
where $H(E)$ is the base 10 logarithm of the fiducial \cite{Haardt12} radiation field and $mE + b$ is the line between the two $(E,H(E))$ coordinate pairs evaluated at $E=1$ Ryd and $18.4$ Ryd.  The value $\alpha_\textrm{UV}=-1.4$ corresponds to the approximate spectral index of $H(E)$ near redshift $z\approx0.5$.

	To fully specify the UV spectral energy distribution, we need a prescription for the intensity of the ionizing radiation.  The ionization state of the optically thin gas, however, is degenerate with respect to the total ionizing intensity and the gas density, which makes it impossible to directly constrain both properties simultaneously.  Instead, it is common to present model results as a function of the ionization parameter \textit{U} = \textit{n}$_{\gamma}$/\textit{n}$_{\textrm{H}}$, where \textit{n}$_{\gamma} =$ the number density of ionizing photons, which scales with the total ionizing intensity of the radiation for a fixed spectral shape.  In our parameterization of the UV radiation field the spectral shape is not fixed, however, so we must first normalize by the total ionizing intensity.  This poses a problem for the physical interpretation of \textit{n}$_{\gamma}$, and by extension, \textit{U}, since the total ionizing intensity is sensitive to the distribution of X-ray photons in a given SED while the ionization state of the gas is not, since the tail of the H~\textsc{i} photoionization cross-section is negligible at such high energies.  As a solution to this problem, we instead define a pseudo-ionization parameter $\textit{Q}_\Gamma$ based on the H~\textsc{i} photoionization rate $\Gamma_{\textrm{H I}}$ rather than on the photon number density:\footnote{Note, unlike \textit{U}, $\textit{Q}_\Gamma$ is not dimensionless.}
\begin{equation}\label{eqn:gammaU}
	\textit{Q}_\Gamma \equiv \frac{\Gamma_{\textrm{H I}}}{\textit{n}_{\textrm{H}}}.
\end{equation}
Like \textit{n}$_{\gamma}$, $\Gamma_{\textrm{H I}}$ also scales linearly with the total ionizing intensity for a fixed spectral shape, and similarly needs to be normalized.  The normalized rate is:
\begin{equation}\label{eqn:gammatildeU}
\tilde{\Gamma}_{\textrm{H I}} = \textit{f}(\alpha_{\textrm{UV}},\beta_{\textrm{He II}}) \cdot \Gamma_{\textrm{H I}},
\end{equation}
which depends on the normalizing function
\begin{equation}\label{eqn:gamma_normalization}
\textit{f}(\alpha_{\textrm{UV}},\beta_{\textrm{He II}}) = \frac{\Gamma_{\textrm{H I}}^0}{\Gamma_{\textrm{H I}}(\alpha_{\textrm{UV}},\beta_{\textrm{He II}})}.
\end{equation}
For consistency with the fiducial values of $\alpha_{\textrm{UV}}$ and $\beta_{\textrm{He II}}$ (i.e., those which reproduce the fiducial UV ionizing radiation field), we set $\Gamma_{\textrm{H I}}^0$ = $\Gamma_{\textrm{H I}}^{\textrm{HM12}}(\textit{z})$, where $\Gamma_{\textrm{H I}}^{\textrm{HM12}}(\textit{z})$ is the redshift-dependent H~\textsc{i} photoionization rate of the UVB from \cite{Haardt12}.  Compared to alternative UVB models (e.g., see \citealt{Fumagalli17}, Fig. 11), $\Gamma_{\textrm{H I}}^{\textrm{HM12}}(\textit{z})$ would seem to be an excellent choice as a lower limit on the true H~\textsc{i} photoionization rate.  Indeed, given the potential for extra radiation intensity from local ionizing sources (e.g., AGN), it makes sense to work with such a lower limit.

\ctable[
	caption={List of Parameters, Priors, and Pre-Computed Ionization Model Grid 
			Details},
	doinside = \footnotesize,
	label={tab:param_info}
	]{lcl@{~}l}{
\tnote[a]{The \textit{i}-th redshift-dependent grid value of $\log n_\mathrm{H}(z)$ is given by $-3.846\times\ln (e^{1.638} - i\times e^{-2.047}) + \ln(1+z)^3$ for $i=0,1,...,29$.}
\tnote[b]{$\log_{10}T$ is only relevant for P+C models.}
\tnote[c]{See Section~\ref{sec:metals_and_abn} regarding why $\mathrm{[\alpha/H]}$ instead of $\mathrm{[Z/H]}$.}
\tnote[d]{The \textit{i}-th grid value of $[\alpha/\mathrm{H}]$ is $3.7\times\ln (i\times e^-2.8344 + e^{-0.811})$ for $i=0,1,...,13$.}
\tnote[e]{In P+C models these parameters are applied as post-\textsc{cloudy} adjustments to shorten the computation time.  Real-time processing of non-solar gas composition produces slight differences (typically $\lesssim0.01$ dex) in the modeled electron number density and ionic column densities.}
\tnote[f]{\textit{N}\textsubscript{H} is not gridded because the solution scales linearly with \textit{N}\textsubscript{H} for an optically thin gas (see Section~\ref{sec:NHscaling}).}
\tnote[g]{$\sigma^2_{\ln P,true}$ is a hyperparameter and $s^2_{\ln P}$ is a property of the other parameters.  Both relate to pressure equilibrium in multiphase models (see Section~\ref{sec:pressure}).}
}{
																														\FL
Parameter								& Bounds		& Step			& Prior Reference							\ML
$\log_{10} n_\mathrm{H}(z)\ ({\rm cm}^{-3})$ & $(-6.3,-1.3)|^{z=0}$ & \tmark[a] & Eqn.~\ref{eqn:prior_density}			\NN
$\log_{10}T \ (K)$\tmark[b]				&$(4.0,6.4)$	& $0.1$		& Eqns.~\ref{eqn:prior_temp},~\ref{eqn:uniform}		\NN
$\mathrm{[\alpha/H]}$\tmark[c]			&$(-3.0,0.7)$	& \tmark[d] & Eqns.~\ref{eqn:prior_metals},~\ref{eqn:uniform}		\NN
$\alpha_{\mathrm{UV}}$					&$(-2.5,-0.28)$	& $0.317$		& Eqns.~\ref{eqn:prior_auv},~\ref{eqn:truncnorm}	\NN
$\beta_{\mathrm{He II}}$				& $(0,3.0)$		& $1.0$		& Eqns.~\ref{eqn:prior_HeII},~\ref{eqn:uniform}		\NN
$\mathrm{[C/\alpha]}$\tmark[e]			&$(-0.8,0.8)$	& $0.4$		& Eqns.~\ref{eqn:prior_abn},~\ref{eqn:uniform}		\NN
$\mathrm{[N/\alpha]}$\tmark[e]			&$(-1.2,0.4)$	& $0.4$		& Eqns.~\ref{eqn:prior_abn},~\ref{eqn:uniform}		\NN
$\mathrm{[\alpha/Fe]}$\tmark[e]			&$(-1.2,0.4)$	& $0.4$		& Eqns.~\ref{eqn:prior_abn},~\ref{eqn:uniform}		\NN
$\log_{10} N_\mathrm{H}\ ({\rm cm}^{-2})$\tmark[f] & $(0,\gtrsim23)$ & --- & Eqns.~\ref{eqn:prior_NH},~\ref{eqn:gamma}	\NN
$\sigma^2_{\ln P,true}$\tmark[g] 		&$(0,3.5)$ 		& ---		& Eqn.~\ref{eqn:priorfull}							\NN
$s^2_{\ln P}$\tmark[g] 					& \multicolumn{2}{c}{Latent Parameter}  & Eqn.~\ref{eqn:priorfull}				\LL
}

\subsubsection{Gas Density and Temperature}
\label{sec:density}

	The UV intensity normalization discussed above affects the gas density as well, since the two quantities are related by Equation~\ref{eqn:gammaU}.  In conjunction with Equation~\ref{eqn:gammatildeU}, the adjusted density corresponding to the normalized H~\textsc{i} photoionization rate is:
\begin{equation}\label{eqn:adjusted_density}
	\tilde{\textit{n}}_{\textrm{H}} = \frac{\tilde{\Gamma}_{\textrm{H I}}}{\textit{Q}_\Gamma} = \tilde{\Gamma}_{\textrm{H I}} \left( \frac{\textit{n}_{\textrm{H}}}{\Gamma_{\textrm{H I}}} \right) = \textit{f}(\alpha_{\textrm{UV}},\beta_{\textrm{He II}}) \cdot \textit{n}_{\textrm{H}}.
\end{equation}
The unadjusted density, $\textit{n}_{\textrm{H}}$, in conjunction with an unnormalized H~\textsc{i} photoionization rate, $\Gamma_{\textrm{H I}}$, describes the ionization state of the gas ($\textit{Q}_\Gamma$).  Equation~\ref{eqn:adjusted_density} shows that upon normalizing the total ionizing intensity, now characterized by $\tilde{\Gamma}_{\textrm{H I}}$, we can recover the same ionization parameter $\textit{Q}_\Gamma$ only if we also multiply the density $\textit{n}_{\textrm{H}}$ by $\textit{f}(\alpha_{\textrm{UV}},\beta_{\textrm{He II}})$.  If one takes the fiducial $\Gamma_{\textrm{H I}}^0$ as a lower limit, then $\textit{f}(\alpha_{\textrm{UV}},\beta_{\textrm{He II}})$ becomes a lower limit by Equation~\ref{eqn:gamma_normalization}, as suggested above, and the adjusted density is also a lower limit: $\tilde{\textit{n}}_{\textrm{H}} \lesssim \textit{n}_{\textrm{H}}^{\textit{true}}$.  Note that treating the density as a lower limit propagates to a lower limit on the derived gas pressure and an upper limit on the inferred cloud size, two properties discussed below.

	In multiphase models, the \textit{relative} density between individual phases irradiated by a common UV ionizing radiation field can be uniquely specified, unlike\footnote{Alternatively, if one has strong prior information on $\Gamma_{\textrm{H I}}^0$, then $\Gamma_{\textrm{H I}}^0$ could be included within a Bayesian model as a ``nuisance'' parameter.  This would enable a full specification of the gas density.  For example, one could use measurements of the UVB value for $\Gamma_{\textrm{H I}}^0$ as a prior, but only if one has high confidence that the absorber is not partially ionized by local galaxy/AGN sources.  We do not take this approach given the limited number of constraints on $\Gamma_{\textrm{H I}}^{\textrm{UVB}}$ and the uncertainty in the local environment.} the \emph{absolute} gas density, which depends on one's choice of $\Gamma_{\textrm{H I}}^0$ (to normalize the radiation field).  Throughout this work, we assume that kinematically aligned absorption components are exposed to a common radiation field.  Under this assumption, we can directly compare gas densities, pressures, and cloud sizes across ionization phases inferred by multiphase models for a given absorber (or multiple absorbers if they share a common set of UV parameters) without worrying about the true normalization of the gas density.

	For single-phase models, the only constraints on the density are the boundaries of the parameter interpolation grid, so we place a uniform prior on the logarithm of the density across the accessible region of parameter space.  In multiphase models, however, the prior also enforces a density ranking among the phases (of a given absorber) so that a particular model phase always takes densities with the same ranking index $i$.  For example, in a three-phase model, the first phase is always assigned the highest density, the second phase gets the intermediate density, and the lowest density goes to the third phase.  When the ionization mechanism (PIE-only or P+C) is consistent across all phases, then this prior is essentially a bookkeeping exercise with no physical implications.  When both PIE-only and P+C phases are intermixed, however, the inferred gas densities \textit{are} impacted by the ranking prior.  The prior is still warranted in this case though, since we generally expect that higher temperature (P+C) gas has a lower density than cooler (PIE-only) gas.  The exception to this prior would be if high density, warm-hot gas were embedded in a cooler photoionized cloud, but the large implied pressure gradient should make any such state short-lived.  Without this phase ranking prior, the model typically prefers the exception, but for non-physical reasons: the density is poorly constrained at higher temperatures and the cloud size prior (see below) prefers higher densities (smaller clouds).  Equation~\ref{eqn:prior_density} gives the probability density of the ranking prior.

	The \textsc{cloudy} P+C models additionally include the gas temperature as a free parameter.  This is particularly useful for modeling hot gas phases in a multiphase absorber.  We place a uniform prior on the `hot' phase temperature $\log\textit{T}$ between the boundaries of the \textsc{cloudy} interpolation grid (Table~\ref{tab:param_info}).  Note that at the lowest gas temperatures the P+C model is largely photoionized, and collisional ionization only becomes dominant at higher temperatures.  In the latter regime, the density dependence of the heating and cooling processes cancel, so the gas density has little effect on the model.  Even at temperatures of $\approx$10\textsuperscript{5.5}~K, however, the density dependence is not completely erased.  A two-phase P+C model with a ranked density structure will elicit a ranked temperature structure (highest temperature corresponding to lowest density) without any informative temperature prior.  If the temperature were independent of density, this would not occur.

\subsubsection{Gas Metallicity and Composition}
\label{sec:metals_and_abn}
	In addition to the standard metallicity parameter normalized to \cite{Asplund09} solar abundances, we also include a set of parameters governing the relative abundances of individual elements.  For the absorbers examined here, we tagged all the $\alpha$-elements to directly follow the metallicity parameter (i.e., at solar relative abundance ratios), while the carbon, nitrogen, and iron abundances were allowed to vary with respect to the $\alpha$ abundances.  All other species were required to follow solar relative abundances except in one absorber, which also demands a non-solar aluminum abundance (see below).  The motivation for this individual treatment of C, N, and Fe is their differing nucleosynthetic origins; the relative abundances of these species are observed to depart from their solar values in various contexts \citep[e.g.][]{McWilliam97,Fabbian09,Lehner13,Cooke17}.  To all metallicity and relative abundance parameters we applied non-informative, uniform priors across the range of the precomputed \textsc{cloudy} grid.

	We detect aluminum in Absorber~0.6A, and, in exception to the scheme above, we find that an adequate fit requires tuning the [Al/$\alpha$] relative abundance.  We include [Al/$\alpha$] as an extra parameter in our models for Absorber~0.6A and assign a truncated normal prior $p_\textrm{Trunc. Normal}([\textrm{Al}/\alpha],0,0.2,-0.5,0.5)$. 

We do not explicitly attempt to include parameters that allow for depletion of some elements by incorporation into dust, but an underabundance of Fe in the gas phase could be interpreted as a result of dust depletion instead of as the result of nucleosynthetic origins.  In the absorption systems studied in this paper, we do not detect many of the refractory elements that are highly sensitive to the presence of dust, so it can be difficult in some cases to distinguish between nucleosynthesis and dust depletion.  Our approach is simply to explore the relative abundances allowed by the models, which can then be interpreted and discussed in various contexts.

\subsubsection{ N{\normalfont (H)} Scaling}
\label{sec:NHscaling}
	All our absorbers fall in the optically thin regime $\textit{N}(\textrm{H}~\textsc{i})\lesssim10^{17}$ where the ionic column densities scale linearly with \textit{N}(H~\textsc{i}).  This allows us to precompute a single grid of ionization models at a generic, though optically thin, value of \textit{N}(H~\textsc{i}) (we chose log~\textit{N}(H~\textsc{i})=$10^{15.5}$) and to scale the column densities to fit any of our absorbers.  Once the grid has been computed, there is no reason to prefer \textit{N}(H~\textsc{i}) as a scaling parameter over any other ionic column density.  Instead, we take the total hydrogen column ($\textit{N}_\textrm{H} = \textit{N}(\textrm{H}~\textsc{i})+\textit{N}(\textrm{H}~\textsc{ii})$) as a more useful scaling parameter since it allows us to establish a direct prior on the physical extent of the absorption ``cloud''.  Each modeled ionization phase is assigned a scaling parameter, so that the column densities of the \textit{p}-th phase $\textit{N}(\textit{X}^{\textrm{ion}}_{\textit{p}})$ are set by scaling the precomputed column densities $\textit{N}(\textit{X}^{\textrm{ion}}_{\textrm{grid}})$:
\begin{equation}
	\textit{N}(\textit{X}^{\textrm{ion}}_{\textit{p}}) = \frac{\textit{N}_\textrm{H}^\textit{p}}{\textit{N}_\textrm{H}^\textrm{grid}} \textit{N}(\textit{X}^{\textrm{ion}}_{\textrm{grid}}).
\end{equation}

	The total hydrogen column density is closely related to the line-of-sight path-length, or ``cloud size'' \textit{l}\textsubscript{abs} of an absorber, so we encode some reasonable assumptions about the cloud size as a prior on $\textit{N}_\textrm{H}^\textit{p}$.  We assume that absorbers are smaller than $\approx$100 kpc so that these absorption systems, which are often clearly composed of several components and multiple phases, can plausibly fit within a galaxy halo.  At low redshifts, these assumptions are supported by CGM studies that have probed \textsl{single} galaxies with multiple sightlines to background QSOs \citep[e.g.,][]{Keeney13,Bowen16}; these studies indicate that the size of metal-enriched CGM clouds is $\ll$ 100 kpc.  At intermediate and high redshifts, several groups have similarly used the multiple sightlines to images of gravitationally lensed QSOs to show that metal absorbers change on scales smaller than 100 kpc \citep{Rauch01,Chen14}.  \citet{Rauch01} probed the spatial density distribution of C~\textsc{iv} absorbers along pairs of sightlines to three gravitationally lensed quasars (\textit{z}\textsubscript{QSO}=\{2.32, 2.72, 3.62\}) and found little difference in the multi-sightline comparison of C~\textsc{iv} column density or velocity shear at scales less than a few hundred parsecs and increasingly greater differences on larger scales, with simple estimates suggesting the absorption likely extends over $\approx$10~kpc regions.  More recently, \cite{Rubin18} calculated a coherence scale of $\gtrsim$2~kpc for strong (W\textsubscript{2796}$>$0.1~$\textrm{\AA}$) Mg~\textsc{ii} absorbers.  We model these assumptions with a Gamma distribution prior over the multiphase path length \textit{l}\textsubscript{abs}, which we define as the total path length through all the discretely modeled gas phases \textit{p} that compose a given absorber.  In this work we explore $\textit{n}_{\textrm{phases}}\leq3$ and we approximate the path length through a single phase \textit{p} by its modeled hydrogen column density to number density ratio.
\begin{equation}\label{eqn:cloudsize}
l_\textrm{abs} = \sum_{p=1}^{\textit{n}_\textrm{phases}} \frac{\textit{N}_\textrm{H}^{p}}{\langle \textit{n}_\textrm{H}^{p} \rangle}
\end{equation}
where $\langle \textit{n}_\textrm{H}^{p} \rangle = f_\textrm{v} \textit{n}_\textrm{H}^{p}$.  Since the hydrogen number density volume filling factor $f_\textrm{v} \leq 1$ is unknown, we proceed by setting it to unity and treating Equation~\ref{eqn:cloudsize} as a lower limit on the multiphase path length.

	A convenient property of the Gamma distribution ensures that the sum $X+Y$ of two independent Gamma variables, $X\sim\mathrm{Gamma}(k_1,\theta)$ and $Y\sim\mathrm{Gamma}(k_2,\theta)$, will also be a Gamma variable $X+Y\sim\mathrm{Gamma}(k_1+k_2,\theta)$.  Since we treat the total path-length as a Gamma variable, we can model path-lengths of individual phases (in a multiphase absorber) as independent Gamma variables by assigning a common scale parameter $\theta_\textrm{path-length}$.  Furthermore, if we consider the hydrogen column density of an individual phase $\textit{N}_\textrm{H}^p$ as the path-length of the phase $l_\textrm{abs}^p$ scaled by its density $\textit{n}_\textrm{H}^p$,
\begin{equation}
	\textit{N}_\textrm{H}^p = \textit{n}_\textrm{H}^p l_\textrm{abs}^p ,
\end{equation}
then the scaling property of the Gamma distribution\footnote{If $X\sim\mathrm{Gamma}(k,\theta)$, then for $c>0$: $cX\sim\mathrm{Gamma}(k,c\theta)$.} allows us to recast the prior on $\textit{l}_\textrm{abs}^\textit{p}$ as a Gamma distribution over $\textit{N}_\textrm{H}$, i.e., $\textrm{P}(\textit{N}_\textrm{H}^p|\textit{n}_\textrm{H}^p) = \mathrm{Gamma}(k^p,\textit{n}_\textrm{H}^p \theta_\textrm{path-length})$.  We find that setting $k^p=0.2$ and $\theta_\textrm{path-length}=15$~kpc produces good results, though we do not attempt to rigorously optimize these values (e.g., through sensitivity testing or by setting hyperpriors.)

\subsubsection{Gas Thermal Pressure}
\label{sec:pressure}
	The multiphase nature of our models allows us to investigate differences in the physical conditions between individual model phases.  One important property is the variation in thermal pressures among the \textit{n}\textsubscript{phases} model phases, which can act as an indicator for thermal pressure equilibrium (TPE) or lack thereof across the modeled absorber.  
The question of whether or not an absorber exists in TPE is important because it equates to a test for hydrostatic equilibrium in the ambient medium under two assumptions: every phase in the absorber 1) exhibits the same ratio of thermal vs. non-thermal pressure support and 2) follows a similar spatial distribution.  Regarding the first point, \cite{Lochhaas20} conducted a detailed investigation of halo pressures in two simulated galaxies and found that thermal pressure support dominates at radii larger than 0.1\textit{R}\textsubscript{\textit{vir}}.  In support of the second assumption, one could argue that the simplest explanation for the observed kinematic correlations discussed in Section~\ref{sec:kinematic_alignment} and exemplified in Figures~\ref{v_align1630}-\ref{v_align1338} is that the gas in different ionization phases \textit{does} follow a similar spatial distribution.  

	We aim to address two questions concerning TPE and, by extension, hydrostatic equilibrium: 1) is a TPE solution \textit{possible}, and 2) \textit{how likely} is TPE?  Suppose we treat the set of \textit{n} (where typically \textit{n}=2 or~3) thermal pressures, which we derive for each sampling of the posterior of our multiphase absorber models, as samples of an underlying log-normal pressure distribution with variance $\sigma_\textit{P,true}^2$.  This assumes that predominantly stochastic processes determine the density and temperature distributions in the underlying medium.  If the underlying medium is a galaxy halo, then an approximately log-normal thermal pressure distribution is supported by \cite{Lochhaas20}.  At the other extreme, we cannot discount the possibility that a single clump of gas plunging through the halo might exhibit non-thermal pressure variations owing to interface physics (e.g., shocks and instabilities).  In this case, the methodology presented here will likely underestimate the probability of hydrostatic equilibrium.  For a log-normal thermal pressure distribution, the \textit{logarithmic} pressure in each phase, $\ln\textit{P}_\textit{i}$, samples a normal distribution with $\sigma_{\ln\textit{P,true}}^2$, and hence the sample variance of the set of \textit{n} phase pressures $s_{\ln\textit{P}}^2=\mathrm{Var}\{ \ln\textit{P}_1,...,\ln\textit{P}_\textit{n}  \}$ has the known sampling distribution
\begin{equation}\label{eqn:chi2}
\frac{(n-1) s_{\ln\textit{P}}^2}{\sigma_{\ln\textit{P,true}}^2} \sim \chi^2(n-1).
\end{equation}
Equation~\ref{eqn:chi2} implicitly defines \textit{P}($\sigma_{\ln\textit{P,true}}^2 | s_{\ln\textit{P}}^2$), i.e., the probability that the underlying log-normal pressure distribution has a variance $\sigma_{\ln\textit{P,true}}^2$ given the sample variance $s_{\ln\textit{P}}^2$ calculated from the set of model phase pressures.  To make a probabilistic interpretation about TPE one needs to condition \textit{P}($\sigma_{\ln\textit{P,true}}^2 | s_{\ln\textit{P}}^2$) on the data, i.e., the column densities \textbf{\textit{N}}.  The posterior predictive distribution for $\sigma_{\ln\textit{P,true}}^2$
\begin{equation}\label{eqn:pressurevariancepostpred}
\textit{P}(\sigma_{\ln\textit{P,true}}^2 | \textbf{\textit{N}}) = \int \textit{P}(\sigma_{\ln\textit{P,true}}^2 | s_{\ln\textit{P}}^2) \textit{P}(s_{\ln\textit{P}}^2 | \textbf{\textit{N}}) \mathrm{d}\theta
\end{equation}
provides a straightforward way for this to be accomplished.  Here $\textit{P}(s_{\ln\textit{P}}^2 | \textbf{\textit{N}})$ is the posterior distribution over $s_{\ln\textit{P}}^2$, which in turn is computed from the posterior distributions of the phase densities and temperatures ($\ln\textit{P}_i = \ln\textit{n}_i + \ln\textit{T}_i$ for $i=1,...,\textit{n}$).  If we define TPE as a constraint on the underlying pressure dispersion of the modeled gas phases, i.e., $\sigma_{\ln\textit{P,true}}^2 < \alpha_\textit{TPE}$ for some threshold value $\alpha_\textit{TPE}$, then a multiphase TPE solution is 1) \textit{possible} if $\textit{P}(\sigma_{\ln\textit{P,true}}^2 < \alpha_\textit{TPE} | \textbf{\textit{N}})>0$ and 2) \textit{likely} if $\textit{P}(\sigma_{\ln\textit{P,true}}^2 < \alpha_\textit{TPE} | \textbf{\textit{N}})>0.5$.  We take $\alpha_\textit{TPE}=0.18$, which is the value of $\sigma_{\ln\textit{P,true}}^2$ when the full width at half maximum (FWHM) of its corresponding \textit{log-normal} distribution is half its mean.  The reader can assess how our interpretation of TPE might change with different values of $\alpha_\textit{TPE}$ based on the information presented in Section~\ref{sec:pressure_gradients}.

\begin{figure*}
\centering
\includegraphics[width=\textwidth]{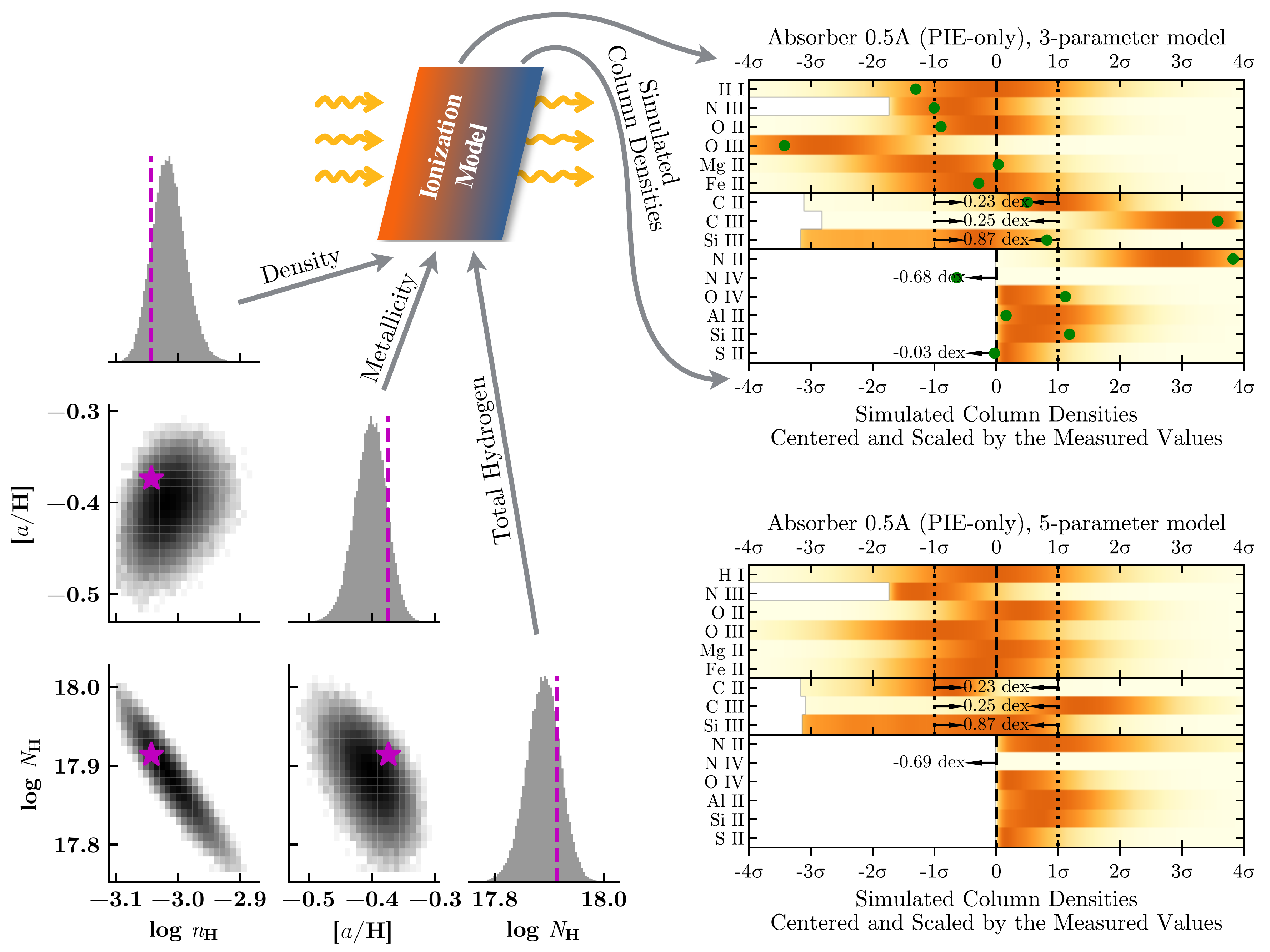}
\caption{A comparison between PP-plots generated for an inadequate 3-parameter single-phase, PIE-only model of Absorber~0.5A (\textit{upper-right}) and a better (see text) 5-parameter version (\textit{lower-right}).  The whole diagram also illustrates how the upper-right PP-plot is related to the posterior distribution of the 3-parameter model.  The magenta stars and dashed lines in the ``corner plot'' (\textit{left}) highlight a random parameter vector sampled via MCMC from the posterior of the 3-parameter model.  Using the ionization model specified by these parameters, we simulate a vector of mock column densities, located as green dots in the upper-right PP-plot.  The horizontal axis for the simulated columns has been standardized by subtracting the actual, measured column density values (i.e., those listed in Table~\ref{tab:z051850_meas}) and scaling by their measurement uncertainties.  Thus it measures the number of (measurement error) standard deviations by which the simulated column densities are in excess of the measured values.  We repeat this procedure 10$^5$ times for independent draws from the posterior to obtain the posterior predictive distributions for individual ions, which we plot as color-coded histograms (horizontal bars) in the upper-right PP-plot.  See the text and Figure~\ref{fig:demo_pdf} for a detailed description of the color mapping and the procedure for generating simulated column densities.  Simulated column densities are only plotted on the constrained side of limits, but we numerically label the logarithmic offset between the peak of the posterior predictive distribution and the measured limit if the peak lies on the the unconstrained side of the limit.  For double bounded limits, the region interior to $\pm1\sigma$ scales logarithmically and the interior numeric gives the logarithmic column density difference between $\pm1\sigma$.  In some cases (e.g., \textsc{N~iii}), there are sharp edges on the left side of the color-coded histograms; this happens when the model reaches the column density = 0 level.}
\label{fig:z051850_example}
\end{figure*}

	The disadvantage of the posterior predictive distribution is that it precludes setting a prior constraint on $\sigma_{\ln\textit{P,true}}^2$ directly.  To allow for such a prior we first set a prior on $s_{\ln\textit{P}}^2$ and make $\sigma_{\ln\textit{P,true}}^2$ a hyperparameter of $\textit{P}(s_{\ln\textit{P}}^2)$.  This allows us to set a hyperprior on $\sigma_{\ln\textit{P,true}}^2$ and use the marginal posterior distribution $\textit{P}(\sigma_{\ln\textit{P,true}}^2 | \textbf{\textit{N}})$ of the hyperparameter $\sigma_{\ln\textit{P,true}}^2$ as a replacement for the posterior predictive distribution expressed in Equation~\ref{eqn:pressurevariancepostpred}.  This approach has the added advantage of requiring us to specify a prior on $s_{\ln\textit{P}}^2$, a derived parameter of the model that would not normally require a prior.  In the absence of an explicit prior, $s_{\ln\textit{P}}^2$ would align with the implicit prior \textit{induced} by the complex intersection of priors on explicit model parameters (indeed, implicit priors are unavoidable for all derived model parameters).  Setting an explicit prior on a derived parameter first requires canceling the induced one.  \cite{Handley19} prove that the maximum entropy prior, i.e., the one that assumes the least information, is achieved with a canceling step where the joint probability of the explicit model parameters is divided by the induced prior on the derived quantity of interest, in this case $s_{\ln\textit{P}}^2$ (see their Equation~3).  We use kernel density estimation (KDE) to capture the prior induced on $s_{\ln\textit{P}}^2$ in 10$^5$ random uniform samples of the explicit parameters.\footnote{The thermal pressure variance depends on just the temperature and density, but in PIE-only models the temperature is itself a derived quantity of the model determined by all the \textsc{cloudy} ionization modeling parameters.  Hence, in general the induced prior depends on all the bounded parameters included in Table~\ref{tab:param_info}.}  After canceling, we multiply by our desired explicit prior, in this case \textit{P}($s_{\ln\textit{P}}^2 | \sigma_{\ln\textit{P,true}}^2$), which is implicitly defined by the inverse of Equation~\ref{eqn:chi2}:
\begin{equation}\label{eqn:inv_chi2}
\frac{\sigma_{\ln\textit{P,true}}^2}{(n-1) s_{\ln\textit{P}}^2} \sim \mathrm{Inv}\text{-}\chi^2(n-1).
\end{equation}
Now, $\sigma_{\ln\textit{P,true}}^2$ is a hyperparameter of the prior on $s_{\ln\textit{P}}^2$, and we are free to assign a hyperprior to it as intended.  Further details are provided in the appendix (see Equation~\ref{eqn:priorfull}).

	Several considerations affect the bounds, weight, and shape of our prior on $\sigma_{\ln\textit{P,true}}^2$.  First, we restrict $\sigma_{\ln\textit{P,true}}^2$ to $0 \leq \sigma_{\ln\textit{P,true}}^2 \leq 3.5^2$.  This range corresponds to the finite support of our KDE for $s_{\ln\textit{P}}^2$, outside of which \textit{P}($s_{\ln\textit{P}}^2 | \sigma_{\ln\textit{P,true}}^2$) quickly becomes negligible.  Next, we set the shape of the prior as a mixture of a Uniform and a Half-Normal distribution and weight them so that $\textit{P}(\sigma_{\ln\textit{P,true}}^2 < \alpha_\textit{TPE} | \textbf{\textit{X}}) = \textit{P}(\sigma_{\ln\textit{P,true}}^2 > \alpha_\textit{TPE} | \textbf{\textit{X}}) = 0.5$, i.e., the prior probabilities of TPE and non-TPE are equal.  This results in a high probability density within the narrow TPE threshold and a low probability density across the much wider non-TPE region (see Section~\ref{sec:pressure_gradients}).

\begin{figure}
\centering
	\begin{subfigure}{0.45\textwidth}
	\centering
	\includegraphics[width=3.0in]{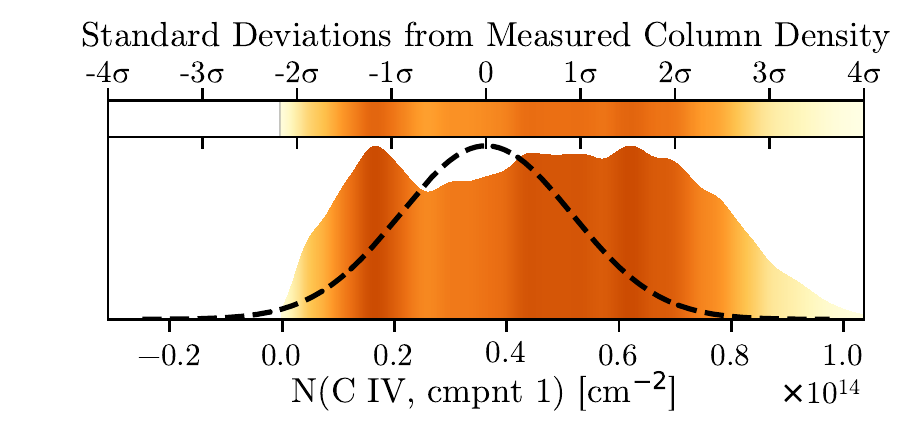}
	\end{subfigure}
	\begin{subfigure}{0.45\textwidth}
	\centering
	\includegraphics[width=3.0in]{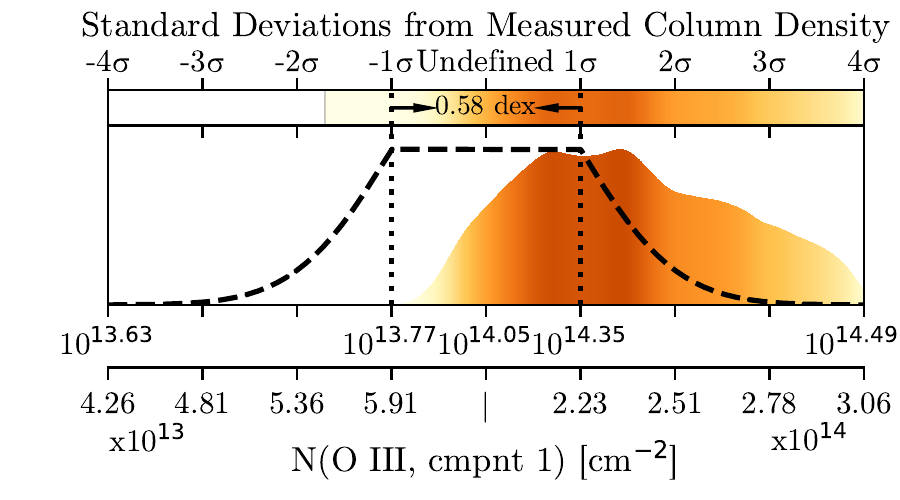}
	\end{subfigure}
	\begin{subfigure}{0.45\textwidth}
	\centering
	\includegraphics[width=3.0in]{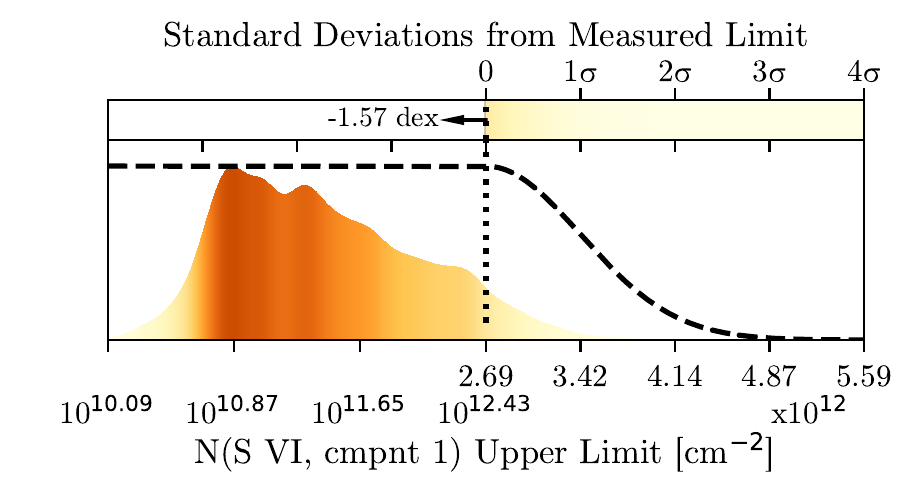}
	\end{subfigure}
\caption{Expanded probability densities for arbitrary C~\textsc{iv} (a), O~\textsc{iii} (b), and S~\textsc{vi} (c) examples.  These are intended to clarify the interpretation of the compact-bars in posterior predictive plots, such as the right side of Figure~\ref{fig:z051850_example} and similar figures throughout this paper and how they compare model-predicted column densities to measurements and limits.  The color distribution encodes peaks and valleys from the smoothed histogram of scaled column density residuals onto a compact color scale.  Dashed black lines illustrate the shape of the assumed uncertainty distribution corresponding to the (mock) column density measurements and limits.  Each panel includes both raw (\textit{bottom}) and scaled (\textit{top}) column density coordinates along the pair of horizontal axes.  The black arrows in the lower two examples are explained in Figure~\ref{fig:z051850_example}.  Column densities on the unconstrained side of the S~\textsc{vi} histogram ($<$0$\sigma$) are logarithmically binned and are ignored by the colorbar.  The O~\textsc{iii} example includes a joint constraint from both a lower and upper limit simultaneously.  In this case, the measured column density has a uniform chance of falling between $\pm$1$\sigma$ (vertical black dotted lines).  Inside these bounds the simulated column densities are logarithmically binned, while the values outside this region are linearly binned.  The histograms have been normalized to reflect a consistent probability value at any given location along the axis, although the relative area under the curve is not conserved due to the differential stretching of the horizontal axis.}
\label{fig:demo_pdf}
\end{figure}


\section{MULTIPHASE IONIZATION MODELING}
\label{sec:results}

	In this section, we investigate the ionization structure of the absorbers presented in Section~\ref{sec:data} using the Bayesian multiphase ionization modeling scheme we describe in Section~\ref{sec:scheme} and Appendix~\ref{app:likelihood}.  Our goal is to find the most probable, physically motivated, ionization equilibrium model that plausibly fits \emph{all} the data while minimizing the number of model parameters.  Beginning with a model consisting of a single phase of photoionized gas, we incrementally add discrete gas phases or alter the ionization mechanism (either PIE-only or P+C, see Section~\ref{sec:scheme}) of a given phase until we achieve an adequate fit.  We use posterior predictive \textit{p}-values to reject models that cannot plausibly reproduce the measured column densities, and we compute Bayes factors to weight and select the ``best'' (i.e., most probable) of the remaining models.  We show that a single-phase model cannot reproduce the measured column densities in four of the five discrete absorption components that we probe in detail, and we explore various two- and three-phase alternatives.  Crucially for what follows, we refer to each resolved velocity component as a unique ``absorber,'' reserving the term ``absorption system'' for groups of absorbers at a common systemic redshift along a given line-of-sight. While this is somewhat different from typical usage in the literature, it is helpful in this paper because a given component can consist of multiple phases that are modeled together and it is convenient to refer to this entity of kinematically aligned phases as an ``absorber''.

	We will use Absorber~0.5A (PG1522+101, see Fig.~\ref{fig:z051850_vpfits}) to illustrate some of our analysis methods in the next section. For convenient reference in this discussion, we summarize in Table~\ref{tab:selection_metrics_z051850} some of the details of the models we will consider for this case.


\subsection{Model Selection}
\label{sec:residual_checks}
	In Section~\ref{sec:model_selection_metrics} we presented two quantitative model selection statistics, Bayes factors and PP \textit{p}-values.  Here we introduce posterior predictive plots (PP-plots) as graphical aids in evaluating the quality of our models.  These PP-plots are designed to visually demonstrate the significance of discrepancies between measured and simulated columns for individual ions.

	Figure~\ref{fig:z051850_example} introduces this type of plot using real data and models for Absorber~0.5A.  The panels on the left side of the figure show marginal posterior distributions, plotted in the typical ``corner plot'' fashion, of parameters from a 3-parameter model derived for this absorber.  Parameter vectors randomly sampled from the model posterior are passed to the (potentially multiphase) ionization model to derive modeled column densities.  On the upper-right side of Figure~\ref{fig:z051850_example}, we plot orange color-coded histograms of simulated column densities drawn from the 3-parameter posterior predictive distributions (see Appendix~\ref{app:ppp} for details) of the observed species.  The histograms' horizontal axes are centered at the \textsl{measured} column density values corresponding to each observed species and scaled by the corresponding measurement $1\sigma$ uncertainty.  The variation in color intensity along each horizontal bar maps to the shape of the respective histogram (of simulated column densities), with the darkest regions corresponding to peak prediction (probability) density.  In an ideal fit the darkest orange region should be centered at zero and predominantly fall within the $\pm1\sigma$ measurement uncertainties.  Figure~\ref{fig:demo_pdf} provides three examples to illustrate how the color maps to probability density in various scenarios (including limits).

	Figure~\ref{fig:z051850_example} also illustrates the process of generating simulated column densities from the posterior predictive distribution.  A single simulated column density begins with a draw from the posterior distribution (left side of Figure~\ref{fig:z051850_example}), which is then passed through the ionization model and converted into a vector of column densities (one for each species of interest).  While the model-predicted column densities are indeed useful (e.g., for interpreting fractional degrees of ionization, as we do below), they lack a meaningful standard of comparison.  Simulated column densities, however, do have a clear basis of comparison, i.e., they should look ``similar'' to the true measured ones.  In Appendix~\ref{app:ppp}, for example, we introduce one possible metric to quantify this similarity in order to compute PP \textit{p}-values.  To sample the posterior predictive distribution of a given species, we draw one simulated column density from the ZTN error model (see Section~\ref{sec:formalism} and Appendix~\ref{app:likelihood}) that takes a model-predicted column density and the measurement uncertainty of the corresponding measured column as its location and scale parameters.  We repeat this step 10$^5$ times, each time using a different model-predicted column density derived as a random draw from the model posterior.  The resulting simulated column densities are drawn from the posterior predictive distribution.

\begin{figure}
\centering
\includegraphics[width=3.2in]{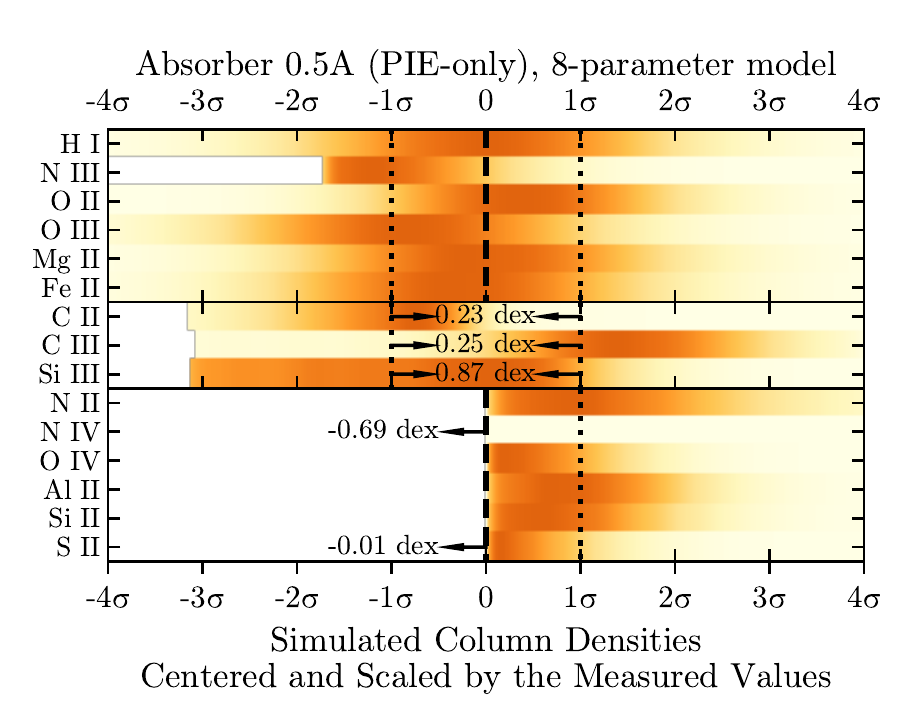}
\caption{Scaled residuals for an 8-parameter single-phase, PIE-only model of Absorber~0.5A.  This model varies the same density, metallicity, and total hydrogen column density parameters as the 3-parameter model (see Figure~\ref{fig:z051850_example}) plus the C/$\alpha$ and N/$\alpha$ relative abundance parameters added by the 5-parameter model (also in Figure~\ref{fig:z051850_example}), as well as the $\alpha$/Fe, $\alpha_\textrm{UV}$, and $\beta_\textrm{He\ II}$ parameters.}
\label{fig:z051850_8par_residual}
\end{figure}

	Inspection of the upper-right panel of Figure~\ref{fig:z051850_example} reveals that for Absorber 0.5A, the 3-parameter PIE model's posterior predictive distributions for N~\textsc{ii} and C~\textsc{iii} both \textsl{overestimate} their respective measured columns by $\approx3$$\sigma$ of the measurement uncertainty, while the prediction for O~\textsc{iii} skews to the left by $\approx3$$\sigma$ (i.e., \textsl{underestimating} the measured columns).  These discrepancies do not extend to the posterior predictive distributions for the remaining detected species (where we count C~\textsc{ii} and Si~\textsc{iii} as detections), nor to the Al~\textsc{ii}, Si~\textsc{ii}, S~\textsc{ii}, N~\textsc{iv}, and O~\textsc{iv} columns, which are all less than or approximately equal to their respective upper limits\footnote{Note that the portion of the posterior predictive histograms lying on the unconstrained side of limits are not plotted in our PP-plots, though the full histogram is used to determine the correct color mapping on the constrained side of the limits, as exemplified in Figure~\ref{fig:demo_pdf}.}.  The bottom-right panel of Figure~\ref{fig:z051850_example} shows the PP-plot for a slightly more complex, 5-parameter model of the same absorber.  This PP-plot improves noticeably in predictive consistency compared to the three-parameter model.  An 8-parameter model, shown in Figure~\ref{fig:z051850_8par_residual}, failts to improve over the 5-parameter model, however, despite its additional model flexibility.  The degree to which one of these competing models is preferred over the others can be discerned using the quantitative selection metrics outlined in Section~\ref{sec:model_selection_metrics}, but the advantage of the PP-plots is their ability to pinpoint patterns underlying the quantitative assessments.

\ctable[
	caption={Selection Metrics for Models of Absorber 0.5A},
	doinside = \footnotesize,
	width=\columnwidth,
	label={tab:selection_metrics_z051850}
	]{r@{--}l@{~}c@{~}cr@{$\pm$}lcc}{
\tnote[a]{The model identifier; the left side follows the shorthand convention discussed in Section~\ref{sec:formalism}.  The right side lists any elements for which we assigned a free (relative abundance) parameter.}
\tnote[b]{One checkmark indicates $\alpha_\textrm{UV}$ was a free parameter in the model, two checkmarks means the same for $\beta_\textrm{He II}$, and models with both parameters fixed (at -1.4, 1, respectively) get no checkmark.  Note that checkmarks do not convey the same meaning in analogous tables for the other absorbers.}
\tnote[c]{The number of free parameters in the model.}
\tnote[d]{Logarithmic Bayes factors, i.e., the ratio of each model's marginal likelihood $\log Z_i$ to that of the highest ranked model $\log Z_{max}$ (in this case, P$_\textrm{5par}$, with C, N varied): $\log K_i = \log (Z_i/Z_{max})$.  Models meeting our selection criterion, $\log K>-1.5$, lie above the midline.}
\tnote[e]{Posterior predictive \textit{p}-values. Models with PP \textit{p}-values $<$0.05 are rejected.}
\tnote[f]{``Rej'' indicates that the model is rejected.}
}{
\FL
\multicolumn{2}{c}{Model ID\tmark[a]} & UV par.\tmark[b] & $N_\textrm{pars}$\tmark[c] & \multicolumn{3}{c}{Selection Metrics} & Note\tmark[f] \NN
\cmidrule(r){5-7}
\multicolumn{2}{c}{} & & & \multicolumn{2}{c}{$\log_{10} K$\tmark[d]} & PP$p$\tmark[e] \ML

  PPa & C,N   &              &  7 &  +0.10 & 0.02 & 0.312 &     \NN
  PPa & C,N   &              &  7 &  +0.06 & 0.05 & 0.319 &     \NN
   Pa & C     &              &  4 &   0.00 & 0.04 & 0.151 &     \NN
   Pa & C,N   &              &  5 &  -0.07 & 0.01 & 0.328 &     \NN
   Pa & C,N,Fe&              &  6 &  -0.78 & 0.03 & 0.275 &     \NN
   Pa & C,Fe  &              &  5 &  -0.97 & 0.01 & 0.125 &     \NN
  PCa & C,N,Fe& \checkmark\checkmark & 11 &  -1.07 & 0.05 & 0.322 &     \NN
   Pa & C     & \checkmark   &  5 &  -1.13 & 0.02 & 0.118 &     \NN
  PPa & C,N,Fe& \checkmark\checkmark & 10 &  -1.22 & 0.01 & 0.255 &     \NN
   Ca & C,N   &              &  6 &  -1.35 & 0.04 & 0.297 &     \NN
   Pa & C,Fe  & \checkmark   &  6 &  -1.35 & 0.01 & 0.120 &     \NN
 \cmidrule(r){5-6}
   Pa & C,N,Fe& \checkmark\checkmark &  8 &  -1.54 & 0.03 & 0.241 &     \NN
  PPa &       &              &  5 &  -1.63 & 0.05 & 0.029 & Rej \NN
  PPa & C,N,Fe& \checkmark\checkmark & 10 &  -1.65 & 0.03 & 0.253 &     \NN
  PPa &       &              &  5 &  -1.86 & 0.02 & 0.032 & Rej \NN
  PPa &       &              &  6 &  -1.93 & 0.03 & 0.035 & Rej \NN
   Ca & C,Fe  &              &  6 &  -2.14 & 0.06 & 0.117 &     \NN
   Pa & C,N   &              &  5 &  -2.49 & 0.21 & 0.033 & Rej \NN
   Ca & C,N,Fe& \checkmark\checkmark &  9 &  -2.52 & 0.65 & 0.236 &     \NN
   Pa &       &              &  3 &  -2.73 & 0.02 & 0.003 & Rej \NN
   Pa & N     &              &  4 &  -2.86 & 0.07 & 0.012 & Rej \NN
   Pa & C,N   &              &  5 &  -3.36 & 0.55 & 0.009 & Rej \NN
   Pa &       & \checkmark   &  4 &  -3.62 & 0.40 & 0.004 & Rej \NN
   Pa &       & \checkmark\checkmark &  5 &  -3.68 & 0.46 & 0.004 & Rej \NN
   Ca & C     &              &  4 &  -3.83 & 0.56 & 0.007 & Rej \LL
}

\subsection{The Single Phase Model}
\label{sec:multiphase_required}

	Among the five absorbers in this study, only Absorber~0.5A (PG1522+101, see Fig.~\ref{fig:z051850_vpfits}) is adequately modeled with a single gas phase, but this absorber also has the fewest metal detections.  Its spectral profiles show robust low-ion absorption but no trace of high-ionization stages.  Despite the simplicity of this absorber, it is poorly fit by a basic 3-parameter model, where the density, overall metallicity, and total hydrogen column density are the only parameters.  The PP-plot in the top-right of Figure~\ref{fig:z051850_example} shows this model, and we can see why it falls short, as discussed above. The PP \textit{p}-value of 0.003 for this model (see Table~\ref{tab:selection_metrics_z051850}) is less than our adopted $\alpha$=0.05 significance threshold for rejection.  If we introduce some additional model flexibility, allowing the carbon and nitrogen relative abundances to vary, the PP-plot improves significantly (Figure~\ref{fig:z051850_example}, \textsl{bottom-right}).  We compare this to the 8-parameter model shown in Figure~\ref{fig:z051850_8par_residual}, which produces almost identical posterior predictive distributions despite the introduction of three further free parameters (the iron relative abundance plus two UV parameters for eight total parameters).

	While neither the 5- nor 8-parameter model can be rejected on the basis of their (satisfactory) PP \textit{p}-values (0.328 and 0.241 respectively), the 5-parameter model \textit{seems} to be strongly preferred over the 8-parameter model by a Bayes factor of $10^{1.47}$ (i.e., $10^{-0.07}/10^{-1.54}$ the reciprocal of the ratio implicit in Table~\ref{tab:selection_metrics_z051850}), though this is slightly less than our decision threshold of $10^{1.5}$ -- see Section~\ref{sec:model_selection_metrics}.  The comparison is ill-advised in this case, however, because Bayes factors are ideal for comparing mutually exclusive models, whereas the 5-parameter model is really an instance of the 8-parameter model with three fixed parameters \textit{arbitrarily} selected from a continuous parameter space in the fuller model.\footnote{If we could make a strong argument that our particular choice of fixed parameters is in some way a special combination when reducing from eight to five parameters, then perhaps Bayes factors would be appropriate for comparing these two models.}  Instead, it is preferable to evaluate the 5-parameter model that results from marginalizing over the three extra parameters of the 8-parameter model, which is what we choose to do in extracting our inferences about this absorber.  In this particular case, the choice between models would hardly impact the parameteric inferences, since there are no significant inconsistencies between the two models with respect to the marginal distributions of their five common parameters.  Note, however, that comparing between models with P+C vs. PIE-only ionization mechanisms is an appropriate use of the Bayes factor, since these are exclusive models.  In fact, we find that the P+C 6-parameter model (which adds the gas temperature as a free parameter to the PIE-only 5-parameter model) is disfavored relative to the PIE-only model by a Bayes factor of $\approx10^{-1.28}$, though this is not large enough to be decisive.  Interestingly, the P+C model prefers cooler temperatures anyway; in this temperature regime, photoionization dominates over collisional ionization and the P+C model is very similar to the PIE-only model.

Looking further into the plethora of model variations listed in Table~\ref{tab:selection_metrics_z051850}, several things stand out.  First, all the models rejected by their PP~\textit{p}-value (the two rightmost columns) are likewise rejected on account of their Bayes factors (4th column, below the midline).  Second, the viable models appear to include a mix of single-phase and two-phase models.  This is somewhat misleading, because we intuitively assign a higher prior probability to the single-phase models, which would boost the posterior odds of the single-phase models relative to the two-phase ones.  This model prior is not accounted for by Table~\ref{tab:selection_metrics_z051850}.  Hence, we listed the best single-phase model as the reference model (i.e., where log$_{10}$ \textit{K}=0) even though there are a pair of two-phase models that are ranked slightly higher on the basis of their Bayes factors alone.  The constrast between two-phase and single phase models is also misleading because the posterior distributions of the two-phase models for this absorber do not deviate significantly from that of the single-phase models (which themselves are very similar).  For example, the two-phase model includes two density parameters, one of which has the same distribution as the single-phase density, while the other has a flat (uniform) distribution across the explorable parameter space implying that it is unconstrained by the data.

A third interesting insight from Table~\ref{tab:selection_metrics_z051850} is that the more dramatic improvements in the models for Absorber~0.5A come from varying the relative abundance parameters for C and/or N in lieu of the UV radiation field parameters (indicated by checkmarks in the second column) rather than vice versa.  This seemingly stands in contrast to the models for the other four absorbers (discussed below) which show much higher sensitivity to $\alpha_\textrm{UV}$ in particular.  Upon closer inspection, however, we find that the fiducial value of $\alpha_\textrm{UV}=-1.4$ assumed in models where we fix the parameter happens, by chance, to be close to the optimal one.  Forcing $\alpha_\textrm{UV}$ to be harder or softer (taking values of -1 or -2 respectively) in a 5-parameter model (with C/$\alpha$ and N/$\alpha$ free) results in much poorer fits, with PP \textit{p}-values $<$0.1.  Note that the 8-parameter model naturally arrives at this conclusion, with a peak in the marginal distribution for $\alpha_\textrm{UV}$ near the fiducial value.  Thus, even this simple, single-phase absorber offers some constraint on the UV radiation field.

\ctable[
	caption={Selection Metrics for Models of Absorbers~0.4A \&~0.4B},
	doinside = \footnotesize,
	width = \columnwidth,
	label={tab:selection_metrics_z041761}
	]{r@{--}l@{~}c@{~}cr@{$\pm$}lcc}{
\tnote[a]{The model identifier follows the shorthand convention discussed in Section~\ref{sec:formalism}.}
\tnote[b]{A checkmark indicates that each absorber (0.4A \&~0.4B) was modeled with an independent set of UV parameters.}
\tnote[c--f]{These tablenotes follow Table~\ref{tab:selection_metrics_z051850}, except that here the highest ranked model is a ``joint UV'' PPPa--CCCb model.}
}{
\FL
\multicolumn{2}{c}{Model ID\tmark[a]} & Ind. UV\tmark[b] & $N_\textrm{pars}$\tmark[c] & \multicolumn{3}{c}{Selection Metrics} & Note\tmark[f] \NN
\cmidrule(r){5-7}
\multicolumn{2}{c}{} & & & \multicolumn{2}{c}{$\log_{10} K$\tmark[d]} & PP$p$\tmark[e] \ML

 PPPa &  CCCb &              & 27 &   0.00 & 0.41 & 0.625 &     \NN
 PCCa &  CCCb &              & 29 &  -0.31 & 0.40 & 0.541 &     \NN
 PPCa &  CCCb &              & 28 &  -0.55 & 0.40 & 0.611 &     \NN
 CCCa &  CCCb &              & 30 &  -1.31 & 0.78 & 0.447 &     \NN
 PPPa &  CCCb &  \checkmark  & 29 &  -1.45 & 0.78 & 0.613 &     \NN
 \cmidrule(r){5-6}
 PCCa &  PCCb &              & 28 &  -1.70 & 0.56 & 0.381 &     \NN
 PCCa &  PPPb &              & 26 &  -1.85 & 0.37 & 0.468 &     \NN
 PCCa &  CCCb &  \checkmark  & 31 &  -2.00 & 1.30 & 0.537 &     \NN
 PPCa &  PCCb &              & 27 &  -2.47 & 0.43 & 0.368 &     \NN
 PPPa &  PCCb &              & 26 &  -2.59 & 0.88 & 0.363 &     \NN
 PCCa &  PPCb &              & 27 &  -2.96 & 0.51 & 0.487 &     \NN
 PPCa &  CCCb &  \checkmark  & 30 &  -3.40 & 1.01 & 0.591 &     \NN
  PCa &  CCCb &              & 26 &  -3.51 & 0.66 & 0.356 &     \NN
  PPa &   PPb &  \checkmark  & 22 &  -3.52 & 0.39 & 0.307 &     \NN
  PCa &  CCCb &  \checkmark  & 28 &  -3.62 & 0.28 & 0.349 &     \NN
  PCa &  PCCb &              & 25 &  -3.72 & 0.30 & 0.148 &     \NN
  PCa &   PPb &              & 21 &  -4.27 & 0.23 & 0.254 &     \NN
 PPPa &  PPPb &              & 24 &  -4.43 & 0.66 & 0.317 &     \NN
  PCa &  PPPb &              & 23 &  -4.68 & 0.16 & 0.213 &     \NN
 PPCa &  PPCb &              & 26 &  -5.06 & 0.78 & 0.274 &     \NN
  PCa &  PPCb &              & 24 &  -5.35 & 0.37 & 0.225 &     \NN
  PPa &   PCb &  \checkmark  & 23 &  -5.37 & 0.42 & 0.219 &     \NN
 PPCa &  PPPb &              & 25 &  -5.55 & 1.16 & 0.292 &     \NN
  PPa &   PPb &              & 20 &  -6.27 & 0.18 & 0.105 &     \NN
 PPPa &  PPCb &              & 25 &  -6.38 & 0.36 & 0.311 &     \NN
  PCa &   PPb &  \checkmark  & 23 &  -8.17 & 0.50 & 0.318 &     \NN
  PPa &   PCb &              & 21 &  -8.31 & 0.39 & 0.081 &     \NN
  PCa &   PCb &  \checkmark  & 24 &  -8.34 & 0.63 & 0.234 &     \NN
  PCa &   PCb &              & 22 & -10.76 & 0.78 & 0.116 &     \NN
   Pa &    Pb &  \checkmark  & 16 & -21.50 & 0.10 & 0.000 & Rej \NN
   Pa &    Pb &              & 14 & -46.73 & 0.01 & 0.000 & Rej \LL
}

\ctable[
	caption={Selection Metrics for Models of Absorbers~0.6A \&~0.6B},
	doinside = \footnotesize,
	width = \columnwidth,
	label={tab:selection_metrics_z068605}
	]{r@{}c@{}l@{~}c@{~}cr@{$\pm$}lcc}{
\tnote[a]{The model identifier follows the shorthand convention discussed in Section~\ref{sec:formalism}.}
\tnote[b]{A checkmark indicates that each absorber (0.6A \&~0.6B) was modeled with an independent set of UV parameters.}
\tnote[c--f]{These tablenotes follow Table~\ref{tab:selection_metrics_z051850}, except that here the highest ranked model is a ``joint UV'' PCCa--CCCb model.}
}{
\FL
\multicolumn{3}{c}{Model ID\tmark[a]} & Ind. UV\tmark[b] & $N_\textrm{pars}$\tmark[c] & \multicolumn{3}{c}{Selection Metrics} & Note\tmark[f] \NN
\cmidrule(r){6-8}
\multicolumn{3}{c}{} & & & \multicolumn{2}{c}{$\log_{10} K$\tmark[d]} & PP$p$\tmark[e] \ML

 PCCa & - &  CCCb &              & 30 &   0.00 & 0.27 & 0.273 &     \NN
 CCCa & - &  CCCb &              & 31 &  -0.69 & 0.39 & 0.255 &     \NN
 PPPa & - &  CCCb &              & 28 &  -0.69 & 0.17 & 0.324 &     \NN
 PCCa & - &  CCCb &  \checkmark  & 32 &  -0.71 & 0.74 & 0.270 &     \NN
 PPCa & - &  CCCb &              & 29 &  -0.98 & 0.34 & 0.310 &     \NN
 PPPa & - &  CCCb &  \checkmark  & 30 &  -1.46 & 0.31 & 0.225 &     \NN
 \cmidrule(r){6-7}
  PPa & - &  CCCb &              & 26 &  -1.94 & 0.45 & 0.178 &     \NN
  PPa & - &  CCCb &  \checkmark  & 28 &  -2.75 & 0.41 & 0.155 &     \NN
 CCCa & - &  CCCb &  \checkmark  & 33 &  -2.79 & 0.31 & 0.217 &     \NN
 PPCa & - &  CCCb &  \checkmark  & 31 &  -2.98 & 1.02 & 0.253 &     \NN
 PCCa & - &  PCCb &              & 29 &  -3.17 & 0.46 & 0.140 &     \NN
 CCCa & - &  PCCb &              & 30 &  -3.22 & 0.55 & 0.146 &     \NN
 PPPa & - &  PCCb &              & 27 &  -3.68 & 0.49 & 0.145 &     \NN
 PPCa & - &  PCCb &              & 28 &  -4.36 & 0.38 & 0.143 &     \NN
  PCa & - &  CCCb &              & 27 & -13.65 & 0.25 & 0.021 & Rej \NN
  PPa & - &   PCb &  \checkmark  & 24 & -16.32 & 0.25 & 0.001 & Rej \NN
  PCa & - &   PCb &  \checkmark  & 25 & -16.83 & 0.26 & 0.001 & Rej \NN
  PPa & - &   PCb &              & 22 & -20.98 & 0.11 & 0.000 & Rej \NN
  PCa & - &   PCb &              & 23 & -23.68 & 1.02 & 0.000 & Rej \NN
   Pa & - &    Pb &  \checkmark  & 17 & -62.01 & 0.07 & 0.000 & Rej \NN
   Pa & - &    Pb &              & 15 & -73.13 & 0.06 & 0.000 & Rej \ML
      &   &  CCCb &              & 16 &   0.00 & 0.27 & 0.225 &     \NN
 \cmidrule(r){6-7}
      &   &  PCCb &  \checkmark  & 15 &  -2.30 & 0.39 & 0.317 &     \NN
      &   &  PPCb &  \checkmark  & 14 & -10.18 & 0.17 & 0.012 & Rej \NN
      &   &  PPPb &  \checkmark  & 13 & -14.30 & 0.17 & 0.000 & Rej \NN
      &   &  PCb  &  \checkmark  & 12 & -14.55 & 0.34 & 0.000 & Rej \NN
      &   &  PPb  &  \checkmark  & 11 & -14.92 & 0.31 & 0.000 & Rej \NN
      &   &  CCb  &  \checkmark  & 13 & -15.37 & 0.74 & 0.000 & Rej \ML
 PPPa &   &       &  \checkmark  & 14 &   0.00 & 0.27 & 0.261 &     \NN
 PPCa &   &       &  \checkmark  & 15 &  -0.37 & 0.39 & 0.232 &     \NN
 \cmidrule(r){6-7}
 PCCa &   &       &  \checkmark  & 16 &  -1.71 & 0.17 & 0.200 &     \NN
 PPa  &   &       &  \checkmark  & 12 &  -3.11 & 0.34 & 0.005 & Rej \NN
 PCa  &   &       &  \checkmark  & 13 &  -4.76 & 0.31 & 0.003 & Rej \NN
 CCCa &   &       &  \checkmark  & 17 &  -5.39 & 0.74 & 0.046 & Rej \NN
 CCa  &   &       &  \checkmark  & 14 &  -7.86 & 0.45 & 0.005 & Rej \LL
}

\begin{figure*}
\centering
\begin{subfigure}{.32\textwidth}
	\centering
	\includegraphics[width=1.0\textwidth]{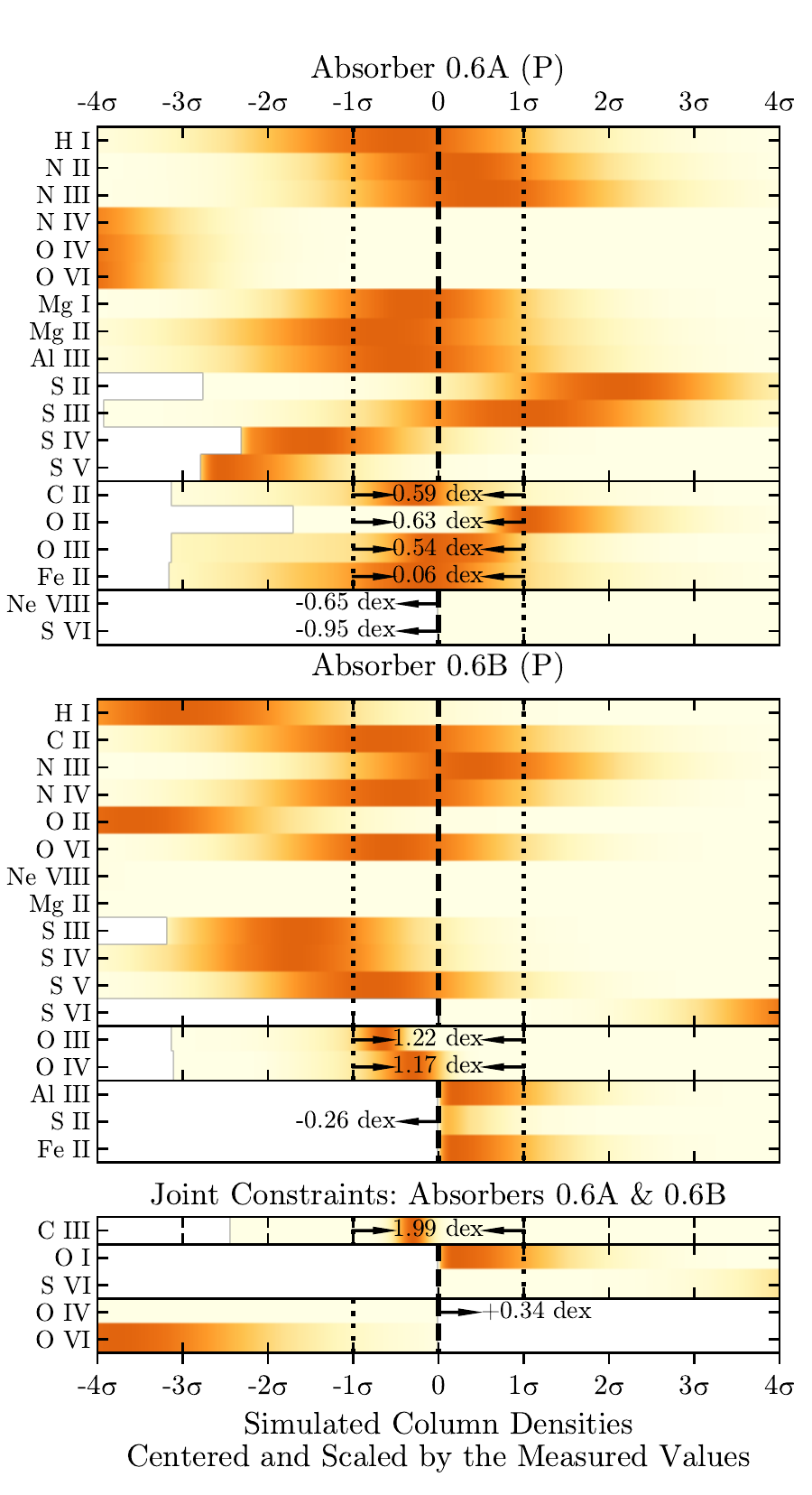}
	\caption{}
	\label{fig:z068605_singlephase}
\end{subfigure}%
\begin{subfigure}{.32\textwidth}
	\centering
	\includegraphics[width=1.0\textwidth]{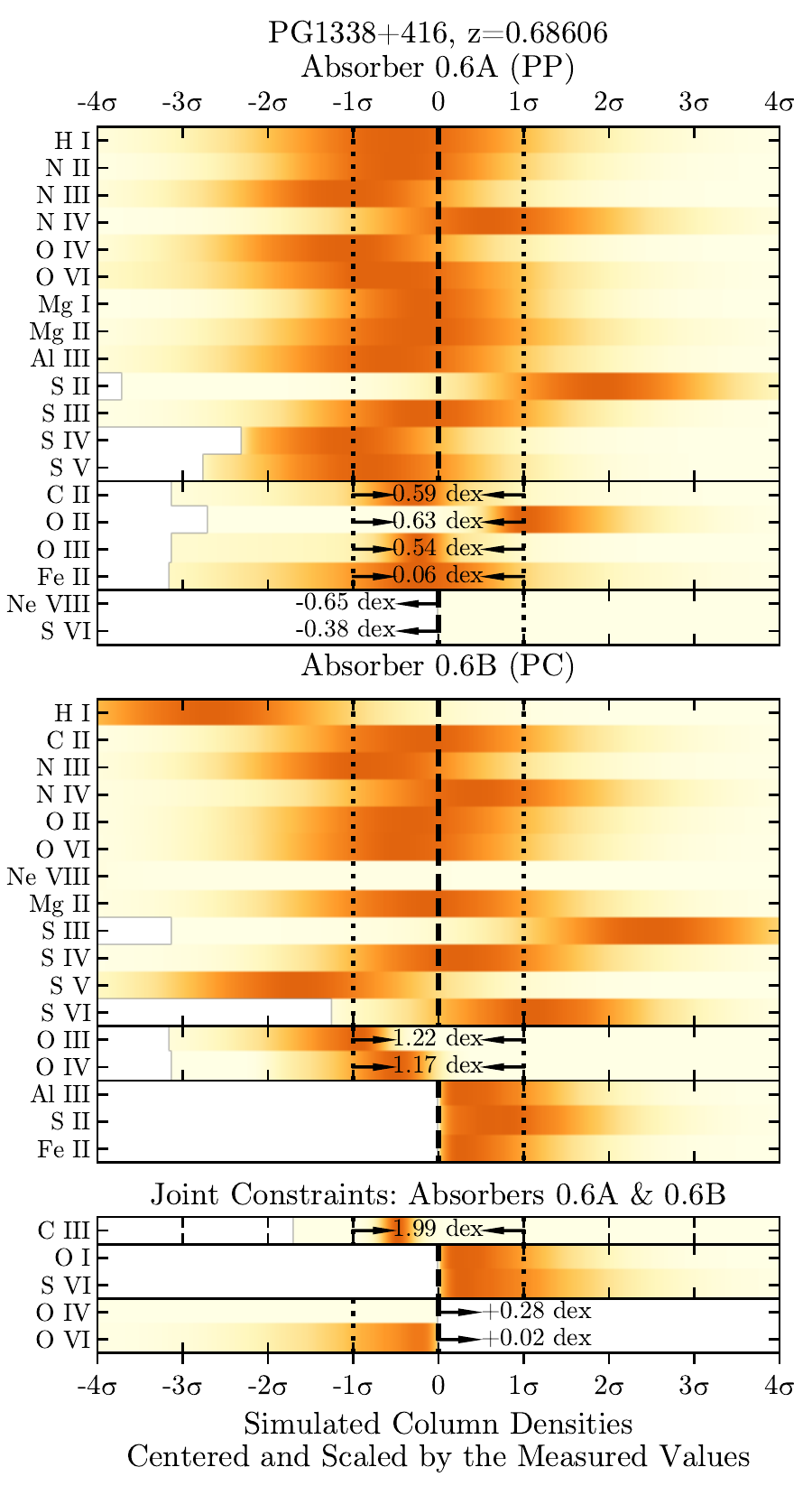}
	\caption{}
	\label{fig:z068605_twophase}
\end{subfigure}
\begin{subfigure}{.32\textwidth}
	\centering
	\includegraphics[width=1.0\textwidth]{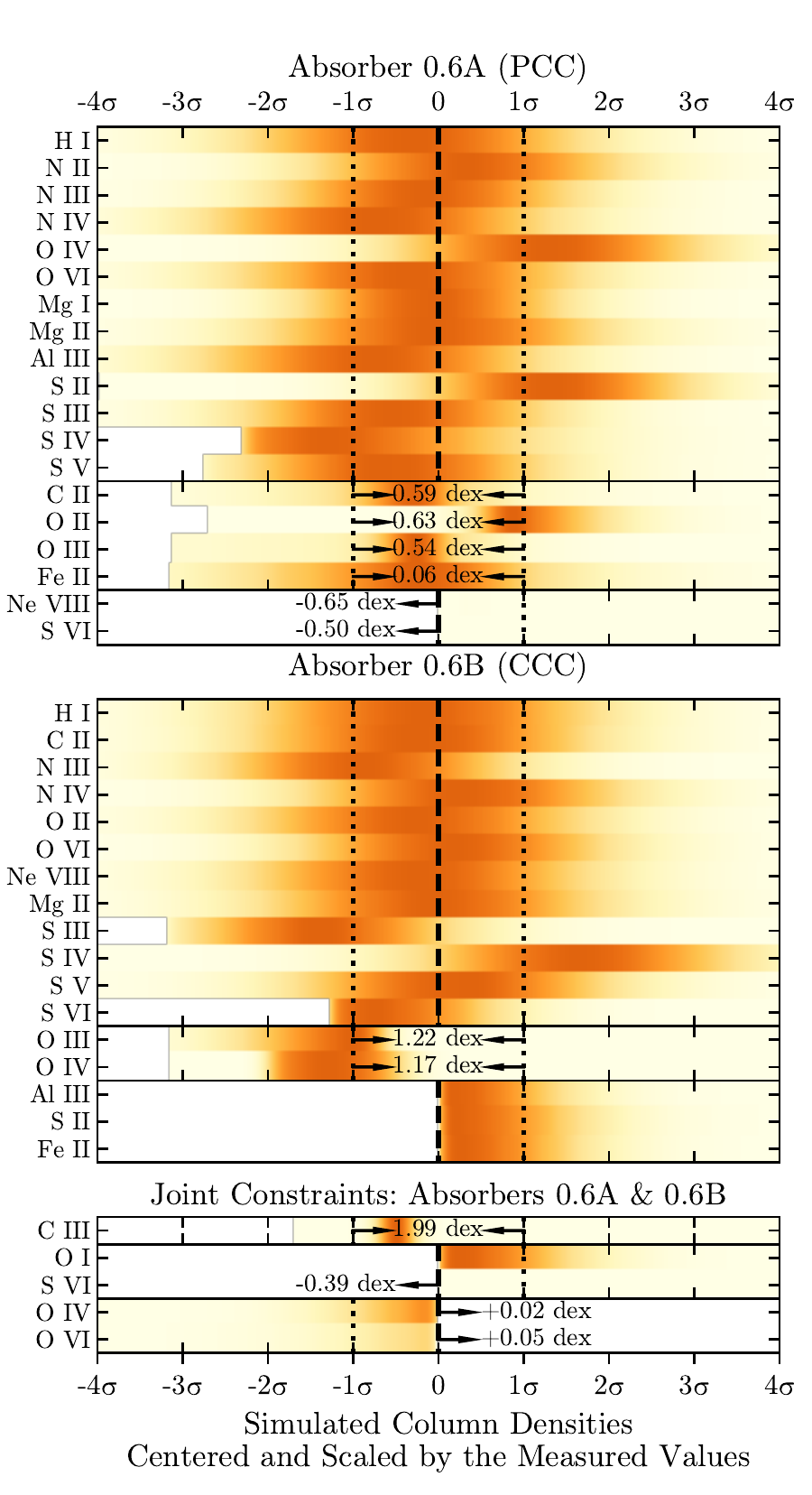}
	\caption{}
	\label{fig:z068605_threephase}
\end{subfigure}
\caption{PP-plots for ions/limits detected in Absorber~0.6A (\textit{top panels}), Absorber~0.6B (\textit{middle panels}), and where the measurement could not be resolved into either absorber (\textit{bottom panels}) are shown for selected models of increasing complexity (from left to right).  These types of plots are explained in detail in Figure~\ref{fig:z051850_example} and Section~\ref{sec:residual_checks}.  The single-phase model (\textit{a}) cannot adequately account for more than half of the species in Absorbers~0.6A and~0.6B.  A two-phase model (\textit{b}), which also assigns independent UV parameters to the two absorbers does better, yet still misses H~\textsc{i}, S~\textsc{iii}, S~\textsc{v}, and Ne~\textsc{viii} in Absorbers~0.6B.  The addition of a third phase for Absorber~0.6B (\textit{c}) results in a statistically acceptable model (PP \textit{p}-values and Bayes factors for these models are listed in Table~\ref{tab:selection_metrics_z068605}) without requiring independent UV parameters for each absorber.  The letters `P' and `C' in the panel titles reference the number of phases used and their ionization mechanism.  See Section~\ref{sec:formalism} for a review of our convention for model specification.}
\label{fig:z068605_PPplot}
\end{figure*}

\begin{figure*}
\centering
\begin{subfigure}{.32\textwidth}
	\centering
	\includegraphics[width=1.0\textwidth]{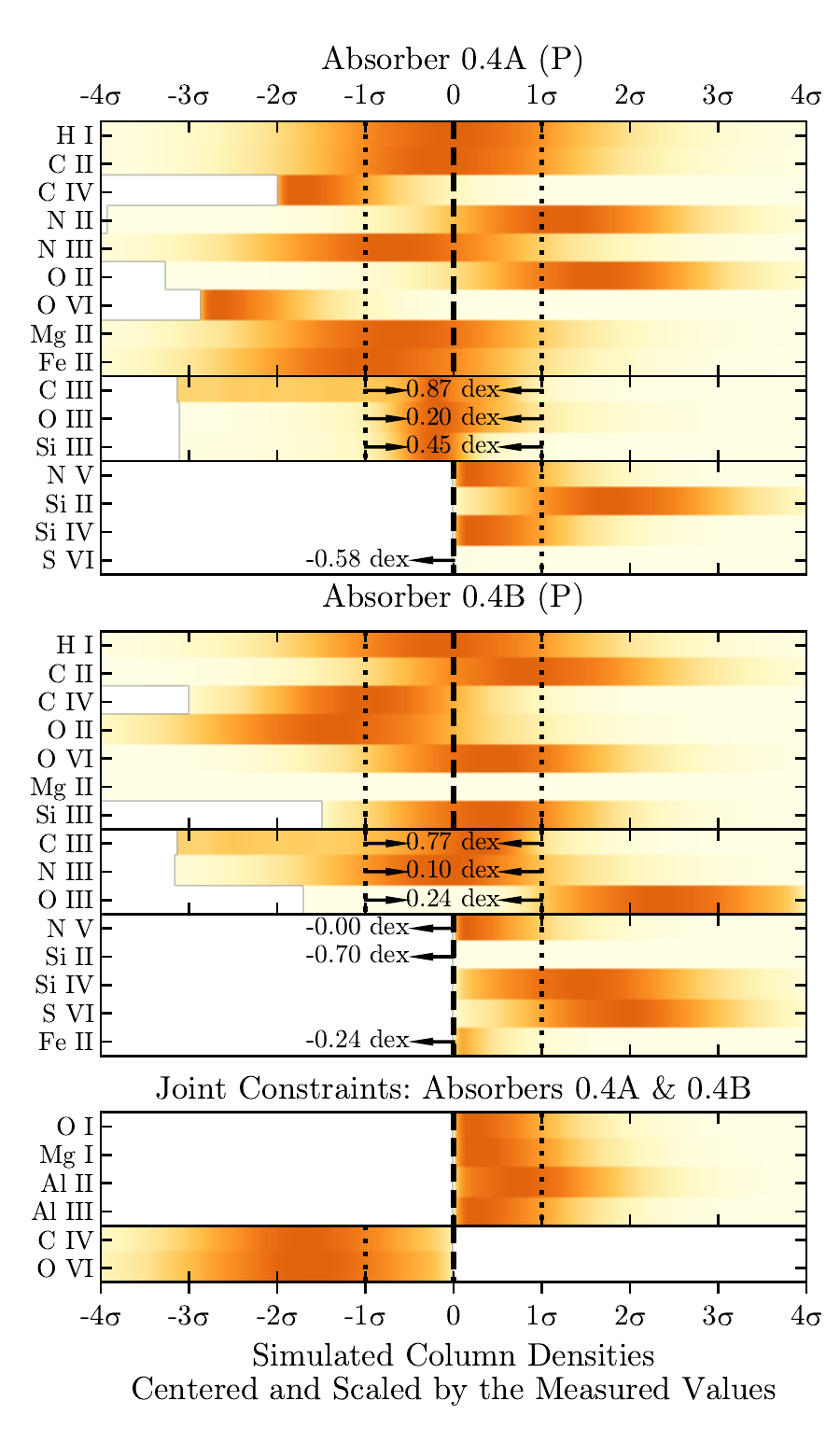}
	\caption{}
	\label{fig:z041761_singlephase}
\end{subfigure}%
\begin{subfigure}{.32\textwidth}
	\centering
	\includegraphics[width=1.0\textwidth]{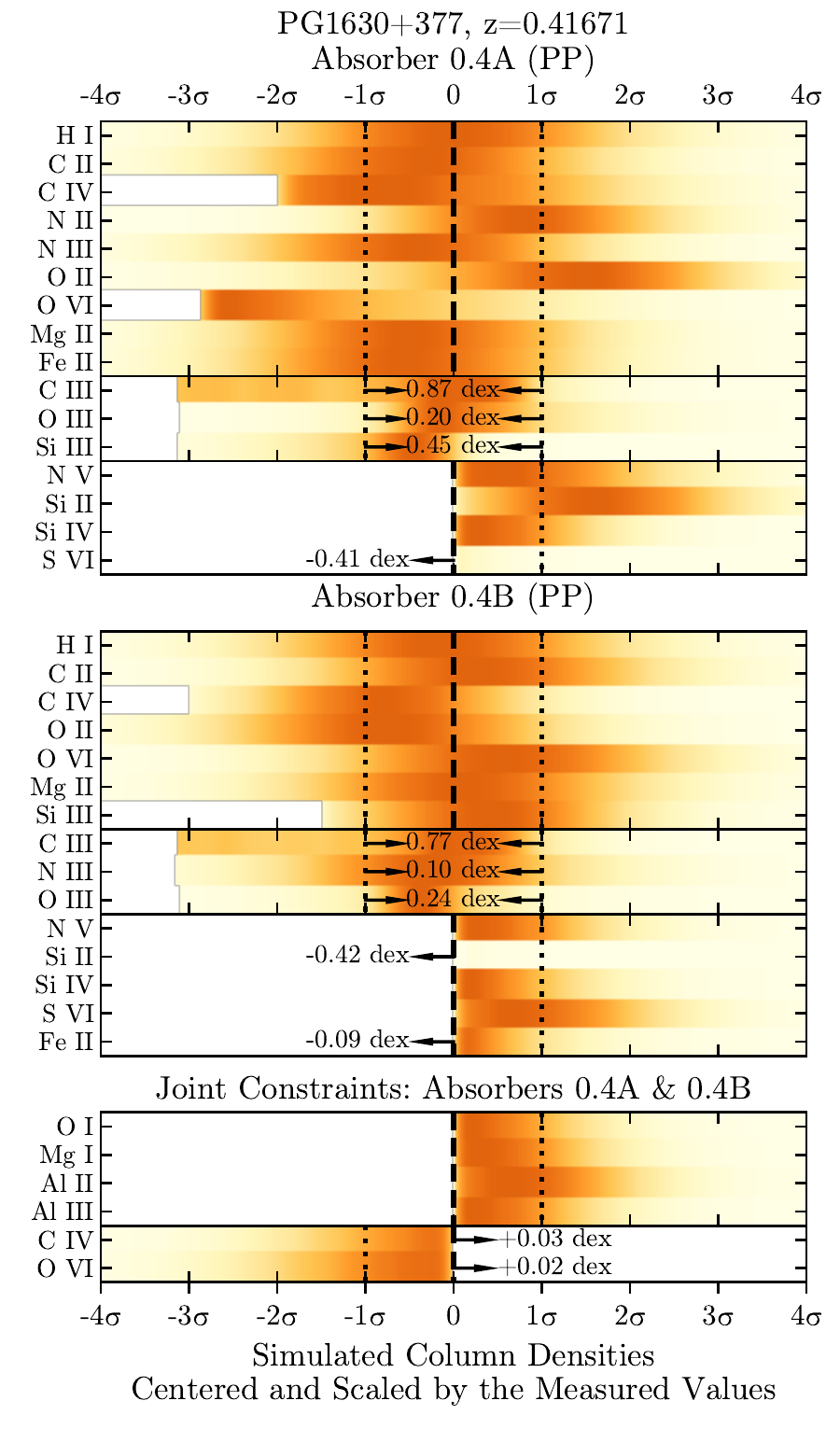}
	\caption{}
	\label{fig:z041761_twophase}
\end{subfigure}
\begin{subfigure}{.32\textwidth}
	\centering
	\includegraphics[width=1.0\textwidth]{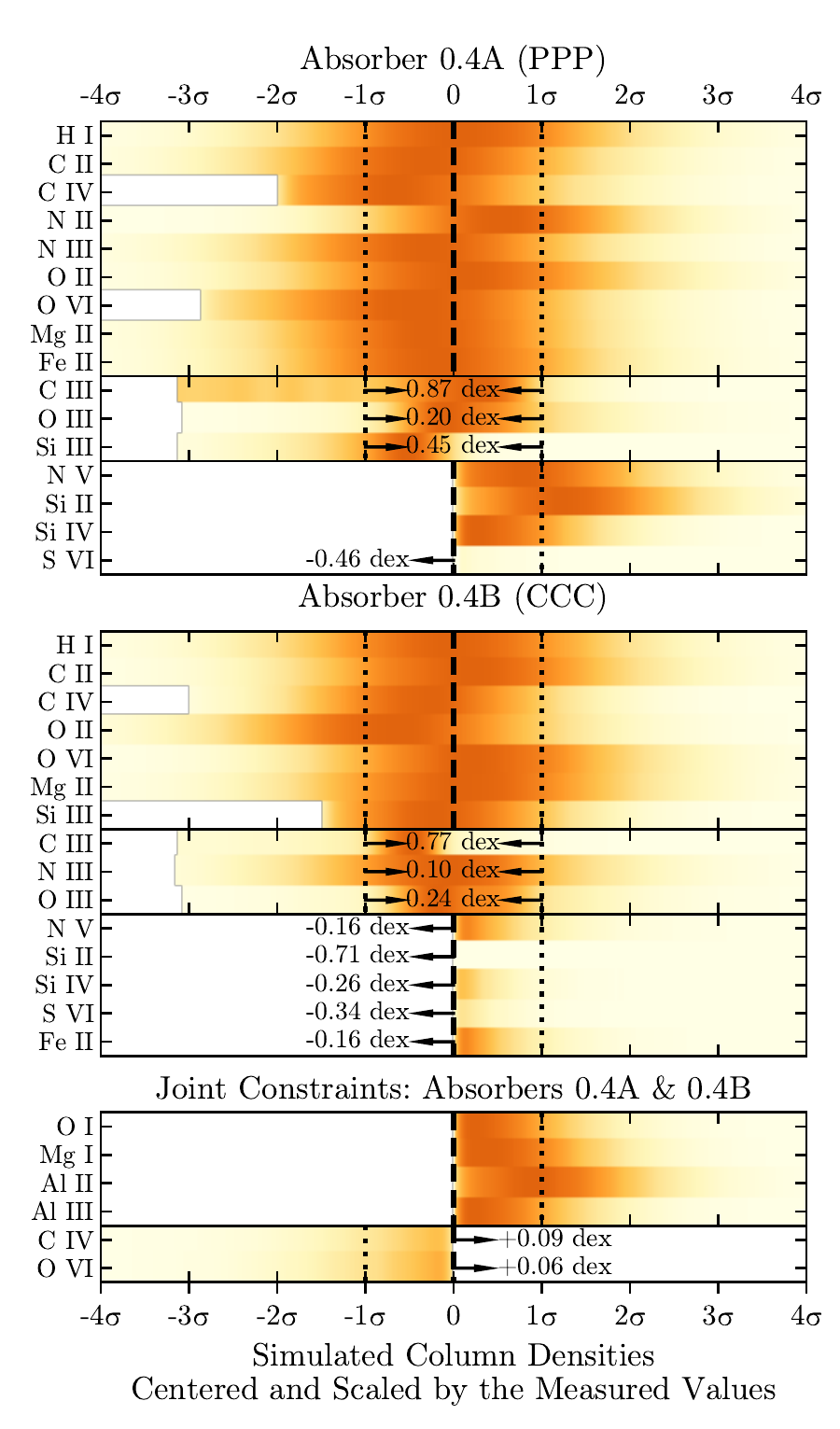}
	\caption{}
	\label{fig:z041761_threephase}
\end{subfigure}
\caption{PP-plots for Absorbers~0.4A \&~0.4B modeled with an increasing number of PIE-only phases in each panel (from left to right).  The meaning of the top, middle, and bottom panels is analogous to those in Figure~\ref{fig:z068605_PPplot}.  Also as in Figure~\ref{fig:z068605_PPplot}, the single-phase model clearly fails owing to $\gtrsim 2 \sigma$ discrepancies between the simulated and observed O~\textsc{ii}, Si~\textsc{ii}, C~\textsc{iv}, and O~\textsc{vi} column densities in Absorber~0.4A and Mg~\textsc{ii}, O~\textsc{iii}, and S~\textsc{vi} in Absorber~0.4B (\textit{a}).  In this case (as opposed to Figure~\ref{fig:z068605_PPplot}), a two-phase model (\textit{b}) provides posterior predictive distributions that are reasonably consistent with the measured columns, while a third phase offers only subtle improvements (\textit{c}).  We note that the mixed two-phase model (PIE-only plus P+C) shown here performs better for O~\textsc{vi} than a PIE-only two-phase model for Absorber~0.4A (not shown), though a PIE-only three-phase model also suffices.  See Figure~\ref{fig:z051850_example} and Section~\ref{sec:residual_checks} for details to aid in deciphering these PP-plots.}
\label{fig:z041761_PPplot}
\end{figure*}

\subsection{Additional Phases}
\label{sec:twophase}

	In contrast to Absorber~0.5A (described above), the remaining four absorbers require \textit{multi}phase models with at least two distinct phases per absorber. Tables~\ref{tab:selection_metrics_z041761} and \ref{tab:selection_metrics_z068605} summarize the models we have tried for the these absorbers, including the PP \textit{p}-values for each model. 

\subsubsection{Multiphase Models for PG1338+416 Absorbers 0.6A and 0.6B} 

We first discuss the absorption system we have studied in the PG1338+416 spectrum.  Figure~\ref{fig:z068605_PPplot} tracks the improvement for 1-, 2-, and 3-phase models of Absorbers~0.6A and 0.6B (see Figure~\ref{fig:z068605_vpfits}) through their respective PP-plots.  Single-phase models with all parameters free (16 total--eight for each absorber in the plotted PIE-only model) fail for both absorbers (Figure~\ref{fig:z068605_singlephase}).  A single PIE-only phase dramatically underproduces the high-ions N~\textsc{iv}, O~\textsc{iv}, S~\textsc{v}, and O~\textsc{vi} in Absorber~0.6A, resulting in a PP \textit{p}-value of 0.00023\footnote{Computed from just the simulations and observations represented in the top panel of Figure~\ref{fig:z068605_singlephase}.}, which falls well below our rejection threshold discussed in Section~\ref{sec:model_selection_metrics}.  A single-phase model for Absorber~0.6B performs even more poorly, only managing to fit a handful of species.  To ensure that our conclusions are driven by the observations themselves and not modeling systematics (e.g., such as those arising from uncertainties in the Voigt profile and ionization model atomic parameters), we reran our single-phase models after artificially boosting the measured column density errors to a minimum of 0.1 dex (with the exception of \textit{N}\textsubscript{H I}).  The slight improvements brought about by loosening these constraints fall far short of what is needed to salvage the single-phase models, implying that our conclusion about the need for multiphase models is robust.

	To improve upon these single-phase ionization models, we explored adding a second and even a third model phase to each of Absorbers 0.6A \& 0.6B.  Each new phase introduces at least two (three if using P+C) additional parameters to govern its density, total hydrogen column density, and (for P+C phases) temperature.  In light of the observed kinematic alignment across various ionization stages (Section~\ref{sec:kinematic_alignment}), which implies some similarity in the spatial distributions of the different gas phases, we assume that a common UV radiation field illuminates all the model phases associated with a given absorber.  Thus additional phases share the UV parameters of the first model phase.  Since both Absorbers 0.6A \& 0.6B are distinct velocity components of a single absorption system at redshift \textit{z}=0.68606, we explore two classes of models.  One with unique pairs of UV parameters \textsl{per absorber} and another ``joint UV'' model, with a single pair of UV parameters shared across all absorbers in this PG1338+416 system.

	With regard to the metallicity and relative abundance parameters, we considered two choices: 1) vary the metallicity parameters independently in every model phase, for every modeled absorber or 2) assign a single set of parameters to cover all the phases modeled for a given absorber, i.e., enforce a homogeneous gas composition in each absorber.  The second choice is reasonable and realistic if the circumgalactic gas is well mixed.  If the CGM is not well mixed, then situations such as a cool, metal-enriched clump of gas embedded in a hot, lower-metallicity halo could occur.  In practice, however, the metallicity of the high-ions is difficult to constrain and, at least for the absorbers considered here, allowing the metallicity to vary independently in each phase does not improve the model result.  This conclusion is borne out by the Bayes factors, which effectively reject variable-metallicity models in favor of those having a homogeneous metallicity across all the gas phases.  These simpler models achieve equally good fits with fewer parameters.  Moreover, we find that variable-metallicity models tend to yield metallicity results that are quite similar to those from homogeneous (well-mixed) models anyway.  Hence the models we present here all include a single set of shared gas composition parameters across all phases in a given absorber.  Parameterized in this manner, each multiphase model only adds an extra two or three parameters per additional phase (per absorber) beyond those required by the full single-phase model.

	Jointly modeling Absorbers~0.6A \&~0.6B with two phases each noticeably improves the predictive consistency for both absorbers as shown in Figure~\ref{fig:z068605_twophase}, though H~\textsc{i} and Ne~\textsc{viii} still pose a problem for the model.  The posterior distributions of the parameters associated with each of these two absorbers are not independent in this model on account of the joint constraints\footnote{These were derived from equivalent measurements independently of any VP measurements for the same species.} on the C~\textsc{iii}, O~\textsc{i}, O~\textsc{iv}, O~\textsc{vi}, and S~\textsc{vi} column densities.  While the top panel of Figure~\ref{fig:z068605_twophase} (Absorber 0.6A) suggests a reasonably good model based on its posterior predictive power (only the weak S~\textsc{ii} detection is off by $\gtrsim$ 1$\sigma$), the model's failure to reproduce H~\textsc{i}, S~\textsc{iii}, S~\textsc{v}, and Ne~\textsc{viii} for Absorber~0.6B (middle panel) accounts for its low Bayes factor (10$^{-20.98}$) and consequently its low ranking in Table~\ref{tab:selection_metrics_z068605}.  The key difference between 0.6A and 0.6B, as we discuss below, is the detection of Ne~\textsc{viii} in 0.6B.  The Ne~\textsc{viii} is hard to explain without a third phase.

	Adding a third model phase to cover the high-ions in Absorber~0.6B (Figure~\ref{fig:z068605_threephase}) largely resolves the problems with the two-phase model.  The model in Figure~\ref{fig:z068605_threephase} also includes a third phase for Absorber~0.6A.  The extra phase does not significantly improve this absorber's PP~plot (top panel), yet the model ranks higher in Table~\ref{tab:selection_metrics_z068605} because it satisfies the stringent pressure equilibrium prior, whereas the two-phase model cannot (see below, Table~\ref{tab:TPE_probabilities}).  The three-phase model presented here also differs from the two-phase one by sharing the same set of UV parameters for both absorbers (what we call ``joint UV'') rather than assigning independent sets of free UV parameters to each (i.e., ``independent UV'').  We find that our ``joint UV'' models all rank near or higher than our ``independent UV'' models.

\subsubsection{Multiphase Models for PG1630+377 Absorbers 0.4A and 0.4B}

	The pair of absorbers in PG1630+377 at $\textit{z}\textsubscript{\textrm{abs}}=0.41760$ (0.4A and 0.4B) similarly require multiphase models, as demonstrated by Figure~\ref{fig:z041761_PPplot}.  As was the case with the PG1338+416 Absorbers~0.6A and 0.6B, artifically boosting the column density errors does not resolve the failures of the single-phase models.  Modeling these two absorbers with two phases each, however, \textit{does} produce a nearly satisfactory PP-plot (Figure~\ref{fig:z041761_twophase}), though the model struggles to produce enough O~\textsc{vi} for Absorber~0.4A.  As with the models for the $\textit{z}\textsubscript{\textrm{abs}}=0.68606$ absorbers, a third phase improves not only the PP~plot (Figure~\ref{fig:z041761_threephase}), but also the prospects of thermal pressure equilibrium (see below, Table~\ref{tab:TPE_probabilities}).  The small number of high-ion detections in this absorber (just C~\textsc{iv} and O~\textsc{vi}, although \textsc{O~iii} may be almost as much of a ``high ion'' as \textsc{C~iv}, see Figure~\ref{fig:ion_potentials}) precludes strong constraints on a third phase in either velocity component, but Table~\ref{tab:selection_metrics_z041761} suggests that models fare better when including one.

	These two absorbers (0.4A and 0.4B) underscore how the number and degree of ionization of detected species play a pivotal role in dictating the optimal number of model phases.  The real significance in the number of model phases stems from the degree to which our inferences depend on them.  In general, we find good agreement between two and three-phase models (for all our multiphase absorbers) where their parameters overlap.  This observation, coupled with the large number of acceptable competing model variations listed in Tables ~\ref{tab:selection_metrics_z051850}, ~\ref{tab:selection_metrics_z041761}, and~\ref{tab:selection_metrics_z068605}, leads us to merge competing models via a simple Bayesian model averaging procedure: the average model is obtained by drawing samples from the posterior of each of the competing models in proportion to (i.e., weighted by) their Bayes factors (listed in the same tables).  Since, as we noted above, the ``joint UV'' models typically outperform the ``independent UV'' alternatives, we choose to include only the ``joint UV'' models in our Bayesian average.  This simplistic approach glosses over several nuances related to Bayesian model averaging \citep[see, for example,][]{Hoeting99}, which are of minor importance to us given the large overlap in the posteriors of the competing models.  Where significant inferential discrepancies do exist between classes of models, we acknowledge them below.

\begin{figure*}
\centering
	\begin{subfigure}{.33\textwidth}
	\centering
	\includegraphics[width=1.0\textwidth]{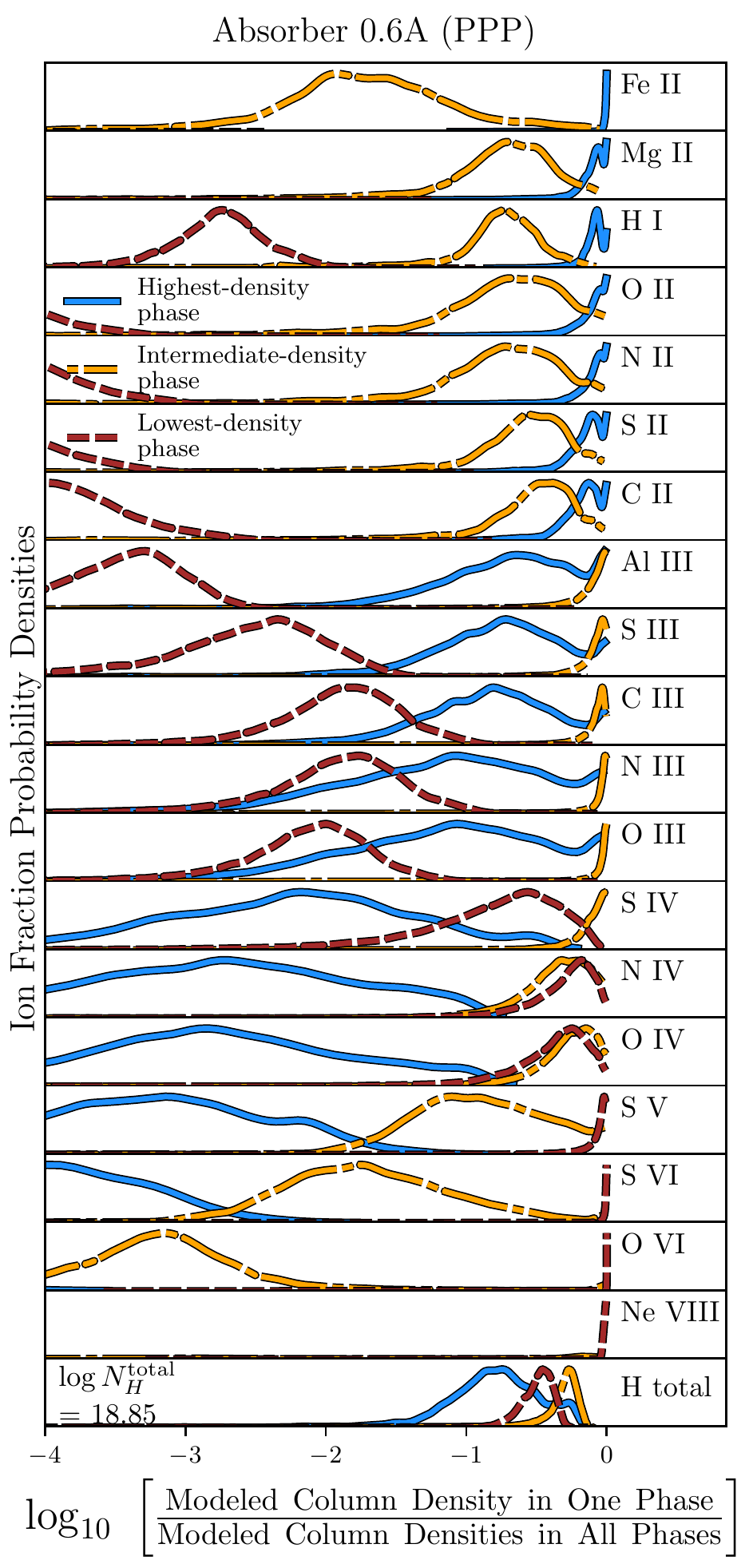}
	\caption{}
	\label{fig:abs06aPPP_fracplot}
	\end{subfigure}
	\begin{subfigure}{.33\textwidth}
	\centering
	\includegraphics[width=1.0\textwidth]{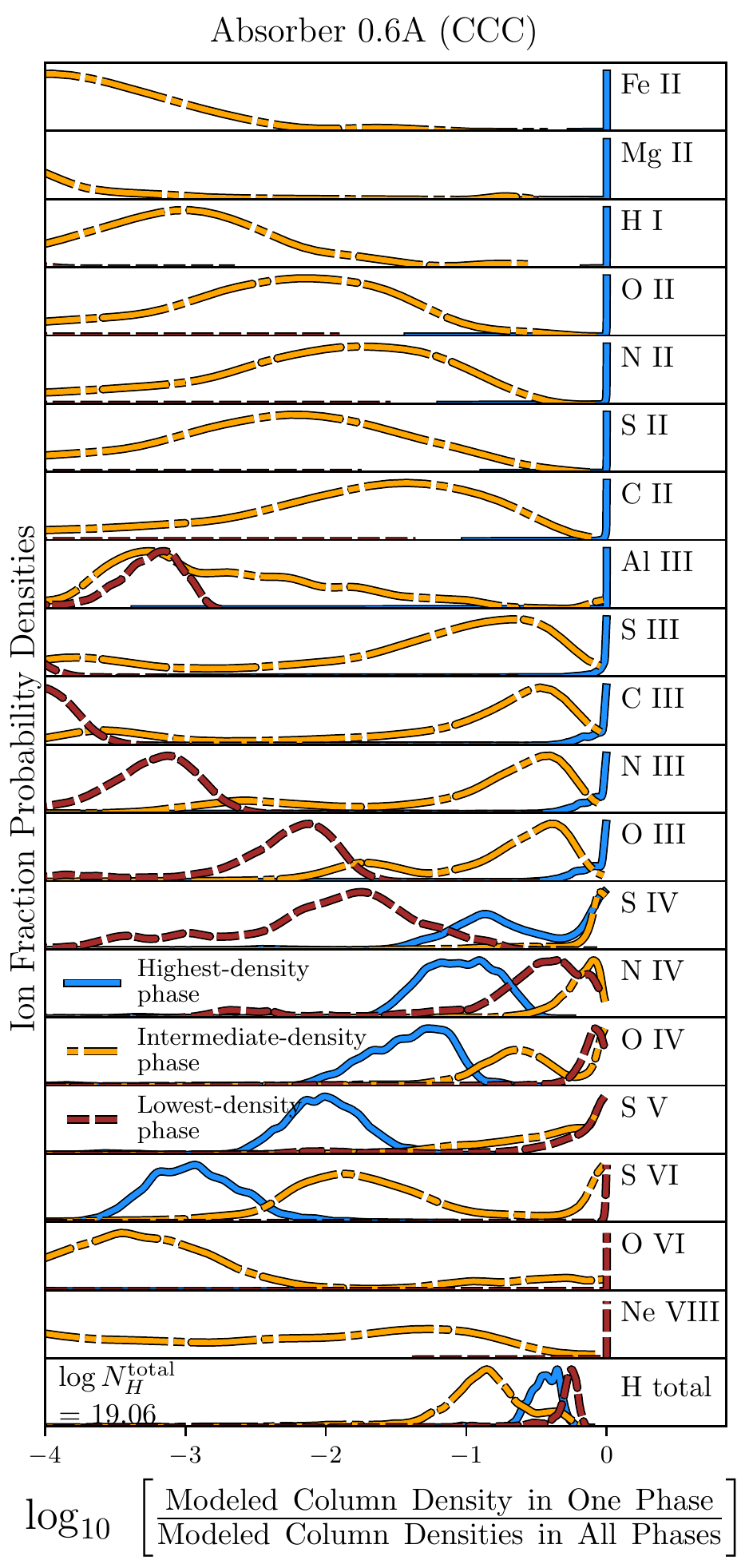}
	\caption{}
	\label{fig:abs06aCCC_fracplot}
	\end{subfigure}
	\begin{subfigure}{.33\textwidth}
	\centering
	\includegraphics[width=1.0\textwidth]{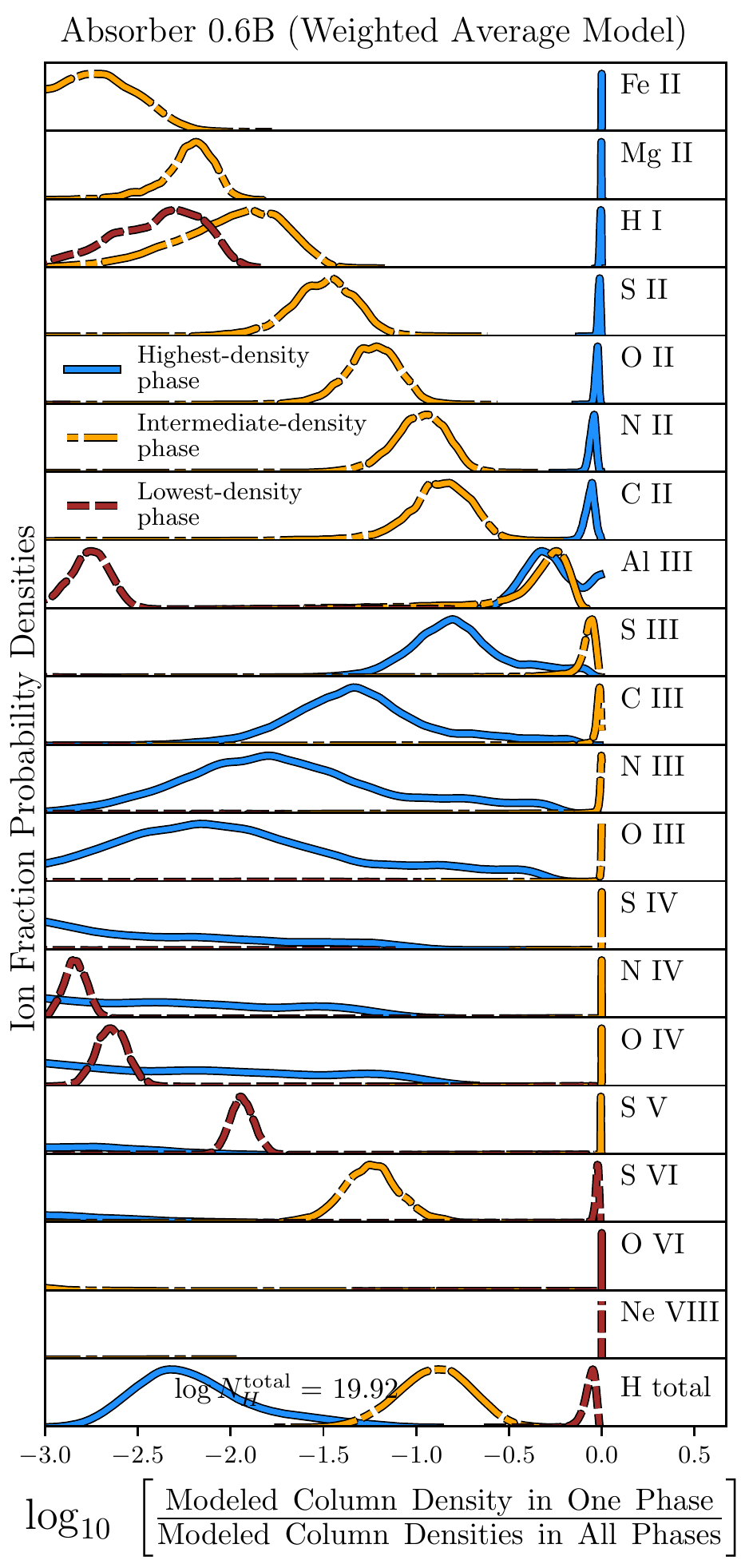}
	\caption{}
	\label{fig:abs06b_fracplot}
	\end{subfigure}
\caption{Fractional, model-predicted column densities of selected ions for Absorbers~0.6A (PPP model) (\textit{left}), 0.6A (CCC model) (\textit{middle}), and 0.6B (\textit{right}).  The curves represent the (unnormalized) probability that a given fraction of the total model-predicted log column density resides in a particular phase (labeled in the legend).  The fractional contributions of all phases in a given panel sum to unity (zero on the log scale).}
\label{fig:fracplot_comparison}
\end{figure*}

\subsection{Ion Distribution Among the Phases}
\label{sec:structure}

	Despite assumptions that are often made in the literature,\footnote{Commonly in the literature, simplifying assumptions are made such as assigning \underline{all} of the \textsc{H~i} to the lowest ionization phase \citep[unless the \textsc{H~i} absorption profile shows evidence of a broad component, e.g.,][]{Tripp01,Richter06}. However, more nuanced efforts to partition the \textsc{H~i} have appeared in more recent papers. For example, \cite{Stern16} have proposed a model in which the \textsc{H~i} is distributed across multiple phases.} many ions can arise in a range of physical conditions, and more than one phase in a multiphase model can contribute some appreciable fraction of a specific ion's total column density.  Here we investigate how these fractions vary from ion to ion, and we consider some insights about the ions we have studied.  Figure~\ref{fig:fracplot_comparison} presents the constraints from our models on the fractions of ions that arise in a low-ionization phase vs. a high-ionization phase for two of our aborbers.  In both panels, the solid blue curves depict the relative probability that the model phase having the highest gas density contains a given fraction (specified as a logarithm along the horizontal axis) of the total modeled column density.  The dot-dashed orange curves and dashed brown curves plot the same for the intermediate and highest-density phases, respectively.  Said another way, the blue, orange, and brown relative probability densities (i.e., $p_\textrm{blue}$, $p_\textrm{orange}$, and $p_\textrm{brown}$) reference the same posterior sample wherever $10^{x_\textrm{blue}} + 10^{x_\textrm{orange}}  + 10^{x_\textrm{brown}}= 1$ is true for any set of \textit{x}-axis values $x_\textrm{blue}$, $x_\textrm{orange}$, and $x_\textrm{brown}$ corresponding to the labeled curves of matching colors.  The total fraction is always unity.  The colored curves are \textit{relative} probabilities since they have been normalized by their peak probability to maximize visibility.  The relative height of or area under the various colored curves is otherwise meaningless.  The curves in Figure~\ref{fig:fracplot_comparison} depict the fraction of each ion that arises in a given phase; e.g., if a curve peaks at an x-axis value $\approx$ 0, then $\approx$ 100\% of the ion column density originates in that phase.

	In addition to confirming our expectation that the highest density phase produces the dominant fraction of low-ions while the majority of high-ions are located in the lowest density phase(s), Figure~\ref{fig:fracplot_comparison} also reveals two important results.

	The first is that the low- and high-ionization gas phases are not always neatly separable.  For example, $\approx20\%$ of the neutral hydrogen in the PPP model for Absorber~0.6A (Figure~\ref{fig:abs06aPPP_fracplot}) is produced by the intermediate density phase, while the lowest density phase only contains $\approx80\%$ of the total.  In fact, this model predicts that there are ten other species that have $\gtrsim20\%$ contributions from two distinct phases (i.e., C~\textsc{ii}, N~\textsc{ii/iv}, O~\textsc{ii/iii/iv}, and S~\textsc{ii/iii/iv}).  The fraction of neutral hydrogen in a given phase though, is particularly important.  While H~\textsc{i} is often assumed to be entirely associated with a separable cold phase, this example shows that this does not have to be true.
	The fraction of a given species' column density that originates in a non-dominant phase (e.g., the fraction of C~\textsc{iii} produced by the lowest ionization phase in Figure~\ref{fig:abs06aPPP_fracplot}) is sensitive to the ionization mechanism.  Contrast the PIE-only model for Absorber~0.6A refered to above (Figure~\ref{fig:abs06aPPP_fracplot}) with a P+C model for the same absorber (Figure~\ref{fig:abs06aCCC_fracplot}).  Whereas a substantial fraction of the PPP model's low ion columns come from an intermediate ionization phase, the CCC model attributes essentially all the low ions to just the lowest density phase.  The production of the intermediate phase is evidently shifted more toward the higher-ions.  This makes sense, because the P+C models typically probe higher temperatures than the PIE-only ones.
	Finally, we also find some absorbers, such as Absorber~0.6B (Figure~\ref{fig:abs06b_fracplot}), where the overlap between phases is largely negligible for most ions.  Even in this case, however, the delineation between low-ions and mid-ions is not always clear cut.  Taken as a whole, all the panels in Figure~\ref{fig:fracplot_comparison} demonstrate that there is not a straightforward way to reliably determine the relative contribution from multiple phases \emph{a priori} for the majority of ions, so models that account for all ionization stages are quite valuable.

\begin{figure*}
\includegraphics[width=6.9in]{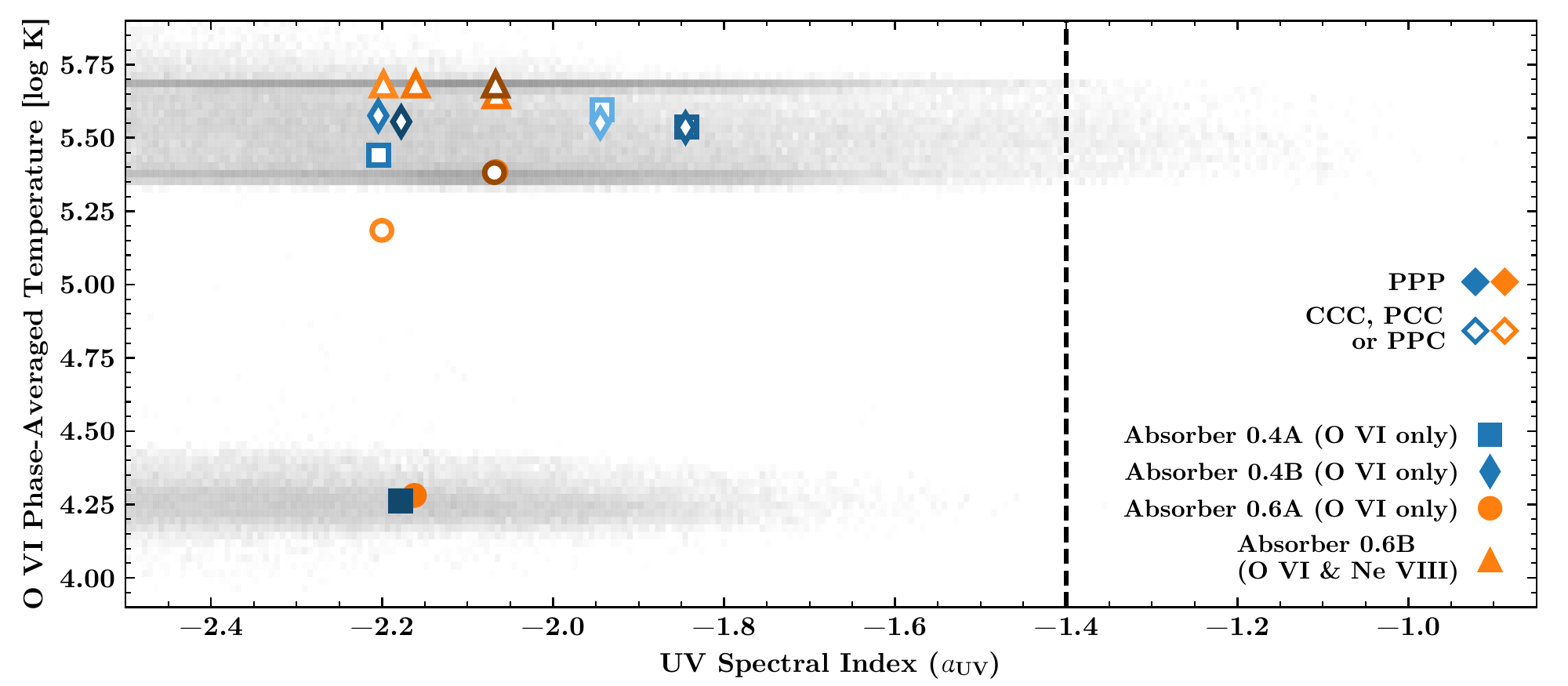}
\caption{An illustration of the ionization regimes preferred for the high-ions by the best models.  This demonstrates that most of the successful models, including all the ones constrained by a Ne~\textsc{viii} detection, produce O~\textsc{vi} at high temperatures ($\log_{10} (T_\textrm{O VI}/K) \approx 5.55$ K).  The vertical axis is the linear average of the gas temperature across all model phases, weighted by the fraction of (modeled) O~\textsc{vi} in each phase.  The horizontal axis tracks the UV spectral index parameter, with lower values indicating softer UV radiation.  The dotted vertical line is placed at $\alpha_\textrm{UV}$=-1.4, the slope of the \protect\cite{Haardt12} UVB at these redshifts ($z\approx0.5$).  The log-scaled, gray 2-D histogram bins samples from the Bayesian weighted average model (see Section~\ref{sec:twophase}), while the plotted points denote mean values for each of the models with Bayes factors $>10^{-1.5}$ in Tables~\ref{tab:selection_metrics_z041761} and~\ref{tab:selection_metrics_z068605}.  The middle-right legend indicates our use of closed symbols to denote purely photoionized (PIE-only) models and open symbols to denote any partially collisionally ionized (P+C) models.  The symbol shapes and colors refer to particular absorbers and absorption systems, respectively.  Finally, the color shading reflects the relative probability of the various models (in each color group), with darker symbols denoting the most probable models according to their Bayes factors.}
\label{fig:ionization_regimes}
\end{figure*}

	The second observation is that intermediate-ions such as C~\textsc{iii}, N~\textsc{iii}, and O~\textsc{iii} can be predominantly produced in the same phase as ``higher''-ions N~\textsc{iv}, O~\textsc{iv}, and S~\textsc{v} (Figure~\ref{fig:abs06b_fracplot}), in the same phase as low-ions C~\textsc{ii}, N~\textsc{ii}, and O~\textsc{ii} (Figure~\ref{fig:abs06aCCC_fracplot}), or in their own unique phase (Figure~\ref{fig:abs06aPPP_fracplot}).  These same figures indicate that the sets of ``low'' and ``high'' ions might be similarly indeterminate.  In particular, if the model has three phases, then the ``high'' ions may or may not include N~\textsc{iv}, O~\textsc{iv}, and S~\textsc{iv/v}.  These observations suggest that the second ionization phase in our models (the first being low-ion dominated) is distinct from both the ``cold'' and hot phases described by a two-phase halo paradigm.  Furthermore, these observations and those of the preceeding paragraph support our hypothesis in the introduction, namely, that analyses based on, e.g., fitting \textsc{C~ii} and \textsc{C~iv}\footnote{The \textsc{C~iv} ion, which is commonly detected in ground-based QSO observations, would be categorized similarly to N~\textsc{iv} and O~\textsc{iv} in Figure~\ref{fig:fracplot_comparison} on account of their similar ionization potentials and photoionization rates (see Figure~\ref{fig:ion_potentials}).} with a single-phase ionization model likely suffer from serious systematic errors, and in general, caution is warranted when only limited constraints on absorber ionization are available.

	Since our models involve summing column densities across phases at varying temperatures, it is important to consider whether the superposition of Voigt profiles implied by the pairs of column densities and temperatures  ($\textrm{N}_\textrm{H\ I}^{i}$ and $\textrm{T}^{i}$ for the \textit{i}th phase) is consistent with the spectra.  We carried out this consistency check for the models presented in Figures~\ref{fig:z068605_threephase} and~\ref{fig:z041761_threephase} and found that modeled spectra do not exceed the observed absorption profiles.  In most cases, the modeled absorption lines have noticeably smaller linewidths, which we interpret as evidence for some degree of turbulent or non-thermal line broadening that is beyond the scope of the models in this paper.  We did not analyze these linewidth differences further here, though we are currently working to develop an integrated Voigt profile and ionization fraction model that can encompass the extra linewidth information in a rigorous manner.

\subsection{Highly Ionized Gas}
\label{sec:OVI_ionization}

The question of whether photoionization can realistically produce O~\textsc{vi} and even Ne~\textsc{viii} in CGM/IGM absorbers has been a recalcitrant question in the literature, and the answer is still being debated \citep[e.g.,][]{Tripp01,Tripp08,Tripp11,Savage05,Savage10,Savage14,Muzahid12,Meiring13,Werk16,Hussain17,Pachat17,Rosenwasser18,Burchett19}. While the ionization state in our PIE-only models depends solely on the UV radiation field and gas density, the P+C models allow for tuning the gas heating independently of the incident radiation.  When tuned to produce lower ionization states, e.g., C~\textsc{ii} and Si~\textsc{ii}, the additional flexibility afforded by the P+C ionization mechanism becomes redundant, since photoionization on its own can readily produce these low-ions.  However, as we consider increasingly more ionized species, e.g., \textsc{C~iv}, \textsc{N~iv}, O~\textsc{iv}, and higher ions, the number of ionizing photons with sufficient energy probably decreases rapidly, and an additional source of gas heating and ionization may be required.  PIE-only models must compensate for the low density of high-energy photons (see Figure~\ref{fig:ion_potentials}) either by increasing the ionization parameter (either through lowering the gas density or via increasing the intensity of the radiation field) \emph{or} by varying the shape of the ionizing UV spectrum (for example, harder ionizing UV slopes will generally produce a larger O~\textsc{vi}/C~\textsc{iv} ratio than softer slopes).

Figure~\ref{fig:ionization_regimes} shows how the ionization mechanism and radiation field are intertwined in our best multiphase models.  The log-scaled, gray 2-D histogram shows the joint probability of all the multiphase models in this paper across $\textit{T}_\textrm{O VI}$--$\alpha_\textrm{UV}$ space, where $\textit{T}_\textrm{O VI}$ is the average phase temperature weighted by the fraction of O~\textsc{vi} in each modeled phase.  The histogram reveals a well-defined photoionization branch ($\textit{T}_\textrm{O VI} \approx 10^{4-4.5}$) as well as collisionally ionized, warm-hot ($\textit{T}_\textrm{O VI} \gtrsim 10^{5.25}$) structures.  The blue and orange symbols mark point estimates for individual models.

Three patterns in Figure~\ref{fig:ionization_regimes} are particularly notable.  First, all the models with favorable Bayes factors (\textit{colored symbols}) lie along the collisional ionization branch, except for one photoionization model each of Absorbers~0.4A and 0.4B.  Thus, we conclude that O~\textsc{vi}, which we detect in all four multiphase absorbers, can be readily explained by a collisional ionization origin.  The two satisfactory models with photoionized O~\textsc{vi}, however, indicate that the detection of O~\textsc{vi} is not itself a sufficient cause to discriminate between ionization mechanisms.

Second, the two absorbers that do not produce any favorably ranked photoionization models exhibit a higher ratio of high- to low-ions than the other two multiphase models.  A brief examination of Figures~\ref{fig:z041761_vpfits} and~\ref{fig:z068605_vpfits} corroborates this statement.  The low-ions (e.g., C~\textsc{ii}, Mg~\textsc{ii}) are significantly weaker and the mid/high-ions (e.g., C~\textsc{iii/iv}, O~\textsc{iii/ovi}) much stronger in Absorber~0.4B (no satisfactory photoionization model) than in Absorber~0.4A (which has a good photoionization model).  A similarly striking pattern emerges for Absorber~0.6B, which has a much stronger high-ion to low-ion ratio than Absorber~0.6A, as also evidenced by a strong Ne~\textsc{viii} line detected in the former absorber.

Our stringent pressure equilibrium prior may play an outsized role in precluding photoionization models for Absorber~0.4B.  Indeed, Table~\ref{tab:selection_metrics_z041761} lists a model (PCCa-PPPb) with an excellent PP~\textit{p}-value and a Bayes factor not far outside the rejection threshold, but with zero probability of thermal pressure equilibrium (Table~\ref{tab:TPE_probabilities}).  For the Ne~\textsc{viii} absorber (0.6B) however, a photoionization model fails badly (see the Bayes factor for the PPPb model in Table~\ref{tab:selection_metrics_z068605}) because strong Ne~\textsc{viii} columns are simply impossible to produce in a photoionized gas phase that obeys our modest cloud size prior, even with a multiphase model.  In addition to Ne~\textsc{viii}, Absorber~0.6B also exhibits very strong high-ion absorption in numerous other species for which we have coverage (i.e., N~\textsc{v}, O~\textsc{iv/vi}, Ne~\textsc{viii}, and S~\textsc{iv/v/vi}).  Even ignoring O~\textsc{vi} and Ne~\textsc{viii}, the other high-ions in this absorber, which exist in an intermediate ionization phase (see Figure~\ref{fig:abs06b_fracplot}), exhibit similarly derived temperatures that on their own are indicative of collisional ionization (e.g., $\textit{T}_\textrm{O\ IV} \approx \textit{T}_\textrm{S\ IV}\approx$10$^{5.0}$ K).

Third, the two collisional ionization bands that appear in the gray histogram in Figure~\ref{fig:ionization_regimes}, i.e., at $\textit{T}_\textrm{O VI}$$\approx$10$^{5.35}$ K and 10$^{5.7}$ K, should be interpreted cautiously.  Although they undoubtedly represent collisionally ionized O~\textsc{vi} (and other high-ions), their narrow thermal precision (i.e., dispersion along the vertical axis) may be an artifact of the discrete nature of our models.  In the likely scenario that the absorber temperatures actually follow a continuous multiphase structure, the phase-averaged temperatures in Figure~\ref{fig:ionization_regimes} could mildly underestimate the distribution of true (hot) gas temperatures.  Another concern is the lack of constraints at the high energy edge of the model.  It is possible, for example, that the quasar sightlines probe some additional phase of hotter gas (e.g., the putative hot, virialized halo), and detection of this phase would require species such as O~\textsc{vii} at shorter wavelengths than those covered by COS.  Under this scenario it is possible that a comprehensive model constrained at all wavelengths could split the O~\textsc{vi} and Ne~\textsc{viii} columns between the collisionally ionized branch in Figure~\ref{fig:ionization_regimes} and some higher temperature branch, thereby forcing an (upward) adjustment to the $\textit{T}_\textrm{Ne VIII}$$\approx$10$^{5.7}$ K temperature in Figure~\ref{fig:ionization_regimes}.


\section{Physical Conditions}
\label{sec:inferences}

\subsection{Pressure Gradients}
\label{sec:pressure_gradients}

\begin{figure}
\includegraphics[width=\columnwidth]{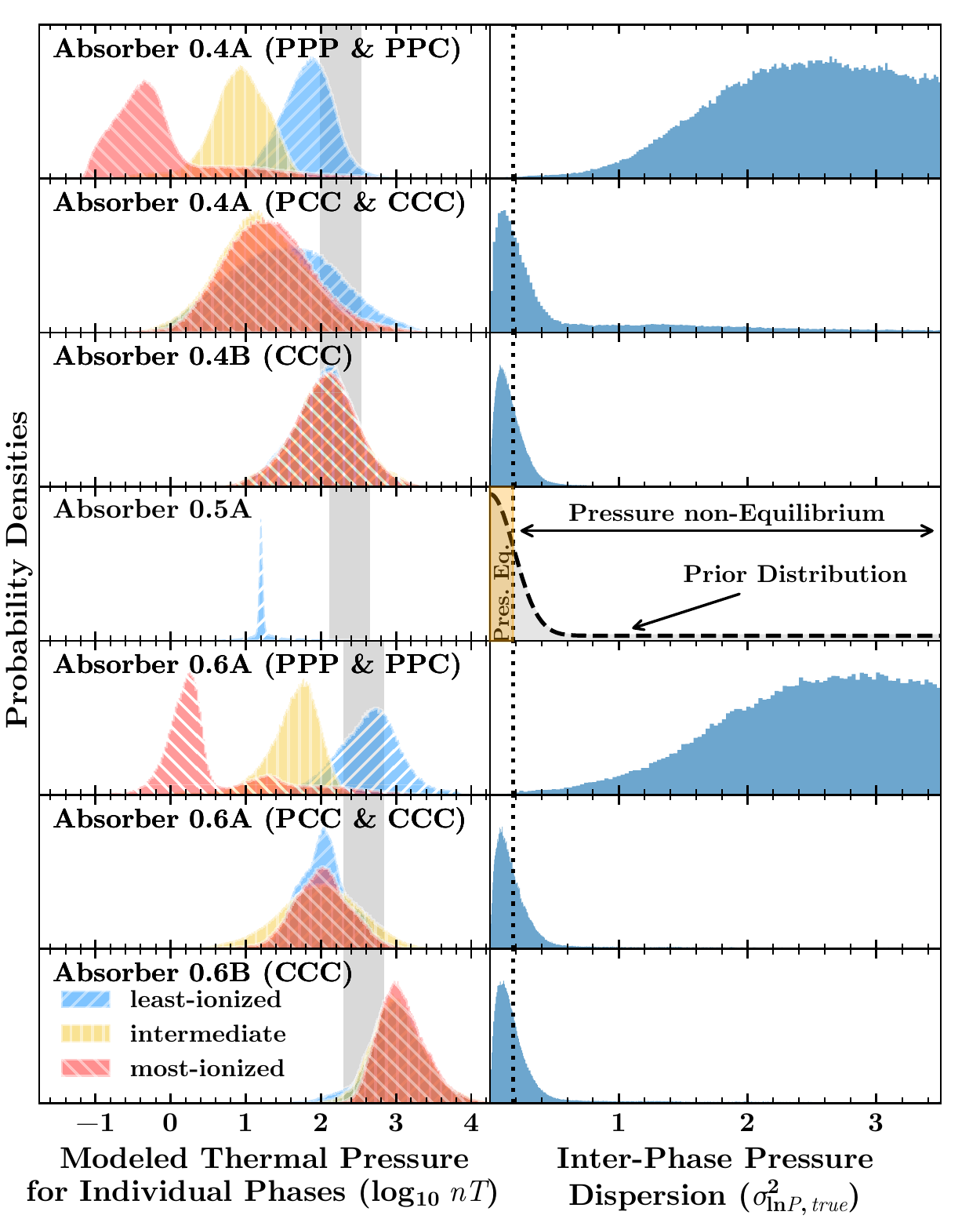}
\caption{\textsl{Left panels:} Bayesian model averaged thermal pressures (\textit{nT}) in individual ionization phases of each absorber.  The blue, yellow, and red hatched histograms represent individual phases from those having the lowest degree of ionization (highest density) to the highest (lowest density), respectively.  For comparison, vertical gray bands represent the range of hydrostatic halo pressures predicted by \protect\cite{Maller04} between 0.2--1\emph{R}\textsubscript{vir} for a 10$^{\textrm{11.5}} \textrm{M}_{\odot}$ halo (higher halo masses would push the gray bands toward higher pressures).  \textsl{Right panels:} the distribution of $\sigma^2_{\ln \textit{P,true}}$, which we use to interpret the probability of thermal pressure equilibrium as discussed in the text (see also Section~\ref{sec:parameters}).  The middle panel on the right side illustrates our prior on $\sigma^2_{\ln \textit{P,true}}$.}
\label{fig:CGM_pressure}
\end{figure}

	An important utility of our multiphase models is their ability to track thermal pressure ($P_{\rm therm}/k = nT$) across individual gas phases.  If the absorbers are in hydrostatic equilibrium within a hot and virialized halo, then the thermal pressures of individual phases should be nearly equal modulo the assumptions discussed in Section~\ref{sec:parameters} (e.g., assuming non-thermal pressure support is negligible).  Indeed, as illustrated in Figure~\ref{fig:CGM_pressure}, we can find models that provide good solutions with pressure balance between the phases for all four multiphase absorbers, though we also find models with satisfactory Bayes factors where the individual phases are not in thermal pressure equilibrium.  These latter models could indicate that more complex models of the CGM may be required, possibly including time-dependent physics, or this could indicate that non-thermal pressure support (e.g., from turbulence, cosmic rays, and/or magnetic fields) is important.\footnote{To be clear, our analysis of \emph{thermal pressure} equilibrium vs non-equilibrium probabilities is based on models that assume \textit{ionization} equilibrium, although we briefly speculate about the potential relevance of exploring non-equilibrium models.  Any subsequent discussion of models that a}  In this section, we present a detailed analysis of the thermal gas pressures required by the models that takes advantage of our Bayesian framework.

In Section~\ref{sec:parameters} we laid out a framework for evaluating 1) whether thermal pressure equilibrium (TPE) is a possible solution for our absorbers, and 2) how likely any such TPE solutions are, i.e., $P(\textrm{TPE})$.  Our approach introduces a hyperparameter to model the thermal pressure dispersion of the absorbing medium at any scale (e.g., a single gas cloud or whole galaxy halo), assuming the pressure is log-normally distributed throughout the medium.  This hyperparameter, $\sigma^2_{\ln \textit{P,true}}$, governs the prior probability of the \textit{modeled} pressure dispersion, i.e., the sample variance $s^2_{\ln \textit{P}}$ of the set of thermal pressures produced by a multiphase model for a given absorber.  To ensure our posterior inferences are properly calibrated, we place a hyperprior on $\sigma^2_{\ln \textit{P,true}}$ that is designed to give equal prior volume/weight to equilibrium and non-equilibrium solutions alike, and to give greater preference to lower pressure dispersions.  This prior is illustrated in the middle-right panel of Figure~\ref{fig:CGM_pressure} along with a vertical dotted line marking the threshold between TPE and non-TPE models (i.e., $\alpha_{TPE}$ from Section~\ref{sec:parameters}).  Posterior samples in the narrow region left of the dotted line are consistent with our definition for TPE, while samples rightward are inconsistent with TPE.

Figure~\ref{fig:CGM_pressure} clearly shows that TPE solutions exist for each of the multiphase absorbers, although we also find excellent models for Absorbers~0.4A and 0.6A that do not achieve TPE.  Interestingly, there appears to be a clear distinction between the pressure solutions from models with two or more PIE-only phases (PPP and PPC) compared to models with two or more P+C phases.  The key difference here is that while the P+C models can assume any range of temperatures between 10$^4$ K and 10$^{6.4}$ K, the PIE-only phases are generally limited to $\textit{T}\lesssim10^{4.3}$ K (e.g., see Figure~\ref{fig:ionization_regimes}).  Consequently, those models with no more than one PIE-only phase have enough flexibility to balance their thermal pressures against the range of inter-phase densities required to fit the measured data.  As summarized in Table~\ref{tab:TPE_probabilities}, for three of the four multiphase absorbers, we find favorably ranked models (based on their Bayes factors listed in Tables~\ref{tab:selection_metrics_z041761} and~\ref{tab:selection_metrics_z068605}) where TPE is not only possible but probable (these models indicate $>$50\% probability of TPE, i.e., the cumulative posterior probability that $\sigma^2_{\ln \textit{P,true}} < \alpha_{TPE}$ is greater than 0.5, which is readily apparent upon inspection of the right side of Figure~\ref{fig:CGM_pressure}.)

\ctable[
	caption={Thermal Pressure Equilibrium Probability for Multiphase Models},
	doinside = \footnotesize,
	width = \columnwidth,
	label={tab:TPE_probabilities}
	]{@{}r@{--}l@{~}c@{~}c@{~}c@{~}c@{~}r@{--}l@{~}c@{~}c@{~}c@{}}{
\tnote[a,b]{These tablenotes are identical to those in Tables~\ref{tab:selection_metrics_z041761} and~\ref{tab:selection_metrics_z068605}.}
\tnote[c]{This table includes only multiphase models from Table~\ref{tab:selection_metrics_z041761} (for Absorbers~0.4A and~0.4B) and Table~\ref{tab:selection_metrics_z068605} (for Absorbers~0.6A and~0.6B), in the same order as listed therein (i.e., best to worst, from top to bottom).  The midlines, which delineate between favored and disfavored models, are likewise located between the same models as in Tables~\ref{tab:selection_metrics_z041761} and ~\ref{tab:selection_metrics_z068605}.}
\tnote[d]{Percent probability of thermal pressure equilibrium, $P(\sigma_{\ln P,true}<0.21)$, for each multiphase absorber (see Section~\ref{sec:parameters}).}
}{
\multicolumn{5}{c}{Absorbers~0.4A and~0.4B} & & \multicolumn{5}{c}{Absorbers~0.6A and~0.6B} \NN
\cmidrule[\heavyrulewidth](r){1-5} \cmidrule[\heavyrulewidth](r){7-11}
\multicolumn{2}{c}{Model ID\tmark[a,c]} & Joint & \multicolumn{2}{c}{$P(\textrm{TPE})\tmark[d]$} & & \multicolumn{2}{c}{Model ID\tmark[a,c]} & Joint & \multicolumn{2}{c}{$P(\textrm{TPE})\tmark[d]$} \NN
\cmidrule(r){4-5}\cmidrule(r){10-11}
\multicolumn{2}{c}{} & UV\tmark[b] & 0.4A & 0.4B & & \multicolumn{2}{c}{} & UV\tmark[b] & 0.6A & 0.6B \NN
\cmidrule(r){1-5}\cmidrule(r){7-11}

PPPa & CCCb &            &  0.0 & 70.9 & & PCCa & CCCb &            & 59.8 & 64.3 \NN
PCCa & CCCb &            & 43.8 & 67.3 & & CCCa & CCCb &            & 69.9 & 64.5 \NN
PPCa & CCCb &            &  0.4 & 68.8 & & PPPa & CCCb &            &  0.0 & 64.5 \NN
CCCa & CCCb &            & 29.9 & 68.6 & & PCCa & CCCb & \checkmark & 62.3 & 62.9 \NN
PPPa & CCCb & \checkmark &  0.0 & 66.2 & & PPCa & CCCb &            &  0.2 & 59.1 \NN
\cmidrule(r){4-5}
PCCa & PCCb &            & 45.4 & 52.9 & & PPPa & CCCb & \checkmark &  0.0 & 62.6 \NN
\cmidrule(r){10-11}
PCCa & PPPb &            & 60.2 &  0.0 & & PPa  & CCCb &            &  0.0 & 58.9 \NN
PCCa & CCCb & \checkmark & 40.0 & 69.4 & & PPa  & CCCb & \checkmark &  0.0 & 65.1 \NN
PPCa & PCCb &            &  0.3 & 65.6 & & CCCa & CCCb & \checkmark & 74.1 & 64.0 \NN
PPPa & PCCb &            &  0.0 & 62.1 & & PPCa & CCCb & \checkmark &  0.0 & 62.2 \NN
PCCa & PPCb &            & 58.3 &  0.0 & & PCCa & PCCb &            & 52.5 &  4.2 \NN
PPCa & CCCb & \checkmark &  1.1 & 63.4 & & CCCa & PCCb &            & 69.1 &  4.3 \NN
PCa  & CCCb &            & 18.6 & 69.9 & & PPPa & PCCb &            &  0.0 &  4.8 \NN
PPa  & PPb  & \checkmark &  0.1 &  0.0 & & PPCa & PCCb &            &  0.4 &  6.7 \NN
PCa  & CCCb & \checkmark & 22.4 & 66.7 & & PCa  & CCCb &            & 15.7 & 65.2 \NN
PCa  & PCCb &            & 21.7 & 60.4 & & PPa  & PCb  & \checkmark &  0.0 & 65.1 \NN
PCa  & PPb  &            & 34.7 &  0.0 & & PCa  & PCb  & \checkmark &  2.3 & 67.2 \NN
PPPa & PPPb &            &  0.0 &  0.0 & & PPa  & PCb  &            &  0.0 & 54.5 \NN
PCa  & PPPb &            & 24.9 &  0.1 & & PCa  & PCb  &            & 23.7 & 40.4 \NN
PPCa & PPCb &            &  0.3 &  0.0 & & \multicolumn{2}{c}{}  &  &      &      \NN
PCa  & PPCb &            & 32.8 &  0.0 & & \multicolumn{2}{c}{}  &  &      &      \NN
PPa  & PCb  & \checkmark &  0.0 &  5.4 & & \multicolumn{2}{c}{}  &  &      &      \NN
PPCa & PPPb &            &  0.0 &  0.0 & & \multicolumn{2}{c}{}  &  &      &      \NN
PPa  & PPb  &            &  0.0 &  0.0 & & \multicolumn{2}{c}{}  &  &      &      \NN
PPPa & PPCb &            &  0.0 &  0.0 & & \multicolumn{2}{c}{}  &  &      &      \NN
PCa  & PPb  & \checkmark & 14.0 &  0.0 & & \multicolumn{2}{c}{}  &  &      &      \NN
PPa  & PCb  &            &  0.1 &  0.1 & & \multicolumn{2}{c}{}  &  &      &      \NN
PCa  & PCb  & \checkmark & 13.0 &  3.8 & & \multicolumn{2}{c}{}  &  &      &      \NN
PCa  & PCb  &            & 25.0 &  0.3 & & \multicolumn{2}{c}{}  &  &      &      \NN
\cmidrule[\heavyrulewidth](r){1-5}
\cmidrule[\heavyrulewidth](r){7-11}
}

Our evidence for TPE in what are most likely CGM absorbers (based on their \textit{N}\textsubscript{H I} and metallicity, see below) contrasts with the prior conclusions of \cite{Werk14}, who analyzed absorption in the CGM of the COS-Halos \textit{L}\textsubscript{*} galaxies (see also, e.g., \citealt{McQuinn18}).  Their work compared gas densities inferred from single-phase, photoionization models with the predictions of analytic halo models of a single and two-phase hydrostatic equilibrium CGM, finding no agreement with either one.  A close examination of Figure~\ref{fig:CGM_pressure} and Table~\ref{tab:TPE_probabilities}, however, reveals that our \textsl{predominantly photoionized}, multiphase models actually show significantly different thermal pressures in the various phases, consistent with the results of \cite{Werk14}.  Moreover, it is notable that the PPP and PPC (non-TPE) models for Absorbers~0.4A and~0.6A have favorable Bayes factors, because the shape of our pressure prior (middle panel on the right in Figure~\ref{fig:CGM_pressure}) severely disfavors non-TPE solutions.  Were these models to only weakly disfavor pressure equilibrium, then we might expect a non-negligible fraction of their posterior volume for $\sigma^2_{\ln \textit{P,true}}$ to lie within the threshold for TPE.  Instead, they altogether reject such a scenario.  In fact, even the poor-performing models (for all absorbers) in Table~\ref{tab:TPE_probabilities}, reveal a bifurcation in their TPE probabilities, with those that are predominantly photoionized consistently having little or no success satisfying the TPE prior on $\sigma^2_{\ln \textit{P,true}}$ compared to those with at least two partially collisionally ionized, P+C phases.  The reason is simply that these models cannot produce a large enough spread in temperature between the relatively cool, photoionzed gas phases to balance their inter-phase differences in gas density.

The left side of Figure~\ref{fig:CGM_pressure} displays the (Bayesian model averaged) posterior distribution of logarithmic thermal pressures for each absorber/phase combination.  Recall that these pressures scale with the intensity of the UV radiation field owing to the degeneracy between the gas density and the UV intensity (Section~\ref{sec:density}).  A consequence of our interpreting the \cite{Haardt12} UVB as a lower limit on the radiation field intensity (as discussed in Section~\ref{sec:UVB}) is that the absolute pressures we compute are likewise lower limits.  The \textit{relative} pressures of individual phases (modeled for a given absorber), however, are well characterized, so they should scale upward in lock step if one postulates an elevated (above the \cite{Haardt12} baseline) UVB intensity.

We compare our modeled thermal pressures in Figure~\ref{fig:CGM_pressure} to the two-phase analytic halo model of \cite{Maller04}, which is the same model used as a point of reference by \cite{Werk14}.  Strikingly, we find that the \emph{highest} pressure phase in each of our multiphase absorbers is \emph{less} than or approximately equal to the minimum hydrostatic equilibrium pressure predicted by \cite{Maller04} for a 10$^{\textrm{11.5}} \textrm{M}_{\odot}$ halo.  If the TPE models are correct, then this could indicate that our absorbers are located in sub-\textit{L}\textsubscript{*} mass halos.  For the non-TPE models, this represents the high end of the pressures in individual phases.  The higher temperature phases require even lower pressures.  We chose to compare against a 10$^{\textrm{11.5}} \textrm{M}_{\odot}$ halo since that is roughly the threshold in halo mass above which halos are expected to remain virialized to the present epoch \citep[e.g.,][]{Keres05,Keres09}.  For example, the hydrodynamical simulations of \cite{Fielding17} for a few idealized halos suggest that non-thermal pressure support can become important within several Gyr of virialization for halos with \emph{M}\textsubscript{Halo}$<$10$^{\textrm{11.5}} \textrm{M}_{\odot}$.  Such halo sizes correspond to relatively low-mass present day galaxies, with stellar masses of \emph{M}\textsubscript{$\star$}=10$^{\textrm{9-10}} \textrm{M}_{\odot}$ corresponding roughly to \emph{M}\textsubscript{Halo}=10$^{\textrm{11.0-11.5}} \textrm{M}_{\odot}.$\footnote{Based on the halo-to-stellar mass ratios of \citealt{Guo10}.}

	A speculative explanation for the low thermal pressures suggested by some of our models may be that these absorbers exist in low mass, non-virialized halos or some other context where thermal pressure equilibrium is not attained.  For example, a disk cloud that is ejected into a halo might be overpressurized for some time as it moves out into the halo.  
In such a halo, it might be possible for cool gas ejected from the disk to dissipate as it expands into the low-pressure CGM.  As the gas clouds expand, their density would drop, causing the ionization parameter to rise and leading to a greater degree of gas ionization until it melds with the ambient halo.

	Either low overall halo pressures or significant support from non-thermal pressures could cause the non-TPE results shown in Figure~\ref{fig:CGM_pressure}.  \cite{Ji19}, for example, find that cosmic rays added to the FIRE-2 simulations can lower the thermal pressure for the low-ion dominated gas phase by more than an order of magnitude relative to the thermal pressure without cosmic rays.  According to their simulations, however, we should expect to find a thermal pressure gradient pointing in the opposite direction of what we find for the PPP and PPC models of Absorbers~0.4A and~0.6A.   

\subsection{Cloud Sizes}
\label{sec:cloud_size}

\begin{figure}
	\includegraphics[width=\columnwidth]{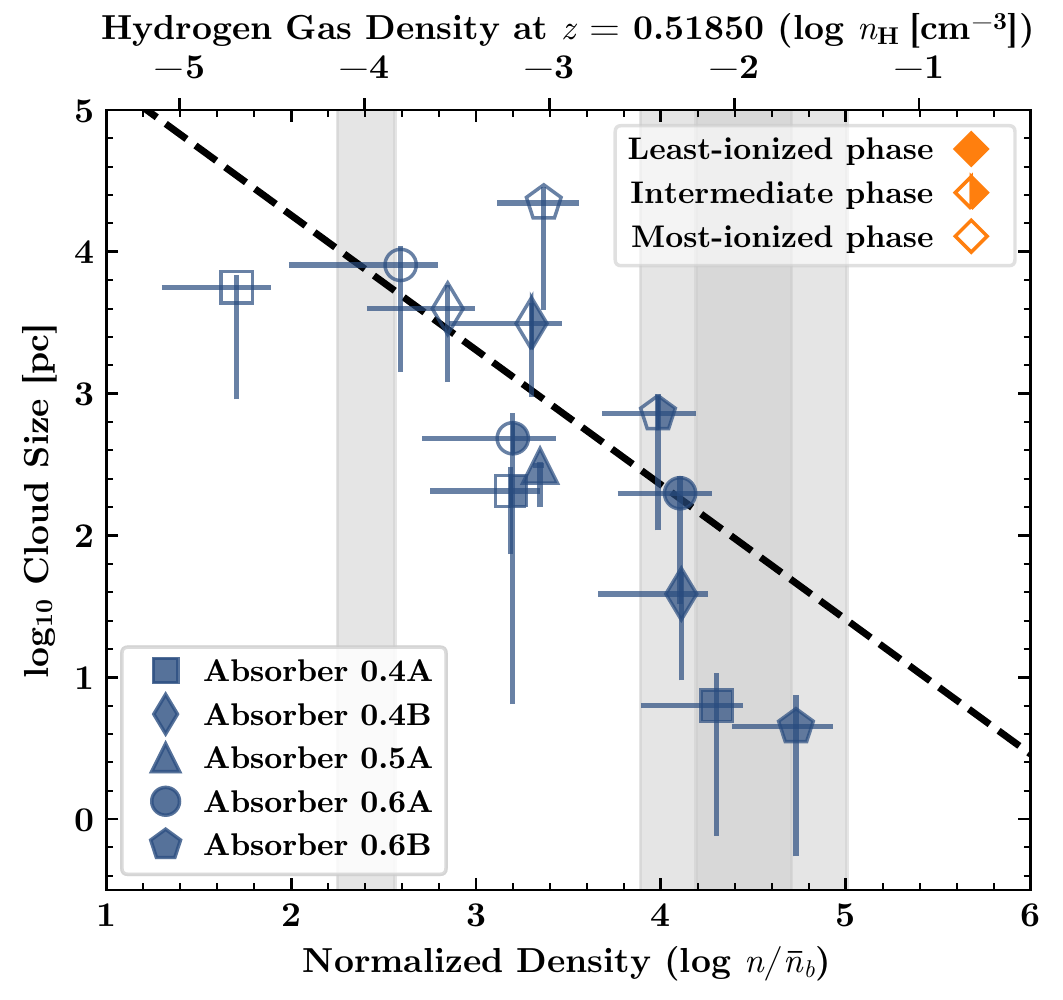}
	\caption{Cloud scale lengths $\mathnormal{l} \geq N_{\rm H}$/\emph{n}$_{\rm H}$ of individual model phases vs. the hydrogen gas density normalized by the cosmic mean baryon number density (taken here as $\bar{n}_b=1.85\times10^{-7} (1+z)^{3}$).  These are extracted from the Bayesian weighted average model (see Section~\ref{sec:twophase}).  The top axis provides a conversion to hydrogen number density (i.e., the parameter used in our models) at redshift \emph{z}=0.51850; one can shift the units of the top axis by -0.09 or +0.14 to convert to redshift \emph{z}=0.41760 or \emph{z}=0.68606, respectively.  Shapes indicate phases pertaining to the averaged-models of specific absorbers, and filled/semi-filled/open symbols denote low/intermediate/high-ion dominated phases within each absorber.  For comparison, the gray bands cover the range of densities predicted for a 10$^{\textrm{11.5}} \textrm{M}_{\odot}$ halo by the two-phase analytic halo model of \protect\cite{Maller04}.  The isolated, leftmost band shows the predicted hot-phase density between 0.2--1\emph{R}\textsubscript{vir} while the wider, rightmost bands show the predicted cool-phase density at the same radii for temperatures T=10\textsuperscript{4} K (\emph{thin band}) and T=10\textsuperscript{3.7--4.3} K (\emph{wide band}).}
	\label{fig:TR_dist}
\end{figure}

	The ratio of the total hydrogen column density to the hydrogen number density provides a scale for the physical extent of a given modeled gas phase, which we can use to estimate absorber sizes (Equation~\ref{eqn:cloudsize}).  In Figure~\ref{fig:TR_dist}, we present such ``cloud'' size estimates, broken down by phase and absorber, for our Bayesian model averages (see Section~\ref{sec:twophase}).  Note that all our models include a prior discouraging cloud sizes larger than 10 kpc (but with a long tail, see Section~\ref{sec:NHscaling}).  Also shown for comparison is the powerlaw size vs. density relation from the multiphase CGM modeling by \cite{Stern16}.  Their multiphase approach differs fundamentally from ours in that their ionic column densities are produced by gas following a powerlaw density distribution, while our approach treats the measured column densities as composites of ionization phases at discrete, constant densities.\footnote{Other key differences include our parameterization of the UVB and relative abundances, and our focus on modeling individual velocity components.  Additionally, they adopted 0.2 dex uncertainties for all their measurements, while most of our measurements are at least three times more restrictive.}

	Examining individual model phases, we find that the cloud size correlates with the degree of ionization, e.g., the phase having the largest scale length is also the one that produces the dominant fraction of high-ions.  The intermediate (in ionization) phases of the Bayesian weighted average models for Absorbers 0.4A, 0.4B, 0.6A, and 0.6B are also intermediate in cloud size, accounting for approximately 6\%, 47\%, 5\%, and 3\%, of the respective absorber's total modeled path length, assuming a volume filling factor of unity in every phase, (i.e., $\textit{f}_\textrm{v}=1$ in Equation~\ref{eqn:cloudsize}).  For dense, low-ion dominated phases these numbers shrink to approximately 0.1\%, 0.5\%, 2.3\%, and 0.03\%, respectively.  The squares of these 1-D distances roughly approximate the relative probability of observing each of the two or three phases in arbitrary lines-of-sight.  For example, if our absorbers are structured in a layered spherical shell geometry with cold gas embedded within successively hotter and volumetrically larger gas phases, then our models would imply an O~\textsc{vi} cross section that is $\gtrsim$300 and $\gtrsim$3000 times larger than the associated (i.e., kinematically aligned) O~\textsc{iii} and Mg~\textsc{ii} cross sections, respectively.  While higher ionization stages do appear more frequently in blind redshift surveys \citep[see, e.g., Fig.~10 in][]{Tripp08} their enhanced detection frequency relative to that of the lower ionization stages is much less than the multiple orders of magnitude discrepancy anticipated by our simple geometrical argument.  Various theories have been proposed to solve this conundrum.  For example, the lower ionization phases could be dispersed throughout the larger high-ion phase as a fine ``mist'' \citep[e.g.,][]{McCourt18,Liang19,Gronke20b}, which would effectively increase the lower-ions' cross-section.  Indeed, \cite{Werk19} find empirical evidence of small ($\le 10$ pc) clumps of low-ions embedded in much larger ($\ge$ 1 kpc) coherent mid-/high-ion structures.  Alternatively, cool gas could form as a shocked, thin shell surrounding a larger hot bubble \citep[e.g.,][]{Thompson16}.  It is also possible that cold clouds form in situ in complex, multiphase outflows from galaxies \citep[e.g.,][]{Gronke20a}. The overall sizes of our absorbers are small enough to allow any of these scenarios.  This contrasts with other studies in the literature \citep[e.g.,][]{Hussain17}, which infer significantly larger cloud sizes that would be in tension with some of the theoretical studies above.  Our smaller sizes partly result from our imposed prior penalty on large cloud sizes.  This does not lessen our conclusion, however, that viable models of highly ionized clouds can be constructed that do not require tremendously large clouds.

\subsection{The Ionizing Radiation Field}
\label{sec:UVB_inferences}

	Models of the extragalactic ionizing flux background are fundamental to our understanding of the IGM/CGM and other topics, such as reionization, yet observational constraints on the shape and intensity of the UV background are scarce.  Not only are models of the \emph{mean} extragalactic radiation field \citep[e.g.,][]{Haardt96,Haardt01,Haardt12,FaucherGiguere09} intrinsically uncertain, but in the vicinity of AGN and star forming galaxies, their uncertainty is compounded by the potential for local additions to the background ionizing flux \citep[see][]{Upton-Sanderbeck18}.  The column density ratios of many ions commonly detected in CGM/IGM absorbers are highly sensitive to the UV radiation field (see, for example, Figure~\ref{fig:UVBplotA}).  Consequently, line-rich QSO absorption systems with excellent quality spectra present an important tool for constraining \emph{local} variations in the ionizing UV radiation field. 

\begin{figure}
	\begin{subfigure}{.45\textwidth}
	\centering
	\includegraphics[width=3.3in]{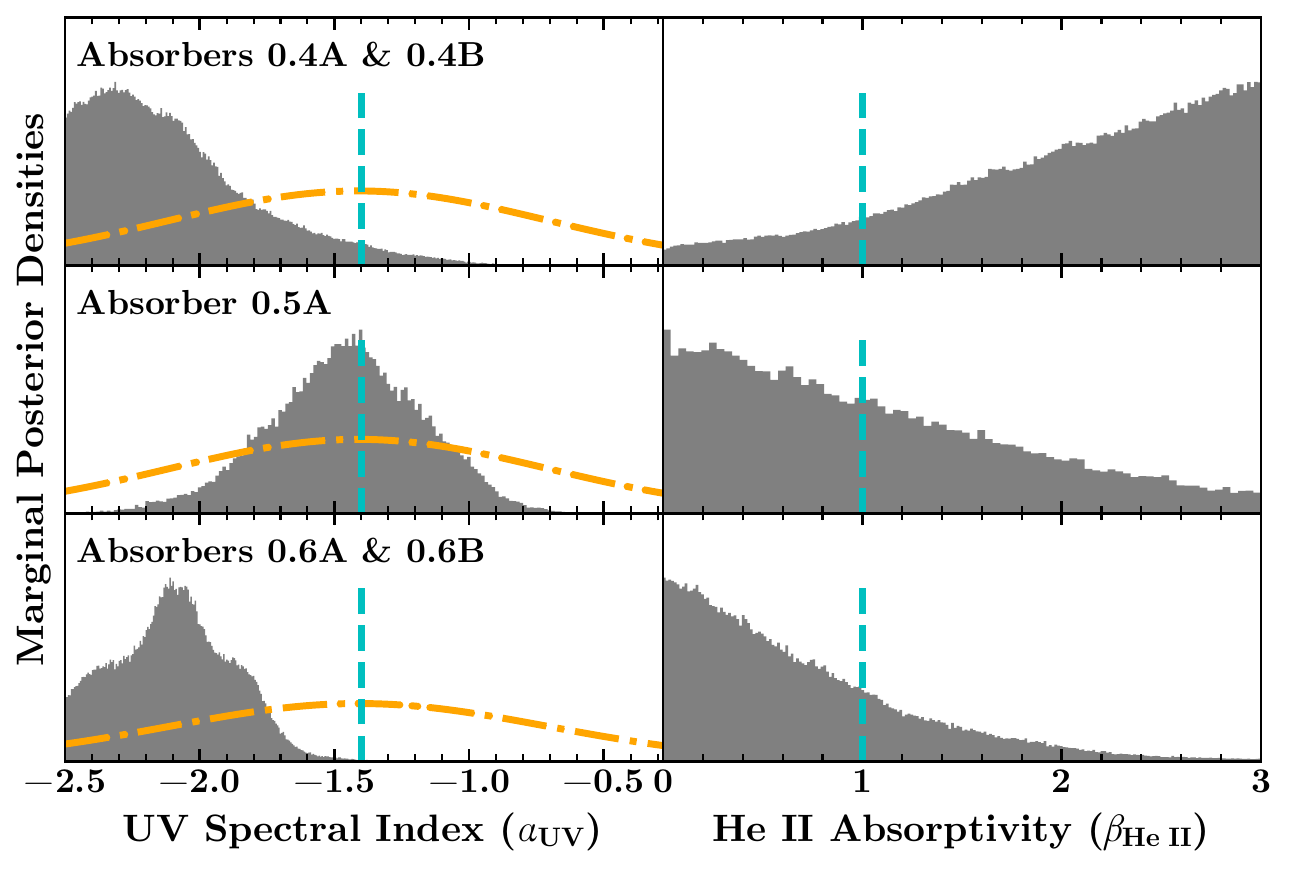}
	\caption{}
	\label{fig:uv_marginals}
	\end{subfigure}
	\begin{subfigure}{.45\textwidth}
	\centering
	\includegraphics[width=3.3in]{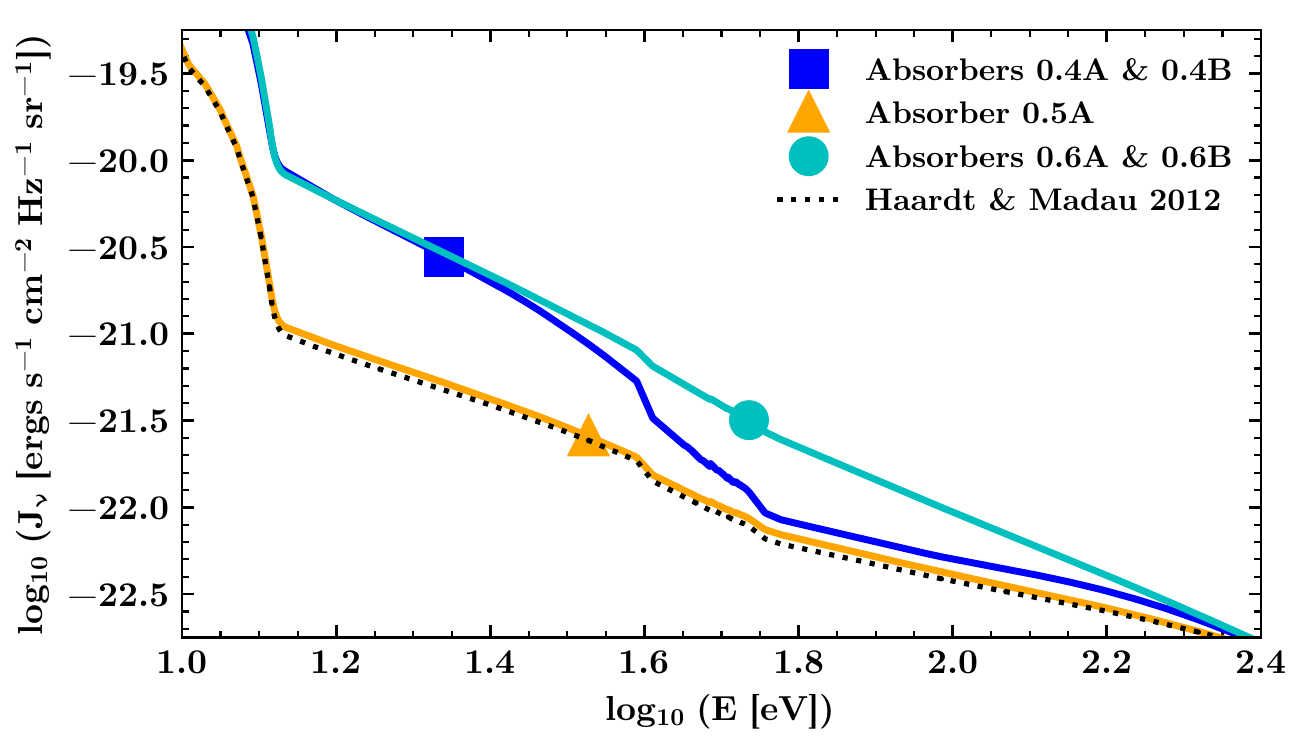}
	\caption{}
	\label{fig:uv_sed_realization}
	\end{subfigure}
\caption{ \textsl{(Top)} Full probability densities for the UV parameters inferred from our Bayesian model-averaged posteriors.  The vertical dashed cyan lines mark the fiducial values, and the orange dash-dotted curves illustrate the prior density for $\alpha_\textrm{UV}$.  Note that the distributions plotted for Absorber~0.5A were derived from the subset of acceptable models that include $\alpha_\textrm{UV}$ and/or $\beta_\textrm{He\ II}$ as free parameters (see Table~\ref{tab:selection_metrics_z051850}).  \textsl{(Bottom)} Realizations of the most probable spectra derived from the posteriors shown in the top panel.  The black dotted line illustrates the fiducial \protect\cite{Haardt12} spectrum at redshift $z$=0.68606.  For plotting purposes, we normalized all the UV spectral energy distributions by $\Gamma_\textrm{H I}^\textrm{HM12}$(\textit{z}=0.68606) (details in Section~\ref{sec:parameters}) to minimize redshift-dependent intensity variations.}
\label{fig:UVB_dist_updated}
\end{figure}

	To extract information about the UV radiation field from our absorbers, we parameterize the ionizing UV spectrum in our models as a modified powerlaw, with a variable slope and He~\textsc{ii} Lyman Limit absorption ``trough'' (see Section~\ref{sec:parameters} for details).  As described in Section~\ref{sec:twophase}, we model the ionizing flux in two ways, 1) with independent UV parameters for each absorber (``independent UV''), and 2) with a shared set of UV parameters covering all absorbers at the same systemic redshift (``joint UV'').  Both approaches can produce models with acceptable Bayes factors (see Tables~\ref{tab:selection_metrics_z041761} and~\ref{tab:selection_metrics_z068605}), but we find that ``joint UV'' models generally score better than the ``independent UV'' versions of the same models.

Figure~\ref{fig:uv_marginals} plots marginal posterior densities for $\alpha_\textrm{UV}$ and $\beta_\textrm{He\ II}$ for each absorption system.  These suggest that the four multiphase absorbers (Absorbers~0.4A, 0.4B, 0.6A, and 0.6B favor a softer UV radiation field than the fiducal one \citep{Haardt12}.  Specifically, the most probable value of $\alpha_\textrm{UV}$ for the absorption system at \textit{z}=0.41760 is $\approx-2.3$, and for the two absorbers at \textit{z}=0.68606 it peaks at $\approx-2.1$.  While we understand that the ionizing UV spectral index of \citep{Haardt12} ($\alpha_\textrm{UV}^\textrm{HM12}$$\approx$-1.4) is itself derived from uncertain models, it is unlikely that the cosmic UV background is as soft as our multiphase absorbers indicate.  One possibility is that a nearby AGN with a comparably soft ionizing flux spectrum might boost the local radiation field ionizing flux density and tilt its spectral slope.  Indeed, the distribution of spectral indices found by \cite{Telfer02} in their subset of radio-loud AGN, with a mean of $\alpha_\textrm{UV}=-1.96\pm0.12$, supports this possibility.  Other studies of composite ultraviolet quasar and AGN spectra, e.g., \cite{Scott04} and \cite{Stevans14}, are also consistent with local values of $\approx-2.3$ to $-2.1$ for individual cases.  Alternatively, the ionizing flux field for these absorbers could have an important contribution from nearby galaxies.  \cite{Upton-Sanderbeck18} describe several potential galactic and circumgalactic sources of ionizing radiation associated with galaxies in addition to the omnipresent UV background.  Radiation from hot stars, for example, boosts the longer UV wavelengths more than the shortest ones, which, when combined with the UVB would help to soften its spectral index.

The second parameter governing the shape of the UV radiation field in our models, $\beta_\textrm{He\ II}$, provides more clues as to the origin of the extra UV flux.  $\beta_\textrm{He\ II}$ adjusts the depth of the He~\textsc{ii} trough at 4 Ryd, which is sensitive to the amount of helium along the path from the radiation source to the absorber.  While the effective helium opacity becomes significant for distant sources, radiation from a local AGN is unlikely to encounter sufficient helium to produce a significant effect.  In contrast, stellar atmospheres generally imprint a strong He edge on stellar spectra.  Thus, if galaxies (stellar sources) are leaking ionizing radiation, we generally expect a deeper He trough in the ionizing UV flux than the He feature of the diffuse UV background alone.  Using these qualitative differences between AGN and stellar sources of ionizing radiation, the posterior distributions of $\beta_\textrm{He\ II}$ from models of Absorbers~0.4A and 0.4B (a strong He~\textsc{ii} trough) may indicate that the extra radiation has a stellar origin, while the models of Absorbers~0.6A and 0.6B (little/no He~\textsc{ii} trough) are consistent with ionizing flux contributions coming from an AGN.

Figure~\ref{fig:uv_sed_realization} offers another perspective on the inferred UV parameters.  The cyan and blue lines reproduce the most probable UV ionizing flux SED for the multiphase absorbers.  Their shapes ($\alpha_\textrm{UV}$) look very similar, except that the absorption trough at 4 Ryd is strong in Absorbers~0.4A and 0.4B and almost non-existant ine Absorbers~0.6A and 0.6B.  Both, however imply an extra component of ionizing flux above the fiducial \cite{Haardt12} estimate (dotted black line).  In constrast, the shape of the single-phase absorber (Absorber~0.5A) broadly aligns with \cite{Haardt12}, based on those models for this absorber where the UV radiation field is allowed to vary.

\subsection{Metallicity and Relative Abundance Ratios}
\label{sec:metallicity_inferences}

\begin{figure}
	\includegraphics[width=\columnwidth]{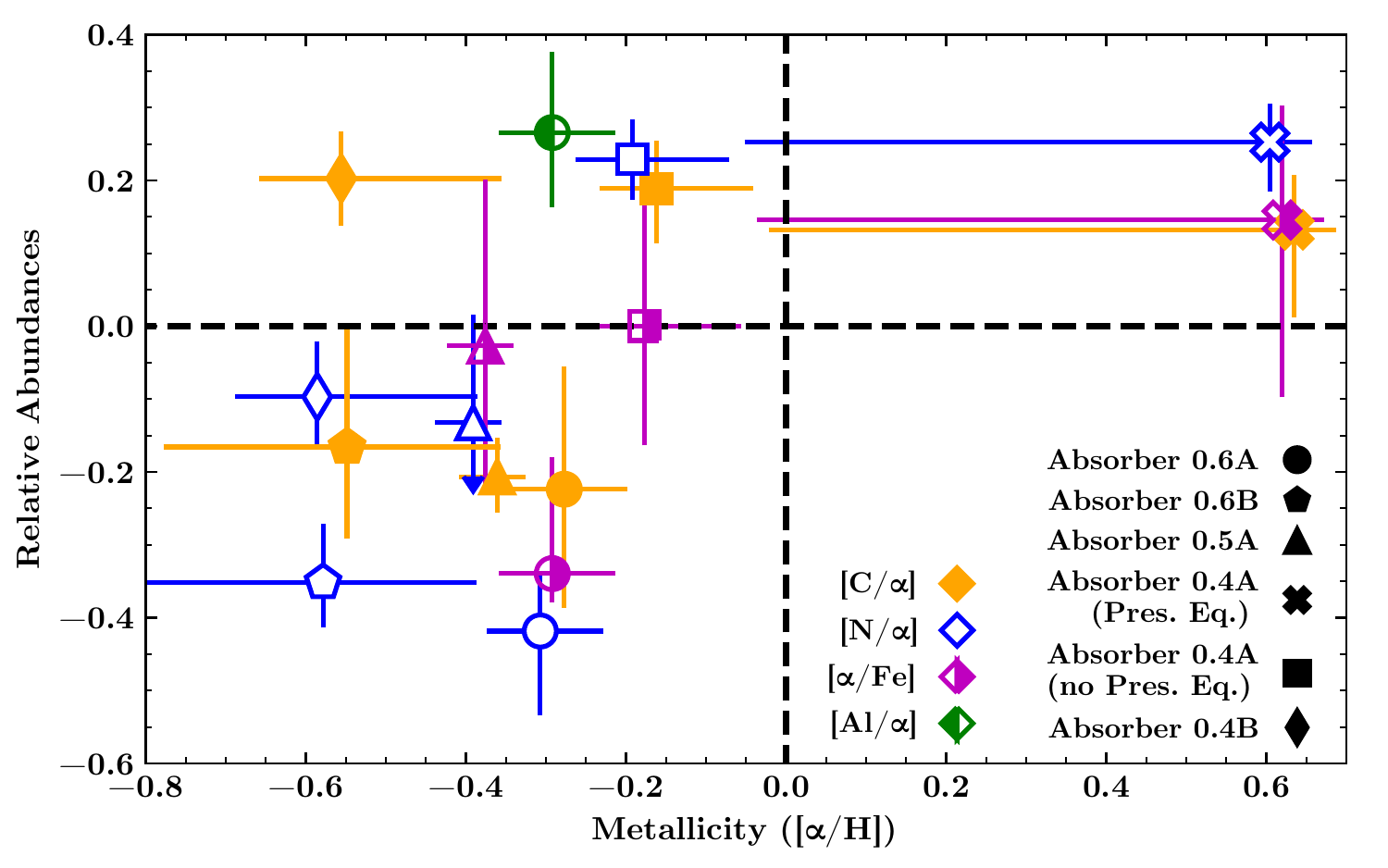}
	\caption{Modeled relative abundance ratios for C, N, Al, and Fe, with respect to the $\alpha$-element abundances, vs. the modeled overall metallicity [$\alpha$/H].  The symbols represent point estimates of the Bayesian model-averaged (see Section~\ref{sec:twophase}) parameter distributions (i.e., marginal posterior densities).  $\alpha$/Fe is not shown for Absorbers~0.4B and~0.6B, where it is poorly constrained.  Likewise, Absorber~0.6A is the only absorber where Al/$\alpha$ could be constrained.  For clarity, we applied a -0.015 dex offset on the horizontal (metallicity) axis to the the blue points ([N/$\alpha$]) and a +0.015 dex offset to the metallicities of the orange points ([C/$\alpha$]).}
	\label{fig:abnd_dist}
\end{figure}

	Our ionization models all include parameters governing the absorbers' metallicity and deviation from solar C/$\alpha$, N/$\alpha$, and $\alpha$/Fe ratios.  Figure~\ref{fig:abnd_dist} plots point estimates of the marginal posterior densities of these parameters.  Recall that all of the non-$\alpha$ elements we use to constrain our models are assigned independent abundance parameters. We use [$\alpha$/H] as an indicator of the overall absorber metallicity.  We chose to keep the metallicity and enrichment parameters constant across all phases within each absorber as discussed in Section~\ref{sec:twophase}.  In three-phase models the metallicity of the highest ionization phase is minimally constrained owing to the limited number of high-ion detections, so the resulting metallicities are primarily driven by the two lowest ionization phases.  

	Unsurprisingly, all five of the absorbers are metal-enriched with $\alpha$-metallicities greater than a quarter solar, a fact that we attribute to the selection of line-rich absorbers for this study.  The metallicities are consistent with the distribution of metallicities measured in previously studied absorber samples with similar neutral hydrogen column densities \citep{Lehner13,Wotta16,Prochaska17}.  We note that a super-solar metallicity is indicated by the thermal pressure equilibrium solution for Absorber~0.4A.  The metallicity for this model really follows a bimodal distribution (hence the large range extending to lower $Z$ shown in Figure~\ref{fig:abnd_dist}) with a weaker secondary peak only slightly more enriched than the non-TPE solution for Absorber~0.4A (the square symbols).  While we cannot rule out such a high metallicity arising in particular instances, the models present acceptable solutions that are less super-solar.

	Our modeled metallicities also lie at the high end of the gas phase O/H metallicity gradient found in the nebular emission of higher-mass (M\textsubscript{$\star$}$>$10\textsuperscript{10} M$_{\odot}$) local galaxies \citep[e.g.,][]{Belfiore17}.  This correlation could suggest that the absorbers formed from enriched gas pushed out (e.g., via stellar/supernova winds or AGN outbursts) of the central regions of such galaxies into the CGM.  That picture becomes less clear, however, when our relative abundance constraints are considered.  Coincident with supersolar O/H, \cite{Belfiore17} also found supersolar N/O ratios, in contrast with our modeling that requires subsolar N/O in all but one case.  One possibility is that at the redshifts of our absorbers (\emph{z}=0.41760, 0.68606, corresponding to lookback times of $\sim$5 Gyr) the secondary production of nitrogen through the CNO-cycle has not yet had time to enrich the gas.  Yet, the gas phase N/O ratio shows no evidence of evolving since \emph{z}$\approx$1--2 in the limited number of galaxies analyzed by \cite{Kojima17} and in a stack of galaxies by \cite{Masters14}, for example.

	Iron is another element with a delayed nucleosynthesis relative to the $\alpha$-elements that provides an interesting point of comparison with three of our absorbers.  Lower metallicity halo stars and globular clusters in the Milky Way offer well-known examples of enhanced $\alpha$/Fe \citep[e.g.,][]{McWilliam97} at low metallicities from the generation of stars formed before Type Ia supernovae had time to boost the iron abundances.  Near solar metallicities, however, the $\alpha$/Fe ratio is observed to decrease to solar levels, which is more consistent with the near-solar $\alpha$/Fe ratios we find for Absorbers~0.4A and~0.5A. Surprisingly, our models prefer a sub-solar $\alpha$/Fe ratio for Absorber~0.6A, i.e, it requires an excess of iron relative to what is expected.  In some models of galactic outflows, the hot outflowing metals are found to separate from the bulk of the ISM mass \citep{Maclow99}, so it is possible that galaxy halos contain pockets of metal-enriched gas with abundance patterns connected to particular metal sources.  A larger sample of CGM abundance ratios will be crucial to further clarify the possible progenitors of these CGM absorbers.

	Absorber~0.6A also exhibits a robust Al/$\alpha$ overabundance relative to solar.  The introduction of an Al/$\alpha$ parameter was necessary to obtain a satisfactory model for this absorber.  Although uncommon, some enhanced Al/$\alpha$ abundances in QSO absorbers have been noted in the literature.  For example, \cite{Ganguly98} reported [Al/$\alpha$]$\approx$0.3 at -0.55$\leq$[\emph{Z}/\emph{Z}\textsubscript{$\odot$}]$\leq$-0.32 for an absorber with similar $N$(\textsc{H~i}) as for the systems in this paper but at \emph{z}=1.94, and \cite{Richter05} found [Al/$\alpha$]$\approx$0.4 for a low-metallicity sub-DLA at \emph{z}=2.187.  Several other examples of enhanced aluminum abundances, predominantly in high redshift, low-metallicity, and/or high-\emph{N}(\textsc{H~i}) QSO absorbers, are noted in \cite{Prochaska02} and \cite{Glidden16}.  These papers offer several speculations regarding the origin of non-solar aluminum relative abundances.  The possibility of producing Al in novae stands out as the most consistent with our ejecta interpretation of the other observed relative abundance trends above, though these references also discuss the possibility that inferred non-solar aluminum relative abundances might originate in part owing to incorrect atomic physics (e.g., ionization and recombination rates).

	Circumgalactic dust has the potential to confuse relative abundance studies since it differentially depletes metals from the gas phase.  While evidence for circumgalactic dust comes from several different lines of inquiry \citep[e.g.,][]{Menard10,Peek15,DeCia16,Frye19}, it is often neglected in analyses of optically thin CGM absorbers where exposure to the UV background is thought to suppress the amount of dust in the gas.  In support of this assumption, \cite{Fumagalli16} found that the physical conditions of a large sample of Lyman Limit Systems (LLSs) were minimally impacted by a model incorporating a dust abundance scaling based on \cite{Jenkins09}, though it should be noted that most of these absorption systems are located at redshifts \emph{z} $\geq$ 2.5 and thus could have lower metallicities and lower dust contents.  While we did not include a dust-specific relative abundance parameterization in our models, we have examined the abundances of species that are expected to be less depleted by dust (e.g., O and S) versus those that have a greater propensity to deplete (e.g., Si, Mg, and Fe), and we did not find any compelling indications of dust depletion patterns (see, e.g., \citealt{Jenkins09}). Indeed, the enhanced Al/$\alpha$ ratio discussed above is also in the wrong direction for a dusty absorber, since aluminum is expected to deplete strongly onto dust grains \citep{Jenkins04}.


\section{SUMMARY}
\label{sec:summary}

	In this work, we developed a Bayesian multiphase ionization modeling scheme with the goal of accounting for all the observed ionization stages simultaneously at the maximum kinematic resolution possible.  To explore the models' dependency on weakly constrained variables, we introduced several new or seldom encountered parameters governing 1) the number of model phases per absorber, 2) the shape of the ionizing radiation field incident on the absorbing gas, and 3) relative abundance ratios of non $\alpha$-elements in the same gas.  We treated each velocity component (referred to throughout as independent ``absorbers'') in an absorption system as being produced by an unresolved composition of gas from multiple different physical states, i.e., ``phases''.  To robustly navigate the enlarged parameter space in search of optimal solutions to our multiphase models, we employed an MCMC sampling procedure and several carefully chosen priors on the model parameters.  These include a novel hyperprior on the thermal pressure, which we use to interpret the probability of thermal pressure equilibrium between the individual phases of our multiphase absorbers.

	To showcase our ionization modeling approach, we selected absorption line measurements and multiphase ionization models for five strong H~\textsc{i} absorbers ($\log \textit{N}_\textrm{H I}$$>$15) identified in three line-rich absorption systems at intermediate redshifts (0.4$<$\emph{z}$<$0.7) in quasar spectra observed as part of the CASBaH program.  The good signal-to-noise, full FUV-to-optical coverage of the background quasar spectra, and the optimal redshifts of the absorbers allowed us to detect several species that are difficult or impossible to observe at low (z $\lesssim$ 0.4) and high (z $\gtrsim$ 1.5) redshifts.  The unusual number and precision of column density measurements gleaned from these absorbers were essential for constraining our complex ionization models.

	Applying multiphase ionization equilibium models to these absorbers, we find that only 1/5 of the absorbers is consistent with a single ionization phase.  The other 4/5 absorbers require \textit{at least two} distinct ionization phases to satisfactorily model all observed ionization stages simultaneously.  In fact, a rigorous model comparison using Bayes factors strongly favors three-phase models for all four multiphase absorbers.  An examination of the individual model phases reveals that for many of the species we detected, including commonly observed ions such as H~\textsc{i}, C~\textsc{ii}, C~\textsc{iii}, Si~\textsc{iii}, and Si~\textsc{iv}, multiple model phases may contribute a non-negligible fraction toward their total ionic column densities, making it impossible to categorize many of these ions as belonging exclusively to a particular gas phase.  In some models, a significant amount (as much as $\approx20$\% or more) of the neutral hydrogen, for example, may originate outside the coolest, densest modeled gas phase that contains $>95$\% of the Mg~\textsc{ii} and Fe~\textsc{ii}.  Interestingly, we also find that singly- and doubly-ionized species often \textit{do not} originate in the same gas phase; ions such as C~\textsc{iii}, N~\textsc{iii}, and O~\textsc{iii} may share more in common (regarding their production) with ``hotter'' ions such as C~\textsc{iv}, N~\textsc{iv}, and O~\textsc{iv} than with low-ions like C~\textsc{ii}, N~\textsc{ii}, and O~\textsc{ii}.  We also examined whether photoionization-only and fixed-temperature photo- plus collisional ionization models (i.e., collisional ionization models where the incident ionizing radiation field is not ignored) are capable of satisfactorily reproducing the measured column densities of all detected species simultaneously.  O~\textsc{vi}, which we detected in all our multiphase absorbers, is readily accounted for by collisional ionization models.  We also found photoionization models that could produce O~\textsc{vi} in two of the absorbers, though not for the two remaining absorbers where we detect strong high-ion column densities, including Ne~\textsc{viii}.

	We discussed several physical properties of our IGM/CGM absorbers, including thermal gas pressures, cloud sizes, deviations from solar metallicity and relative abundances (for C, N, Al, and Fe), and implications for the UV ionizing radiation field impinging on these systems.  For all four multiphase absorbers we found model configurations that can be interpreted as having multiphase pressures consistent with thermal pressure equilibrium, though for two absorbers we also found successful models with inter-phase thermal pressures severely out of equilibrium with each other.  We observed that the key factor for building thermal pressure equilibrium models is the dynamic range of the modeled temperature.  Photoionization models simply cannot achieve the higher temperatures needed to balance lower gas densities of more highly ionized gas phases.  In addition to the question of equilibrium, we also find that all our multiphase models exhibit low overall thermal pressures, and the set of non-pressure equilibrium models fall well below reasonable estimations of the halo pressure based on a two-phase halo analytical model for a 10$^{\textrm{11.5}} \textrm{M}_{\odot}$ halo.  This suggests that the absorbers either reside in relatively low-mass (sub-\textit{L}\textsubscript{*}) halos (we blindly selected our absorbers with respect to their environments) or they have significant non-thermal pressure support.

	Our models also successfully produce absorbers with realistic cloud scale lengths of $\lesssim10-30$ kpc after setting a prior to restrict excessively extended solutions.  The highest density gas phase in each absorber is consistent with sub-kiloparsec scale lengths.  We also inferred the shape of the ionizing radiation field from our models, finding evidence that suggests the presence of a local source of radiation near four of our absorbers, while the other one is consistent with photoionization by the UV background ionizing flux.  Finally, we derive metallicities greater than 25\% solar in all five absorbers and find evidence of significant deviations from solar relative abundances, particularly for N/$\alpha$, in all five.  These case studies serve as a preview of our larger, ongoing CASBaH investigation of physical conditions in IGM/CGM absorbers that will utilize the multiphase ionization modeling scheme developed herein.

\section*{Acknowledgements} The authors gratefully acknowledge Martin Weinberg for many helpful and stimulating discussions about statistical methodologies and the circumgalactic medium. This research is based on observations with the NASA/ESA Hubble Space Telescope, which is operated by the Association of Universities for Research in Astronomy, Incorporated, under NASA contract NAS5-26555. Financial support for HST programs HST-GO-11741 and HST-GO-13846 was provided through grants from the Space Telescope Science Institute under NASA contract NAS5-26555.  N. Katz acknowledges support from NASA grant 80NSSC18K1016.  Some data presented herein were obtained at the W. M. Keck Observatory, which is operated as a scientific partnership between the California Institute of Technology, the University of California, and the National Aeronautics and Space Administration. The Keck Observatory was made possible by the generous financial support of the W. M. Keck Foundation.  The authors recognize and acknowledge the very significant cultural role and reverence that the summit of Maunakea has always had within the indigenous Hawaiian community.  We are most fortunate to have the opportunity to conduct observations from this mountain.  The large grid of ionization models required for this work were precomputed in parallel using the Linux package ``GNU Parallel'' \citep{Tange2011} on resources provided by the University of Massachusetts' Green High Performance Computing Cluster (GHPCC).  Financial support for computational costs was provided in part by a University of Massachusetts Amherst Graduate School Dissertation Research Grant.


\section*{Data Availability} The data underlying this article are presented within the article except for several large interpolation tables, which were generated from publicly available software (\textsc{cloudy}, \url{https://nublado.org}), and the original UV and optical quasar spectra, which are publicly available from the Mikulski Archive for Space Telescopes \url{https://archive.stsci.edu/}, and from the Keck Observatory Archive \url{https://www2.keck.hawaii.edu/koa/public/koa.php}.  The various softwares used to analyze the spectra and compute Bayesian multiphase ionization models were developed by us and are not available for download at present.



\appendix

\section{BAYESIAN MODEL DETAILS}

\subsection{The Likelihood Function}
\label{app:likelihood}

	The likelihood function (Equation~\ref{eqn:like}) for our multiphase model assumes that column densities measured from the spectra are random variates of a zero-truncated normal distribution (ZTN).  What we refer to throughout this paper as the zero-truncated normal (ZTN) distribution is really a simplification of the truncated normal distribution, which is defined as:
\begin{equation}\label{eqn:truncnorm}
	p_\mathrm{Trunc.\ Normal}(x,\mu,\sigma,a,b) = \frac{\phi(\frac{x-\mu}{\sigma})}{\sigma \left( \Phi(\frac{b-\mu}{\sigma}) - \Phi(\frac{a-\mu}{\sigma}) \right)}
\end{equation}
where $\phi(\cdot)$ and $\Phi(\cdot)$ reference the standard normal probability density and cumulative density functions (respectively), which we provide here for completeness:
\begin{equation}\label{eqn:STDNpdf}
	\phi(x) = \frac{1}{\sqrt{2\pi}} e^{-\frac{1}{2}x^2}
\end{equation}
and
\begin{equation}\label{eqn:STDNcdf}
	\Phi(x) = \frac{1}{2}(1 + \mathrm{erf}(\frac{x}{\sqrt{2}})).
\end{equation}
The ZTN probability density corresponds to setting $a,b=0,-\infty$ in Equation~\ref{eqn:truncnorm}, which simplifies to:
\begin{equation}\label{eqn:ZTNpdf}
	f_\mathrm{ZTN}(x,\mu,\sigma) = \frac{\phi(\frac{x-\mu}{\sigma})}{\sigma \left( 1 - \Phi(-\frac{\mu}{\sigma}) \right)}
\end{equation}
and the accompanying cumulative density function is:
\begin{equation}\label{eqn:ZTNcdf}
	F_\mathrm{ZTN}(x,\mu,\sigma) = \frac{\Phi(\frac{x-\mu}{\sigma}) - \Phi(-\frac{\mu}{\sigma})}{1 - \Phi(-\frac{\mu}{\sigma})}
\end{equation}
Our adoption of the ZTN as our column density sampling distribution was motivated by simulations of absorption features of different strengths in simulated counts spectra at various signal-to-noise ratios.  For each test case, we ran 1000 different realizations of the spectrum, each time fitting a Voigt profile (with fixed velocity and \emph{b}-value constrained between 10-25 km s\textsuperscript{-1}) to the absorption feature (even if non-detected) using $\chi^{2}$ minimization.  The histogram of best fitting column densities matches best with a ZTN distribution, even when the mean of the ZTN was fixed to the ``true'' (simulated) column density value.  Note that the assumption of ZTN distributed column densities may break down in more complex scenarios than those considered above, such as multiple adjacent absorption lines (whether from the same system or a contaminating one), or for Voigt-profile fitting measurements with shared (and hence correlated) parameters.  A proper treatment of such covariates has not been considered in the literature, however, and would be beyond the scope of this paper.  For the present, using the simpler (truncated normal) likelihood distribution described above should suffice as an approximation.  

	The full likelihood distribution is the product of independent ZTNs, one for each species-absorber pair (denoted by subscripts $ion,c$).  The width of an individual ZTN (for species $ion$) is set to the measured uncertainty in the column density measurements, i.e., $\sigma_\textrm{ZTN} = \sigma_{ion,c}^{meas}$.  The mean\footnote{To be precise, we use $\mu_\textrm{ZTN}$ to denote the 1st moment of the \textit{non}-truncated normal distribution.} of each ZTN sampling distribution is set by the model-predicted column density for a given set of model parameters, i.e., $\mu_\textrm{ZTN} = N_{ion,c}^{model}$.  Equation~\ref{eqn:like} gives the likelihood of the observed column densities as the product of ZTN probabilities.  The specific form of the ZTN for each term in the product depends on whether the measured column densities belong to the set of $Detections$, $Upper\ Limits$, or $Lower\ Limits$ (i.e., $\{D\}$, $\{UL\}$, or $\{LL\}$ in Equation~\ref{eqn:like}).  Since the column density measurement (kinematic) resolution often differs for various species (for the reasons discussed above), we introduce the concept of \textit{component groups} to handle measurements that span one or more modeled absorbers (i.e., velocity components).  These component groups (CG) are exclusive, meaning they encompass adjacent, but non-overlaping portions of the absorption features.  A CG may contain anywhere from one to all of the measured column density constraints used in the model.  CGs are in essence bookkeeping tools, enabling us to compare a lower (kinematic) resolution measurement to the more finely resolved models, by keeping track of which modeled absorbers (i.e., higher resolution gridpoints of the model) are encompassed by the CG.

	While the likelihood of detection is given by the ZTN probability density function, for upper limits the likelihood is the \textit{integrated} tail probability $P(N_\textrm{ZTN} \leq N_{ion,c}^{limit})$, which is obtained by evaluating the ZTN cumulative distribution function (CDF, given by Equation~\ref{eqn:ZTNcdf}) at the measured limit $N_{ion,c}^{limit}$.  Lower limits are analogous, but correspond to upper tail probabilities, i.e., $P(N_\textrm{ZTN} \geq N_{ion,c}^{limit})$.  Thus the full likelihood distribution is given by (with reference to Equations~\ref{eqn:ZTNpdf} and~\ref{eqn:ZTNcdf}):
\begin{equation}\label{eqn:like}
\emph{L}(\textbf{\emph{N}}|{\bmath{\theta}}) = \scalebox{1.25}{$\displaystyle \prod_{ion}^{N_{\rm ions}} \prod_{c=1}^{N_{\rm CG}}$} 
\begin{cases}
	f_{\rm{ZTN}}(N_{ion,c}^{meas},N_{ion,c}^{model},\sigma_{ion,c}^{meas}), & {\rm if}\ ion \in \{D\}^c \\
	F_{\rm{ZTN}}(N_{ion,c}^{limit},N_{ion,c}^{model},\sigma_{ion,c}^{limit}), & {\rm if}\ ion \in \{UL\}^c \\
	1-F_{\rm{ZTN}}(N_{ion,c}^{limit},N_{ion,c}^{model},\sigma_{ion,c}^{limit}), & {\rm if}\ ion \in \{LL\}^c \\
\end{cases}
\end{equation}
\begin{equation*}
{\rm where}\ N_{ion,c}^{model} = \scalebox{1.25}{$\displaystyle \sum_{i=1}^{N_{\rm abs}(c)\ }$} \scalebox{1.25}{$\displaystyle \sum_{p=1}^{N_{\rm phases}(ion^i)}$} N^{\textsc{cloudy}}_{ion}(\bmath {\theta_{p}^{i}}) \left( \frac{N_{\rm H}^{i,p}}{N_{\rm H}^{0}} \right),
\end{equation*}
\noindent and 
\begin{equation*}
\bmath {\theta_{p}^{i}} = \{ \log{n_{\rm H}^{i,p}}, \log{T^{i,p}}, \alpha_{\rm UV}^{i}, \beta_{\rm He\ II}^{i}, \log{Z^{i}}, [C/\alpha]^{i}, [N/\alpha]^{i}, [\alpha/Fe]^{i} \}.
\end{equation*}
Here, $N_\textrm{ions}$, $N_\textrm{CG}$, $N_\textrm{abs}$, and $N_\textrm{phases}$, refer, respectively, to the number of unique species used to constrain the model, the number of component groups used to represent the measurements for these species, the number of absorbers being jointly modeled (e.g., two absorbers at the same systemic redshift), and the number of phases used by the model for each absorber.  The modeled column densities $N_{ion,c}^{model}$ actually represent the sum over column densities predicted at a more granular level (i.e., by individual absorbers and phases).  These more granular column densities $N^{\textsc{cloudy}}_{ion}(\bmath {\theta_{p}^{i}})$ are obtained by interpolating over our grid (Table~\ref{tab:param_info}) of \textsc{cloudy} ionization model evaluations, with free grid parameters $\bmath {\theta_{p}^{i}}$.  Note that $\log T^{i,p}$ is only included for the P+C model.  Finally, the total hydrogen ratio adjusts the gas content of the absorber with a normalization for the zeropoint of the interpolation grid.

\subsection{Prior Probability Density Functions}
\label{app:priors}

	Section~\ref{sec:scheme} includes a thorough qualitative description of the priors incorporated in our ionization modeling.  Here we present the underlying formulae, using bold symbols throughout to represent parameter sets.  The full prior distribution can be written as:
\begin{equation}\label{eqn:priorfull}
	P(\bmath{\theta}) = \hat{P}(\bmath{\theta}) P(\bmath{\sigma^2}_{\ln P,true}) \frac{P(\bmath{s^2}_{\ln P}|\bmath{\sigma^2}_{\ln P,true})}{P(\bmath{s^2}_{\ln P}|\hat{P}(\bmath{\theta}), P(\bmath{\sigma^2}_{\ln P,true}))}
\end{equation}
where the first term $\hat{P}(\bmath{\theta})$ represents the joint prior distribution over the primary model parameters and the remaining terms in Equation~\ref{eqn:priorfull} describe our thermal pressure prior, including a term for the hyperprior $P(\bmath{\sigma^2}_{\ln P,true})$.  Equation~\ref{eqn:inv_chi2} implicitly defines $P(\bmath{s^2}_{\ln P}|\bmath{\sigma^2}_{\ln P,true})$, the prior on the sample variance of the inter-phase pressures (a derived parameter of the model).  Following \cite{Handley19}, as explained in Section~\ref{sec:scheme}, we also need to divide by $P(\bmath{s^2}_{\ln P}|\hat{P}(\bmath{\theta}), P(\bmath{\sigma^2}_{\ln P,true}))$, that is, the prior induced on $\bmath{s^2}_{\ln P} | \bmath{\sigma^2}_{\ln P,true}$ by the explicit priors $\hat{P}(\bmath{\theta})$.

The prior on the primary model parameters is a joint distribution
\begin{equation}
	\hat{P}(\bmath{\theta}) = P(\bmath{\alpha_{\rm UV}}, \bmath{\beta_{\rm He\ II}}, \bmath{{\rm X}/\alpha}, \bmath{\alpha/{\rm H}}, \log{\bmath{T}}, \log{\bmath{n_{\rm H}}}, \log{\bmath{N_{\rm H}}}),
	\end{equation}
which in our application simplifies to
\begin{multline}\label{eqn:prior_terms}
	\hat{P}(\bmath{\theta}) =P(\bmath{\alpha_{\rm UV}})   P(\bmath{\beta_{\rm He\ II}})   P(\bmath{\alpha/{\rm H}})   P(\bmath{{\rm X}/\alpha}) P(\log{\bmath{T}}) P(\log{\bmath{n_{\rm H}}}) \\ \times\ P(\log{\bmath{N_{\rm H}}} | \log{\bmath{\tilde{n}_{\rm H}}} = \log{\bmath{n_{\rm H}}} + \log{f(\bmath{\alpha_{\rm UV}}, \bmath{\beta_{\rm He\ II}})} )
\end{multline}
though we can imagine inserting various additional conditional dependencies given sufficiently strong prior information, e.g., future prior constraints on relative abundance ratios might be conditional on the metallicity.  Note that the conditional prior for \textit{N}\textsubscript{H} has been formulated as a prior on the cloudsize, which requires that it be cast in terms of the adjusted density $\log{\tilde{n}_{\rm H}}$ using the logarithmic form of Equation~\ref{eqn:adjusted_density}.  The specific expressions for each of the terms in Equation~\ref{eqn:prior_terms} are given by Equations~\ref{eqn:prior_auv}--\ref{eqn:prior_NH}, some of which make reference to the standard probability distributions listed in Equations~\ref{eqn:truncnorm}--\ref{eqn:STDNcdf} and~\ref{eqn:uniform}--\ref{eqn:gamma}.  Ideally we would model every discrete velocity component as a distinct absorber, but in a crowded absorption system it is not always possible to resolve the individual components in every observed species.  In these instances, two or more absorbers must be modeled together with joint constraints on any unresolved species.  In the following, the index $i$ extends over the number $N_{abs}$ of distinct absorbers modeled together.  For the ionizing UV spectral shape parameters, the index $u$ extends over the set of distinct UV parameter pairs (with cardinality $N_{UV}$).  Note that $N_{UV}$$\leq$$N_{abs}$ since we allow for the possibility of sharing UV parameters between models for one or more absorbers (at the same systemic redshift).  Finally, the index $p$ extends over the set of phases in the model for the $i$th absorber.  Thus,
\begin{equation}\label{eqn:prior_auv}
	p(\bmath{\alpha_{\rm UV}}) = \prod_{u}^{N_{UV}} p_\mathrm{Trunc.\ Normal}(\alpha_{\rm UV}^u,\alpha_{\rm UV}^\mathrm{HM12}(z),0.7,-2.5,-0.28)
\end{equation}
where $\alpha_{\rm UV}^\mathrm{HM12}(z)$ is the fiducial \citep{Haardt12} ionizing UV spectral index measured at redshift $z$;
\begin{equation}\label{eqn:prior_HeII}
	p(\bmath{\beta_{\rm He\ II}}) = \prod_{u}^{N_{UV}} p_\mathrm{Uniform}(\beta_{\rm He\ II}^u,0,3)
\end{equation}
and
\begin{equation}\label{eqn:prior_abn}
	p(\bmath{X/\alpha}) = \prod_{element \in \bmath{X}}^{\mathrm{card}(\bmath{X})} \prod_{i}^{N_{abs}(element)} p_\mathrm{Uniform}([element/\alpha]_i, a_{ele}, b_{ele})
\end{equation}
where $\bmath{X}$ includes the elements C, N, and Fe with $(a_{ele}, b_{ele})$=(-0.8,0.8), (-1.2,0.4), (-1.2,0.4) for C/$\alpha$, N/$\alpha$, and Fe/$\alpha$, respectively.  As explained in the main text, we make an exception if $element=\textrm{Al}$, replacing the uniform probability density with $p_\mathrm{Trunc.\ Normal}([element/\alpha]_i, 0, 0.2, -0.5, 0.5)$;
\begin{equation}\label{eqn:prior_metals}
	p(\bmath{\alpha/{\rm H}}) = \prod_{i}^{N_{abs}} p_\mathrm{Uniform}([\alpha/{\rm H}]_i,-3.0,0.7)
\end{equation}
where $[\alpha/{\rm H}]$ represents the [Z/H] metallicity parameter varied in our grid of \textsc{cloudy} models, which is better understood as $[\alpha/{\rm H}]$ after we allow the significant non-$\alpha$ elements to vary freely
\begin{equation}\label{eqn:prior_temp}
	p(\log{\bmath{T}}) = \prod_{i}^{N_{abs}} \prod_{p}^{N^i_{phases,P+C}} p_\mathrm{Uniform}(\log{T_{i,p}},4.0,6.4)
\end{equation}
where $p$ is restricted to those model phases (if any) where P+C is used (since $\log T$ is not a parameter of PIE-only phase models)
\begin{equation}\label{eqn:prior_density}
	p(\log{\bmath{n_{\rm H}}}) = \prod_{i}^{N_{abs}} \prod_{p}^{N^i_{phases}} (N^i_{phases}+1)! \cdot {1}\{ n_{\rm H}^{min} \leq n_{\rm H}^{i,p} < n_{\rm H}^{i,p-1} \leq n_{\rm H}^{max} \}
\end{equation}
where the indicator function only accepts solutions with one-to-one ordering between the density and index of individual phases in each modeled absorber.  Limits on $n_{\rm H}^{min}$ and $n_{\rm H}^{max}$ are imposed by the boundaries of the \textsc{cloudy} ionization model grid (see Table~\ref{tab:param_info}).  The normalizing constant $(N^i_{phases}+1)!$ is simplest to understand if one considers the line segment between $n_{\rm H}^{min}$ and $n_{\rm H}^{max}$ as being split into $N^i_{phases}+1$ smaller segments by the densities $\bmath{n}_{\rm H}^i$, which correspond to the $N^i_{phases}$ phases specified by the model for the $i$th absorber.  These line segments can be arranged in $(N^i_{phases}+1)!$ possible ways, only one of which gives non-zero probability in Equation~\ref{eqn:prior_density}.
\begin{equation}\label{eqn:prior_NH}
	p(\log{\bmath{N_{\rm H}}} | \log{\bmath{\tilde{n}_{\rm H}}}) = \prod_{i}^{N_{abs}} \prod_{p}^{N^i_{phases}} p_\mathrm{Gamma}(\log{\bmath{N}_{\rm H}^{i,p}}, 0.2, \tilde{n}_{\rm H}^{i,p}\cdot 15\ \mathrm{kpc})
\end{equation}
where we use the adjusted density given by Equation~\ref{eqn:adjusted_density}.  The following two equations list the standard Uniform and Gamma probability distributions invoked above for completeness.  Note that we define the Truncated Normal distribution, which is also invoked above, in Equation~\ref{eqn:truncnorm}.
\begin{equation}\label{eqn:uniform}
	p_\mathrm{Uniform}(x,a,b) = \frac{1}{b-a}\; \textrm{for}\ a \leq x < b
\end{equation}
and
\begin{equation}\label{eqn:gamma}
	p_\mathrm{Gamma}(x, k,\theta) = \frac{x^{k-1}e^{-\frac{x}{\theta}}}{\theta^k \Gamma(k)}\; \textrm{for}\ x,k,\theta > 0.
\end{equation}

\subsection{Numerical Estimation of the Marginal Likelihood}
\label{app:marginal_likelihood}
	\cite{Weinberg13a} and \cite{Weinberg13} describe a novel approach for estimating the marginal likelihood from samples of the posterior distribution.  Noting that posterior volumes with the smallest density of samples dominate the uncertainty in the marginal likelihood integral, they show how the variance can be reduced by restricting the marginal likelihood estimation to a small volume centered on the posterior mode.  The marginal likelihood can be written as
\begin{equation}\label{eqn:ML}
	P(\bmath{N}|M) \int_{\Omega_s} dP(\bmath{\theta}|\bmath{N}) = \int_{\Omega_s} P(\bmath{\theta}|M) P(\bmath{N}|\bmath{\theta}, M)
\end{equation}
where $\Omega_s \subset \Omega$ and $\Omega$ represent the full set of posterior samples.  $\Omega_s$ is carefully chosen to minimize the variance in the numerical evaluation of the two integrals.  The integral on the left-hand side is just the fraction $\Omega_s/\Omega$ if evaluated via Monte Carlo sampling.  The right-hand side can be evaluated using the tree-based tessellation VTA algorithm proposed by \cite{Weinberg12} or by naive Monte Carlo resampling of the volume defined by $\Omega_s$ as suggested by \cite{Weinberg13}.

	We make two adjustments to the algorithm.  First, we use a custom iterative procedure to locate a hyperrectangle that circumscribes the posterior mode.  We identify samples with marginal posterior probabilities within a fraction $p$ of the marginal posterior mode that form a solid hyperrectangle containing the mode.  We iteratively adjust $p$ until the number of posterior samples satisfying this condition equals the desired value, $N_{\Omega_s}$.  We use a value of $N_{\Omega_s}$=1000, which offers the best tradeoff in minimizing the variance of both integrals in Equation~\ref{eqn:ML}.  Second, we estimate Equation~\ref{eqn:ML} using the \textit{vegas} adaptive Monte Carlo algorithm \citep{Lepage78} implemented in the \textsc{vegas} Python package\footnote{Version 3.4 https://vegas.readthedocs.io/en/latest/index.html}.  This algorithm relies on both importance sampling and stratified sampling strategies to estimate the integral.  We obtain much more efficient sampling using this algorithm than with using naive Monte Carlo integration.

\subsection{MCMC Tests}
\label{app:tests}

\begin{figure*}
\centering
	\begin{minipage}{.495\textwidth}
		\centering
		\begin{subfigure}{\textwidth}Ä
			\centering
			\includegraphics[width=3in]{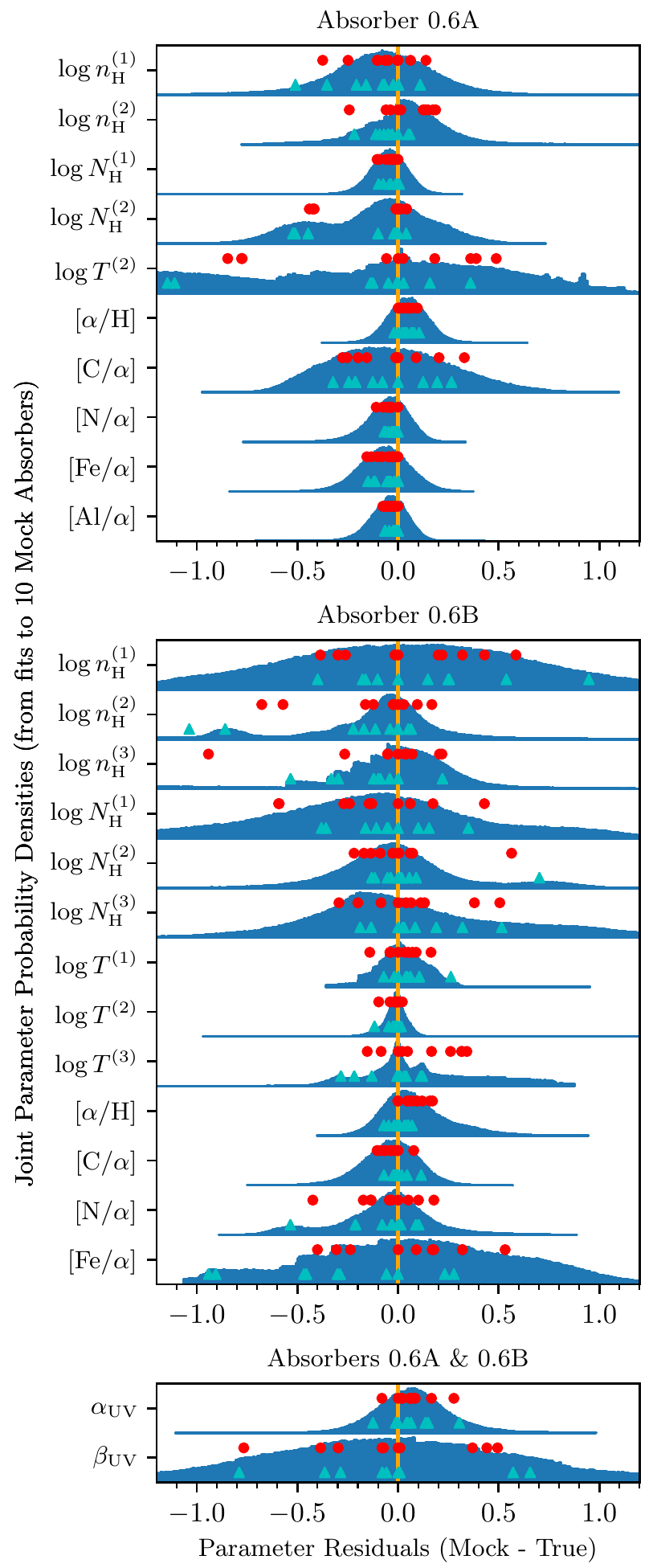}
			\vspace{0.3in}
			\label{fig:z068605_test}
		\end{subfigure}
		\begin{subfigure}{\textwidth}
			\centering
			\includegraphics[width=3in]{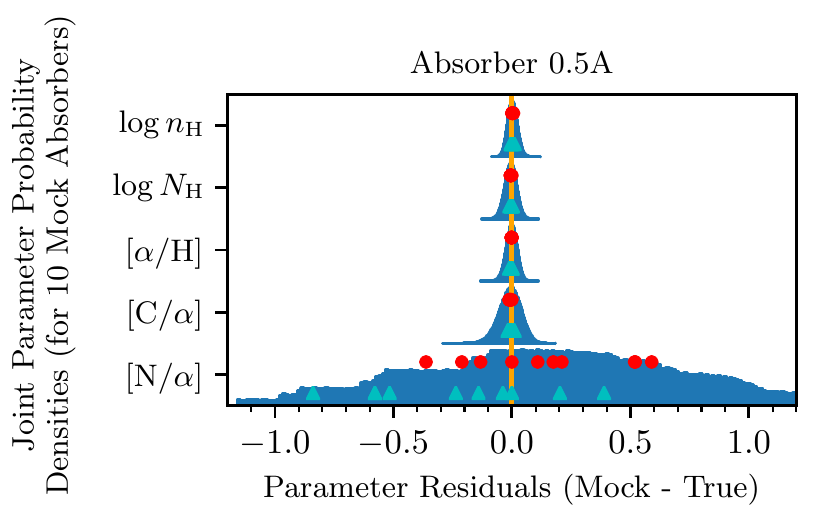}
			\label{fig:z051850_test}
		\end{subfigure}
	\end{minipage}
	\begin{minipage}{.495\textwidth}
		\centering
		\begin{subfigure}{\textwidth}
			\centering
			\includegraphics[width=3in]{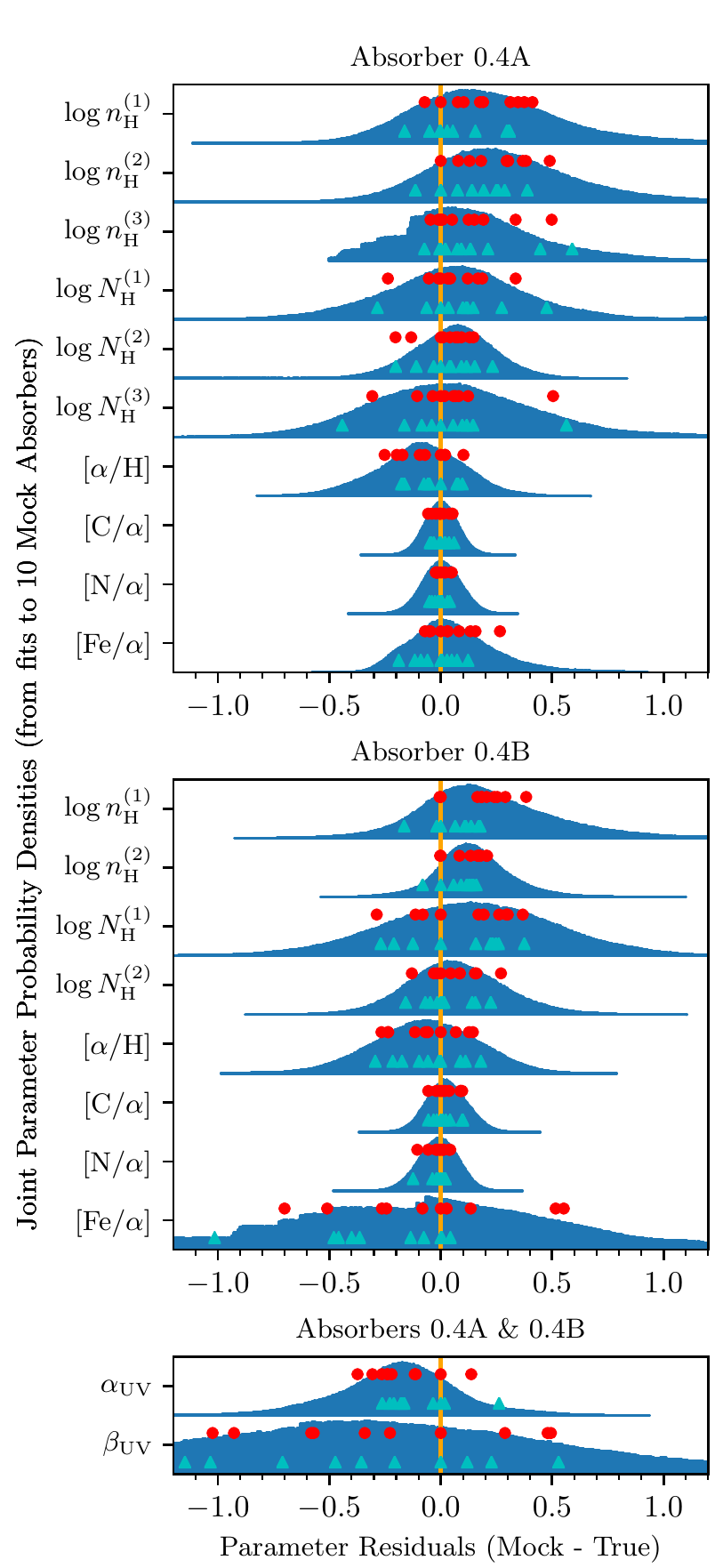}
			\vspace{0.15in}
			\label{fig:z041761_test}
		\end{subfigure}
		\begin{minipage}{0.92\textwidth}
			\centering
			\caption{``Joint parameter residuals'' for fits to mock absorption data (see text for details) designed to test the accuracy and precision of parameteric inference using our Bayesian ionization modeling scheme.  Each subfigure summarizes test results from fitting PPPa-PPb (Mock Absorbers~0.4A \&~0.4B), P$_{5par}$ (Mock Absorber~0.5A), and PCa-CCCb (Mock Absorbers~0.6A \&~0.6B) models to ten mock absorption systems each.  The true parameters for each test model were randomly selected from the posteriors of the same models applied to the real Absorbers' column densities.  The horizontal axes in these plots are all centered on the true value of the parameter.  Blue histograms show the joint probability densities of individual parameters from the groups of mock absorber models after subtracting the true parameter value.  Parameters are grouped by absorber, and superscripts are used to denote distinct phases (``1'' refers to the lowest degree of ionization, etc...).  Red circles and cyan triangles show the mean and median, respectively, of the inferred parameters from the individual mock absorber models.  See the main text for various considerations that should be taken into account when interpreting these results.  \vspace{0.25in}}
			\label{fig:three_systems_test}
		\end{minipage}
	\end{minipage}
\end{figure*}

\begin{figure*}
\centering
\begin{subfigure}{.495\textwidth}
	\centering
	\includegraphics[width=1.0\textwidth]{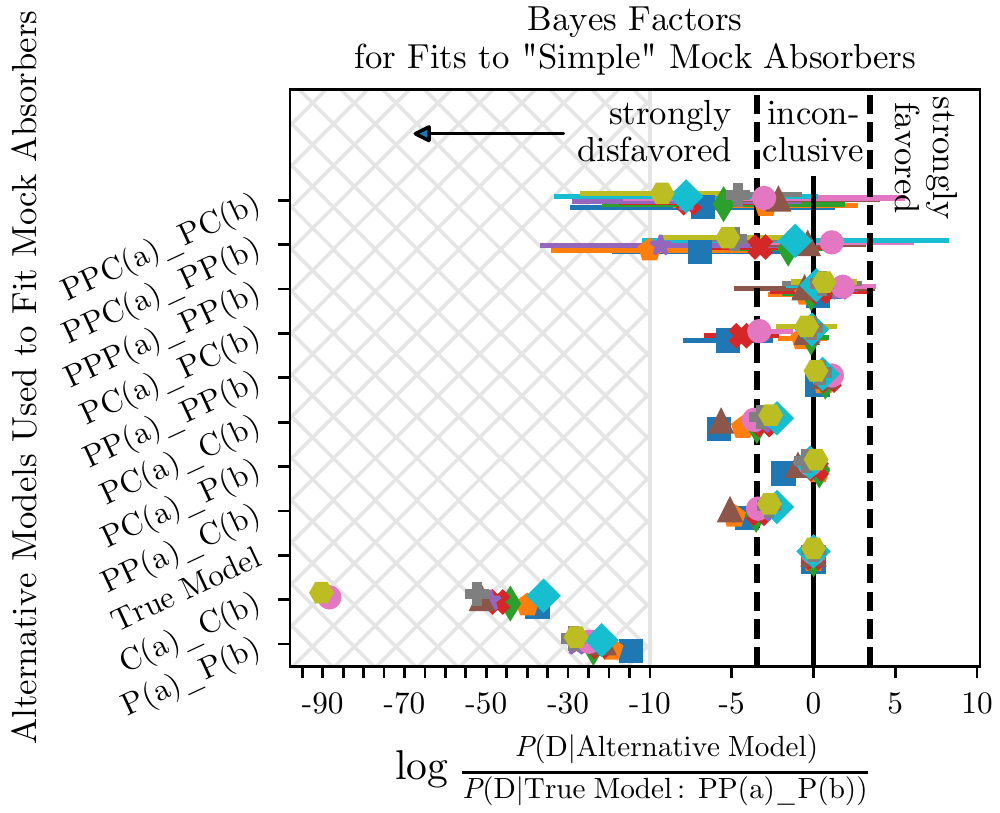}
	\label{fig:simple_bayesratio_test}
\end{subfigure}
\begin{subfigure}{.495\textwidth}
	\centering
	\includegraphics[width=1.0\textwidth]{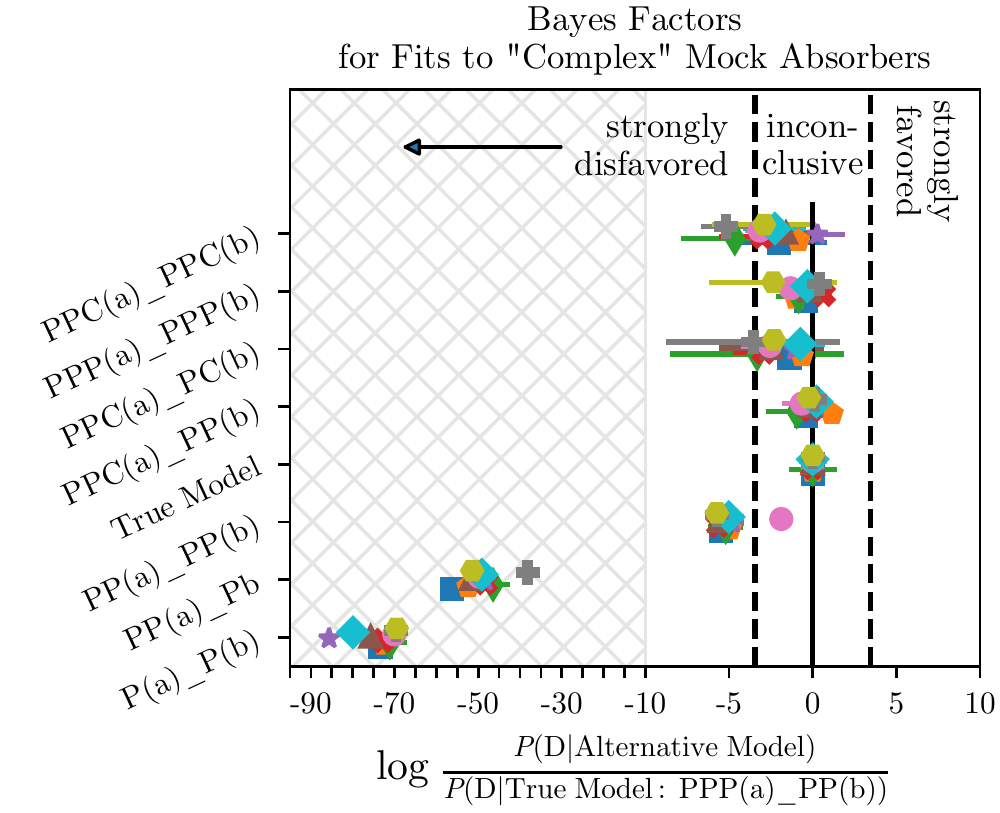}
	\label{fig:complex_bayesratio_test}
\end{subfigure}%
\caption{A comparison of Bayes factors for fits to ten ``simple'' and ten ``complex'' mock absorption systems (see the main text for the distinction between these sets).  The various models used to fit the mock absorbers are labeled along the vertical axis, with labels specifing the number and ionization type of phases (following the convention described in Section~\ref{sec:formalism}) for each component (``a'' and ``b'').  The Bayes factor for each model-mock absorber pair is the ratio of its marginal likelihood to that of the true model.  Colors and symbols redundantly refer to individual mock absorption systems.  The solid vertical line represents alternative models with equal odds as the true model, while the dashed lines indicate a strong preference for or against the alternative model following the interpretation scale of \protect \cite{Jeffreys61}.  The horizontal axis has been linearly scaled by a factor of four in the hatched region to minimize whitespace.  A small vertical jitter is applied to enhance marker visiblity.}
\label{fig:bayesratio_test}
\end{figure*}

	We tested our Bayesian ionization modeling scheme on numerous mock absorbers to 1) estimate the accuracy and precision of our parameter inferences and 2) assess the sensitivity of Bayes factors for model selection.  These properties depend on both the (unknown) physical parameters of the true absorbers and on the number and quality of the absorption line constraints.  Performing global tests over these variables would be computationally challenging and superfluous to our purposes.  Instead, we narrow our scope and test mock absorbers designed to mimic the constraints and inferred parameters of the observed (real) absorbers presented in this study.

	We designed the first set of tests, i.e., for accuracy and precision, as follows.  We begin by randomly drawing ten parameter vectors from the posterior of highest ranked model (by Bayes factors) for each observed absorber and compute the corresponding model-predicted column densities.  Each parameter vector and its associated column densities represents a distinct mock absorber.  Drawing mock absorbers in this manner (i.e., from the posteriors of the models for real absorbers) ensures that the mock absorbers test a similar parameter space as that probed by the real ones, but with sufficient variation.  We also want our mock column densities to have similar constraining power over the models as the real absorbers' column densities.  To achieve this, we 1) record column densities for the same set of species observed in the real absorbers while borrowing the same measurement classifications (detection or upper/lower limit)\footnote{In some cases we had to extend upper and/or lower limits for individual mock absorbers so that the mock column densities would remain inside the limits.}, and 2) apply the measurement uncertainties from the real data to the mock data. 

After creating the mock absorbers, we then apply the same Bayesian ionization model that we used for the true absorbers to each mock absorber, after removing the pressure equilibrium and cloud size priors.

	For each set of ten mock absorbers, we combined posterior samples of the individual models after first subtracting their known parameter vectors that we originally drew from the posteriors of real models; we refer to these as ``joint parameter residuals.''  Figure~\ref{fig:three_systems_test} shows histograms of these joint parameter residuals, along with the mean and median of the parameter residuals for individual mock absorbers.  Using these figures, we can estimate the accuracy of the parameters\footnote{More precisely, we can evaluate the accuracy of \emph{point estimates} of the parameter marginal posterior distributions.} inferred by our models for the real absorption systems.  For example, the modeled UV spectral index ($\alpha_\textrm{UV}$) for Mock Absorbers 0.6A \& 0.6B is overestimated by $\approx$0.08 dex on average, which can be interpreted to mean that the corresponding measurements for the real absorbers (See Figure~\ref{fig:abnd_dist}) most likely overestimate the true (unknown) value by a similar amount.  But Figure~\ref{fig:three_systems_test} also shows an $\approx$35\% chance that the true $\alpha_\textrm{UV}$ is \emph{under}estimated (i.e., by integrating the blue histogram in the region $>$0), so the bias is relatively weak.  Overall, the joint parameter residuals reveal relatively minor bias, in all cases $\lesssim$1$\sigma$, but in some cases with rather large variances.  Note, however, that the interpretation of the variance depends on the parameter.  For example, the iron relative abundance parameter for Mock Absorber~0.6B has far too wide a variance to draw any useful insight from the model.  This comes as no surprise, however, since iron in this absorber is only detected as an upper limit, and hence provides only a weak constraint on the model.

	We also devised a second set of tests to evaluate the sensitivity of Bayes factors for selecting the true model from a set of models with various phase combinations, i.e., including both ionization mechanism (PIE-only vs. P+C) and number (of phases) per velocity component.  Ideally, Bayes factors should show a preference for the true model regardless of its complexity, since Bayes factors naturally penalize both too simplistic and too overly complex models.  In practice, the signal is not always strong enough to be conclusive, or it may be affected by insufficient data, bad priors, or imprecise and/or biased computation of the marginal likelihood.  Hence, we present these tests as an empirical calibration of Bayes factors specific to our application.  For these tests we produced two sets of ten photoionized mock absorption systems, one generated from a complex model, the other from a simpler model.  Specifically, the ``complex'' set of mock absorbers were drawn from the posterior of the PPPa-PPb model of the real Absorbers 0.4A \& 0.4B in the same fashion as that described above for the accuracy and precision test (in fact the same mock absorbers were re-used).  The ten ``simple'' mock absorbers were similarily produced, but drawn from the posterior of a lower complexity PPa\_Pb model of the same real absorbers.\footnote{We use the lower complexity PPa-Pb model simply to narrow the parameter selection space of the ``simple'' mock absorbers.  The accuracy of the model with respect to the real data has no effect on the efficacy of the mock absorbers and tests derived from it.}  We then model the mock absorbers with a suite of ionization models with varying degrees of complexity, including the correct models from which they were derived.  At every intersection of the mock absorbers and trial ionization models, we compute Bayes factors as the ratio of the marginal likelihood for the trial model over that of the true model.  

	Figure~\ref{fig:bayesratio_test} presents logarithms of the Bayes factors from the second test relative to the true model.  The logarithmic ratios are all essentially less than or equal to zero, meaning no model is strongly preferred over the true model.  Further, we can conclusively rule out models with a lower degree of complexity (i.e., fewer phases) than the true model.  This lends credence to our important conclusion regarding the inadequacy of single phase models (which implies the need for multiphase alternatives).  Models with a slightly higher degree of complexity than the true model, however, are not clearly disfavored.  One possible explanation for this ambiguity is that the extra phases in higher-complexity models might only contain a trivial amount of gas relative to the content of the other phases.  The small amount of gas in these extra phases might produce an unecessarily optimized fit, an improvement that is balanced by the penalty of adding parameters.  Under this hypothesis, the higher complexity model minus the trivial phases would look similar to the true model.  However, Occam's razor would have us prefer the less complex model if the Bayes factor does not clearly favor the more complex one.

	Additionally, we note that for these particular mock absorbers the Bayes factors indicate a weak, but consistent preference for purely photoionized models over ones incorporating collisional ionization.  While this trends in the right direction (the true models are purely photoionized), it is not strong enough to be conclusive, likely owing to the paucity of high ion constraints in these mock absorbers (see Table~\ref{tab:z041761_meas} for a list of the species used).  Note that this is not evidence for a universal preference for photoionization over collisional ionization, but only for the particular case of Absorbers 0.4A \& 0.4B.

\subsection{Posterior Predictive p-values}
\label{app:ppp}

	Fundamentally, data (\textit{y}\textsuperscript{\textit{sim}}) simulated from a good model should look similar to the observed data (\textit{y}\textsuperscript{\emph{obs}}).  This can be stated in terms of some test statistic \emph{T}(\textit{y}) that summarizes each dataset: \emph{T}(\textit{y}\textsuperscript{\emph{obs}}) should have a reasonable probability of being drawn from the distribution of \emph{T}(\textit{y}\textsuperscript{\textit{sim}}).  If \emph{T}(\textit{y}\textsuperscript{\emph{obs}}) is an outlier, we reject the generative model.  \emph{p}-values are commonly used to quantify the statistical significance of a measurement with respect to the null hypothesis or null distribution.  The classical p-value is calculated over the distribution of \textit{y}\textsuperscript{\textit{sim}} for fixed (known) $\theta$ as:
\begin{equation}
p_{\textrm{C}} = \mathrm{Pr}(T(\textit{y}^{sim}) \geq T(\textit{y}^{obs} | \theta)).
\end{equation}
Thus, classical \emph{p}-values summarize the discrepancy between the observed data and the values expected by a model with a particular value of $\theta$.  

	In a Bayesian context, the observed data is compared to the posterior predictive distribution, which integrates over the posterior distribution of the (unknown) parameters.  Thus, \textit{T} is typically a function of $\theta$ as well as \textit{y}, i.e., \emph{T}=\emph{T}(\textit{y},$\theta$).  The (Bayesian) posterior predictive \emph{p}-value is thus defined as \citep{Gelman96,Gelman04}:
\begin{equation}
p_{\textrm{B}} = \mathrm{Pr}(T(\textit{y}^{sim},\bmath{\theta}) \geq T(\textit{y}^{obs},\bmath{\theta}) | \textit{y})
\end{equation}
where the probability is taken over the posterior distribution of $\theta$ and the posterior predictive distribution of \textit{y}\textsuperscript{\textit{sim}}, i.e., the joint distribution \emph{p}(\textit{y}\textsuperscript{\textit{sim}},$\theta$|\textit{y}\textsuperscript{obs}).  In practice, one computes \emph{T}(\textit{y}\textsuperscript{\emph{obs}},$\theta$) and \emph{T}(\textit{y}\textsuperscript{\textit{sim}},$\theta$) for each set of parameters sampled from the posterior.  The mock data, \textit{y}\textsuperscript{\textit{sim}}, are randomly drawn from the likelihood distribution, with one replicated dataset for each of the posterior samples.

	The optimal test statistic for computing \emph{p}\textsubscript{B} depends on the model at hand.  Here, we choose to sum the scaled residuals between the data and model column densities,
\begin{equation}
T(y,\theta) = \sum\limits_{i}^{ions} \frac{\left|y_i - N^{model}_i(\bmath{\theta})\right|}{\sigma^{obs}_{i}}.
\end{equation}
Here, $y_i$ represents either the observed or simulated column density for the $i$th ion, $N^{model}_i$ is the model-predicted column density corresponding to a particular set of parameters $\bmath{\theta}$, and $\sigma_i$ denotes the uncertainty in the measurement of the corresponding observed column density (even when $y_i$=$y^{sim}_i$). This statistic was chosen to complement the qualitative goodness-of-fit plots in Section~\ref{sec:results}, since discrepancies between the data and model are more intuitively understood when expressed as residuals.  With this statistic, \emph{p}\textsubscript{B} should be interpreted as a two-tailed probability, with an ideal value at \emph{p}\textsubscript{B}=0.5.  Large residuals drive \emph{p}\textsubscript{B}$\approx$0 and indicate that the observed data are unlikely predictions of the model (i.e., potentially warranting a rejection of the model).  Overly narrow residuals, which produce \emph{p}\textsubscript{B}$\approx$1, are also unlikely to occur if the null hypothesis (the model being tested) is the correct one, so high numbers likely indicate some degree of overfitting.

\begin{figure}
	\includegraphics[width=\columnwidth]{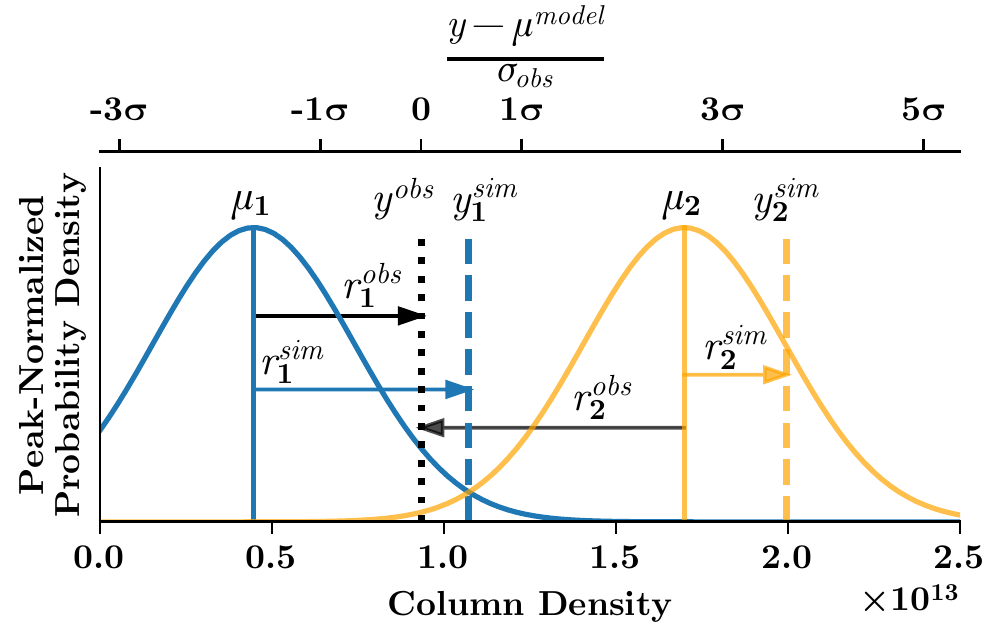}
	\caption{A graphical illustration of scaled residuals.  Model-predicted column densities $\mu_{1}$ and $\mu_{2}$ (blue and orange vertical lines, respectively), corresponding to two arbitrary parameter vectors $\bmath{\theta_{1}}$ and $\bmath{\theta_{2}}$ drawn from the (hypothetical) posterior, become the means of ZTN distributions (identically colored curves) each having the same standard deviation $\sigma_{\textrm{obs}}$).  The distance \emph{r}\textsuperscript{\emph{obs}} (black arrows) between either $\mu_1$ or $\mu_2$ and the observed column density (vertical black dotted line) can be expressed as a column density (bottom axis) or as a scaled residual (top axis, $\mu^{model}\in\{\mu_1,\mu_2,...\}$).  When computing posterior predictive \emph{p}-values, a single simulated column density $\textit{y}^{\textit{sim}}_{i}$ is drawn from each unique ZTN, which is linked to the unique posterior sample $\bmath{\theta_{i}}$.  We illustrate this concept for the two simulated column densities $\textit{y}^{\textit{sim}}_{1}$ and $\textit{y}^{\textit{sim}}_{2}$, denoted by dashed vertical lines that are colored to correspond to the ZTN distributions from which they were drawn.  A scaled residual is likewise computed for each of these simulated data points, with the distance \emph{r}\textsuperscript{\textit{sim}} shown by the respective colored arrows.
}
	\label{fig:scaled_resid_meas}
\end{figure}

	Figure~\ref{fig:scaled_resid_meas} illustrates the concept of scaled residuals.  The likelihood distribution, \emph{L}($\textbf{\emph{N}}|{\bmath{\theta}}$) (Equation~\ref{eqn:like}), gives the probability of the data given the model, so we treat each observed column density as a random sample from a ZTN distribution (Equation~\ref{eqn:ZTNpdf}).  The ionization model sets the mean of this distribution, while its variance is fundamentally determined by the noise in the data, which we approximate as $\sigma^{obs}_i$.  Figure~\ref{fig:scaled_resid_meas} plots two sampling distributions for an arbitrary ion along with the corresponding simulated column densities.  We are interested in the significance of any discrepancy rather than the absolute value of the discrepancy, hence the need for scaling by the standard deviation.

\begin{figure}
	\includegraphics[width=\columnwidth]{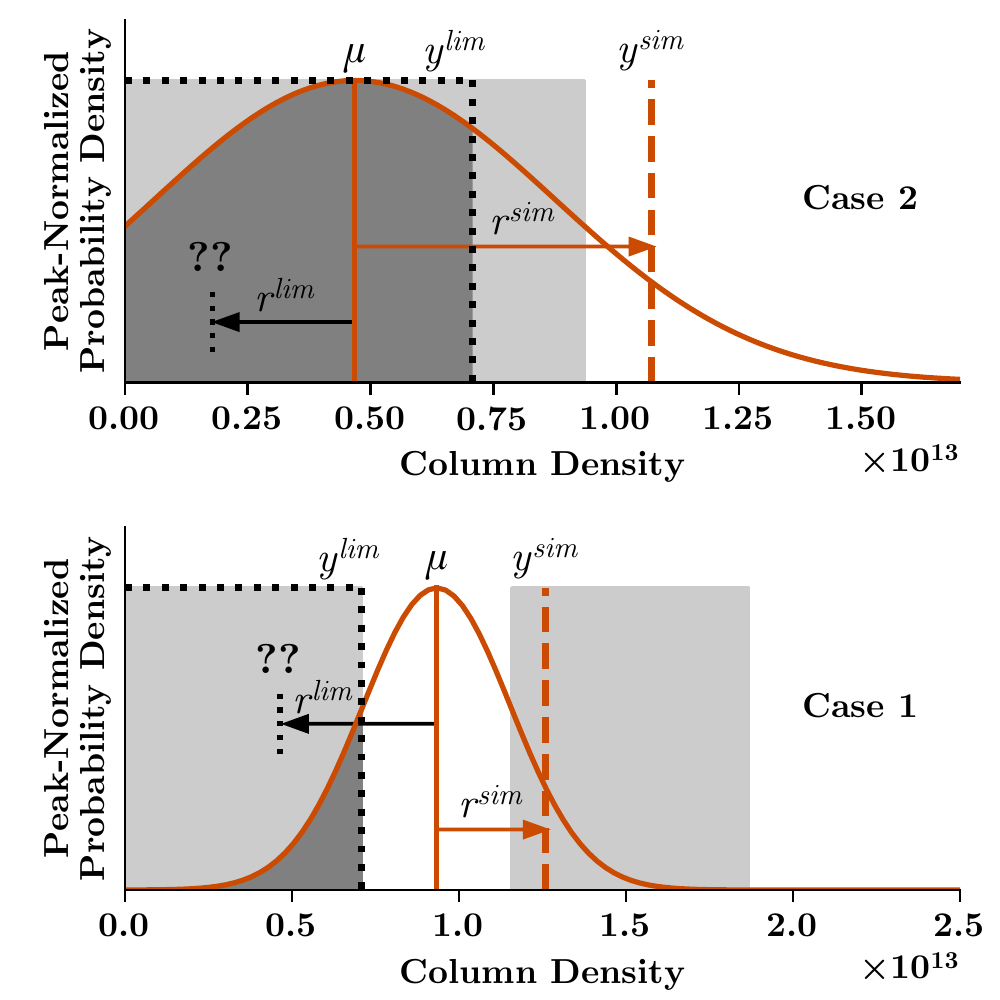}
	\caption{An illustration of column density residuals for lower limits.  In both cases, $\mu$, the mean of the ZTN distribution (red curve), is the model-predicted column density corresponding to a particular set of parameters drawn from the posterior of a given model.  We compute the posterior predictive probability distribution by integrating over all possible values of $\mu$ implied by the posterior.  The red vertical dashed lines represent simulated column densities drawn from the ZTN for two illustrative cases, while the dotted black lines indicate the measured upper limit.  As a necessary step to obtaining a $p$-value, we must determine if the simulated measurement produces a larger residual with respect to $\mu$ than the observed limit.  Case 1 differs from Case 2 in the location of $\mu$ relative to the observed column density limit.  In both cases the residual \emph{r}\textsuperscript{\textit{sim}} is straightforward to measure, while we can only measure an upper limit on \emph{r}\textsuperscript{\emph{lim}}.  In some regions (white/not shaded) it is clear that either \emph{r}\textsuperscript{\textit{sim}}$<$\emph{r}\textsuperscript{\emph{lim}} (Case 1) or \emph{r}\textsuperscript{\textit{sim}}$>$\emph{r}\textsuperscript{\emph{lim}} (Case 2), but in the lightly shaded regions it is not clear which residual is larger.  We overcome this problem by statistically generating realizations of the column density limits.  That is, we draw samples from the posterior predictive distribution in the region allowed by the measured limit (shaded in dark grey).  Note that when \emph{y}\textsuperscript{\textit{sim}}$<$\emph{y}\textsuperscript{\emph{lim}} we treat both \emph{r}\textsuperscript{\emph{lim}} and \emph{r}\textsuperscript{\textit{sim}} as limits and ignore these points when computing the test statistics \emph{T}(\textit{y}\textsuperscript{\emph{lim}},$\bmath{\theta}$) and \emph{T}(\textit{y}\textsuperscript{\textit{sim}},$\bmath{\theta}$).}
	\label{fig:scaled_resid_lims}
\end{figure}

	Extracting scaled residuals for censored data (upper and/or lower column density limits) is not as straightforward as with measurements for detected species.  In these cases the observed data residuals $\left|y^{obs}_i - N_i(\theta)\right|$ are themselves limits, a point illustrated in Figure~\ref{fig:scaled_resid_lims}.  To make a fair comparison with model simulated column densities, any simulated columns that fall outside the observed detection limit are likewise treated as limits, and we ignore these when computing test statistics \emph{T}(y\textsuperscript{\emph{obs}},$\theta$) and \emph{T}(y\textsuperscript{\textit{sim}},$\theta$).  In some cases, however, the posterior predictive distribution will generate y\textsuperscript{\textit{sim}} values that would be considered detections in a real dataset given the signal-to-noise ratio of the observed spectrum.  Since these provide a real constraint on the ionization modeling, we resort to a probabilistic analysis of the observed limits to provide a meaningful comparison to the simulated data.  As noted above in Appendix~\ref{app:likelihood}, each column density limit is interpreted as the tail probability of the ZTN distribution implied by a given set of model parameters.  Thus, we can realize an observed column density from the limits by sampling the relevant tail of the model-implied column density sampling distribution.\footnote{Note that without the tail constraints this would be identical to the sampling procedure used to generate simulated data.}  We then use the realized column density to compute \emph{T}(y\textsuperscript{\emph{obs}},$\theta$) as though it were an actual detection.  This approach does not bias the test statistic since the realized and simulated column densities are ultimately drawn from the same distribution (i.e., before applying the observational limit constraints).  The darkest shaded regions in Figure~\ref{fig:scaled_resid_lims} exemplify the tail regions from which upper limit column density realizations are drawn.  Lower limits are similar but mirrored, while double limits specify sampling intervals rather than tails.


\bsp	
\label{lastpage}
\end{document}